\documentclass[11pt]{article}
\pdfoutput=1
\usepackage[utf8]{inputenc}
\usepackage{multirow}
\usepackage{amsmath, amsfonts, amssymb}
\usepackage{array}
    \newcolumntype{C}[1]{>{\centering\arraybackslash}m{#1}}
\usepackage{comment}
\usepackage{graphicx}
\usepackage{pdflscape}
\usepackage{psfrag}
\usepackage{amsthm}
\usepackage{enumerate}
\usepackage{arydshln}
\usepackage{pifont}
    
\usepackage{soul}
\usepackage{slashed}
\usepackage{subcaption}
\usepackage{mathrsfs}
\usepackage{ytableau}
 \usepackage{a4wide}
  \usepackage{tikz}
  \usepackage{tikz-cd}
  \usetikzlibrary{shapes.geometric}
  \usepackage{tcolorbox}
  \definecolor{dark-gray}{gray}{0.20}
  \definecolor{gray}{gray}{0.30}
  \definecolor{light-gray}{gray}{0.80}
  \definecolor{dark-red}{rgb}{0.7,0,0}
  \definecolor{dark-green}{rgb}{0.1,0.4,0}
  \definecolor{dark-blue}{rgb}{0.3,0.3,0.7}
  \definecolor{light-blue}{rgb}{0.8,0.8,1}
      \definecolor{swamp}{RGB}{240, 199, 197}
      
  \usepackage{pifont}

\usepackage{newunicodechar} 
\usepackage{setspace}

\usepackage{ifthen}
\renewcommand*{\arraystretch}{0.9}
\usepackage{longtable}

\newcommand{\be}{\begin{equation}}
\newcommand{\ee}{\end{equation}}
\newcommand{\eq}[1]{(\ref{#1})}

\def\be{\begin{equation}}
\def\ee{\end{equation}}
\def\bea{\begin{eqnarray}}
\def\eea{\end{eqnarray}}

\newcommand{\R}{\mathbb{R}}

\newcommand{\Z}{\mathbb{Z}}
\AtBeginDocument{

}

\newcommand{\ed}{{\rm d}}

\newcommand{\figref}[1]{figure~\ref{#1}}

\newcommand{\secref}[1]{Section~\ref{#1}}
\numberwithin{equation}{section}

\usepackage{jheppub}

\usepackage{cleveref}

\hypersetup{
	colorlinks=true,
	linkcolor=dark-blue,
	citecolor=dark-red,
	urlcolor=dark-green,
	linktoc=page
}

\theoremstyle{definition}

\theoremstyle{remark}

\crefname{appendix}{Appendix}{Appendices}

\title{An M-theory dS maximum from Casimir energies \\ on Riemann-flat manifolds}

\author{Bruno Valeixo Bento} 
\author{and Miguel Montero}
\affiliation{Instituto de F\'{i}sica Te\'{o}rica IFT-UAM/CSIC,
C/ Nicol\'{a}s Cabrera 13-15, Campus de Cantoblanco, 28049 Madrid, Spain}

\emailAdd{bruno.bento@ift.csic.es}
\emailAdd{miguel.montero@csic.es}

\abstract{We initiate the study of flux compactifications on non-supersymmetric Riemann-flat manifolds (RFM's) with Casimir energy. While curvature and other corrections are suppressed in RFM's, the inclusion of Casimir energies allows one to evade standard dS no-go theorems, and the absence of orientifolds or other singular sources means that the construction is completely captured by ten or eleven-dimensional supergravity. We obtain a fully explicit formula for the Casimir stress-energy in a general RFM, including its ten or eleven-dimensional profile. The Casimir energy localizes in particular loci of the RFM, which we call ``Casimir branes''. The tension of Casimir branes sometimes cancels exactly, due to a spacetime analog of worldsheet Atkin-Lehner symmetry. We use Casimir energies to construct an explicit $dS_5$ maximum solution of a flux compactification of M-theory on a specific 6-dimensional RFM. The resulting solution is scale-separated, has a vacuum energy of $10^{-8}$ in five-dimensional Planck units, the Hubble radius is $10^4$ Planck lengths, and the light fields have masses of order $H$. This is a fully explicit, top-down de Sitter maximum in M-theory, with precisely computable vacuum energy. While the solution is not parametric, it is under very good control: higher derivative and loop corrections to the vacuum energy are suppressed in powers of a small parameter $\delta V/V\sim 10^{-5}$, and M2 and M5-brane instantons are negligible. In short, the solution survives all known corrections. Nevertheless, it might be sensitive to more exotic ones, such as e.g. loops of 11d Planckian virtual black holes if there were a large enough number of them. We also extend the Ewald numerical method for lattice sums to arbitrary dimensions and develop an efficient numerical implementation.}

\begin{document}
\emergencystretch 3em
\hypersetup{pageanchor=false}
\makeatletter
\let\old@fpheader\@fpheader
\preprint{IFT-25-070}

\makeatother

\maketitle

\hypersetup{
    pdftitle={},
    pdfauthor={},
    pdfsubject={}
}

\newcommand{\remove}[1]{\textcolor{red}{\sout{#1}}}

\newcommand{\red}[1][\text{(check)}]{\textcolor{red}{#1}}

\newcommand{\D}[1][\gamma]{\mathbf{D}_{\bf #1}}
\newcommand{\bvec}[1][\gamma]{\vec{b}_{\bf #1}}
\newcommand{\RFM}[2][T]{%
  \ifthenelse{\equal{#1}{T}}%
    {\frac{T^{#2}}{\Gamma}}%
    {\frac{\mathbb{R}^{#2}}{\mathcal{B}}}%
}
\newcommand{\Tr}[2]{{\rm Tr_{\bf #1}}(#2)}
\newcommand{\TrB}[1]{\Tr{B}{#1}}
\newcommand{\TrF}[1]{\Tr{F}{#1}}
\newcommand{\Vcas}{V_{\text{Cas}}}

\section{Introduction and basic idea}
\label{sec:introductions}

One of the most pressing open questions in String Theory is to ascertain beyond doubt whether it can accommodate long-lived accelerated expansion. The foremost reason for this is to connect to observations \cite{SupernovaCosmologyProject:1998vns,SupernovaSearchTeam:1998fmf}. However, independently of any observational implications, whether long-lived accelerating expansion exists or not is an essential question about the non-supersymmetric Landscape that we must elucidate. There are arguments (some of them older than the authors of this paper \cite{Dine:1985he}) to the effect that we do not expect to have accelerating cosmologies near the boundaries of moduli space, where strings are arbitrarily weakly coupled or we have some other control parameter that can be tuned arbitrarily \cite{Ooguri:2006in,Lee:2019wij}. Together with the lack of explicit string constructions, these have led to concrete Swampland conjectures \cite{Palti:2019pca,vanBeest:2021lhn} that acceleration may not exist in asymptotic regions of moduli space \cite{Danielsson:2018ztv,Obied:2018sgi,Ooguri:2018wrx,Andriot:2018wzk,Andriot:2018mav,Bedroya:2019snp}, a statement for which there is now ample systematic evidence (see e.g. \cite{Grimm:2019ixq,Calderon-Infante:2022nxb,Etheredge:2024tok}).

In practice, all of the above means that any stringy attempt at long-lived phases of accelerated expansion is pushed away from the asymptotic boundaries of moduli space. Since it is hard to imagine a fully non-perturbative gravitational setup where we would have any measure of control, most accelerated expansion scenarios end up living on the edge between asymptotic and non-perturbative, where a description in terms of string perturbation theory, or a large internal space may still be expected to hold, but where additional non-asymptotic ingredients are added to the mix, in order to avoid the asymptotic no-go theorems. This is certainly the case of the existing de Sitter scenarios, such as KKLT\footnote{See also \cite{Parameswaran:2010ec} for  heterotic orbifold models that include KKLT-like ingredients, such as hidden sector gaugino condensation, with de Sitter maxima and a supersymmetric Standard Model.} \cite{Kachru:2003aw} and LVS \cite{Balasubramanian:2005zx}, whose fate is still under debate (see e.g. \cite{Gao:2020xqh, Demirtas:2021nlu,Junghans:2022exo,Bena:2022cwb,Gao:2022fdi,Lust:2022lfc,Lust:2022xoq,Hebecker:2022zme,Bena:2022ive,ValeixoBento:2023nbv,McAllister:2024lnt,Kim:2024dnw,Moritz:2025bsi}) in large part because the mix of ingredients required is quite difficult to control. The origin of this complexity can be traced in part to one particular no-go theorem \cite{Gibbons:1984kp,Maldacena:2000mw} which ensures that no de Sitter critical point may be achieved only through the use of classical supergravity ingredients (fluxes and curvature). To circumvent this, existing dS scenarios include stringy sources such as orientifolds, non-perturbative effects, and/or higher-derivative corrections, all of which are beyond the reach of 10 or 11-dimensional supergravity, and are intrinsically complicated, leading to questions such as e.g. the backreaction of the singular sources introduced \cite{Gao:2020xqh} and so on. 

The goal of this paper is to sidetrack all of this discussion, by looking at a different, much simpler way to avoid the no-go theorem of \cite{Maldacena:2000mw}, and which can be fully analyzed using supergravity tools: Casimir energy \cite{Appelquist:1983vs,Arkani-Hamed:2007ryu}, the quantum response of matter fields to a non-supersymmetric background. This idea, pioneered in \cite{DeLuca:2021pej}, exploits the fact that Casimir energy (which can violate the energy condition assumed in \cite{Maldacena:2000mw}) can be computed explicitly and reliably, using standard techniques (see e.g. \cite{Appelquist:1983vs,Birrell:1982ix,Arkani-Hamed:2007ryu}). This can be done for many theories, including 11-dimensional supergravity, as well as in all 10-dimensional supergravities arising as the low-energy limits of String Theory. 
While many other works have explored Casimir energies and their interplay with the cosmological constant problem, both from a phenomenological and fundamental point of view, \cite{DeLuca:2021pej} gives the first concrete and fully top-down construction of a de Sitter minimum in string/M-theory that relies on Casimir energy, together with flux and curvature contributions.

On top of the key ingredient of Casimir energy, the approach of \cite{DeLuca:2021pej} also involves a non-trivial warping throughout the compactification manifold (but much weaker than in the strongly warped throats appearing in KKLT and LVS), due to the class of manifolds considered, which are hyperbolic.\footnote{Hyperbolic manifolds were also used in the minimal dS constructions of \cite{Haque:2008jz,Danielsson:2009ff}, in which smeared D-brane/O-plane sources are used instead of Casimir energies. Negative curvature spaces (together with non-zero Romans mass) were identified as necessary complements to a Type IIA setup with fluxes and localised sources in order to obtain de Sitter solutions. The presence of the localised sources, which was dealt in that reference by means of a smearing approximation, brings in the complexity that is also present in models like KKLT and LVS.\label{ft:IIA-models}} Although warping is a classical phenomenon, fully accounted for in supergravity, it can lead to complicated non-linear equations that are hard to solve explicitly. In this paper, we will instead look at compactifications involving Casimir energy and fluxes on manifolds equipped with a metric that is Riemann-flat on a first approximation; as a result, warping will be very small everywhere, and we will be able to solve and describe the solutions fully explicitly as a perturbative series around the flat metric, where each term will be less important than the preceding one.\footnote{In \cite{Avalos:2023mti,Avalos:2023ldc,Detraux:2024esd} non-supersymmetric heterotic string models with D-term uplifts are studied in the free fermionic formulation, where higher-loop contributions that arise after a stringy Scherk-Schwarz spontaneous supersymmetry breaking seem to allow for positive vacuum energies. Whether these dS solutions survive full stabilisation of the moduli (including the dilaton) requires further study.}
This means that our solutions are small perturbations of Riemann-flat backgrounds that solve the classical equations of motion without sources---neither fluxes nor Casimir energies. As we shall see explicitly, one can add flux and Casimir contributions that are not only small but of the same order, so that treating them simultaneously as perturbations to the flat solution is consistent. We discuss the issue of consistency and control in detail for the explicit 5d de Sitter saddle of Section~\ref{sec:dS5-maximum}.

The main technical result is a fully explicit formula to compute Casimir energies in a general Riemann-flat manifold (RFM). Interestingly, we find that the Casimir energy is a sum of contributions, each of which is localized on a specific submanifold of the RFM, which we also determine explicitly. Since the contributions are localized, we may think of them as  effective ``Casimir branes'';  the Casimir energy may then be replaced by a system of such branes which behaves in a similar way to the usual D-branes and orientifolds of more standard string compactifications. Therefore, the Casimir brane picture allows us to pull back a lot of the intuition from these models and use it effectively for model building with Casimir energies.

Furthermore, the Casimir formula that we obtained allows us to determine not just the contribution of Casimir energies to the lower-dimensional vacuum energy, but the explicit 11d profile of the vacuum energy as well. In other words, we go beyond the ``smearing approximation'' \cite{Acharya:2006ne,Baines:2020dmu}, which is where most analysis of 4d effective field theories derived from string theory stop. We obtain in this way completely smooth sources of negative energy in 11 dimensions;  the ``Casimir branes'' have a finite thickness which we compute explicitly, unlike the singular orientifolds commonly used as a source of negative energy, and whose ten-dimensional backreaction is a source of confusion \cite{Saracco:2012wc,Junghans:2020acz,Marchesano:2020qvg}. To put it in another way, when it comes to computing the backreacted higher-dimensional solution, our non-supersymmetric solutions are in a much better position than e.g. DGKT \cite{DeWolfe:2005uu}.

In some special cases, we will find that the tensions of different Casimir branes all add up to zero, so that the corresponding contribution to Casimir energy vanishes. As we will see in Subsection \ref{sec:alsym}, this phenomenon is a spacetime analog of an exotic mechanism to produce a vanishing cosmological constant in string perturbation theory, dubbed ``Atkin-Lehner symmetry'' \cite{atkin1970hecke,Moore:1987ue,Dienes:1990qh}. The mechanism had so far only been realized in very stringy models with a two-dimensional target space (see also \cite{Satoh:2021nfu} for a recent string-inspired discussion on generalizations of this). The version that we find works in higher dimensions, requires no worldsheet description, and can be reliably analyzed with EFT techniques. The resulting enlargement of the class of quantum gravity models where we can find a symmetry reason for a vanishing vacuum energy may have interesting implications for the construction of top-down models with a vanishing cosmological constant.

Armed with this picture and the explicit Casimir energy formula, we embark on a systematic exploration of the landscape of flux compactifications of RFM's---focusing on the M-theory corner---and look for de Sitter critical points. A major technical obstacle is that the tension of Casimir branes in general depends on moduli in an intricate way, via an infinite sum that converges slowly; we overcome this obstacle by extending the method of Ewald summation (used for similar sums that appear in physical chemistry) to arbitrary dimensions, and constructing an efficient numerical implementation for it.  

Even within the M-theory corner, the landscape of RFM compactifications is vast. In this paper we restrict to compactifications of M-theory down to four and five dimensions, with $G_4$-flux turned on. In an upcoming paper \cite{dS-nogos-MMBB}, we will show that $dS_4$ minima cannot be found from an M-theory compactification on an RFM with only these ingredients due to tadpole restrictions. Here, instead, we will look for $dS$ saddle points (that is, positive saddles of the vacuum energy with at least one unstable direction\footnote{The persistence of unstable directions in de Sitter solutions was previously noted e.g. in \cite{Danielsson:2011au,Danielsson:2012et,Shiu:2011zt,Chen:2011ac} for the Type IIA constructions with O-plane sources like the ones mentioned in footnote \ref{ft:IIA-models} \cite{Haque:2008jz,Danielsson:2009ff}.}). 
These provide a clean and clear example of the qualities of RFM compactifications. Furthermore, recent  DESI results \cite{DESI:2024mwx,DESI:2024kob,DESI:2025zgx,DESI:2025fii} have provided some evidence for dynamical dark energy as opposed to a cosmological constant.
If this is the case, a $dS$ maximum like the ones described in this paper would be phenomenologically favored over a minimum \cite{Agrawal:2018rcg}. Furthermore, the maxima we find are always at enhanced symmetry points, which could help addressing the question of initial conditions that one inevitably faces when describing Dark Energy with a maximum of the potential via e.g. a thermal Coleman-Weinberg-like transition.

After an exhaustive study of 2-moduli $dS_4$ compactifications of M-theory on an RFM of cyclic holonomy, we have found no dS maximum. While $dS_4$ maxima in M-theory may well exist when compactifying on RFM's of non-cyclic holonomy, the focus of our paper is instead a $dS_5$ solution that we found in five dimensions, and which we describe in the following paragraphs.

\subsection*{The main result: A $dS_5\times\mathcal{F}_6$ maximum in M-theory}
We found a concrete Riemann-flat manifold $\mathcal{F}_6=T^6/\mathbb{Z}_8$ such that an M-theory compactification on $\mathcal{F}_6$ with $G_4$ flux along a particular 2-cycle yields a five-dimensional de Sitter solution of M-theory, with vacuum energy
$V_{\text{saddle}} = 4.43\cdot 10^{-8} \,\ell_5^{-5}$.
The solution is supported by a balance of Casimir energy and $G_4$-flux. It is also scale separated, with a hierarchy
$\ell_\text{KK}\sim 4.6\cdot10^{-3} H^{-1}_0$. The scalar sector of the low-energy EFT contains five saxionic scalars, with masses $m_i^2=\{-83.3 \,,\, -5.98 \,,\, -359 \,,\, 6.68\,,\, -22.1\}\cdot H_0^2$, as well as five axions that get negligible masses from non-perturbative effects. See Table \ref{resumen} for more details on the solution and its properties.

To our knowledge, this is the first time a concrete, top-down dS maximum compactification in string or M-theory has been realized in sufficient detail to compute the vacuum energy precisely, as well as the spectrum of low-lying excitations. For contrast, recent work \cite{Chen:2025rkb} has robust arguments for the existence of dS maxima in stringy regimes, but the vacuum is estimated to be only around $10^{-3}$ in Planck units and there is not enough control to reliably compute the vacuum energy or the masses of low-lying fields. Similar control issues affect other proposals of dS maxima within supergravity \cite{Haque:2008jz,Danielsson:2009ff,Andriot:2021rdy,Andriot:2022bnb,Andriot:2022way,Andriot:2024cct,Chen:2025rkb}, and all the more so other constructions of dS minima \cite{Kachru:2003aw,Balasubramanian:2005zx,DeLuca:2021pej}, where the intricate nature of the construction means that some $\mathcal{O}(1)$ guesswork (e.g. in the values of Pfaffian factors, effects of stringy corrections, etc) becomes all but necessary. Furthermore, a controlled $dS_5$ saddle can also be related to a $dS_4$ via the dS/dS correspondence \cite{Alishahiha:2004md,Gorbenko:2018oov}.

The property that sets our solution apart from these proposals is that the Riemann-flat manifold affords us very good control over corrections, which we discuss in detail in Section \ref{sec:ctrl}. As explained above, our explicit Casimir formula provides a completely smooth picture at the 11-dimensional level, which in turn allows us to estimate the classical 11d backreaction explicitly (and would allow for a systematic, rigorous treatment of the full 11d profile in future work). Higher-derivative and loop corrections are similarly suppressed, all in powers of a small parameter $\epsilon\sim 10^{-5}$, related to the smallness of the vacuum energy and the large volume of the internal space. M2 and M5-brane instantons are negligible. Since there are no singular sources (branes, orientifolds), or non-perturbative effects, this means that our solution survives all the known effects that are normally discussed in the context of string compactifications. An explicit non-supersymmetric compactification, constructed in a perturbative expansion around an exact solution of the classical equations of motion, like the one we will find here, also aligns with the considerations in \cite{Sethi:2017phn}, which argued that the calculation of quantum effects around classically time-dependent vacua is fraught with subtleties, and that we lack the proper framework to do so in flux compactifications of String Theory. In that same reference, it was suggested that, whenever applicable, supergravity might be an appropriate substitute, that is, as long as volumes are large and there are no stringy ingredients such as branes, etc. The solutions constructed here are in precisely that regime. 

The question remains whether there could be unknown effects that destabilize the solution. The smallest closed geodesic is only $3.87$ eleven-dimensional Planck lengths long, so one could imagine loops of 11-dimensional virtual black holes around this circle. Although we estimate the effect of a single black hole to be small, based on  general properties of the Casimir energy, we do not really know how many long-lived black holes M-theory has at Planckian energy. If there is an unexpectedly large number of these (larger than what a na\"ive extrapolation of the BH entropy formula would suggest), they could destabilize the solution. Because of this, we cannot tell rigorously whether the solution survives or not, but the larger point is that the Landscape of RFM flux compactifications is a promising arena where we may get closer than ever to a fully controlled string compactification exhibiting positive vacuum energy. 

The paper is organised as follows:\begin{itemize}
\item In \secref{sec:warm-up-dS7} we present a simple example of a $dS_7$ from M-theory on $T^4$ that will serve as a warm-up for the constructions studied in the following sections.
\item In \secref{sec:RFMs}, we briefly introduce Riemann-flat manifolds and their properties, as well as the necessary tools for the compactifications studied here; this includes the mathematical description of RFM's, their invariant metrics and allowed spin structures; we also comment on their classification and focus on the particular case of cyclic RFM's. 
\item \secref{sec:casimir-RFMs} is devoted to the computation of Casimir energies in Riemann-flat manifolds; starting with the Casimir energy of massless free fields, we explain how it can be computed in quotient spaces through the method of images and derive an explicit formula for the Casimir energy on Riemann-flat manifolds; we show in particular how the computation of the Casimir energy on RFM's is always reduced to a similar computation performed on the invariant subspace with respect to each element of the holonomy group of the RFM, which often has dimension lower than the RFM, and introduce the concept of ``Casimir branes''; we also explain how the vacuum energy can sometimes vanish as a result of a spacetime version of an Atkin-Lehner symmetry. We end this section by extending the technique of Ewald resummation to arbitrary dimensions, which allows for an efficient evaluation of the sums over lattices required for the Casimir energies, especially when the invariant subspace has lower dimension. 
\item In \secref{sec:dS5-maximum}, we apply the results of Sections \ref{sec:RFMs} and \ref{sec:casimir-RFMs} to M-theory compactified on a 6-dimensional Riemann-flat manifold, which results in a controlled $dS_5$ solution; we work through this construction in detail, carefully computing the Casimir and flux potentials, analysing the 5d saddle point and discussing the control of this solution. 
\item Finally, in \secref{sec:dS4-maxima} we scan a full family of cyclic 7-dimensional RFM's in the search for a $dS_4$ solution, showing that none of the RFM's whose study can be reduced to a 2-moduli problem through the use of symmetry arguments can provide a saddle for the Casimir potential; we also comment on cases that remain open, all of which require the study of at least three moduli. 
\item We conclude in \secref{sec:conclusions}, where we summarise our results and discuss their relation to current and future work. Several appendices contain details or alternative derivations of results in the main text.
\end{itemize}
 Finally, the following list summarizes some aspects of our notation:

\begin{itemize}

\item We are interested in compactifications of M-theory or other higher-dimensional string theories, with spacetime dimension $D$. The dimension of the compactification manifold will be $k$, and the lower-dimensional theory has dimension $d=D-k$. 
\item We use a mostly minus metric convention, and greek indices denote $D$-dimensional tensor components.
\item The coordinates in the internal $k$-dimensional manifold will be denoted as $\vec{z}$ or $z^i$.

\item When we write an RFM as Euclidean space $\mathbb{R}^k$ modulo a group, that group is called $\mathcal{B}$, i.e. $\text{RFM}=\mathbb{R}^k/\mathcal{B}$. An element of $\mathcal{B}$ is an affine transformation of $\mathbb{R}^k$, of the form $\D[]+\bvec[]$, and will be denoted abstractly as $b\in\mathcal{B}$. The quotient group that captures holonomies in an RFM is called $\Gamma$, and $\gamma\in\Gamma$ is a generic element. Whenever we want to emphasize the image of a given $b$ represented as $\D[]+\bvec[]$, via the projection map to $\Gamma$, we will write $\D +\bvec$. Since we will be looking at cases where $\Gamma$ is cyclic, we will often choose a specific generator of $\Gamma$, which we will call $\mathbf{g}$ (for generator), and work with $\D[g]+\bvec[g]$. The order of a group $\Gamma$ will be denoted as $\vert \Gamma\vert$.

\item We denote the trace of an element $\D[]$ under a given representation $\mathbf{r}$ of $SO(D-2)$ as $\Tr{r}{\D[]}$. We will sometimes collect the traces under all bosonic and fermionic representations using the notation $\TrB{\D[]}$ and $\TrF{\D[]}$, respectively.

\end{itemize}

\noindent\textbf{Note added}: As this work was nearing completion we learned of \cite{DallAgata:2025}, which also discusses Casimir energies in Riemann-flat manifolds in the context of supergravity theories.

\noindent\textbf{Note added in v3}: There was a calculational error (described in more detail in Subsection \ref{subsec:dS5-maximum-RFM}) in the computations leading to the $dS_5$ saddle described in previous versions of the manuscript (we thank M. Aparici for pointing this out to us).  In this version of the paper we have fully rewritten Section \ref{sec:dS5-maximum}, replacing the wrong calculations with corrected ones. Although the resulting $dS_5$ saddle is still under very good control, the quality is smaller than what we reported in previous versions (vacuum energy $10^{-8}$ instead of $10^{-15}$; control parameter $\epsilon\sim 10^{-5}$ instead of $\epsilon\sim 10^{-10}$). This motivates us further to search for better quality RFM vacua, a project which is ongoing. In soon--to--appear work together with M. Aparici, we will present a novel $dS_4$ solution where the control parameters are as good as the ones quoted in previous versions of this paper. No other results of the paper (theoretical framework, other checks, checks in four dimensions, Casimir formulae---indeed, nothing outside Section \ref{sec:dS5-maximum}) are impacted at all. 

\section{Warm-up: A \texorpdfstring{$dS_7$}{dS7} from M-theory on \texorpdfstring{$T^4$}{T4}}
\label{sec:warm-up-dS7}

The key idea in this paper is to use quantum effects to evade no-go theorems like \cite{Maldacena:2000mw} and achieve de Sitter solutions. In this section, we will take a first step towards this goal by explaining the general idea and a simple implementation of it, which will motivate the central topic of this paper---an exploration of the Landscape of Riemann-flat manifolds, or RFM's for short. To evade the no-go theorem of \cite{Maldacena:2000mw} we need sources of stress-energy in the compactification that violate the null energy condition. Popular choices are orientifolds or higher-derivative effects \cite{Grana:2005jc,Denef:2007pq,Blaback:2019zig}, but here we will choose Casimir energies \cite{Appelquist:1983vs}, which we will now review; a good source is \cite{Arkani-Hamed:2007ryu}, or the textbook \cite{Birrell:1982ix}.

 Consider a $D$-dimensional QFT in a background spacetime $\mathcal{M}$ equipped with a metric $g$. This may be a bona fide QFT or merely a low-energy EFT, possibly coupled to gravity; if so, we will also take into account the standard linearized graviton and its standard Fierz-Pauli action and couplings to matter. Other than the linearized fluctuation, which will be treated as just another quantum field on $\mathcal{M}$, the background geometry will be taken to be non-dynamical.  Finally, if the QFT is spin or has any other symmetries, we will also consider $\mathcal{M}$ to be equipped with the appropriate background fields. We will however set to zero any background gauge fields for continuous symmetries (see \cite{Arkani-Hamed:2007ryu,Ibanez:2017kvh,Hamada:2017yji} for a case where this dependence is taken into account in a particular example).
  
The basic observable in any QFT like the one just described is the stress-energy tensor $T_{\mu\nu}$, and its expectation value in the vacuum state,
\begin{equation} \langle  T_{\mu\nu}\rangle_{\mathcal{M}}.\end{equation}
When there is no ambiguity, we will omit the $\mathcal{M}$ subscript. In the general case, absent any symmetries, the vev $\langle  T_{\mu\nu}\rangle$ is a generic, non-zero tensor field on $\mathcal{M}$. It is the quantum stress-energy of the QFT on the given background, and it is called the ``Casimir stress-energy tensor'' or, more vaguely, ``Casimir energy''. In some special cases, when the background $\mathcal{M}$ preserves some symmetries,  the vev $\langle  T_{\mu\nu}\rangle$ can be constrained. For instance, if $\mathcal{M}$ is a maximally symmetric space (such as dS, Minkowski, or AdS spaces) then $\langle  T_{\mu\nu}\rangle=\rho\, g_{\mu\nu}$. If the QFT additionally preserves supersymmetry in a maximally symmetric $\mathcal{M}$, then $\rho=0$; in general, unbroken supercharges can constrain some components of $\langle  T_{\mu\nu}\rangle_{\mathcal{M}}$. 

We will postpone a detailed exposition of how to compute Casimir energies to \secref{sec:casimir-RFMs} and describe now the basic approach that we will follow in this paper. Much like in \cite{DeLuca:2021pej}, we are after dS solutions of Einstein's equations where the effects of the Casimir energy are taken into account. In other words, we seek to solve Einstein's equations
\begin{equation} G_{\mu\nu}=8\pi G \left[T_{\mu\nu}^{\text{Cl}} +\langle  T^{\mu\nu}\rangle_{\mathcal{M}}\right],\label{eeq}\end{equation}
where $T_{\mu\nu}^{\text{Cl}}$ is some classical piece, which in our case will come entirely from fluxes in the internal space. We will then balance the classical and Casimir pieces against each other, to find a backreacted solution of Einstein's equations \eq{eeq} with a positive cosmological constant.

One may wonder how this approach could be self-consistent,  given that $\langle  T^{\mu\nu}\rangle_{\mathcal{M}}$ was computed in a background geometry, on which it now backreacts via \eq{eeq}.  The idea is that, under favorable circumstances, it is possible to treat the right-hand side of \eq{eeq} as a small perturbation, and solve the problem iteratively. More concretely, suppose that the manifold $\mathcal{M}$ admits a Ricci-flat metric $g^{(0)}$ (which therefore solves the vacuum Einstein equations), on top of which the Casimir energy and flux term in \eq{eeq} turn out to be small in Planck units, i.e. 
\begin{equation} T_{\mu\nu}^{\text{Cl}}\sim \langle  T^{\mu\nu}\rangle_{\mathcal{M}}\sim \epsilon\, M_P^D,\quad \epsilon\ll1.\end{equation}
 We can then perturb
 \begin{equation} 
    g^{(0)}\,\rightarrow g^{(0)}+ \epsilon\, g^{(1)}+\ldots,\quad T_{\mu\nu}=  \epsilon\, T^{(0)}_{\mu\nu}+\epsilon^2\, T^{(1)}_{\mu\nu}+\ldots 
    \label{eq:eps-expansion}
\end{equation}
 and expand Einstein's equations as a perturbative series in \eq{eeq}, to obtain a system of equations
 \begin{equation} 
    G^{(n)}_{\mu\nu}= 8\pi G\, T^{(n-1)}_{\mu\nu} \,,\label{pertex}
\end{equation}
 which can be solved iteratively in $\epsilon$. Each of these is a linear differential equation, for which a solution typically exists (and can be ensured by rigorous theorems in the elliptic case \cite{gilbarg2001elliptic,evans2010partial}). Importantly, the right-hand side of \eq{pertex} at order $n$ only depends on quantities computed at order $n-1$, as is standard in perturbation theory. The whole system can then be solved iteratively. For small enough $\epsilon$, one expects this expansion to be convergent, and a solution to the \emph{full} $D$-dimensional backreacted Einstein's equations to be achievable by this procedure. The smallness of the parameter $\epsilon$ also ensures the irrelevance of e.g. higher-derivative corrections (which are proportional to a power of $\epsilon$). 

In this paper, we will only carry out the first level of the perturbative analysis \eq{pertex} at $n=1$. However, as remarked in the Introduction, the fact that the only ingredients in \eq{eeq} are Casimir energies and fluxes, and both are under good control from ten/eleven-dimensional supergravity, means that going to higher $n$ is conceptually straightforward (albeit technically cumbersome).  The point is that a fully explicit 10d/11d uplift of the solutions we propose here is certainly feasible, via the procedure just outlined. To our knowledge, this is not the case for any other proposal for dS minima or maxima in String Theory, including \cite{Andriot:2022bnb,Andriot:2022way,Andriot:2024cct,Chen:2025rkb}.  The fact that the backreaction of our dS solutions can be studied explicitly in eleven dimensions is a remarkable property, which we wish to emphasize. It means that the fate of these vacua can be ascertained with a finite amount of work, and have a small chance of leading to endless discussions. Even some supersymmetric AdS vacua, such as DGKT \cite{DeWolfe:2005uu}, do not have a fully clear 10d uplift, due to the presence of singular, stringy sources as orientifolds \cite{Saracco:2012wc,Junghans:2020acz,Marchesano:2020qvg} that complicate the 10d analysis. Nothing like that can happen for the vacua presented here. In the proposal of \cite{DeLuca:2021pej}, which also uses Casimir energies as a key ingredient to achieve de Sitter, one may also solve for the 11-dimensional backreaction explicitly, since there are no ingredients beyond supergravity; however, unlike for an RFM, there is no analog of the ``unperturbed'' metric $g^{(0)}$, since hyperbolic manifolds do not carry a metric that solves Einstein's equations before Casimir energies are included. As a result, one must solve the eleven-dimensional equations all at once, in contrast to the easier situation in an RFM.
  
We can now particularize our discussion to the compactification ansatz we are interested in. Specifically $\mathcal{M}$ will be of the form
\begin{equation}\mathcal{M}=\mathbb{R}^{D-k}\times \mathcal{N}_k,\end{equation}
where the manifold $\mathcal{N}_k$ is compact and the metric on $\mathcal{M}$ is of the form
\begin{equation} ds_{\mathcal{M}}^2= ds^2_{\Lambda}+ ds_{\mathcal{N}_k}^2,\label{e0comp}\end{equation}
with  $ds^2_{\Lambda}$ being a maximally symmetric Lorentzian metric with cosmological constant $\Lambda$\footnote{In general, we would have to introduce a warping function $e^\omega$ in front of the first term of \eq{e0comp}. However, since our starting metric $g^{(0)}$ will be Riemann-flat and thus unwarped, $\omega$ will not appear until second order in the $\epsilon$ expansion. We will not reach this level in this paper and hence ignore $\omega$ altogether.}.
Due to the maximal symmetry of the non-compact dimensions, the stress-energy tensor vev $\langle  T_{\mu\nu}\rangle_{\mathcal{M}}$ has a similar decomposition to the metric in \eq{e0comp}, so that
\begin{equation}
    \langle  T_{\mu\nu}\rangle_{\R^{D-k}\times \mathcal{N}_k}=  \hat{\rho}\, g_{\mu\nu}^{\mathbb{R}^{D-k},\,\Lambda}+ T^{\mathcal{N}_k}_{\mu\nu} \,.
    \label{e1comp}
\end{equation}
Crucially, the function $\hat{\rho}$ can have either sign; we will be interested in the case $\hat{\rho}<0$, where the corresponding term in the stress-energy tensor violates the $(D-k)$-dimensional version of the null energy condition. The tensor $T^{\mathcal{N}_k}_{\mu\nu}$ only has legs along the compact manifold $\mathcal{N}_k$. In general, $\hat{\rho}$ is related to $T^{\mathcal{N}_k}_{\mu\nu}$ by stress-energy tensor conservation (absent gravitational anomalies) $\nabla_\mu  \langle  T^{\mu\nu}\rangle_{\mathcal{M}}=0$.  For a CFT, there is an additional relation coming from the trace anomaly, but there are no other general relationships between $\hat{\rho}$ and $T^{\mathcal{N}_k}_{\mu\nu}$. However, typically these quantities have to be computed on a case-by-case basis.

As explained above, in this paper we just aim to find the backreacted lower-dimensional cosmological constant at first order, and leave the higher-order analysis to further work. An explicit formula for the cosmological constant at this order  can be obtained by taking the $00$ component of Einstein's equations \eq{eeq} and integrating over the compact space $\mathcal{N}_k$. The resulting system is described by a $(D-k=d)$--dimensional effective field theory, where the flux piece is described by the standard flux potential \cite{Grana:2005jc}, and the Casimir stress-energy is obtained from an effective potential
\begin{equation}
    V_{\text{Cas}} = \int d^kz\, \sqrt{-g_0}\, \hat{\rho} \,,
\end{equation}
by integrating the Casimir stress-energy tensor over the whole of the internal compact space $\mathcal{N}_k$.
Since it will not cause ambiguity for us, we will refer to the quantity $V_{\text{Cas}}$ as the Casimir potential or Casimir energy of the compactification. In general, $V_{\text{Cas}}$ receives contributions from both massless and massive $D$-dimensional fields, but the contributions of massive states are suppressed by factors of $e^{-mR}$ \cite{Arkani-Hamed:2007ryu}, where $m$ is the mass of the field and $R$ is a quantity with units of length that measures the volume of $\mathcal{N}_k$ according to the metric $g^{(0)}$ (so that $R^k\sim\text{Vol}(\mathcal{N}_k)$, possibly up to $\mathcal{O}(1)$ constants). In this work, we will focus on compactifications of 10d and 11d quantum gravities, where the gap to the first massive state is controlled by the higher-dimensional Planck or string scale. As a result, we will ignore contributions of massive states, and focus solely on the Casimir potential generated by the massless, $D$-dimensional fields. Since the Casimir energy is the response function of the QFT in the background spacetime metric $g^{(0)}$, which we take to be Riemann-flat, the only dimensionful parameter $V_{\text{Cas}}$ can depend on in a scale-invariant theory is $R$. Therefore, dimensional analysis fixes
\begin{equation}
    V_{\text{Cas}}\sim -\frac{\mathcal{C}}{R^{d}}\,,
\end{equation}
where the quantity $\mathcal{C}$ (that may depend on dimensionless moduli of $\mathcal{N}_k$) is the Casimir coefficient. The extra minus sign has been introduced for reasons that will become clear below.

The total lower-dimensional scalar potential is then given by
\begin{equation}
    V^{(d)} = V_{\text{Fluxes}}+V_{\text{Cas}}\,,
    \label{vtot}
\end{equation}
and our goal will be to find saddles of \eq{vtot} where $V^{(d)}>0$. The flux term contributes a sum of positive terms, all of which scale as powers of $R$ \cite{Grana:2005jc}; specifically, a $p$-form field flux in the internal space produces a contribution to $V_{\text{Fluxes}}$ scaling like
\begin{equation}
    \frac{n_p^2}{R^{2p-k}} \,.
\end{equation}
After rescaling to the lower-dimensional Einstein frame, one gets a potential schematically of the form
\begin{equation} 
    \frac{V^{(d)}}{M_P^d}\sim\frac{1}{R^{\frac{d}{d-2}k}}\left(\sum_p\frac{n_p^2}{R^{2p-k}} -\frac{\mathcal{C}}{R^{d}} \right).\label{fpot22}
\end{equation}
that is a sum of negative powers of $R$. It follows that, to achieve a saddle with $V^{(d)}>0$, at least one term has to be negative. Since the flux terms are positive definite, our only chance is $\mathcal{C}>0$. Indeed, when $\mathcal{C}>0$ the stress-energy tensor violates the null-energy condition, and can thus evade the no-go of \cite{Maldacena:2000mw}. This is why $\mathcal{C}$ was defined with an extra minus sign in \eq{fpot22}. An analogous role is played by orientifolds in standard flux compactifications \cite{Grana:2005jc}.

\begin{figure}[!htb]
    \centering
    \includegraphics[width=0.6\linewidth]{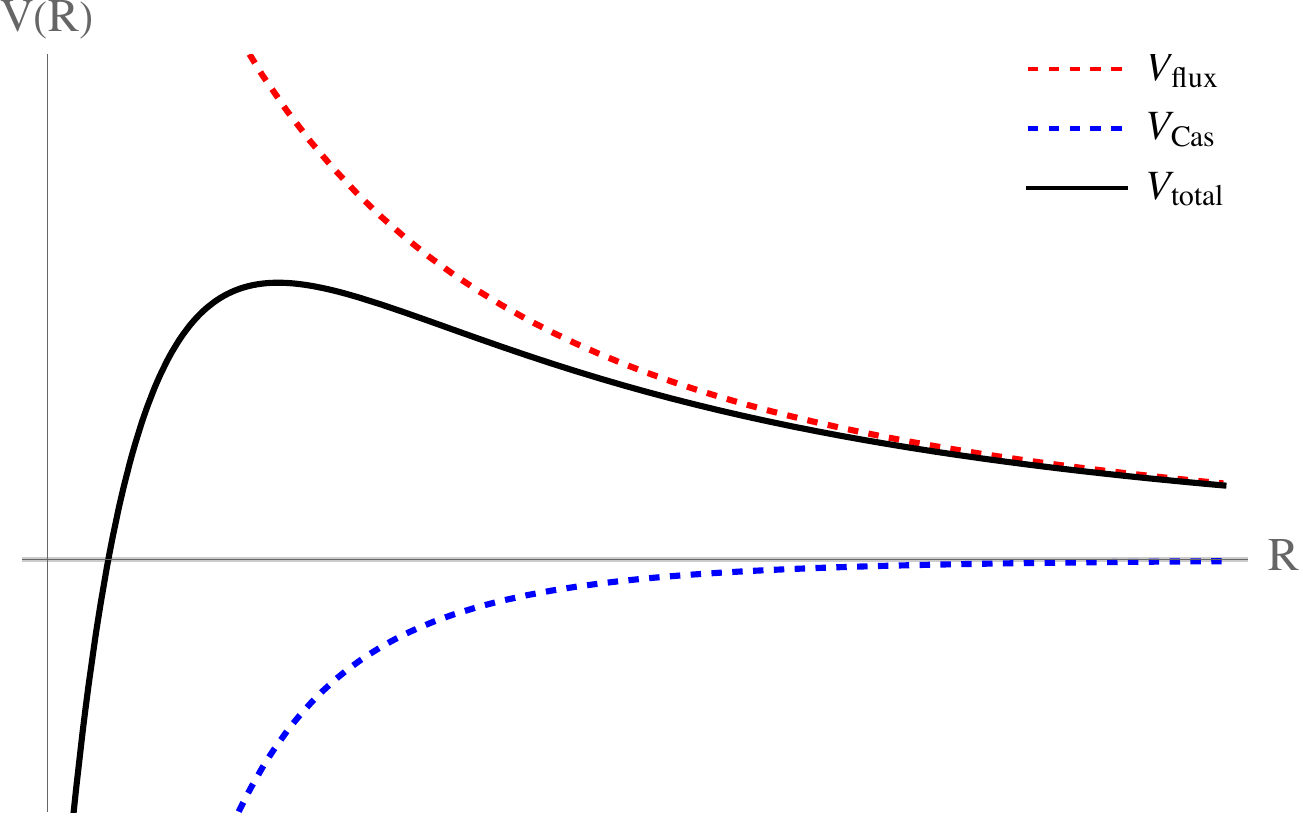}
    \caption{In this paper, we will look for dS maxima solutions where a negative Casimir energy contribution (blue dashed line) and a flux contribution (red dashed line) generate a combined potential (black solid line) for the volume modulus $R$ with a maximum. There is no curvature contribution since the manifold is Riemann-flat. Due to Einstein frame rescalings, the value of the maximum can easily become small in Planck units even for moderately large $R$. Other moduli not included in this picture will also be saddle points due to additional symmetries present at the maximum.}
    \label{fig:key-idea}
\end{figure}

We will look for maxima of \eq{fpot22} where the Casimir contribution is balanced by fluxes, as illustrated in Figure \ref{fig:key-idea}. One could try to add more fluxes to get a dS minimum, analogously to the construction in \cite{DeLuca:2021pej}. Finding a minimum is trickier than it seems at first glance, due to tadpoles and other stringy features, and we leave a detailed discussion to a future publication \cite{dS-nogos-MMBB}. However, finding a maximum does not seem hard at all! We will finish this section by describing one such concrete first attempt, a close relative of the non-susy AdS vacuum in the Appendix of \cite{Luca:2022inb} (see also \cite{Parameswaran:2024mrc}). 

Consider M-theory on $T^4$, threaded by $n_4$ units of $G_4$-flux. The resulting seven-dimensional theory has 10 geometric moduli, coming from the size $R$ of the $T^4$ and its shape moduli. Although the Casimir coefficient $\mathcal{C}$ can in principle depend on all of these, there are special points of moduli space where we can ensure that the dependence of $\mathcal{C}$ on these moduli vanishes at first order, due to symmetries. Specifically, suppose we are at a point in $T^4$ moduli space where the lattice $\Lambda$ defining the $T^4\equiv\mathbb{R}^4/\Lambda$ has a non-trivial symmetry group. For instance, we could take the lattice $\Lambda$ to be the root lattice of the $C_4$ Lie algebra rescaled by $R$. This is the lattice generated by the vectors
\begin{equation}
    (R,-R,0,0),\, (0,R,-R,0),\, (0,0,R,-R)\,\, \text{and}\,\, (0,0,0,2R)
\end{equation}
in $\mathbb{R}^4$ equipped with its standard inner product. Then the $T^4$ and the Casimir energy are invariant under the action of the Weyl group $W(C_4)$, and this action fixes all $T^4$ moduli to a single point. As a result, the gradient $\nabla \mathcal{C}$ with respect to $T^4$ shape moduli transforms nontrivially under $W(C_4)$, and invariance under it means it has to vanish. Thus, at the $C_4$ point in moduli space, we can ignore all shape moduli of $T^4$, and the problem is effectively one-dimensional---we only need to care about the size modulus $R$. The use of symmetries to find saddle points of functions has a long history, but see \cite{Parameswaran:2024mrc} for recent applications (also involving Casimir energy) and \cite{Chen:2025rkb} where a version of this idea including stringy symmetries was applied to the search of dS maxima. 

With these provisions, the scalar potential \eq{fpot22} becomes
\begin{equation}
    \frac{V^{(7d)}}{M_7^7}= \left(\frac{\ell_{11}^{\,9}}{2R^4}\right)^{\frac75}\left[\left(\frac{\pi}{2}\right)^{\frac23}\frac{n_4^2}{2 R^4\ell_{11}^{3}}-\frac{\mathcal{C}}{R^7}\right],\quad \mathcal{C}\approx 8.827 \,.
    \label{f09}
\end{equation}
The factors of $\pi$ in front of the flux piece come from imposing the correct quantization conditions for the M-theory $G_4$ flux; see \cite{Polchinskiv2,Witten:1996md,deAlwis:1996ez,deAlwis:1996hr} and Appendix \ref{app:fluxQ} for details. The value of the Casimir coefficient for M-theory on $T^4$ in \eq{f09} will be calculated in \secref{sec:Casimir-on-RFMs}; for now, we merely remark that it depends on the spin structure on $T^4$, and that the number quoted above corresponds to an antiperiodic spin structure along all cycles of the $T^4$. Finally, we have rescaled to Einstein frame by reducing the 11d integrated potential and expressing it in units of the reduced seven-dimensional Planck's mass 
\begin{equation} 
    M_7^{5}\equiv \frac{\text{Vol}(T^4)}{\ell_{11}^{\,9}}= \frac{2\, R^4}{\ell_{11}^{\,9}} \,,
\end{equation}
where $\ell_{11}$ is the reduced 11-dimensional Planck length. 
The potential \eq{f09} does have a maximum, at
\begin{equation}
    R\approx \frac{2.58\, \ell_{11}}{n_4^{2/3}},\quad \frac{V^{(7d)}}{M_7^7}\approx 6.85\cdot 10^{-6}\, n_4^{42/5} \,.
    \label{f10}
\end{equation}
Even taking the smallest possible value $n_4=1$, the radius is Planckian, with higher values of $n_4$ leading to even smaller radii\footnote{This is to be contrasted with the supersymmetric Freund-Rubin $AdS_7\times S^4$ compactification of M-theory \cite{Freund:1980xh,Aharony:2008ug}, where increasing the 4-form flux leads to bigger radii. In that case, the AdS minimum is achieved by a balance of the $G_4$ flux and a curvature term, which scales as $R^{-\alpha}$ with $\alpha$ smaller than the exponent of the flux term. On the other hand, the Casimir vacua in the main text scale as $R^{-\alpha}$ with $\alpha$ bigger than the flux contribution. This explains the different behaviors. For the same reason, the Casimir vacua are automatically scale-separated, while Freund-Rubin compactifications are not.}. The upshot is that by this simple means we do get a de Sitter maximum. The vacuum energy is only $10^{-6}\,\ell_7^{-7}$, due to a favorable combination of numerical factors. This means that the dS maximum would have a Hubble radius equal to \cite{Chen:2014fqa}
\begin{equation}
    \frac{1}{H_0}=\sqrt{\frac{15}{V^{(7d)}}}\approx 1480 \quad\text{(7d Planck lengths)}\,,
\end{equation}
and hence seven-dimensional gravitational backreaction will not immediately kill the vacuum. 
The hierarchy between KK and Hubble scales is around 234, so it is also a little bit scale separated.  What about 11-dimensional backreaction? The energy density in the flux piece is around $1.7\cdot 10^{-4}\,\ell_{11}^{-11}$, and so any M-theory higher-derivative corrections of the form
\begin{equation} \int \vert G_4\vert^{2n}\label{hder2}\end{equation}
will be suppressed as roughly $10^{-4n}$. Of course, the $\mathcal{O}(1)$ coefficients  in front of \eq{hder2}, about which little is known, are a source of trouble (see however \cite{Grimm:2017okk,Grimm:2017pid,Hyakutake:2005rb,Hyakutake:2007sm,Hyakutake:2006aq}). If they are large enough they could spoil the solution, but on the other hand we expect these to form an asymptotic series, with each coefficient smaller than the next. Finally, any terms involving the 11d Riemann tensor in combination with curvature,
\begin{equation} 
    \int R_{\mu\nu\rho\sigma}^k\vert G_4\vert^{2n} \,, 
    \label{hder3}
\end{equation}
will vanish to leading order in our solution, since the leading metric on $T^4$ is Riemann-flat. They will start contributing at order $\epsilon^2$ in the expansion \eq{eq:eps-expansion} where, as we have seen,  $\epsilon\sim10^{-4}$ for this solution. Therefore, they might be similarly small. Finally, using the explicit expression for the M2-brane tension in Appendix \ref{app:fluxQ}, these can be shown to give negligible contributions.

All in all, we find that the solution \eq{f10} is not obviously killed by any explicitly known effects, and it might turn out to actually survive as a true M-theory vacuum. At the same time, such a small $T^4$ ($R\approx 2.58$) might be corrected by exotic, non-supersymmetric M-theory effects that we know nothing about. Rather than exploring the consistency of the solution any further, we will try to improve on it, by finding similar compactifications of M-theory that lead to larger values of $R$, while at the same time preserving the nice features that lead to increased control---chief among which is a background metric $g^{(0)}$ that is exactly Riemann-flat. The next sections will further develop this approach. 

\section{A review of Riemann-flat Manifolds (RFM's)}
\label{sec:RFMs}
Most of the mileage in the construction of \secref{sec:warm-up-dS7} comes from the fact that the unperturbed metric $g^{(0)}$  in the compactification manifold  $T^4$ is Riemann-flat. There are many more Riemann-flat manifolds (RFM's) than $T^4$, and they share or even enhance many of the nice properties (control under corrections, computable Casimir energy) that we used in \secref{sec:warm-up-dS7}. This section serves as a brief introduction to RFM's and their properties. The subject of RFM's is amply discussed in the mathematical literature, see e.g. \cite{borwein2013lattice}.

A $k$-dimensional Riemann-flat manifold $\mathcal{F}_k$ is equipped with a metric for which the Riemann tensor exactly vanishes, so that the metric is exactly flat. Thus, these manifolds look like Euclidean space in a finite neighbourhood of any point. In fact, the universal cover of any Riemann-flat manifold $\mathcal{F}_k$ is Euclidean space $\mathbb{R}^k$; this means that all RFM's arise as a quotient of $\mathbb{R}^k$ with its standard Euclidean metric,
by a subgroup $\mathcal{B}\subset ISO(k)$ of the Euclidean group in $k$ dimensions (the group of isometries of the standard flat metric in $\mathbb{R}^k$), 
\begin{equation}
    \mathcal{F}_k=\mathbb{R}^k/\mathcal{B} \,.\label{rfmdef}
\end{equation} 
Importantly, the group action is free, meaning that no $b\in\mathcal{B}$ has fixed points. This ensures that the quotient is a manifold and not an orbifold. 

A general isometry in $ISO(k)$ can be written as a rotation followed by a translation,
\begin{equation} 
    b\in\, ISO(k):\, \vec{z}\,\rightarrow  \mathbf{D}\, \vec{z}+ \vec{b}\,,\quad \mathbf{D}\in SO(k) \,,
    \label{ewerrr}
\end{equation}
so every element in $\mathcal{B}$ can be presented in this form. There is a natural normal subgroup of $\mathcal{B}$, corresponding to all those $b\in\mathcal{B}$ with $\mathbf{D}=\mathbf{I}$; this is called the translation subgroup. The corresponding vectors $\vec{b}_i$ generate a lattice $\Lambda\subset\mathbb{R}^k$, that can be identified with the translation subgroup itself. The quotient \eq{rfmdef} producing $\mathcal{F}_k$ can be taken in two steps, where in the first we quotient $\mathbb{R}^k$ by the lattice $\Lambda$ to produce a torus (since we are interested in compact manifolds, we will take the lattice $\Lambda$ to be of full rank $k$),
\begin{equation} 
    T^k=\mathbb{R}^k/\Lambda \,.
\end{equation}
This torus is then quotiented by $\Gamma$, the group of isometries of $T^k$ defined as the quotient of $\mathcal{B}$ by the translation subgroup,
\begin{equation} 
    \Gamma\equiv \mathcal{B}/\sim\,, \quad b\sim b' \quad\text{iff}\quad b'\circ b^{-1}\in \Lambda \,.
\end{equation}
In practice, representatives of an element $\gamma\in\Gamma$ are given by affine transformations of the form \eq{ewerrr} where $\vec{b}$ is taken modulo translations. Furthermore, elements in $\Gamma$ must preserve the lattice $\Lambda$, or otherwise the quotient would be empty. This also means that $\Gamma$ is a \emph{finite} group of isometries of $T^k$.  In mathspeak, one says that these groups fit into an exact sequence of abelian groups,
\begin{equation}
    0\,\longrightarrow \Lambda\,\longrightarrow\, \mathcal{B}\,\longrightarrow \Gamma\,\longrightarrow1 \,,
    \label{bieberles}
\end{equation}
and that the group $\Gamma$ is a subgroup of the automorphism group of the lattice $\Lambda$.  Whenever we want to specify the class $\gamma$ corresponding to a given affine transformation, we will indicate it via a subscript, as in $\mathbf{D}_\gamma\, \vec{z}+ \vec{b}_\gamma$. The group $\Gamma$ is also called the point group or holonomy group of the RFM, since as an abstract group, it is isomorphic to the group of rotation matrices $\mathbf{D}_\gamma$, which are entirely responsible for any holonomy that vectors take upon following a closed curve, since the manifold is Riemann-flat. Thus RFM's have finite holonomy groups and, in fact, they are the only manifolds with this property \cite{eee57c5f-3af2-3ab0-b3c0-3d30ad3fad43}. The fact that $\Gamma$ is isomorphic to a subgroup of $SO(n)$ that preserves the lattice $\Lambda$ (equipped with the standard inner product) means that it is a crystallographic group in $k$ dimensions, about which much is known. Hence one may think of RFM's as ``crystals'', similar to the familiar Brillouin zone of condensed matter systems \cite{ashcroft2011solid}. However, not any crystallographic group gives rise to an RFM---demanding absence of fixed points in the $\Gamma$ action (called torsion-freeness of $\mathcal{B}$ in the math literature) leads to constraints on the pairs $(\mathbf{D}_\gamma,\vec{b}_\gamma)$. For instance, one such restriction is that the fixed-point equation for each generator individually,
\begin{equation} 
    \mathbf{D}_\gamma\, \vec{z}+ \vec{b}_\gamma= \vec{z}+ \vec{l}\,,\quad \vec{l}\in\Lambda \,,
\end{equation}
has no solution. This can be rearranged to give 
\begin{equation} 
    (\mathbf{D}_\gamma-\mathbf{I})\, \vec{z}+\vec{b}_\gamma=0\,\, \text{mod}\,\Lambda \,.
    \label{fpfcond}
\end{equation}
Since the operator on the left-hand side is the projector onto the non-invariant subspace of $\mathbf{D}_\gamma$, this equation will have a solution unless $\vec{b}_\gamma$ has a non-zero component along the invariant subspace of $\mathbf{D}_\gamma$. Thus, $\mathbf{D}_\gamma$ must have a non-trivial invariant subspace, which is not true for general crystallographic actions.

Even with torsion-freeness taken into account, there is still a large Landscape of RFM's. One can prove \cite{eee57c5f-3af2-3ab0-b3c0-3d30ad3fad43} that
\begin{itemize}
\item Two RFM's with defining groups $\mathcal{B}\,,\,\mathcal{B'}$ are isomorphic if and only if there is an $a\in ISO(k)$ such that $\mathcal{B}'=a\circ\mathcal{B} \circ a^{-1}$ (i.e. one group is the same as the other up to a global rotation+translation), and
\item There are only finitely many isomorphism  classes of RFM's in each dimension $k$. 
\end{itemize}
The second theorem tells us that the Landscape of string compactifications in RFM's is finite, and aligns well with expected finiteness properties of the Quantum Gravity Landscape \cite{Heckman:2019bzm,Hamada:2021yxy,Delgado:2024skw}. The classification program for RFM's has been completed in dimensions up to six \cite{Cid2001}; in dimension 7 and higher, it is hindered by the fact that the classification of crystallographic groups themselves is an open problem \cite{Moeck_2019}. M and F-theory compactifications to four dimensions require seven and eight-dimensional manifolds, respectively, so a classification would be highly beneficial for a detailed exploration of this corner of the string Landscape.

Since the above discussion has been a bit abstract, we will now illustrate it with the simplest example of RFM that is not a torus: the Klein bottle (KB). This is customarily presented as the quotient of the complex plane $\mathcal{C}$ by the group of isometries generated by
\begin{equation}
    z\,\rightarrow\, -\bar{z}+\frac{i}{2},\quad z\,\rightarrow\, z+\beta \,. 
\end{equation}
These two actions define the group $\mathcal{B}$ for the KB. Writing the complex number $z=z_1+iz_2$ as a vector $\vec{z}=(z_1,z_2)$, we can recast the above transformations as\begin{equation} 
    b_1:\, \vec{z}\,\rightarrow \left(\begin{array}{cc}-1&0\\0&1\end{array}\right)\vec{z}+\left(\begin{array}{c}0 \\ 1/2\end{array}\right),\quad b_2:\,  \vec{z}\,\rightarrow\,\vec{z}+\left(\begin{array}{c}\beta\\0\end{array}\right),
\end{equation}
which is more reminiscent of \eq{ewerrr}. The translation group is generated by $b_1^2$ (which equals translation by the vector $(0,1)$) and $b_2$, while the group $\Gamma$ is generated by $b_1$ with entries taken modulo 1, and is isomorphic to $\mathbb{Z}_2$. 

The Klein bottle is the only two-dimensional flat manifold with holonomy $\mathbb{Z}_2$, and is uniquely specified by the data $(\Lambda,\Gamma)$. However, in more general examples, this data (with $\Gamma$ read as an abstract group) does not fully specify an RFM, since the extension problem \eq{bieberles} may have more than one solution. An example in 3d is given by $\Lambda=\mathbb{Z}^3$ and $\Gamma=\mathbb{Z}_2$, generated by either of
\begin{equation} \left(\begin{array}{ccc}1&0&0\\0&-1&0\\0&0&1\end{array}\right)\quad\text{or}\quad \left(\begin{array}{ccc}1&0&1\\0&-1&0\\0&0&1\end{array}\right).\end{equation}
These two related RFM's were used in \cite{Montero:2022vva} to identify a discrete $\theta$ angle in an eight-dimensional compactification of M-theory. In general, different $\mathcal{B}$'s with the same $(\Lambda,\Gamma)$ can differ by the action of the matrices $\mathbf{D}_\gamma$ and vectors $\vec{b}_\gamma$. The possibilities for $\mathbf{D}_\gamma$ can be somewhat constrained, using results from representation theory of integer matrices \cite{laffeylectures} as we now explain. First, it is always possible to perform a change of coordinates in the ambient space $\mathbb{R}^k$ such that the lattice $\Lambda$ becomes exactly $\Lambda=\mathbb{Z}^k$, the standard integer lattice. The price one pays for this is that the ambient metric (and hence, the induced metric on the RFM) are no longer standard, but are replaced by some other flat metric $G$. However, now the matrices $\mathbf{D}_\gamma$ preserve the lattice $\mathbb{Z}^k$, and are hence represented by integer matrices of unit determinant---matrices in $SL(k,\mathbb{Z})$. Representation theory of conjugacy classes of such matrices tells us that they are all conjugate to block-diagonal form,
\begin{equation} \mathbf{D}_\gamma\sim\left(\begin{array}{ccc|ccc}
 \mathbf{B}_1& 0 & 0 & \vdots  & \vdots & \vdots \\
0 & \ddots & 0 & \vec{v}_1 & \cdots & \vec{v}_l \\
0 & 0 & \mathbf{B}_i & \vdots  & \vdots & \vdots\\\hline
  0 & \cdots & 0 & \ddots&\vdots &  \\
    \vdots & 0 & \vdots & \cdots & \mathbf{I}_{l\times l} & \cdots \\
   0 &\cdots&0& &\vdots & \ddots\end{array}\right)
\label{simmy}\end{equation}
where the diagonal blocks $\mathbf{B}_i$ are standard and determined by the order of the matrix \cite{laffeylectures}, and $l$ is the dimension of its invariant subspace; we will see a detailed example of this in Section \ref{sec:dS5-maximum}. The only ambiguity is left in the choice of the upper-right entries, which must be analyzed on a case-by-case basis. 

Since the canonical form \eq{simmy} is achieved by conjugation in $SL(k,\mathbb{Z})$, it is not very useful when $\Gamma$ is non-abelian --- there may not exist a similarity transformation that simultaneously puts  all $\mathbf{D}_\gamma$ in the form \eq{simmy}. However, when $\Gamma$ is abelian, and in particular, when $\Gamma=\mathbb{Z}_n$ has a single generator, \eq{simmy} can be used to produce a complete classification of RFM's (see \cite{eee57c5f-3af2-3ab0-b3c0-3d30ad3fad43,945db8a4-b06b-325c-b91e-9c218130485c}). In the remainder of this paper, we will focus solely on this case---Riemann-flat manifolds of cyclic holonomy---as they are significantly easier to study and provide a nice entry point to the RFM landscape. However, we hope to return to the more interesting case of non-cyclic RFM's in the future. 

In the same way that we have just classified the possible $\mathbf{D}_\gamma$'s, it is also possible to classify the vectors $\vec{b}_\gamma$ associated to them. For a given $\mathbf{D}_\gamma$, the affine change of basis $\vec{z}\,\rightarrow \mathbf{U}\,\vec{z}+\vec{u}$ (with inverse $\mathbf{U}^{-1}\,\vec{z}-\mathbf{U}^{-1}\vec{u}$) acts on the transformation as
\begin{equation} 
    \mathbf{D}_\gamma\,\vec{z}+\vec{b}_\gamma\ \rightarrow\   \mathbf{U}\cdot  \mathbf{D}_\gamma\cdot \mathbf{U}^{-1}\,\vec{z}- \mathbf{U}\cdot  \mathbf{D}_\gamma\cdot \mathbf{U}^{-1}\vec{u} + \mathbf{U}\,\vec{b}_\gamma +\vec{u} \,.
\end{equation}
If we require that $\mathbf{D}_\gamma$ remains invariant, for instance for $\mathbf{U}=\mathbf{D}^q_\gamma$ for some integer $q$, the above becomes a transformation of the vector $\vec{b}_\gamma$,
\begin{equation}
    \vec{b}_\gamma\,\rightarrow\, \mathbf{D}^q_\gamma\, \vec{b}_\gamma + (\mathbf{I}-\mathbf{D}_\gamma)\vec{u}= \mathbf{D}^q_\gamma\,\vec{b}_\gamma+\vec{u}_\perp \,,
\end{equation}
where $\vec{u}_\perp$ is the projection of any integer vector onto the subspace orthogonal to the invariant subspace of $\mathbf{D}_\gamma$. In other words, we are free to rotate the transverse component of $\vec{b}_\gamma$ by any power of $\mathbf{D}_\gamma$ and shift it by the transverse projection of any vector.  The set of equivalence classes of $\vec{b}_\gamma$ vectors is the set of solutions of the equation
\begin{equation}\vec{b}_\gamma= \mathbf{D}_\gamma \, \vec{b}_\gamma\,\quad \text{mod}\quad \Lambda_{\vec{u}_\perp},\,\label{eq:shift-vectors}\end{equation}
where $ \Lambda_{\vec{u}_\perp}$ is the lattice generated by all the $\vec{u}_\perp$, and we take into account that the vectors $\vec{b}_\gamma$ satisfy $p\,\vec{b}_\gamma\in \mathbb{Z}^k$, where $p$ is the order of $\gamma$.  In many cases, there is just one equivalence class, and we can set $\vec{b}_\gamma^\perp=0$, but not always, as we will see in Section \ref{sec:dS5-maximum}. 

Another point of physical interest to us is to determine how many geometric moduli the RFM has. These are by definition deformations of the ambient metric $G$ that preserve the Riemann-flat condition. Even in absence of supersymmetry, they correspond to classically massless scalars, that may be lifted by classical or quantum corrections. The torus $T^k$ has
\begin{equation} \frac{k (k+1)}{2}\end{equation}
geometric moduli, including its overall volume. The additional quotient by $\Gamma$ has the appealing effect of projecting some of those. This is because the invariant metrics $G$ satisfy the condition
\begin{equation}
    \mathbf{D}^T_\gamma \cdot\mathbf{G}\cdot \mathbf{D}_\gamma= \mathbf{G}\,,\quad \forall\, \mathbf{D}_\gamma\,,
    \label{ewi}
\end{equation}
which freeze some of the $T^k$ moduli to specific values. For instance, in the case of the Klein bottle discussed above, the complex structure of the covering $T^2$ is forced to have $\text{Re}(\tau)=0$ or $1/2$. 
To find the dimension of the moduli space and the perturbations explicitly, we simply perturb the metric $\mathbf{G}$ in \eq{ewi} by a small symmetric perturbation $\delta\mathbf{G}$. Invariance leads to a linear system,
\begin{equation} 
    \delta\mathbf{G}\cdot  \mathbf{D}_\gamma= (\mathbf{D}^T_\gamma)^{-1} \cdot\delta\mathbf{G}\,.
    \label{eprot0}
\end{equation}
From \eq{eprot0}, we see that the volume is never frozen by the RFM, and that any invariant vector $\vec{v}$ under $\mathbf{D}_\gamma$ will lead to an invariant deformation $\delta\mathbf{G}\propto \vec{v}\otimes\vec{v}$. Since, as explained around equation \eq{fpfcond},  $\mathbf{D}_\gamma$ always has a non-trivial invariant subspace, an RFM always has at least two geometric moduli \cite{bettiol2018teichmuller}\footnote{There are (non-cyclic) examples where this lower bound is saturated, as will be described in \cite{wipber}.}. All the frozen $T^k$ moduli are absent at very low energies, but since their higher KK modes are not projected out, in some sense they should be regarded as having attained masses of order the Kaluza-Klein scale. This will be relevant in Section \ref{sec:dS5-maximum}, where we study the backreaction of an RFM compactification. 

Determining the full KK spectrum for fields of any spin in an RFM is a relatively easy (if cumbersome) task, since fields on an RFM $\mathcal{F}_k$ can be described equivariantly as those field modes on the covering $T^k$ invariant under the $\Gamma$ action. Thus, it suffices to determine the KK spectrum of the $T^k$, and project down to the subspace invariant under $\Gamma$. For the particular case of zero modes of $p$-form fields, this can be done efficiently by algebraic methods. Interestingly, there are examples of non-isomorphic RFM's with identical KK spectrum for scalar fields \cite{10.1215/S0012-7094-92-06820-7} and $p$-forms \cite{miatello2003flatmanifoldsisospectralpforms}, which provide a negative answer to the famous question of whether one can ``hear the shape of a drum'' in higher dimensions.

The final topic we need to review in this section about generalities of RFM's is their spin structures. We will be considering compactifications of supergravity theories on RFM's, and therefore, we must choose boundary conditions for fermionic as well as bosonic fields. Along a non-trivial 1-cycle, fermions can be either periodic or antiperiodic, and such a choice is called a spin structure. Not every manifold admits a spin structure and, interestingly, there are examples of non-spin RFM's in dimensions 4 and above \cite{dekimpe2006,PutryczSzczepanski+2010+323+332}, so we must be careful to choose RFM's that do. While the classification of spin structures for general RFM's is a convoluted topic \cite{lutowski2015spin}, it is possible to provide a concrete characterization for manifolds of cyclic holonomy \cite{Hiss16012008}. In this paper, we will focus on a subclass of RFM's with cyclic holonomy where 
\begin{equation} \mathcal{F}_k=T^{l-1} \times \left(\frac{T^{k-l+1}\times S^1}{\mathbb{Z}_n}\right)\label{circw}\end{equation}
and the group $\mathcal{B}$ is generated by a single element $\D[g]\,\vec{z}+\bvec[g]$, where the last circle factor in \eq{circw} is in the invariant subspace of $\D[g]$. Such manifolds are called mapping tori, and they are featured prominently in topology and the study of global anomalies \cite{bams/1183554722,Witten:1985xe}. A mapping torus can be regarded as a cylinder $T^{k-l+1}\times[0,1]$ where the two ends of the cylinder are identified with a certain diffeomorphism of the $T^{k-l+1}$ fiber. A mapping torus admits a spin structure if and only if there are spinors in the torus cover invariant under the quotienting action or, equivalently, if the gluing diffeomorphism preserves the spin structure of the $T^{k-l+1}$ fiber. 

On the torus $T^{k}$, a spin structure can be specified by the periodicity of spinor fields $\psi(\vec{z})$ under lattice transformations,
\begin{equation} 
    \psi(\vec{z}+\vec{n})=e^{2\pi i\,\vec{h}\cdot\vec{n}}\, \psi(\vec{z}) \,.
    \label{mim}
\end{equation}
The diffeomorphism $\D[g]$ acts on spinor fields of $T^{k-l+1}$ as
\begin{equation}
    \psi(\vec{z})\,\rightarrow \mathcal{D}_{\bf g}\psi(\D[g]\,\vec{z}\,),\end{equation}
where $\mathcal{D}_{\bf g}$ is one of the two spin lifts of $\D[g]$ (related by a sign). Using \eq{mim}, we get that 
\begin{equation}
    \mathcal{D}_{\bf g}\psi(\D[g]\,(\vec{z}+\vec{n}))= e^{2\pi i\,\vec{h}\cdot\D[g]\vec{n }} \,\mathcal{D}_{\bf g}\psi(\D[g]\,\vec{z}) \,,
\end{equation}
so to preserve the spin structure, the vector $\vec{h}$ must satisfy the consistency condition\footnote{This can also be derived directly from the condition \eq{mecon} that one-dimensional projective representations of any group (such as spinors) represent elements in the same conjugacy class by the same phase. Consider a one-dimensional representation of $\mathcal{B}$ where translations by $\vec{n}$ are represented as $e^{2\pi i\,\vec{h}\cdot\vec{n}}$. The conjugation $b_\gamma\, T_{\vec{k}}\,b_\gamma^{-1}$, where $T_{\vec{k}}$ is translation by $\vec{k}$ corresponds to a pure translation, since
\begin{equation*}
    \D[g](\D[g]^{-1}\vec{z}- \D[g]^{-1}\bvec[g]+\vec{k})+\bvec[g] = \vec{z}+\D[g]\cdot\vec{k} \,.
\end{equation*}
Demanding that the phase representing the above is independent of $b_\gamma$ amounts to  \eq{mimi}. }
\begin{equation} 
    (\mathbf{D}^T_g-\mathbf{I})\cdot 
        \vec{h}\in\mathbb{Z}^{k-l+1} \,.
    \label{mimi}
\end{equation}
On top of this, there is one more condition, related to the component $\vec{h}_{S^1}$ of the vector $\vec{h}$ along the base. Suppose $\D[g]$ is of order $p$. Then, acting with the generator $p$ times yields a translation along the vector $p\,\vec{b}_{\bf g}$, which therefore must be in the lattice $\mathbb{Z}^k$. On the other hand, in the spin representation the chosen spin lift satisfies $\mathcal{D}^p_{\bf g}=(-1)^{s_{\bf g}} \mathbf{I}$, i.e. it is the identity up to a sign. Therefore, the choice of spin structure on the covering $T^k$ along the direction $p\, \vec{b}_{\bf g}$ is correlated with the choice of spin lift, and we must have
\begin{equation} 
    s_\mathbf{g}\equiv 2\,p\, \vec{h}\cdot \vec{b}_{\bf g}\quad \text{mod}\,\,2\mathbb{Z} \,. 
    \label{swer2}
\end{equation}
Note that for even order groups $\Gamma$, we have $s_\mathbf{g} = 0$ regardless of the choice of spin lift, constraining the vector $\vec{h}$ directly. 
Since due to \eq{fpfcond} the vector $\bvec[g]$ necessarily has a component along the direction of $\vec{h}_{S^1}$, equation \eq{swer2} correlates the spin structures on the base and fiber of the covering torus. The combined system of equations given by \eq{mimi} and \eq{swer2} always has a solution, which correlates with the fact that all cyclic, mapping tori RFM's admit a spin structure \cite{Hiss16012008}. Finally, the conditions \eq{mimi} and \eq{swer2} can also be derived by demanding that the Casimir energy of fermionic fields as will be computed in Section \ref{sec:casimir-RFMs} is a well-defined function on the RFM. 

The techniques reviewed in this section allow one to determine the classical low-energy EFT that arises from dimensional reduction on any RFM. We are now ready to tackle the main point of this paper---the calculation of quantum effects (Casimir energies) in an RFM.

\section{Casimir energy in RFM's}
\label{sec:casimir-RFMs}

We now explain the central theme of this paper, which is the derivation of an explicit formula to compute Casimir energies in RFM's. The general features of Casimir energies were already discussed in Section \ref{sec:warm-up-dS7}. Here we will zoom in on the simpler case of free fields and the method of images to compute the Casimir energy for free fields in quotient spaces; finally, we will particularize this general result to the case of RFM's. There are many references for the methods we will use here, including \cite{Appelquist:1983vs,Birrell:1982ix,Arkani-Hamed:2007ryu}.
 
\subsection{Casimir energy of massless free fields} \label{sec:casimir-massless-fields}
As discussed in Section \ref{sec:warm-up-dS7}, absent any symmetries, there is no general expression for the Casimir stress-energy tensor $\langle T_{\mu\nu}\rangle_{\mathcal{M}}$. However, for the particular case of free fields, a concrete answer may be obtained by the method of point-splitting regularization \cite{Birrell:1982ix}. Free fields are directly relevant to compactifications of ten-dimensional strings or M-theory, since the 10 or 11d low-energy supergravities are all free at energies much below the string or Planck scales.  For this reason, although a generalization to massive fields is not complicated (see \cite{Birrell:1982ix,Arkani-Hamed:2007ryu}), we will focus on Casimir energies of massless fields only. The contributions of these fields always dominate the Casimir energy in the limit of very large compactification manifolds. In this subsection and the following we provide explicit expressions for $\langle  T_{\mu\nu}\rangle$ and $\hat{\rho}$, in the case of a massless, free field $\phi$ of arbitrary spin $s\leq2$. Since the discussion at this level will be formal, we omit Lorentz and internal symmetry indices; $\phi$ could represent a scalar, spinor, Rarita-Schwinger field, or vector. All such free fields are described by quadratic actions, of the schematic form  
\begin{equation} 
    S = \int \sqrt{-g}\,\mathcal{L}
    = \frac12 \int \sqrt{-g}\, \phi\cdot \mathcal{D}\cdot \phi
    = 0 \,,
    \label{eomff2}
\end{equation}
leading to an equation of motion 
\begin{equation} 
    \mathcal{D}\phi=0 \,. 
\end{equation}
Here, $\mathcal{D}$ is an appropriate linear differential operator of order at most two, which also couples to the background metric and gauge fields. Standard examples are the covariant derivative for vector or scalar fields, the Dirac or Rarita-Schwinger operators for fermions, and the Fierz-Pauli Lagrangian for spin two fields. From \eq{eomff2}, the stress-energy tensor can be computed, as 
\begin{equation}
    T_{\mu\nu}\equiv \frac{-2}{\sqrt{-g}}\frac{\delta S}{\delta g^{\mu\nu}}= \phi \cdot\left[ \frac{\delta\mathcal{D}}{\delta g^{\mu\nu}} - \frac12 g_{\mu\nu}\mathcal{D}\right]\phi \,.
    \label{setgen}
\end{equation}
The crucial feature of \eq{setgen} is that, for a free theory, $T_{\mu\nu}$ is quadratic in the field $\phi$ and its derivatives and, therefore, its expectation value is directly related to the differential operator in \eq{setgen} acting on the two-point function. As usual in QFT, the na\"ive evaluation of $\langle T_{\mu\nu}\rangle$ in terms of $\langle \phi(x)\phi(x)\rangle$ produces divergences due to the contact singularity on the two-point function. The idea of the point-splitting method is to regularize this by separating the two points in the propagator slightly, and then taking the limit as the separation vanishes. This leads to a regularized expression for the stress-energy tensor at any point $x$ in the manifold, as 
\begin{equation} 
    \langle T_{\mu\nu}(x)\rangle= \lim_{y\rightarrow x}  \left[ \frac{\delta\mathcal{D}}{\delta g^{\mu\nu}} - \frac12 g_{\mu\nu}\mathcal{D}\right] \langle \phi(x)\,\phi(y)\rangle 
    \label{pspl}
\end{equation}
where the derivatives in $\mathcal{D}$ are taken with respect to either $x$, $y$, or a combination of both---these only correspond to different choices of regularization scheme. The regularization \eq{pspl} works for any free field, of any spin. As an example, for a massless, free scalar field, with stress-energy tensor 
\begin{equation} 
    T^{\text{sc}}_{\mu\nu} = \frac12\left(2\nabla_\mu \phi\,\nabla_\nu \phi - g_{\mu\nu}\nabla^\rho\phi\,\nabla_\rho\phi\right),
\end{equation}
the point-splitting formula \eq{pspl} becomes
\begin{equation} 
    \langle T^{\text{sc}}_{\mu\nu}\rangle\,\rightarrow\, \frac12\lim_{x\rightarrow y}\left(\nabla_\mu \nabla'_\nu+\nabla'_\mu\nabla_\nu  - g_{\mu\nu}\nabla^\rho\nabla'_\rho\right)G(x,y) \,,
    \label{f00}
\end{equation}
in terms of the scalar Green's function $G(x,y)$. As we will see in the next subsection, UV divergences will cancel automatically in certain spaces when using this regularization, including the ones of interest. To compute the components of the stress-energy tensor along the internal dimensions, one needs to evaluate \eq{pspl} for all massless fields of any spin in the theory under consideration. However, if one is only interested in the components along the non-compact directions---in other words, the value of the Casimir energy $\hat{\rho}$---as we are in this paper, then a small trick allows us to treat all fields of any spin simultaneously, using only \eq{f00}.
To do this, we will momentarily compactify all spatial directions on a torus of side $L$ and compute $T_{00}$. When one does this, the Lorentz symmetry of the theory is broken completely, there is no spin since there is no Lorentz group in zero dimensions, and the operator $\mathcal{D}_{ab}$ can be diagonalized into a basis of internal polarizations. From this point of view, all that remains are scalar fields. The (0+1)-dimensional Lagrangian therefore becomes a sum of free fields, with  $g_\phi$ (the number of physical degrees of freedom of the field $\phi$) copies of a free scalar for bosonic fields in higher dimensions, or a sum of free fermions for fermionic fields. Since fermion and boson contributions only differ by a sign, the Casimir energy can then be computed as 
\begin{equation} 
    \hat{\rho}=-\langle T_{00}\rangle= -\sum_{\text{Polarizations}}(-1)^F\langle T^{\text{sc}}_{00}\rangle \,,
    \label{eqw3}
\end{equation}
where $\langle T^{\text{sc}}_{\mu\nu}\rangle$ is the expectation value of the stress-energy for a scalar field, and the contribution is positive for bosonic fields while negative for fermionic fields. Finally, nothing in \eq{eqw3} depends on $L$ (save for a trivial volume factor), so we can take $L\rightarrow\infty$ and still use the formula \eq{eqw3}. 

The Casimir energy of arbitrary spin fields is then computed using \eq{f00}; as we will see in the next subsection, UV divergences will cancel automatically in certain spaces when using this regularization, including the ones of interest, yielding a completely finite result.  There is just one important caveat: the formula \eq{f00} depends on a scalar propagator $G(x,y)$, but in general this is not unique. Although we have managed to reduce the calculation of Casimir energy of an arbitrary spin field to a sum of scalar fields corresponding to the different polarizations, each polarization may transform differently under the Lorentz symmetry of the ten-dimensional theory, and therefore involve a different scalar field propagator (i.e. the propagator of a scalar field with twisted boundary conditions along the cycles of the covering torus). The source of the ambiguity lies in the fact that the covariant derivative $\nabla$ seen by each polarization may be coupled to an additional background connection, for an internal or spacetime symmetry, and thus different polarizations are affected differently. We will see an example of this in the next subsection. 

\subsection{Free fields in a quotient space: the method of images}
\label{sec:method-of-images}

Armed with a way to regularize and compute $\hat{\rho}$, we now focus our attention on the computation where $\mathcal{N}_k$ is of the form
\begin{equation}\mathcal{N}_k=\frac{\tilde{\mathcal{N}}_k}{\mathcal{B}},\label{manfin}\end{equation}
and where $\mathcal{B}$ is a group of isometries of the manifold $\tilde{\mathcal{N}}_k$\footnote{This can be generalized to the case where $\mathcal{B}$ includes internal symmetries by including the transformation law for each charged field, but we will not study these cases in this paper.}, which we assume to act freely (so that there are no fixed points and $\mathcal{N}_k$ is indeed a manifold). The particular case of the Riemann-flat manifolds studied in Section \ref{sec:RFMs}  has $\tilde{\mathcal{N}}_k=\mathbb{R}^k$, and will be the subject of Subsection \ref{sec:Casimir-on-RFMs}; here, we explain how to compute the propagator and hence free energy for a general manifold of the form \eq{manfin}.  There is a technique to compute the propagator of a scalar field on $\mathcal{N}_k$, given one on $\tilde{\mathcal{N}}_k$ which is $\mathcal{B}$-symmetric, known as the method of images \cite{feynman_lectures_vol2}. Since the underlying equation of motion is linear, one can construct a single-valued propagator on  $\mathcal{N}_k$ by starting with a propagator $G_{\tilde{\mathcal{N}}_k}(x,y)$ on the covering space and  summing over the images of any point under the group $\mathcal{B}$,
\begin{equation} 
    G_{\mathcal{N}_k}(x,y)=\sum_{b\in\mathcal{B}} G_{\tilde{\mathcal{N}}_k}(x,b(y)) \,.
    \label{psplits}
\end{equation}
The resulting expression is manifestly single-valued as a function of $y$, and also as a function of $x$, since 
\begin{equation} G_{\mathcal{N}_k}(b(x),y)=\sum_{b\in\mathcal{B}} G_{\tilde{\mathcal{N}}_k}(b(x),b(y))=\sum_{b\in\mathcal{B}} G_{\tilde{\mathcal{N}}_k}(x,b^{-1}\circ b(y)),\label{inveq}\end{equation}
where we have used that the parent propagator $G_{\tilde{\mathcal{N}}_k}(x,y)$ respects the isometry group $\mathcal{B}$.

The method of images combines very nicely with the point-splitting regularization of the previous subsection, since the $y\rightarrow x$ divergence of the two-point function in the sum \eq{psplits} comes entirely from the contribution where  $b$ is the identity in the sum over $\mathcal{B}$. This contribution is, in turn, identical to the original propagator on the covering space $\tilde{\mathcal{N}}_k$, and does not see the $\mathcal{B}$ quotient at all. As such, this term of the sum in \eq{inveq} (or in the corresponding sum in \eq{f00}) inherits any properties that the propagator or Casimir energy might have on $\tilde{\mathcal{N}}_k$. For instance, if the physical theory is supersymmetric, and the background $\tilde{\mathcal{N}}_k$ preserves supersymmetries, such that there is a Bose-Fermi cancellation on $\tilde{\mathcal{N}}_k$, such cancellation will also happen in the identity term of \eq{inveq}. But since all divergencies in \eq{f00} come from this term alone, we reach the conclusion that \emph{the Casimir energy {\rm\eq{eqw3}} of a supersymmetric theory on $\mathcal{N}_k$ will be  manifestly UV-finite} whenever $\tilde{\mathcal{N}}_k$ preserves supersymmetry, even if no supercharges are preserved by $\mathcal{N}_k$. One could perhaps say that the supersymmetries of $\tilde{\mathcal{N}}_k$ live on $\mathcal{N}_k$ as some form of ``non-invertible'' supersymmetry \cite{Heckman:2024obe}, an idea that deserves further exploration.

We also note in passing that, even on a general $\mathcal{N}_k$ not of the form \eq{manfin}, some universal divergences are expected to cancel in the Casimir energy formula \eq{eqw3}---for instance, those that correspond to a higher-dimensional cosmological constant (which in the point-splitting regularization correspond to the ultralocal limit where one replaces the compactification manifold by $\mathbb{R}^D$)---but in general, we do expect to have divergences in the one-loop Casimir energy, associated to unprotected counterterms of the higher-dimensional theory that might contribute to the lower-dimensional vacuum energy. See \cite{Burgess:2023pnk} for a more detailed exposition. The protection due to the quotient, however, is much stronger, in particular for the case of RFM's, where $\tilde{\mathcal{N}}_k$ is maximally supersymmetric flat space.

At any rate, once we remove the divergent identity contribution, the propagator is finite,
\begin{equation} 
    G^{\text{ren.}}_{\mathcal{N}_k}(x,y)=\sum_{b\neq{\mathbf{I}}\in\mathcal{B}} G_{\tilde{\mathcal{N}}_k}(x,b(y))\,,
    \label{psplitsren}
\end{equation}
corresponding to a finite expression for the vacuum energy via \eq{eqw3}. 

The regularized propagator \eq{psplitsren} is not the only one that can be constructed via the method of images. For instance, if $\phi$ is a complex scalar field (as the polarizations of a higher-dimensional field would be), any expression of the form
\begin{equation} 
    G^{\text{ren.}}_{\mathcal{N}_k,\varphi}(x,y)=\sum_{b\neq{\text{Id.}}\in\mathcal{B}} e^{i\varphi(b)}G_{\tilde{\mathcal{N}}_k}(x,b(y)) \,,
    \label{psplitsren2}
\end{equation}
where $e^{i\varphi(b)}$ is a one-dimensional representation of $\mathcal{B}$, is a valid propagator on $\mathcal{N}_k$.\footnote{For $\phi$ real, this phase is at most a sign.} However, the resulting expression $G^{\text{ren.}}_{\mathcal{N}_k,\varphi}(x,y)$ is in general not a well-defined function on $\mathcal{N}_k$, as the phases spoil the proof of invariance in equation \eq{inveq}. Rather, it becomes the appropriate expression for a propagator of a field  $\phi$  which is a section of a non-trivial  line bundle over $\mathcal{N}_k$. When studying spinning fields as described in the previous subsection, we could decompose them into different polarizations, each  described by a complex scalar field which will transform in non-trivial line bundles over $\mathcal{N}_k$; the associated phases will be an example of the phenomenon, discussed in Subsection \ref{sec:casimir-massless-fields}, that different polarizations may involve different propagators.

Even though \eq{psplitsren2} is not a single-valued function on $\mathcal{N}_k$, the Casimir energy that we derive from it should be. As we will see below, the Casimir energy is obtained after taking two derivatives of $G^{\text{ren.}}_{\mathcal{N}_k,\varphi}(x,x)$, so this function should be single-valued as a function of $x$. For any $b_0\in\mathcal{B}$,
\begin{align} 
    G^{\text{ren.}}_{\mathcal{N}_k,\varphi}(b_0(x),b_0(x)) &= \sum_{b\neq{\text{Id.}}\in\mathcal{B}} e^{i\varphi(b)}G_{\tilde{\mathcal{N}}_k}(b_0(x),b\circ b_0(y)) \nonumber \\
    &= \sum_{b\neq{\text{Id.}}\in\mathcal{B}} e^{i\varphi(b)}G_{\tilde{\mathcal{N}}_k}(x,b_0^{-1}\circ b\circ b_0(y)) \,,
    \label{psplitsren3ymedio}
\end{align}
so the Casimir energy will be single-valued if 
\begin{equation} 
    e^{i\varphi(b_0^{-1}\circ\, b\,\circ\, b_0(y))}= e^{i\varphi(b)}\,,\quad\forall b_0\,,b\,.
\label{mecon}\end{equation}
This condition, which is automatically true for any one-dimensional representation of $\mathcal{B}$, tells us more generally that the most general shift in the propagator must be given by a one-dimensional function of conjugacy classes -- a character-- of $\mathcal{B}$. We will see momentarily that this is indeed the case, even when the representation under $\mathcal{B}$ is not one-dimensional.

Another, equivalent, point of view comes from generalizing \eq{psplitsren} to fields transforming in higher-dimensional representations of  $\mathcal{B}$. Since $\mathcal{B}$ is a subgroup of isometries, the transformation properties of a field $\phi^a$ are induced by its $d$-dimensional transformation properties under the Lorentz group. This generalizes \eq{psplitsren2} to
\begin{equation} 
    G^{\text{ren.}\, ab}_{\mathcal{N}_k,}(x,y)=\sum_{b\neq{\mathbf{I}}\in\mathcal{B}} R_{c}^a(b)\, G^{cb}_{\tilde{\mathcal{N}}_k}(x,b(y))\,,
    \label{psplitsren3}
\end{equation}
where $R_{c}^a(b)$ is a higher-dimensional representation of $\mathcal{B}$. When using the formula for the Casimir energy \eq{eqw3}, the sums over polarizations should be replaced by traces over the matrices $R_{c}^a(b)$---in both cases, the resulting expressions are equivalent, and we learn that we need to include traces of the matrices $R_{c}^{a}$ when inserting \eq{psplitsren3} into \eq{eqw3}.

Finally, we will be interested in some cases where the group $\mathcal{B}$ acts on the fields only projectively. In this paper, where all concrete examples will involve purely geometric actions, this will only be the case for fermionic fields, and the projectivity corresponds to the familiar fact that spinors represent a $2\pi$ rotation as multiplication by $-1$. In cases like this, all the results of this paper are still valid, and we can use an expression such as \eq{psplitsren3}, but with the group $\mathcal{B}$ replaced by a central extension $\hat{\mathcal{B}}$ (which is represented linearly), fitting into a short exact sequence
\begin{equation}1\,\longrightarrow \mathcal{Z}\,\longrightarrow\, \hat{\mathcal{B}}\,\longrightarrow\, \mathcal{B}\,\longrightarrow 1\,.\end{equation}
The group $\mathcal{Z}$ in the extension will be $\mathbb{Z}_2$ in the examples considered in this paper (corresponding to the fact that Spin is a $\mathbb{Z}_2$ extension of SO), but in general it can be more convoluted\footnote{For instance, for type IIB compactifications, it could involve subgroups of the $SL(2,\mathbb{Z})$ duality group \cite{Tachikawa:2018njr}.}.

\subsection{Casimir energy formula for RFM's}
\label{sec:Casimir-on-RFMs}

We are now ready to explain how to compute the Casimir energy for RFM's, which will be the main technical result of this paper. More concretely, we wish to compute the quantity $\hat{\rho}$ and its integral for the RFM's described in Section \ref{sec:RFMs}, using the method of images outlined in the previous subsection. We denote a point in $\mathbb{R}^k$ (the universal cover of the $k$-dimensional RFM $\mathcal{F}_k$) as a $k$-dimensional vector $\vec{z}$. Points in $\mathcal{F}_k$ will then be equivalence classes of these under $\mathcal{B}$. Consider a field $\phi$, which transforms under $\mathcal{B}$ as
\begin{equation} 
    \phi^a(\vec{z})\,\rightarrow\, R^a_c(b)\,  \phi^c(b(\vec{z}))
    = R^a_c(b)\, \phi^c(\mathbf{D}_\gamma\, \vec{z} + \vec{b}_\gamma+\vec{n})\,,
    \label{cas-2-0}
\end{equation}
where we have used the decomposition \eq{ewerrr} and the fact that isometries of $\mathbb{R}^k$ act linearly. Using \eq{psplitsren3} together with \eq{f00}, and plugging it into \eq{eqw3}, one obtains a sum over images of derivatives of flat-space propagators, \begin{equation} 
    \hat{\rho}(\vec{z})= -\frac{\Gamma\left(\frac{D}{2}\right)}{2\pi^{D/2}}\sum^{\sim}_{\substack{\gamma\in\Gamma\\\vec{n}\in\mathbb{Z}^k}} \frac{\text{Tr}(\mathbf{R})}{\vert\vec{z}- (\mathbf{D}_\gamma\,  \vec{z} + \vec{b}_\gamma+\vec{n})\vert^D},
    \label{precas}
\end{equation}
where we have replaced the sum over $\mathcal{B}$ by a sum over the lattice $\Lambda=\mathbb{Z}^k$ and the group $\Gamma$ defined in Section \ref{sec:RFMs}. The tilde over the sum means that the identity in $\mathcal{B}$ is excluded from the sum (although, for a supersymmetric theory, we can put it back in and it will not make a difference, as explained in Subsection \ref{sec:method-of-images}). The trace $\text{Tr}(\mathbf{R})$ replaces the sum over polarizations in \eq{eqw3}; it is the same sum, written in a basis-independent way as explained near the end of Subsection \ref{sec:method-of-images}. An important point is that since Casimir energy is a sum over physical degrees of freedom, the matrices $\mathbf{R}=R^a_c$ are in Lorentzian signature---they represent elements of the group $SO(D-1,1)$. This automatically takes into account a proper counting of physical polarizations for any field, since under the decomposition
\begin{equation} SO(D-1,1)\,\rightarrow\, SO(D-k-1,1)\times SO(k)\end{equation}
the $D$-dimensional field $\phi^a$ decomposes  into representations,
\begin{equation} 
    \mathbf{r}_{SO(D-1,1)}\,\rightarrow\, \sum_i \mathbf{r}^i_{SO(D-k-1,1)}\otimes \tilde{\mathbf{r}}^i_{SO(k)} \,,
\end{equation}
and since $\mathcal{B}$ only ever involves an action in a subgroup of the $SO(k)$ factor, the trace can be grouped in terms of the corresponding $(D-k)$--dimensional irreps, as
\begin{equation} 
    \text{Tr}_{ \mathbf{r}_{SO(D-1,1)}}=\sum_i  g_{\mathbf{r}^i_{SO(D-k-1,1)}}\text{Tr}_{\tilde{\mathbf{r}}^i_{SO(k)}}
\end{equation}
where again $g_{\mathbf{r}}$ is the number of physical polarizations of a $(D-k)$--dimensional massless field in representation $\mathbf{r}$ of the lower-dimensional Lorentz group (for instance, a four-dimensional massless vector has only two polarizations and so on). Alternatively, since there are only massless fields in higher dimensions, we can work with the massless little group of the $D$-dimensional theory, which is $SO(D-2)$. Since we will focus on M-theory compactifications, for the concrete examples in Sections \ref{sec:dS5-maximum} and \ref{sec:dS4-maxima} we will have to evaluate these traces for the 11-dimensional supergravity fields---graviton, gravitino, and 3-form. Details on how to do this can be found in Appendix \ref{ap:traces}.

 Equation \eq{cas-2-0} admits further simplification. The matrix $\mathbf{R}$ representing a pure translation must be proportional to the identity, since translations are a central element in $\mathcal{B}$ and we are allowing for a projective representation. Bloch's theorem \cite{ashcroft2011solid} tells us that the most general such representation is of the form 
\begin{equation} 
    T_{\vec{n}}\,\rightarrow\, e^{2\pi i\,\vec{h}\cdot\vec{n}} \cdot \mathbf{I}\,,
    \label{projrep}
\end{equation}
where $T_{\vec{n}}$ is a translation by integer vector $\vec{n}$ and $\vec{h}$ is a vector defining the representation. As mentioned at the end of Subsection \ref{sec:method-of-images}, in this paper projective representations will only appear for fermions, via the spin lift of $SO$ elements in $\mathcal{B}$. As a result, the phase must be a sign, and the vectors $\vec{h}$ must have all integer or half-integer coordinates; in fact, $\vec{h}$ is the vector that quantifies spin structures in the RFM (or its covering $T^k$) introduced in \secref{sec:RFMs}.

Using this, we can finally recast \eq{precas} as 
\begin{equation} 
    \hat{\rho}(\vec{z})= -\frac{\Gamma\left(\frac{D}{2}\right)}{2\pi^{D/2}}\sum^{\sim}_{\substack{\gamma\in\Gamma\\\vec{n}\in\mathbb{Z}^k}} \frac{\Tr{\bf r}{\D}\, e^{2\pi i\,\vec{h}\cdot\vec{n}} }{\vert\vec{z}- (\mathbf{D}_\gamma\,  \vec{z} + \vec{b}_\gamma+\vec{n})\vert^D}\,.
    \label{precas2}
\end{equation}
This is a concrete, UV-finite expression for $\hat{\rho}(\vec{z})$ for a given free field in a representation $\textbf{r}$. It takes the form of an explicit lattice sum over $\mathbb{Z}^k$, together with a sum over the finite group $\Gamma$. The phase factors $e^{2\pi i\,\vec{h}\cdot\vec{n}}$ account for the presence of non-trivial spin structures for fermionic fields (or, more generally, for non-trivial bundles for discrete symmetries on $T^k$). Finally, there is an overall $(-)$ sign if the field is fermionic. As advertised in the previous subsection, the formula is finite, and UV-finite for a supersymmetric theory. In fact, at one loop, this  is a property of the spectrum only, so any theory for which the number of bosonic and fermionic degrees of freedom match in $D$ dimensions would yield a finite one-loop Casimir energy when compactified on Riemann-flat manifolds, even if it was non-supersymmetric. 

Although we have derived an explicit formula for $\hat{\rho}$, we are more interested in its integral over the compact manifold $\mathcal{F}_k$; only after integration will we obtain the effective potential in $(D-k)$ dimensions that can be used to search for dS maxima, as in Section \ref{sec:dS5-maximum}. This will be the subject of the next subsection, but we finish this one by noticing that the explicit expression for $\hat{\rho}$, as well as the implicit expressions for all components of the stress-energy tensor that we obtained in Subsection \ref{sec:casimir-massless-fields}, provide us with a fully explicit, $D$-dimensional result for the Casimir stress-energy tensor, which can be used as a starting point in any detailed study of backreaction and control issues in these solutions. That such an explicit $D$-dimensional backreaction is available means that including higher-order corrections in this analysis is a matter of calculation, and no ambiguities (e.g. due to subtleties in field profiles around localized sources, as in \cite{Saracco:2012wc,Junghans:2020acz,Marchesano:2020qvg}) can arise. We hope to return to this exciting but probably grueling question in the future.

\subsection{Explicit formula for RFM Casimir energy \& ``Casimir branes''}\label{casformula}

The lower-dimensional Casimir energy (the effective potential coming from the Casimir term) is given by an integral of $\hat{\rho}(\vec{z})$ in \eq{precas2} over the compact space $\mathcal{F}_k$. In other words, we have
\begin{equation} 
    \Vcas = \int_{\mathcal{F}_k} d^k\vec{z}\, \sqrt{G}\,  \hat{\rho}(\vec{z}) 
    = -\frac{\Gamma(s)}{2\pi^{s}} \frac{1}{\vert\Gamma\vert}\sum^{\sim}_{\substack{\gamma\in\Gamma\\\vec{n}\in\mathbb{Z}^k}}  \int_{[0,1]^k} d^k\vec{z}\,\sqrt{G}\,  \frac{\Tr{\bf r}{\D}\, e^{2\pi i\,\vec{h}\cdot\vec{n}} }{\vert\vec{z}- (\mathbf{D}_\gamma\,  \vec{z} + \vec{b}_\gamma+\vec{n})\vert^D} \,.
    \label{vcasfed}
\end{equation}
Here, we have replaced the integral over the manifold $\mathcal{F}_k$ by an integral over the covering torus $T^k=\mathbb{R}^k/\mathbb{Z}^k$ (so that the fundamental domain over which we integrate is $[0,1)^k$), equipped with a general metric $G_{ij}$. The factor of $1/\vert \Gamma\vert$, where $\vert \Gamma\vert$ is the order of the finite group $\Gamma$ of Section \ref{sec:RFMs}, accounts for the fact that the torus is covering $\mathcal{F}_k$ a total of $\vert \Gamma\vert$ times. The sum is taken over all $\gamma\in\Gamma$ and $\vec{n}\in\Z^k$ such that the denominator does not vanish, as indicated by the $\sim$ over it. In other words, the sum excludes the term for which $|\Vec{z}-\Vec{f}_\gamma(\Vec{z})+\Vec{n}|^2 = 0$. Due to the fixed-point-freeness condition of the group $\mathcal{B}$, such a divergence only happens for $\gamma=\mathbf{I}$, $\vec{n}=0$.  The vector $\vec{h}$ encodes twisted boundary conditions of various fields, that transform as a phase under translations, as described in the previous subsection\footnote{Whenever the vector $\vec{h}\neq\vec{0}$, there is a covering torus $\tilde{T}^k\rightarrow T^k$ on which the bundle specified by the vector $\vec{h}$ is trivial, and hence we can take $\vec{h}=0$ on the covering torus. In other words, by passing to a central extension of the group $\mathcal{B}$, we can always take $\vec{h}=0$. As a concrete example of this, in the case of $S^1$ with antiperiodic boundary conditions, we have $h=1/2$ on the $S^1$ identified as $x\sim x+L$; but the boundary condition is periodic under the transformation $x\sim x+2L$, which defines a double cover circle.}. 

The core problem in determining Casimir energies in Riemann-flat manifolds is then the evaluation of sum-integrals of the form
\begin{equation} 
    \mathcal{E}(\gamma)\equiv -\frac{\Gamma(s)}{2\pi^{s}}\frac{1}{\vert\Gamma\vert} \sum_{\vec{n}\in\Z^k}^\sim\int_{[0,1]^k}dV\, \frac{e^{2\pi i \vec{h} \cdot\vec{n}}}{\vert (\mathbf{I}-\mathbf{D}_\gamma)\vec{z}+\vec{b}_\gamma+\vec{n}\vert^{2s}} \,,
    \label{csum}
\end{equation}
where the lattice is taken to be $\mathbb{Z}^k$ with inner product given by $G_{ij} n^i n^j$, we have introduced $s\equiv D/2$, and the induced volume element is
\begin{equation} 
    dV = \sqrt{G}\, dz^1\wedge\ldots \wedge dz^k \,.
\end{equation}
We can then rewrite \eq{vcasfed} as
\begin{equation} 
    \Vcas = \sum_{\textbf{r}}\sum_{\gamma\in\Gamma} \Tr{\bf r}{\D}\,\mathcal{E}(\gamma) \,.
\end{equation}
Each $\mathcal{E}(\gamma)$ therefore corresponds to the contribution of a given ``twisted sector''\footnote{Not to be confused with the usual worldsheet twisted sectors that arise in perturbative string orbifolds. An RFM is a free orbifold and therefore all worldsheet twisted sectors are very massive (consisting entirely of winding states) in any perturbative string construction involving them.}, where the propagators are shifted from those in flat space. The case of flat space, $\gamma=\mathbf{I}$, corresponds exactly to the Casimir energy of the covering torus.  Notice also that this is the only term where there is a divergence, for $\vec{n}=0$; in other words, for $\gamma\neq\mathbf{I}$, the sum over $\vec{n}$ in \eq{csum} has no restrictions.

The basic technical challenge that we solve in this paper is to evaluate the sum-integral \eq{csum} in full generality. In this subsection, we will explain how to perform the integral in \eq{csum}, producing an explicit result in terms of an infinite sum, and discuss the physical interpretation of the result. Subsection \ref{sec:numerical-sums} then outlines an efficient way to evaluate this sum numerically. 

Since evaluating the integral in \eq{csum} is a cornerstone for the results of this paper, and we need to be sure of its validity, we have done it in three independent ways. We now present the most straightforward derivation, based on Mellin transform techniques, and relegate the other two cross-checks (one from an explicit evaluation of the integral and the other from the expressions developed in Section \ref{sec:numerical-sums}) to Appendix \ref{directspace}.

To evaluate the $\vec{z}$ integral in the expression for $\mathcal{E}(\gamma)$, we will use the so-called Mellin transform\footnote{The Mellin transform, and more generally Mellin space, takes a fundamental role in conformal field theories, the AdS/CFT correspondence and QFT in AdS, see e.g. \cite{Mack:2009mi,Penedones:2010ue,Gopakumar:2016cpb,Fitzpatrick:2011ia,Nizami:2016jgt}. Mellin space is analogous to Fourier space, except scale invariance now takes the role that belongs to translation symmetry in the more familiar Fourier case.} \cite{borwein2013lattice}
\begin{equation}
    M_s[f] \equiv \frac{1}{\Gamma(s)}\int_0^\infty \ed t ~ t^{s-1} f \,,
\end{equation}
and in particular the identity
\begin{equation}
    \frac{1}{|(\mathbf{I} - \D)\,\vec{z} - \bvec + \Vec{n}|^{2s}} = \frac{1}{\Gamma(s)}\int_0^\infty \ed t ~ t^{s-1} e^{-|(\mathbf{I} - \D)\,\vec{z} - \bvec + \Vec{n}|^{2}t}.
\end{equation}
This implies that our sum is simply the Mellin transform of a sum of exponential terms of the form $e^{-|(\mathbf{I} - \D)\,\vec{z} - \bvec + \Vec{n}|^{2}t}$.
After permuting integrals and sums, we get
\begin{align}
   \mathcal{E}(\gamma) &= -\frac{\Gamma(s)}{2\pi^{s}}\cdot
    \frac{\sqrt{G}}{|\Gamma|} \frac{1}{\Gamma(s)}\int_0^\infty \ed t ~ t^{s-1} \int_{[0,1]^k}\ed^k\vec{z} ~  {\sum_{\Vec{n}\in\Z^k}} e^{-|(\mathbf{I}-\D)\Vec{z} + \bvec + \Vec{n}|^{2}t + 2\pi i\, \Vec{h}\cdot\Vec{n}} \,.
    \label{eq:Casimir-potential-generic-sum-non-diagonal-metric}
\end{align}
We then complete the square,
\begin{align}
    -|\Vec{n}+\Vec{\alpha}|^2 t + 2\pi i\, \Vec{h}\cdot\Vec{n}
    &= -(\Vec{n}+\Vec{\alpha})^T(t\, \mathbf{G})(\Vec{n}+\Vec{\alpha}) + 2\pi i\, \Vec{h}\cdot\Vec{n} \nonumber \\ 
    &= -(\Vec{n}+\Vec{\alpha} - \frac{\pi i}{t}\mathbf{G}^{-1}\Vec{h})^T(t\,\mathbf{G})(\Vec{n}+\Vec{\alpha} - \frac{\pi i}{t}\mathbf{G}^{-1}\Vec{h}) \nonumber \\
    &\phantom{=}~ - 2\pi i\,\Vec{\alpha}\cdot\Vec{h} - \frac{\pi^2}{t}\Vec{h}^T\mathbf{G}^{-1}\Vec{h} \,,
\end{align}
with $\vec{\alpha} = (\mathbf{I}-\D)\Vec{z} + \bvec$, so that the sum can be written as
\begin{align}
    {\sum_{\Vec{n}\in\Z^k}}' e^{-|\Vec{\alpha} + \Vec{n}|^{2}t + 2\pi i\, \Vec{h}\cdot\Vec{n}} 
    &= e^{- 2\pi i\,\Vec{\alpha}\cdot\Vec{h} - \frac{\pi^2}{t}\Vec{h}^T\mathbf{G}^{-1}\Vec{h}}{\sum_{\Vec{n}\in\Z^k}} e^{-|\Vec{n}+\Vec{\alpha} - \frac{\pi i}{t}\mathbf{G}^{-1}\Vec{h}|^2 t} \,.
    \label{eq:vector-sum}
\end{align}
Now consider the general form of Poisson resummation \cite{cdadf853-2bd9-3849-9928-547f47a288f0}, 
\begin{equation}
    \sum_{\vec{n}\in\Lambda} f(\vec{n}) = \sum_{\vec{q}\in\Lambda^*} \mathcal{F}[f](\vec{q}) \,,
\end{equation}
where $\mathcal{F}[f]$ is the Fourier transform of $f$ and $\Lambda^*$ the lattice dual to $\Lambda$,
\begin{equation}
    \Lambda^* \equiv \{\vec{q}\in\R^k \,|\, \vec{n}\cdot\vec{q}\in\Z \,,\,\forall\vec{n}\in\Lambda \} \,,
\end{equation} 
together with the following property of multidimensional Fourier transforms \cite{osgood2019lectures}
\begin{equation}
    \mathcal{F}[f(\mathbf{A}\Vec{z}+\vec{b})](\vec{\xi}) = \frac{e^{2\pi i\,\vec{b}\cdot (\mathbf{A}^{-1})^T\vec{\xi}}}{|\det \mathbf{A}\,|}\mathcal{F}[f(\Vec{z})]((\mathbf{A}^T)^{-1}\vec{\xi}) \,.
\end{equation}
Since the metric $\mathbf{G}$ can always be written as a Graham matrix $\mathbf{G}=\mathbf{M}^T\mathbf{M}$ (simply by finding an orthonormal basis), we have that 
\begin{equation}
    \vec{v}^T(t\,\mathbf{G})\vec{v}=(\sqrt{t}\,\mathbf{M}\vec{v})^T (\sqrt{t}\,\mathbf{M}\vec{v}) \,,
\end{equation} 
so that we may write
\begin{equation}
    \mathcal{F}[f((\sqrt{t}\,\mathbf{M})(\Vec{n}+\vec{\sigma}))](\vec{\xi}) = \frac{e^{2\pi i\,\vec{\sigma}\cdot \vec{q}}}{|\det \sqrt{t}\,\mathbf{M}\,|}\cdot \mathcal{F}[f(\Vec{n})]((\sqrt{t}\,\mathbf{M}^{T})^{-1}\vec{q}) \,, 
\end{equation}
with $\vec{\sigma} = \Vec{\alpha} - \frac{\pi i}{t}\mathbf{G}^{-1}\Vec{h}$. The Fourier transform is then
\begin{align}
    \mathcal{F}\left[e^{-|\Vec{n}+\Vec{\alpha} - \frac{\pi i}{t}\mathbf{G}^{-1}\Vec{h}|^2 t}\right]
    &= \frac{\pi^{\frac{k}{2}}}{t^{\frac{k}{2}}|\det \mathbf{M}\,|}\cdot e^{2\pi i (\vec{\alpha} - \frac{\pi i}{t}\mathbf{G}^{-1}\vec{h})\cdot\vec{q}}\cdot e^{-\frac{\pi^2 \vec{q}^T \mathbf{G}^{-1}\vec{q}}{t}} \\
    &= \frac{\pi^{\frac{k}{2}}}{t^{\frac{k}{2}}\sqrt{G}} \cdot e^{-\frac{\pi^2}{t}(\vec{q}^T \mathbf{G}^{-1}\vec{q} - 2\,\mathbf{G}^{-1}\vec{h}\cdot\vec{q})} \cdot e^{2\pi i \,\vec{\alpha}\cdot\vec{q}} \,.
\end{align}
The  space coordinate $\vec{z}$ now only appears in the simple exponential term $e^{2\pi i \,\vec{\alpha}\cdot\vec{q}}$,  and thus the spatial integrals over $\vec{z}$ can be performed directly, term by term.
We can then write \eqref{eq:Casimir-potential-generic-sum-non-diagonal-metric} as
\begin{align}
  \mathcal{E}(\gamma) = -\frac{\Gamma(s)}{2\pi^{s}}\cdot 
    \frac{\sqrt{G}}{|\Gamma|} \frac{\pi^{\frac{k}{2}}}{\Gamma(s)}\int_0^\infty \ed t ~ \frac{t^{s-\frac{k}{2}-1}}{\sqrt{G}} 
    {\sum_{\Vec{q}\in\Z^k}} e^{-\frac{\pi^2}{t}(\vec{q}-\vec{h})^T \mathbf{G}^{-1}(\vec{q}-\vec{h}) + 2\pi i \,\bvec\cdot(\vec{q}-\vec{h})} 
    \cdot \mathcal{I}_k(\D),
    \label{eq:Casimir-potential-general-sum-pre-integral}
\end{align}
where we have introduced the integral
\begin{align}
    \mathcal{I}_k(\D) 
    = \int_{[0,1]^k} \ed^k \vec{z} ~&e^{2\pi i \,[(\mathbf{I} - \D)\vec{z}]\cdot(\vec{q}-\vec{h})} 
    = \int_{[0,1]^k} \ed^k  \vec{z} ~e^{2\pi i \,\vec{z}\cdot[(\mathbf{I} - \D)^T\cdot(\vec{q}-\vec{h})]} \,.
    \label{b012}
\end{align}
The result of all these steps is now clear: the integral over the compact space (more precisely, over the covering torus $T^k$) can be performed explicitly,
\begin{equation}
    \mathcal{I}_k(\D)  
    = \prod_{j=1}^{k} \frac{e^{2\pi i \,[(\mathbf{I} - \D)^T\cdot(\vec{q}-\vec{h})]_j} - 1}{[(\mathbf{I} - \D)^T\cdot(\vec{q}-\vec{h})]_j}  \,.
\end{equation}
Recalling the consistency condition \eq{mimi}, namely that 
\begin{equation}
    (\mathbf{I} - \D)^T\cdot\vec{h}\in\Z^k \,,
\end{equation}
which limits the possible choices of $\vec{h}$ compatible with a consistent spin structure, that $\vec{q}\in\Z^k$ and that $\D\in GL(k,\Z)$, we conclude that the components $[(\mathbf{I} - \D)^T\cdot(\vec{q}-\vec{h})]_j$ are always integers. Consequently, each factor in this product is only non-zero if $[(\mathbf{I} - \D)^T\cdot(\vec{q}-\vec{h})]_j = 0$, in which case it simply gives 1, 
\begin{equation}
    \mathcal{I}_k(\D)  
    = \prod_{j=1}^{k} \delta_{0,[(\mathbf{I} - \D)^T\cdot(\vec{q}-\vec{h})]_j}  \,.
\end{equation}
In other words, the integral is only non-vanishing along the invariant subspace of $\D^T$ defined by 
\begin{equation}
    (\mathbf{I} - \D)^T (\vec{q}-\vec{h}) = 0.
    \label{eq:invariant-subspace-dual-lattice}
\end{equation}
A solution to this equation takes the form $\vec{q}=\vec{\kappa}+\vec{\eta}$, where $\vec{\kappa}$ is a generic vector of the sublattice $\Lambda^\parallel$ defined as the intersection of $\Z^k$ with the invariant subspace of $\D^T$, and $\vec{\eta}\in\Z^k$ is any integer vector that satisfies 
\begin{equation}
    (\mathbf{I} - \D)^T\vec{\eta} = (\mathbf{I} - \D)^T\vec{h} \,.
    \label{eq:eta-condition}
\end{equation}
Note that there might not be a solution to this equation; in particular, one cannot generically choose $\vec{h}$ itself since its components may be half-integers. Whenever no solution for $\vec{\eta}\in\Z^k$ exists, the sum vanishes identically---we will keep track of this condition by including in the final result a factor $\hat{\delta}_{\vec{h}}$ which evaluates to $1$ if a solution exists and to $0$ otherwise. The factor $\hat{\delta}_{\vec{h}}$ is related to a field theory version of Atkin-Lehner symmetry \cite{atkin1970hecke,Moore:1987ue,Dienes:1990qh}; we elaborate on this in Subsection \ref{sec:alsym}. In any case, after imposing \eq{eq:eta-condition} the sum over $\vec{q}\in\Z^k$ is thus restricted to a sum over $\vec{\kappa}\in\Lambda^\parallel$,
\begin{align}
   \mathcal{E}(\gamma) &= -\hat{\delta}_{\vec{h}}\,\frac{\Gamma(s)}{2\pi^{s}}\cdot 
    \frac{\sqrt{G}}{|\Gamma|}\frac{\pi^{\frac{k}{2}}}{\Gamma(s)}\int_0^\infty \ed t ~ \frac{t^{s-\frac{k}{2}-1}}{\sqrt{G}} 
    {\sum_{\Vec{\kappa}\in\Lambda^\parallel}} e^{-\frac{\pi^2}{t}(\vec{\kappa}-\vec{\beta})^T \mathbf{G}_\parallel^{-1}(\vec{\kappa}-\vec{\beta}) + 2\pi i \,\bvec^\parallel\cdot(\vec{\kappa}-\vec{\beta})} \,,
    \label{eq:Casimir-potential-general-sum-integrated}
\end{align}
where $\vec{\beta}\equiv\vec{h}-\vec{\eta}\in\Lambda^\parallel\otimes\mathbb{Q}$.
In effect, the Casimir sum collapses into a sum over the invariant subspace of $\D^T$, which has dimension $k' \leq k$. Notice that the invariant subspace is never empty, since otherwise the corresponding element $\gamma\in\Gamma$ would have fixed points, as explained in Section \ref{sec:RFMs}. Finally, the inner products are now restricted onto this invariant subspace, with induced metric $G_\parallel$ and norm $\vert \cdot \vert_\parallel$. The vector $\bvec^\parallel$ is the projection of $\bvec$ onto this same subspace. 

Continuing with our derivation, at this stage, one can perform Poisson resummation back to position space from this reduced sum over the $k'$-dimensional lattice $\Lambda^\parallel$, 
\begin{equation}
    \mathcal{E}(\gamma) = -\hat{\delta}_{\vec{h}}\,\frac{\Gamma(s)}{2\pi^{s}}\cdot
    \frac{\sqrt{G}}{\vert\Gamma\vert} \frac{\pi^{\frac{k-k'}{2}}}{\Gamma(s)}\int_0^\infty \ed t ~ t^{s - \frac{k-k'}{2} - 1} \frac{\sqrt{G_\parallel}}{\sqrt{G}}\sum_{\vec{\xi}\in\Xi} e^{-|\vec{\xi} + \bvec^\parallel|^{2}_\parallel t + 2\pi i \,\Vec{\beta}\cdot\vec{\xi}} \,.
    \label{eq:reduce-Casimir-sum-pre-Mellin}
\end{equation}
obtaining a sum over a lattice $\Xi$, dual to $\Lambda^\parallel$. Since 
\begin{equation} (\mathbf{D}^T_\gamma\,\vec{\kappa})\cdot \vec{\xi}= \vec{\kappa}\cdot (\mathbf{D}_\gamma\,\vec{\xi}),\label{mam}\end{equation}
the dual lattice $\Xi$ is naturally identified with those vectors in the invariant subspace of $\mathbf{D}_\gamma$ which have integer inner products with every $\kappa\in\Lambda^\parallel$,
\begin{equation}
    \Xi = (\Lambda^\parallel)^* \equiv \{\vec{\xi}\in\R^k  \,\vert\, \vec{\xi}\cdot\vec{\kappa}\in\Z,\, \mathbf{D}_\gamma\vec{\xi}=\vec{\xi} \,,\,\forall\vec{\kappa}\in\Lambda^\parallel \} \,,
\label{eqxi}\end{equation}
where the $\cdot$ denotes the standard pairing between a space and its dual. We will now show that $\Xi$ as defined above can be identified with the orthogonal projection of the defining lattice $\mathbb{Z}^k$ onto the invariant subspace of $\mathbf{D}_\gamma$ with respect to the $G_{ij}$ inner product. To show this, notice that, since $\Lambda^\parallel$ is a primitive sublattice of $\mathbb{Z}^k$ (as it is defined by a linear equation), there is a basis $\{\vec{\kappa}_i\}$, $i=1,\ldots k$ of $\mathbb{Z}^k$ where the $k'$ first vectors form a basis of $\Lambda^\parallel$ (see e.g. \cite{spyros}). Consider the corresponding dual basis $\{\vec{\xi}_i\}$. The first $k'$ vectors of this basis satisfy $\vec{\xi}_i\cdot \vec{\kappa}_j=\delta_{ij}$, and therefore satisfy all conditions to be a basis of $\Xi$ in \eq{eqxi}, except for the fact that they might not lie on the invariant subspace of $\mathbf{D}_\gamma$. However, any vector $\vec{\xi}$ can be decomposed as
\begin{equation}\vec{\xi}=\vec{\xi}^\perp+\vec{\xi}^{\,\text{Inv}},\end{equation}
where $\vec{\xi}^{\,\text{Inv}}$ is the orthogonal projection onto the invariant subspace and $\vec{\xi}^\perp$ is orthogonal to it. Since 
\begin{equation} \vec{\kappa}\cdot \vec{\xi}= \vec{\kappa}\cdot \vec{\xi}^{\,\text{Inv}}\end{equation}
for any $\kappa\in\Lambda^\parallel$, we may replace the vectors $\vec{\xi}$ by their orthogonal projection to construct a basis, and thus $\Xi$ is indeed the projection of $\mathbb{Z}^k$ onto the invariant subspace of $\mathbf{D}_\gamma$. 

Finally, performing the Mellin transform in \eq{eq:reduce-Casimir-sum-pre-Mellin}, we find 
\begin{equation}
    \Vcas = \sum_{\textbf{r}}\sum_{\gamma\in\Gamma} \Tr{\bf r}{\D}\,\mathcal{E}(\gamma)
    \,,\quad
    \mathcal{E}(\gamma) = -\hat{\delta}_{\vec{h}}\,\frac{\Gamma(s_\gamma)}{2\pi^{s_\gamma}}\cdot
    \frac{\sqrt{G_\parallel}}{\vert\Gamma\vert} \sum_{\vec{\xi}\in\,\Xi_\gamma} \frac{e^{2\pi i \,\Vec{\beta_\gamma}\cdot\vec{\xi}}}{|\vec{\xi} + \bvec^\parallel|^{2s_\gamma}_\parallel} \,,
    \label{eq:VC-invariant-subspace}
\end{equation}
where $s_\gamma = s - \frac{k-k'}{2}$, the lattice $\Xi_\gamma$ and the vector $\vec{\beta}_\gamma$, as well as the projected metric $\mathbf{G}_\parallel$ and inner product $|\cdot|_\parallel$, depend on the element $\gamma\in\Gamma$. 
 
Equation \eq{eq:VC-invariant-subspace} is the final expression we were looking for. The lower-dimensional Casimir energy is expressed as an infinite sum, with no integrals left, where each $\gamma\in\Gamma$ gives a definite contribution $\mathcal{E}(\gamma)$. Notice that for the identity element in $\Gamma$, the invariant subspace is the full lattice $\Z^k$ and the sum stays $k$-dimensional. 

The fact that $\Xi_\gamma$ is just the invariant lattice of $\mathbf{D}_\gamma$ allows for a nice physical interpretation of our final result for $\mathcal{E}(\gamma)$ \eq{eq:VC-invariant-subspace}. This expression is a lattice sum just like the one that would be used to compute Casimir energies on a torus, only that it is localized on the invariant subspace of $\D$. Since the volume element of this subspace is precisely $\sqrt{G_\parallel}$, we find that  $\mathcal{E}(\gamma)$ behaves exactly as the contribution to the effective potential that would come from an effective $k'$-dimensional ``Casimir brane'' wrapped on the invariant subspace of $\D$, of tension given by \eq{eq:VC-invariant-subspace}. In fact, already from \eq{csum}, before performing the $\vec{z}$ integral, we can tell that the region of highest energy density in $\mathcal{E}(\gamma)$ comes from the invariant subspaces, which is where the denominator becomes largest. What \eq{eq:VC-invariant-subspace} tells us is that this qualitative feature survives a detailed treatment of the problem.

\begin{figure}[!htb]
\centering
\begin{subfigure}{0.48\textwidth}
    \includegraphics[width=\textwidth]{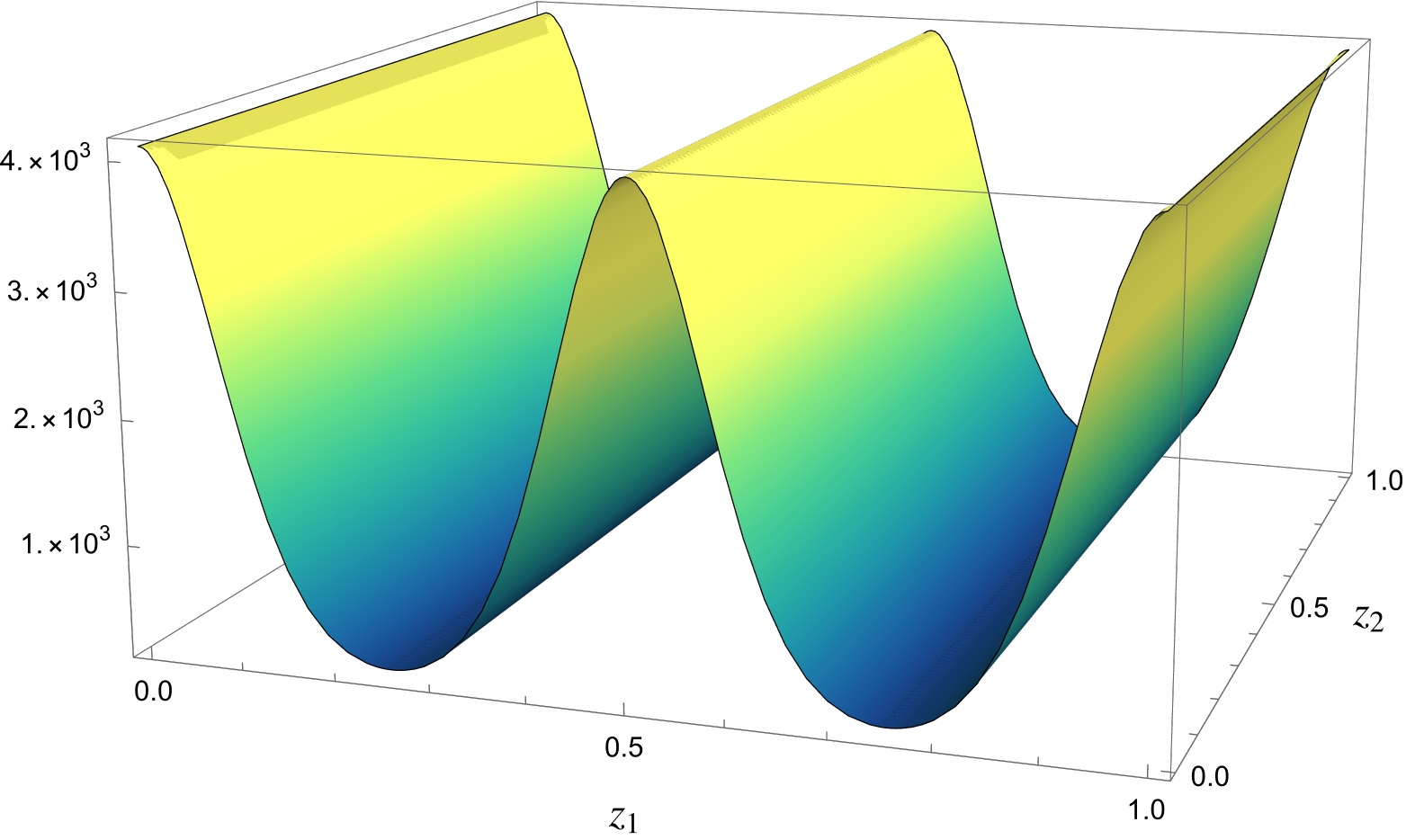}
    \caption{$T^2/\Z_2$ (Klein bottle)}
    \label{fig:CBrane-KB}
\end{subfigure}
\hfill
\begin{subfigure}{0.48\textwidth}
    \includegraphics[width=\textwidth]{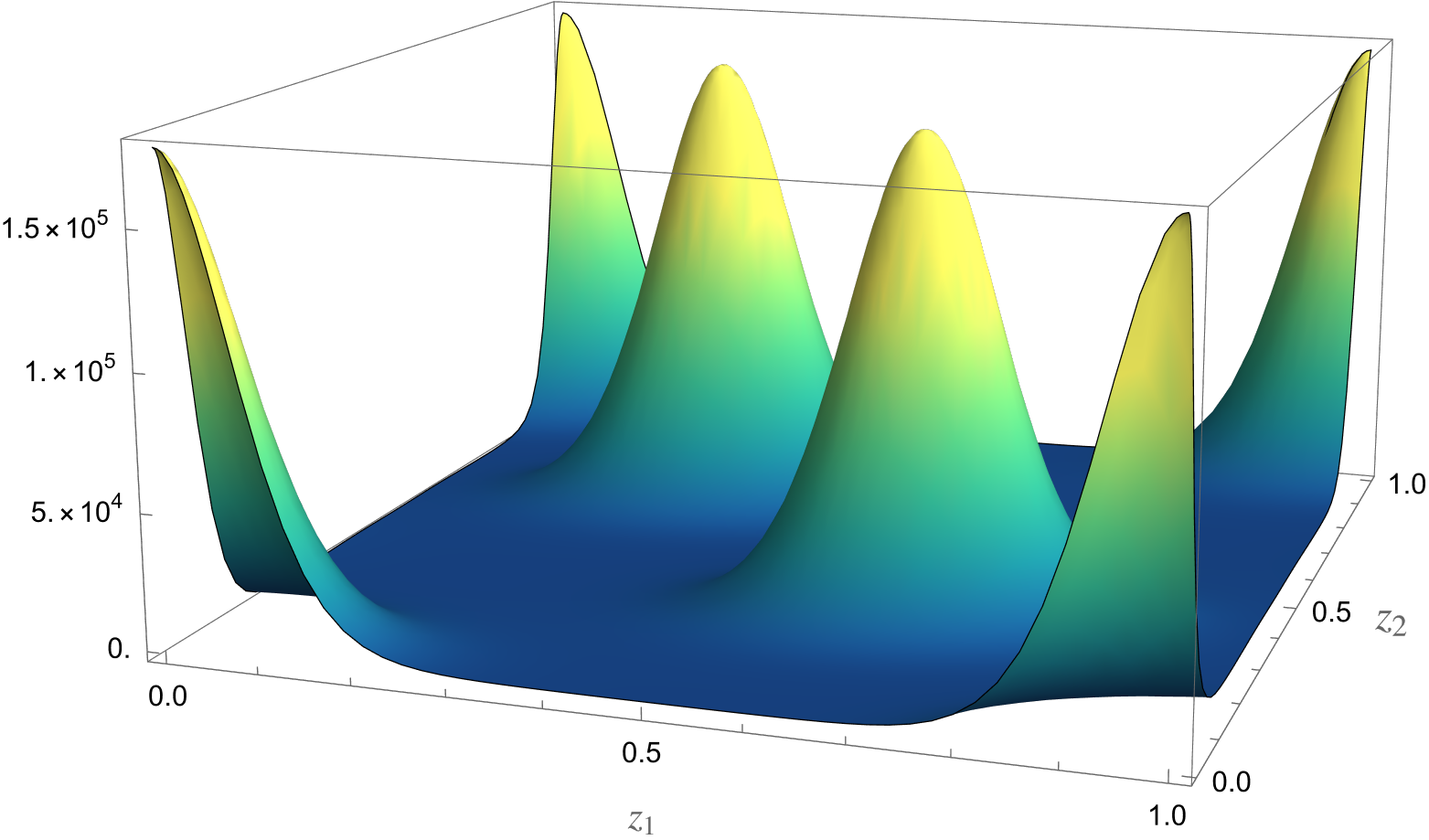}
    \caption{$T^3/\Z_3$}
    \label{fig:CBrane-Z3}
\end{subfigure}
        
\caption{Casimir branes for Klein bottle and 3d RFM.}
\label{fig:CBranes}
\end{figure}

The usefulness of the Casimir brane picture is that the total Casimir energy can be understood as a sum over contributions corresponding to different effective ``Casimir branes'' wrapped on different submanifolds of the RFM. For instance, the $\D=\mathbf{I}$ contribution corresponds to a space-filling Casimir brane; a $\D$ with a one-dimensional invariant subspace corresponds to a $(d+1)$-dimensional Casimir brane; and so on (see Figure \ref{fig:CBranes} for an explicit example). This allows much of the intuition that one has from D-brane model building to be imported to settings with Casimir energies---in particular, it provides a $D$-dimensional picture of the backreaction in the internal space---although one must keep in mind that (unlike ordinary D-branes and orientifolds) the tension of a Casimir brane depends on the precise submanifold on which it is wrapped. In fact, its overall scalling is fixed by the size of the submanifold it wraps. Furthermore, unlike D-branes, but just like orientifolds, Casimir branes can have a negative tension, which is why they evade the no-go theorem of \cite{Maldacena:2000mw} and can provide classical saddle points. 

A point of curiosity is the $\hat{\delta}_{\vec{h}}$ factor. When it evaluates to zero, the Casimir energy vanishes, but as we will see in examples below, in general this happens because there are several Casimir branes whose tensions exactly cancel! Due to the symmetries of the problem, one is forced to introduce localized objects of equal and opposite tension. This is reminiscent of what happens in e.g. orientifold compactifications with 16 supercharges \cite{Ibanez:2012zz}, but now the phenomenon is happening at one-loop and in a non-supersymmetric context. In Section \ref{sec:alsym} we explain this vanishing via a spacetime version of Atkin-Lehner symmetry \cite{atkin1970hecke,Moore:1987ue,Dienes:1990qh}. We believe this warrants further study, since having a symmetry reason for the vanishing of (some terms of) a one-loop Casimir energy by what looks like an extreme fine-tuning of the tension of Casimir branes might give us a way to engineer anomalously small vacuum energies and address aspects of the cosmological hierarchy problem.

In this paper we will not rely heavily on the Casimir brane picture, but we believe it may be a useful tool going forward. Furthermore, the fact that the RFM Casimir energy takes the form of a sum of branes may suggest the existence of a dual picture where the Casimir branes are identified with contributions of more standard branes or orientifolds wrapping non-supersymmetric cycles. For the time being we will ignore these interesting questions and finish this subsection describing the checks we performed to verify \eq{eq:VC-invariant-subspace}.

On top of the three separate derivations for \eq{eq:VC-invariant-subspace}, provided here and in Appendix \ref{directspace}, we have also checked it numerically, by first performing the sum in \eq{csum} using the method of Ewald summation described in Subsection \ref{sec:numerical-sums} and then integrating over $\vec{z}$ numerically, for two simple cases in two and three dimensions which we describe below.
A direct numerical check of the formula for higher dimensional sums quickly becomes prohibitive, since the evaluation of the integrand takes longer time.  

\begin{figure}[!htb]
    \centering
    \includegraphics[width=0.6\linewidth]{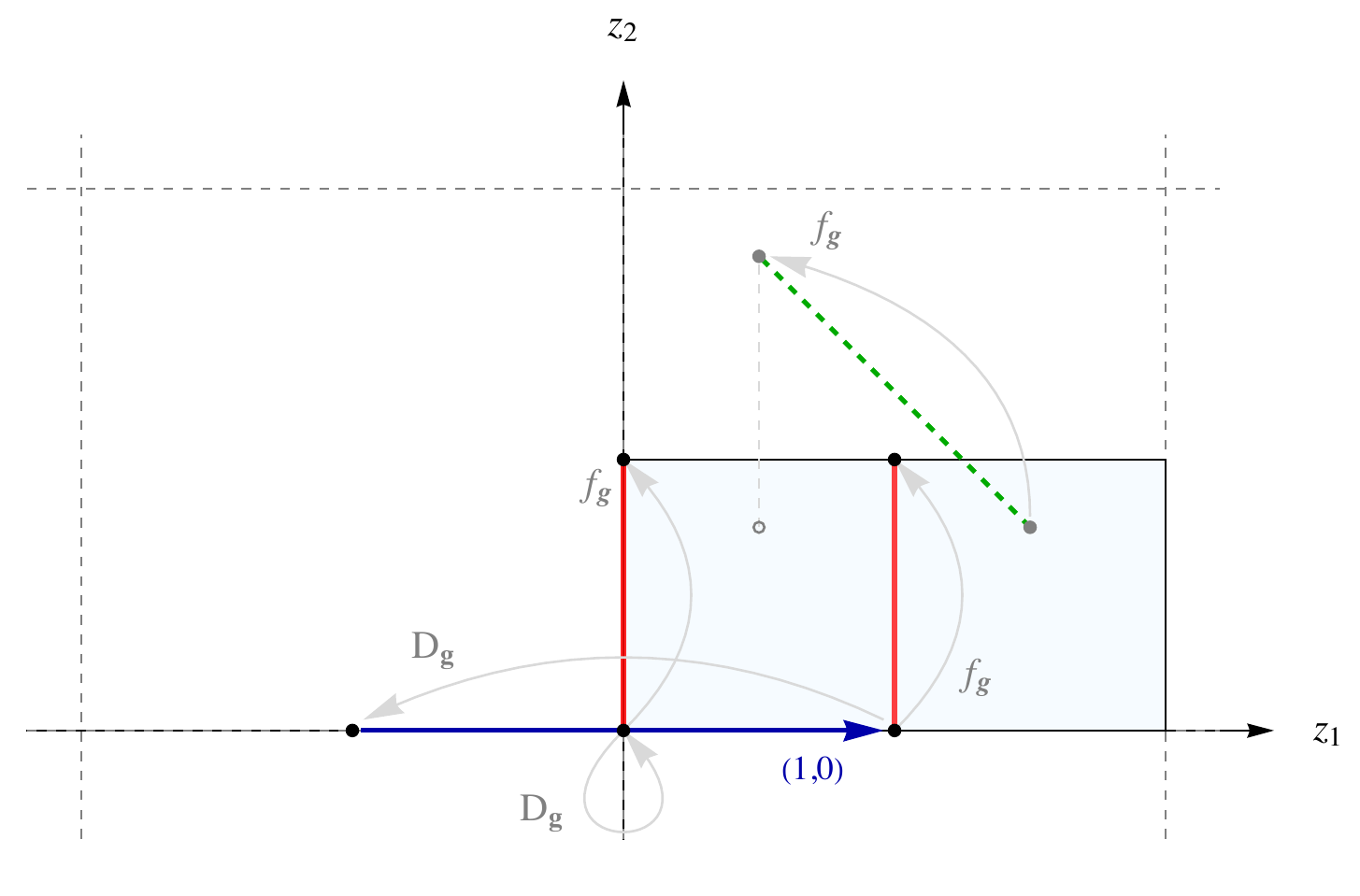}
    \caption{Lattice of the Klein bottle $T^2/\Z_2$, with invariant (up to $T^2$ identifications---blue, bold arrow) points $z_1 \in\{0,\frac12\}$ corresponding to the locations of the Casimir branes (red, bold). Note that Casimir branes connect pairs of identified points that are the closest and therefore contribute the most to the Casimir energy. For comparison we show a point that is not invariant under $\D[g]$ and is thus further apart from its images under $f_{\bf g}(\vec{z}) = \D[g]\,\vec{z} + \bvec[g]$ (green, dashed).}
    \label{fig:KB-lattice}
\end{figure}

The first example we study is the Klein bottle. This is the simplest 2-dimensional RFM $(k=2)$, corresponding to the quotient $T^2/\Z_2$ with the $\Z_2$ group generated by 
\begin{equation}
    \D[g] = \left(
        \begin{array}{*2{C{1.05em}}}
         -1 & 0 \\
         0  & 1
        \end{array}
    \right) \,, \quad
    \bvec[g] = \left(
        \begin{array}{c}
         0 \\ \frac12 
        \end{array} 
    \right) \,.
\end{equation}
The most general metric on the Klein bottle is
\begin{equation}
    \mathbf{G} = R^2\left(\begin{array}{cc}
         x & 0  \\
         0 & x^{-1} 
    \end{array}\right) \,,
\end{equation}
and the subspace left invariant by $\D[g]$ is one-dimensional $(k'=1)$,
\begin{equation}
    (\mathbf{I} - \D[g])\,\vec{z} = \left(\begin{array}{c}
         2z_1  \\ 0  
    \end{array}\right) \overset{!}{=} 0 ~\text{mod}~\Z^2 
    \quad\implies\quad
    z_1 \in\left\{0,\frac12\right\} \,,\, z_2\in[0,1) \,.
\end{equation}
This gives us the positions of the Casimir branes, located at $z_1 \in\{0,\frac12\}$ and wrapping the $z_2$ direction (\figref{fig:CBranes}). Note that the conditions \eq{mimi} and \eq{swer2} force any consistent choice of $\vec{h}$ to satisfy 
\begin{align}
    2h_1\in\Z \,,\quad s_g = 2h_2\,\text{mod}\,2\Z \,. 
\end{align}
Hence we are free to choose any $h_1$, which will fix a specific Pin$^+$ lift\footnote{The Klein bottle is not an orientable space, and so it does not admit a spin structure. There is however a non-orientable generalization of this, called ``Pin'' structures, that works the same in practice. There are two kinds, Pin$^+$ and Pin$^-$, of which Pin$^+$ is in some sense closest to an ordinary spin structure. Since we are just giving this example for illustrative purposes, we skip the subtleties associated to the construction of a Pin$^+$ structure on the Klein bottle. See e.g. \cite{Witten:2015aba} for further information.}, i.e. $\mathcal{D}_\mathbf{g}^2 = (-1)^{2h_2}\mathbf{I}$. We must take this into account when taking traces over fermionic representations.
For this element, we then have $s_\mathbf{g} = s - \frac12$, $\mathbf{G}_\parallel = (R^2\,x^{-1})$ and $\vec{\beta}_\mathbf{g} = \vec{h} - \vec{\eta}_\mathbf{g}$, with $\vec{\eta}_\mathbf{g}$ any vector in $\Z^2$ satisfying \eq{eq:eta-condition}, which reads $2\eta_1 = 2h_1$. We conclude that the contribution from $\D[g]$ is only non-vanishing if $h_1 = 0~\text{mod}~\Z$, i.e. for untwisted boundary conditions around the $z_1$ direction.

The contribution from $\D[g]$ is then
\begin{equation}
    \mathcal{E}(\mathbf{g}) 
    = -\frac{\Gamma(s_\mathbf{g})}{2\pi^{s_\mathbf{g}}}\cdot
    \frac{x^{s_\mathbf{g}-\frac12}}{2R^{2s_\mathbf{g} - 1}} \sum_{\xi\in\Z} \frac{e^{2\pi i \,h_2\xi}}{|\xi + \frac12|^{2s_\mathbf{g}}} 
    = -\frac{\Gamma(s_\mathbf{g})}{2\pi^{s_\mathbf{g}}}\cdot
    \frac{x^{s_\mathbf{g}-\frac12}}{R^{2s_\mathbf{g} - 1}} (2^{2s_\mathbf{g}} - 1)\zeta(2s_\mathbf{g}) \delta_{h_2,0} \,. 
\end{equation}
The fact that the invariant subspace is 1-dimensional allows us to obtain analytical results for the Klein bottle. Moreover, we find that the Casimir energy vanishes whenever the boundary condition around any of the two cycles is twisted, i.e. only for $\vec{h}=(0,0)$ do we find non-vanishing Casimir energy.
For $s=\frac{11}{2}$, which would be appropriate for the study of M-theory on the Klein bottle, we have 
\begin{equation}
    \mathcal{E}(\mathbf{g}) = -\frac{124\pi^5}{945}\frac{x^{9/2}}{R^9} \approx -\frac{40.2\,x^{9/2}}{R^9} \,. 
\end{equation}
This result exactly matches the direct numerical integration of \eqref{csum}, which we also performed separately. Plotting the integrand, we can explicitly see the regions where the Casimir energy is localised, i.e. the Casimir branes located at $z_1 \in\{0,\frac12\}$ and wrapping the $z_2$ direction (\figref{fig:CBranes}). 

In M-theory one must include the contributions of the graviton, 3-form and gravitino; as we just argued, any choice of boundary conditions other than $\vec{h}=\{0,0\}$ will result in a vanishing contribution from $\D[g]$. Finally, we need the traces of $\D[g]$ in these three representations to weigh the respective contributions in the sum \eq{eq:VC-invariant-subspace}. Following the discussion in Appendix \ref{ap:traces} with $\vec{\theta}=\{\pm\pi,0,0,0\}$, we find
\begin{equation}
    \Tr{\mathbf{44}}{\D[g]} = 16 \,,\quad
    \Tr{\mathbf{84}}{\D[g]} = 0 \,,\quad 
    \Tr{\mathbf{128}}{\D[g]} = 0 \,,
\end{equation}
regardless of the choice of spin lift $\mathcal{D}_\mathbf{g}$. On the other hand, the identity term $\mathbf{I}$ contributes as 
\begin{equation}
    \Vcas^{(\mathbf{I})} = -\frac{\Gamma(s)}{2\pi^{s}}\cdot
    \frac{128\,x^{2s}}{2R^{2s-2}} \sum_{\vec{n}\in\,\Z^2} \frac{1 - e^{2\pi i \,\Vec{h}\cdot\vec{n}}}{|x^2n_1^2 + n_2^2|^{2s}} \,. 
\end{equation}
Due to the presence of $x$ inside the sum, this term must be summed numerically. For comparison, when $x=1$, these can be summed analytically to get 
\begin{align}
    \Vcas^{(\mathbf{I})} &= 0 \,,\quad\text{for}\quad \vec{h}=(0,0),\\
    \Vcas^{(\mathbf{I})} &=-\frac{\Gamma(s)}{2\pi^{s}}\cdot
    \frac{128\,x^{2s}}{2R^{2s-2}}\cdot 4\beta(s)[\zeta(s)-\eta(s)] \approx -\frac{0.56}{R^9} \, \,,\quad\text{for}\quad \vec{h}=\Big(\frac12,0\Big).
\end{align}

\begin{figure}[!htb]
    \centering
    \includegraphics[width=0.4\linewidth]{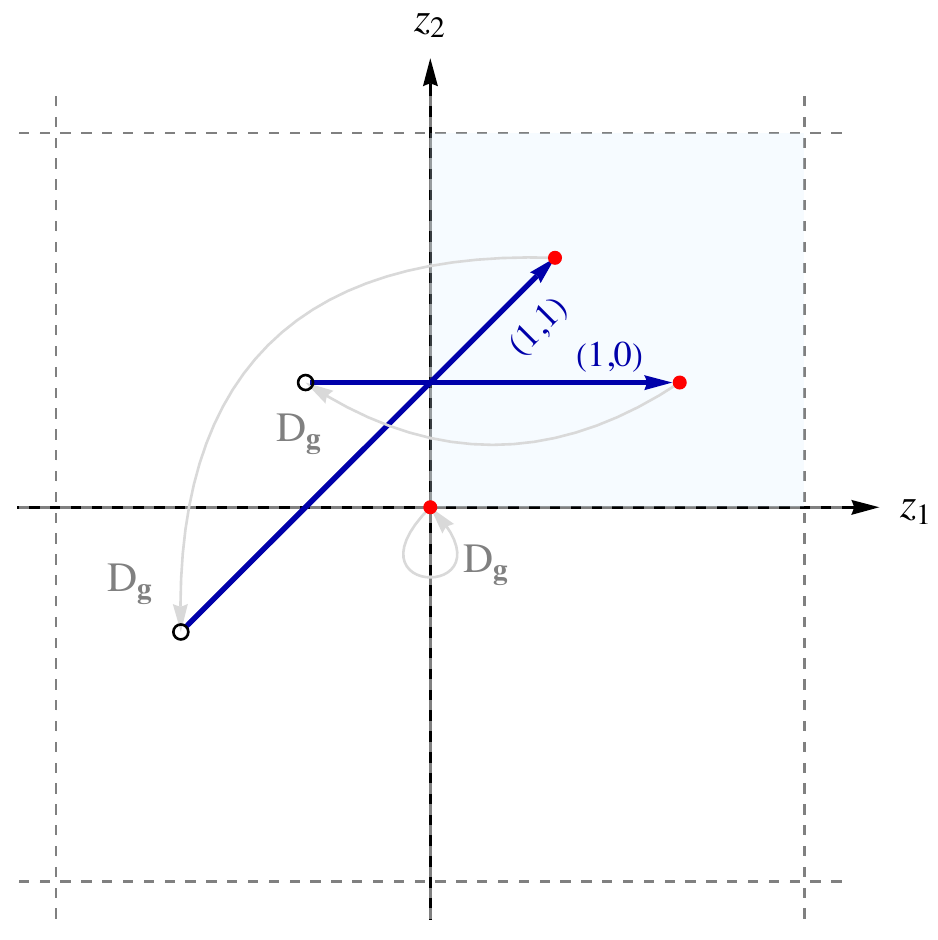}
    \caption{Lattice of the $T^3/\Z_3$ RFM at $z_3 = 0$, with invariant (up to $T^3$ identifications---blue, bold arrows) points $(z_1,z_2)\in\{(0,0),(\frac13,\frac23),(\frac23,\frac13)\}$ corresponding to the locations of the Casimir branes (red, filled points). Note that these are \emph{not} fixed points since each action of $\D[g]$ is accompanied by a shift by $\bvec[g]$ along $z_3$, not shown here (see \figref{fig:3d-RFM-lattice-3d}).}
    \label{fig:3d-RFM-lattice}
\end{figure}
\begin{figure}[!htb]
    \centering
    \includegraphics[width=0.65\linewidth]{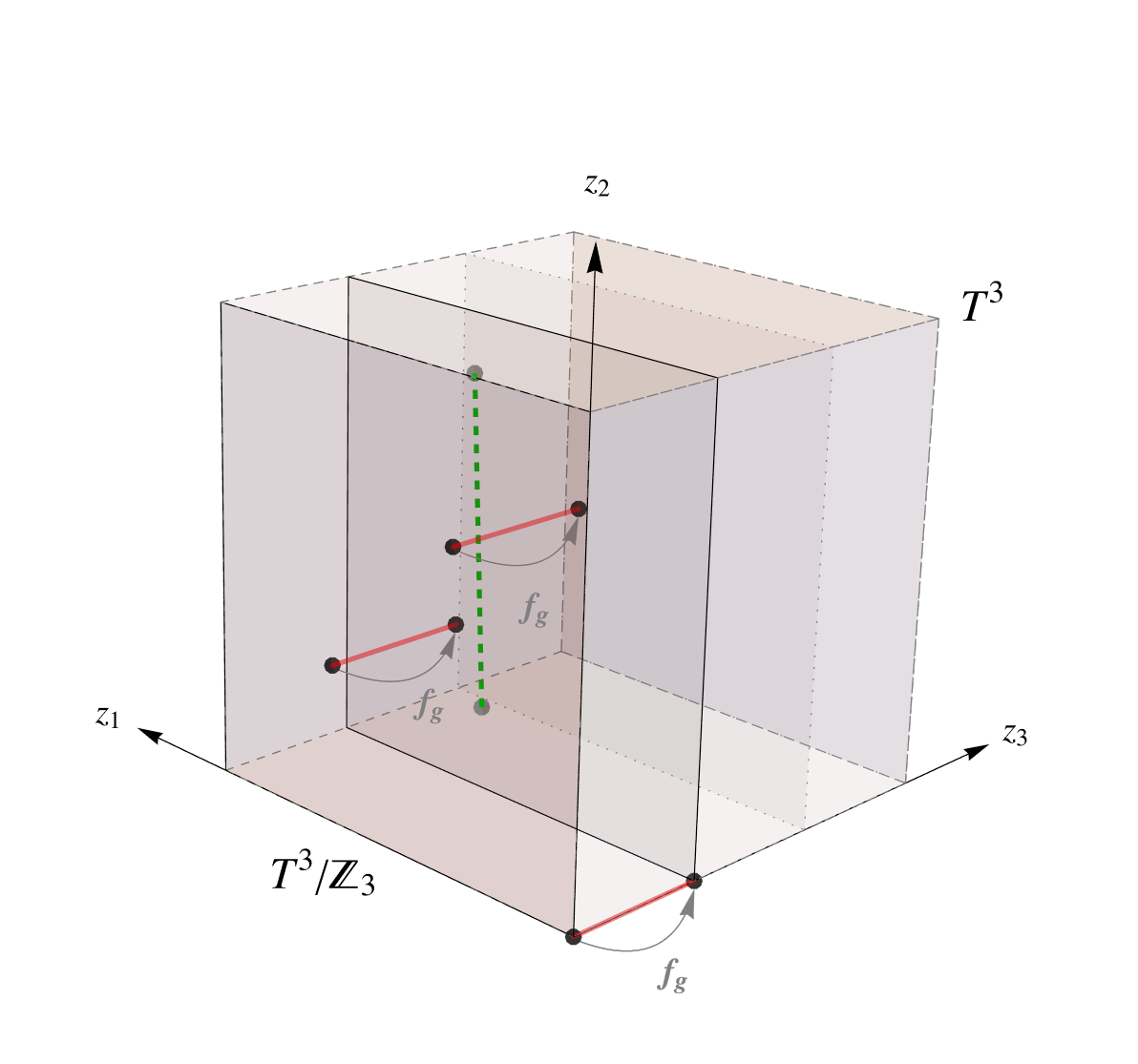}
    \caption{Casimir branes on $T^3/\Z_3$ RFM corresponding to the points that remain invariant under $\D[g]$ (red, bold lines). Note that the Casimir branes connect the pairs of points identified by the $\Z_3$ symmetries that are the closest to each other and therefore contribute the most to the Casimir energy. We show for comparison a point that is not invariant under $\D[g]$ so that its image is further apart and will thus contribute less to the Casimir energy (green, dotted line).}
    \label{fig:3d-RFM-lattice-3d}
\end{figure}

Let us consider a second example, the Riemann-flat manifold $T^3/\Z_3$ with the $\Z_3$ group generated by 
\begin{equation}
    \D[g] = \left(
        \begin{array}{*2{C{1.05em}}|*1{C{1.05em}}}
         0 & -1 & 0 \\
         1 & -1 & 0 \\ \hline 
         0 & 0  & 1 
        \end{array}
    \right) \,, \quad
    \bvec[g] = \left(
        \begin{array}{c}
         0 \\ 0 \\ \frac13 
        \end{array} 
    \right) \,.
    \label{eq:Z3-action}
\end{equation}
This manifold was used in \cite{Montero:2022vva} to construct new compactifications of string theory to seven dimensions and sixteen supercharges involving discrete $\theta$ angles.
The most general metric on this RFM is
\begin{align}
    \mathbf{G} = R^2\, x^{2/3} \begin{pmatrix}
         2 & -1 & 0 \\
         -1 & 2 & 0 \\ 
         0 & 0  & x^{-2} 
        \end{pmatrix} \,,
\end{align}
and the subspace invariant under $\D[g]$ is again 1-dimensional, 
\begin{equation*}
    (\mathbf{I} - \D[g])\,\vec{z} = \left(\begin{array}{c}
         z_1 + z_2 \\ -z_1 + 2z_2 \\ 0    
    \end{array}\right) \overset{!}{=} 0 ~\text{mod}\,\Z^3 
\end{equation*}
\begin{equation}
    \implies\quad
    (z_1,z_2) \in\left\{(0,0),\left(\frac13,\frac23\right),\left(\frac23,\frac13\right)\right\} \,,\, z_3\in[0,1) \,,
\end{equation}
which specifies the locations of the Casimir branes (see figures \ref{fig:CBranes}, \ref{fig:3d-RFM-lattice}, and \ref{fig:3d-RFM-lattice-3d}).

For the element $\D[g]$, we have $s_\mathbf{g} = s - 1$, $\mathbf{G}_\parallel = (R^2\,x^{-4/3})$, and $\vec{\beta}_\mathbf{g} = \vec{h} - \vec{\eta}_\mathbf{g}$, with $\vec{\eta}_\mathbf{g}$ any vector in $\Z^3$ satisfying \eq{eq:eta-condition}. For the choices of $\vec{h}$ allowed by the consistency condition \eq{mimi}, which forces $h_1,h_2 = 0\,\text{mod}\,\Z$, any vector $\vec{\eta}_\mathbf{g}$ satisfies \eq{eq:eta-condition}; in particular we can choose $\vec{\eta}_\mathbf{g} = \vec{0}$. Finally, we must also have
\begin{equation}
    s_g = 2h_3\,\text{mod}\,2\Z \,,
\end{equation}
so that a choice of $h_3$ is tied to a choice of specific spin lift $\pm\mathcal{D}_\mathbf{g}$ satisfying $\mathcal{D}_\mathbf{g}^3 = (-1)^{2h_3}\mathbf{I}\,$. As before, we must take this into account when taking traces over fermionic representations.

The contribution from $\D[g]$ is then
\begin{equation}
    \mathcal{E}(\mathbf{g}) 
    = -\frac{\Gamma(s_\mathbf{g})}{2\pi^{s_\mathbf{g}}}\cdot
    \frac{x^{\frac43 s_\mathbf{g}-\frac23}}{3R^{2s_\mathbf{g} - 1}} \sum_{\xi\in\Z} \frac{e^{2\pi i \,h_3\xi}}{|\xi + \frac13|^{2s_\mathbf{g}}} \,.
\end{equation}
When $h_3=0$, for untwisted boundary conditions, we find
\begin{equation}
    \mathcal{E}(\mathbf{g}) 
    = -\frac{\Gamma(s_\mathbf{g})}{2\pi^{s_\mathbf{g}}}\cdot
    \frac{x^{\frac43 s_\mathbf{g}-\frac23}}{3R^{2s_\mathbf{g} - 1}} \left[\zeta\left(2s_\mathbf{g},\frac13\right) + \zeta\left(2s_\mathbf{g},\frac23\right)\right] 
    \approx -\frac{221.4\,x^{\frac{16}{3}}}{R^8}\,.
\end{equation}
As for the Klein bottle, having a 1-dimensional invariant subspace allows us to obtain this analytical result (see Appendix \ref{ap:lattice-sums-1d-inv-space}); we have used again $s=\frac{11}{2}$ as in M-theory for the numerical check. This result matches exactly the direct numerical integration of \eqref{csum}. Plotting the integrand, we can explicitly see the regions where the Casimir energy is localised, i.e. the Casimir branes located at $(z_1,z_2)\in\{(0,0),(\frac13,\frac23),(\frac23,\frac13)\}$ and wrapping the $z_3$ direction (\figref{fig:CBranes}).
Alternatively, the choice $h_3 = \frac12$ gives 
\begin{equation}
    \mathcal{E}(\mathbf{g}) 
    = -\frac{\Gamma(s_\mathbf{g})}{2\pi^{s_\mathbf{g}}}\cdot
    \frac{x^{\frac43 s_\mathbf{g}-\frac23}}{3R^{2s_\mathbf{g} - 1}} \,\frac{1}{2^{2s_\mathbf{g}}}\left[\zeta\left(2s_\mathbf{g},\frac16\right)-\zeta\left(2s_\mathbf{g},\frac13\right) - \zeta\left(2s_\mathbf{g},\frac23\right) + \zeta\left(2s_\mathbf{g},\frac56\right)\right] \,.
\end{equation}
Funnily enough, we see that the contribution of twisted sectors to the Casimir sums (i.e. the tensions of Casimir branes) can often be computed analytically, even when the untwisted term (corresponding to the parent $T^3$ contribution) cannot!
The traces of $\D[g]$ over the M-theory representations are 
\begin{equation}
    \Tr{\mathbf{44}}{\D[g]} = 20 \,,\quad
    \Tr{\mathbf{84}}{\D[g]} = 21 \,,\quad 
    \Tr{\mathbf{128}}{\D[g]} = 40 \,.
\end{equation}
This 3d RFM contains one more non-trivial element, namely $\D[g]^2$. Repeating the exact same steps for this element, we find that the result with $h_3=0$ is the same as for $\D[g]$. However, when $h_3=\frac12$ we have $\mathcal{E}(\mathbf{g}^2) = - \mathcal{E}(\mathbf{g})$. This is not a coincidence; it follows from the fact that $\mathbf{g}^2 = \mathbf{g}^{-1}$. In Appendix \ref{ap:lattice-sums-1d-inv-space} we show that for cyclic RFM's with 1-dimensional invariant subspaces, we always find 
\begin{equation}
    \mathcal{E}(\mathbf{g}^j) = -\mathcal{E}(\mathbf{g}^{p-j}) \,, 
\end{equation}
where $p=|\Gamma|$, the order of the cyclic group. On the other hand, the explicit formulas for the traces in Appendix \ref{ap:traces} together with the fact that the eigenvalue arguments $\theta_i$ are $p^{\rm th}$ roots of unity, i.e. $p\,\theta_i = 2\pi\Z$, imply that the traces of $\D[g]^j$ and $\D[g]^{p-j}$ are the same over bosonic representations.

We can compare our results with those of \cite{Acharya:2020hsc}, where the partition function of Type IIB string theory on $T^3/\Z_3$ is computed. In the supergravity limit---where our results apply---the contribution from all degrees of freedom and all elements of $\Z_3$ (i.e. the value of $\Vcas$ in \eqref{eq:VC-invariant-subspace}) is given in (3.7) and (3.8) of \cite{Acharya:2020hsc}, with the $T^2$ fibre much smaller than the $S^1$ base, for $h_3=0$ and $h_3=\frac12$ respectively. Note that the traces over the bosonic and fermionic representations are
\begin{align}
    &&&&& \TrB{\D[g]} = 41 \,,\quad &&\TrB{\D[g]^2} = 41 \,,\quad &&\TrB{\mathbf{1}} = 128 \,, && \\
    &&&&& \TrF{\D[g]} = 40\,(-1)^{s_g} \,,\quad &&\TrF{\D[g]^2} = 40\,(-1)^{s_g} \,,\quad &&\TrF{\mathbf{1}} = 128\,(-1)^{s_g} \,, &&
\end{align}
with the spin lift $s_g$ fixed by the choice of $h_3$, such that $s_g = 2h_3\,\text{mod}\,2\Z$ \eqref{swer2}. We then find
\begin{subequations}
    \begin{align}
    \Vcas\big|^{h_3=0}_{\rm IIB} &= -\frac{\Gamma(4)}{2\pi^4}\cdot\frac{2\cdot 81}{3R_1^7} \left[\zeta\left(8,\tfrac13\right) + \zeta\left(8,\tfrac23\right)\right] \,, \\ 
    \Vcas\big|^{h_3=\frac12}_{\rm IIB} &= -\frac{\Gamma(4)}{2\pi^4}\cdot\frac{2}{3R_1^7} \left[\zeta\left(8,\tfrac12\right)+\tfrac{1317}{32}\left[\zeta\left(8,\tfrac13\right) - \zeta\left(8,\tfrac23\right)\right] - \tfrac{5}{32}\left[\zeta\left(8,\tfrac16\right)+ \zeta\left(8,\tfrac56\right)\right]\right] \,,
\end{align}
\end{subequations}
which matches the result of \cite{Acharya:2020hsc} up to an overall factor of 2, after the rescaling $R_1\to\sqrt{2}R_1$. This overall factor is likely due to a specific choice of conventions.

\begin{figure}[!htb]
\centering
\begin{subfigure}{0.48\textwidth}
    \includegraphics[width=\textwidth]{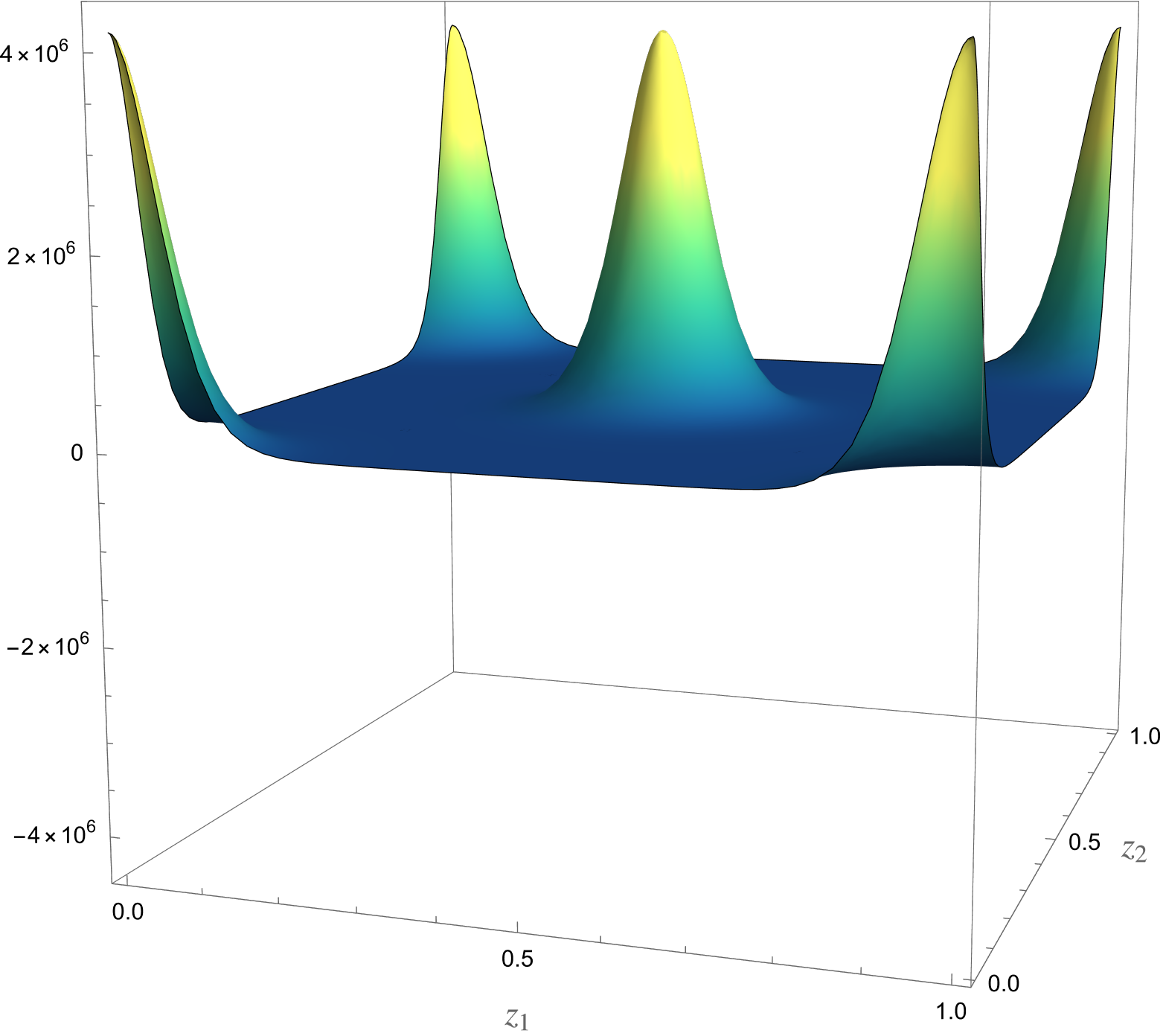}
    \caption{Periodic $\vec{h}=(0,0,0)$.}
    \label{fig:CBrane-T3Z4-boson}
\end{subfigure}
\hfill
\begin{subfigure}{0.48\textwidth}
    \includegraphics[width=\textwidth]{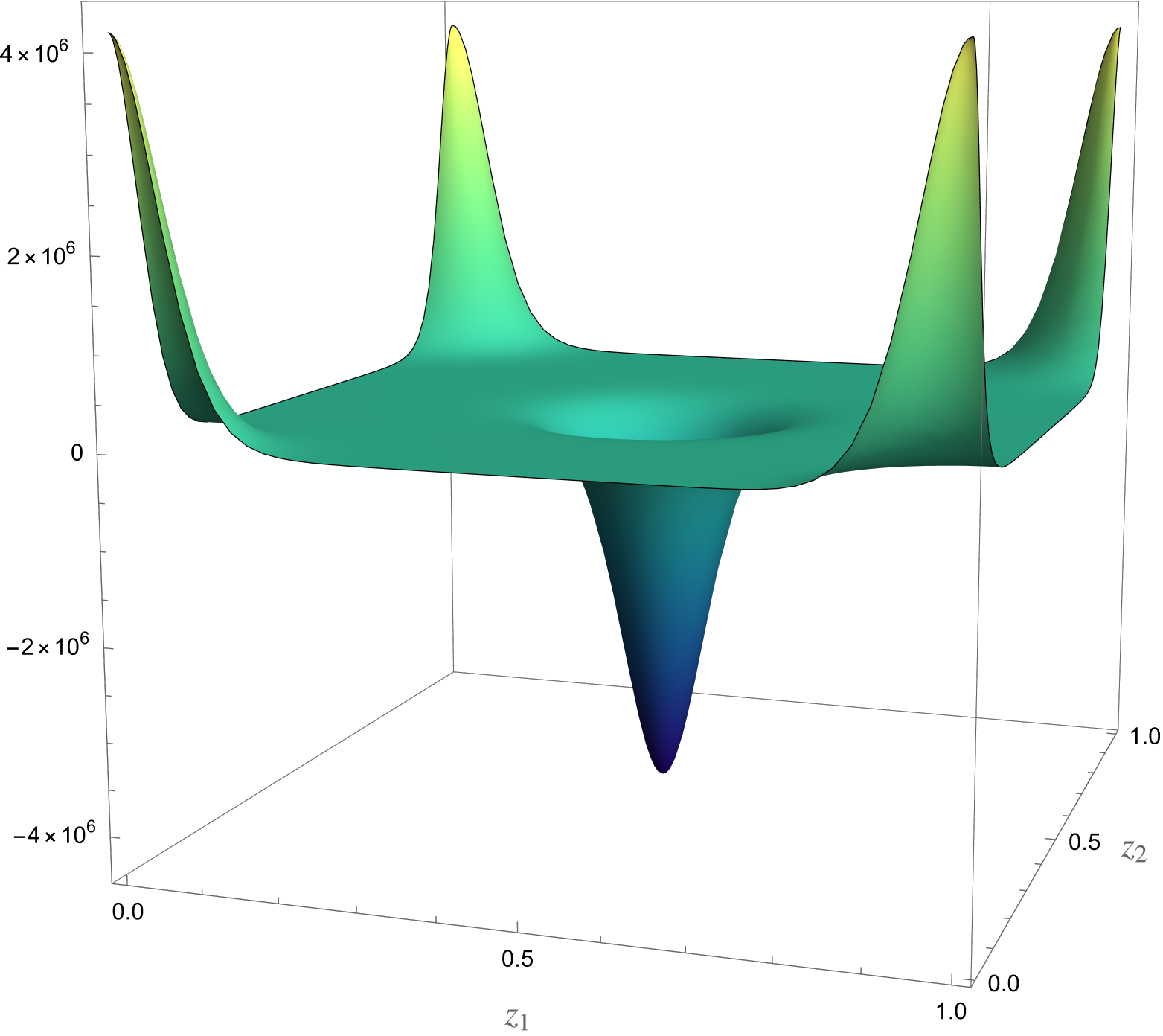}
    \caption{Anti-periodic $\vec{h}=(\frac12,\frac12,0)$.}
    \label{fig:CBrane-T3Z4-fermion}
\end{subfigure}
        
\caption{Casimir energy density from a field with periodic (anti-periodic) boundary conditions around the fibre. In both cases we identify clearly the Casimir branes at $(z_1,z_2) \in \{(0,0),(\frac12,\frac12)\}$; for periodic boundary conditions, the two Casimir branes have positive tension, but for anti-periodic boundary conditions they have opposite tension, which results in a cancellation for the total Casimir energy. We will discuss this phenomenon in detail in Section \ref{sec:alsym}.}
\label{fig:CBranes-T3Z4}
\end{figure}

Consider an alternative 3-dimensional RFM with holonomy group $\Z_4$, namely the manifold $T^3/\Z^4$ with the group generated by 
\begin{equation}
    \D[g] = \left(
        \begin{array}{*2{C{1.05em}}|*1{C{1.05em}}}
         0 & 1 & 0 \\
         -1 & 0 & 0 \\ \hline 
         0 & 0  & 1 
        \end{array}
    \right) \,, \quad
    \bvec[g] = \left(
        \begin{array}{c}
         0 \\ 0 \\ \frac14 
        \end{array} 
    \right) \,.
    \label{eq:Z4-action}
\end{equation}
Our condition \eq{mimi} allows for the choices $\vec{h} = (0,0,h_3)$ and $\vec{h}=(\frac12,\frac12,h_3)$; furthermore we must satisfy the additional condition \eq{swer2},
\begin{equation}
    s_\mathbf{g} = 2h_3 \,\text{mod}\,2\Z \,.
\end{equation}
For the $\Z_4$ action \eq{eq:Z4-action}, we have $\mathcal{D}_\mathbf{g}^4 = -\mathbf{I}$ and thus $s_\mathbf{g} = 1$, which constrains $h_3 = \frac12$. According to \eq{eq:eta-condition}, only the choice $\vec{h} = (0,0,\frac12)$ gives a non-zero Casimir energy. We can clearly see this effect by plotting the integrand for both choices: when the field has twisted boundary conditions around the fibre, regions of positive energy density cancel against regions of negative energy density (figure \ref{fig:CBranes-T3Z4}). We can interpret this result as the cancellation between two Casimir branes of opposite tension. For this RFM the Casimir branes are indeed localised at $(z_1,z_2) \in \left\{\left(0,0\right),\left(\frac12,\frac12\right)\right\}$ and wrapping $z_3$; the Casimir brane at $(z_1,z_2) = (\frac12,\frac12)$ has negative tension when anti-periodic boundary conditions are chosen.

\begin{figure}[!htb]
\centering
\begin{subfigure}{0.4\textwidth}
    \includegraphics[width=\textwidth]{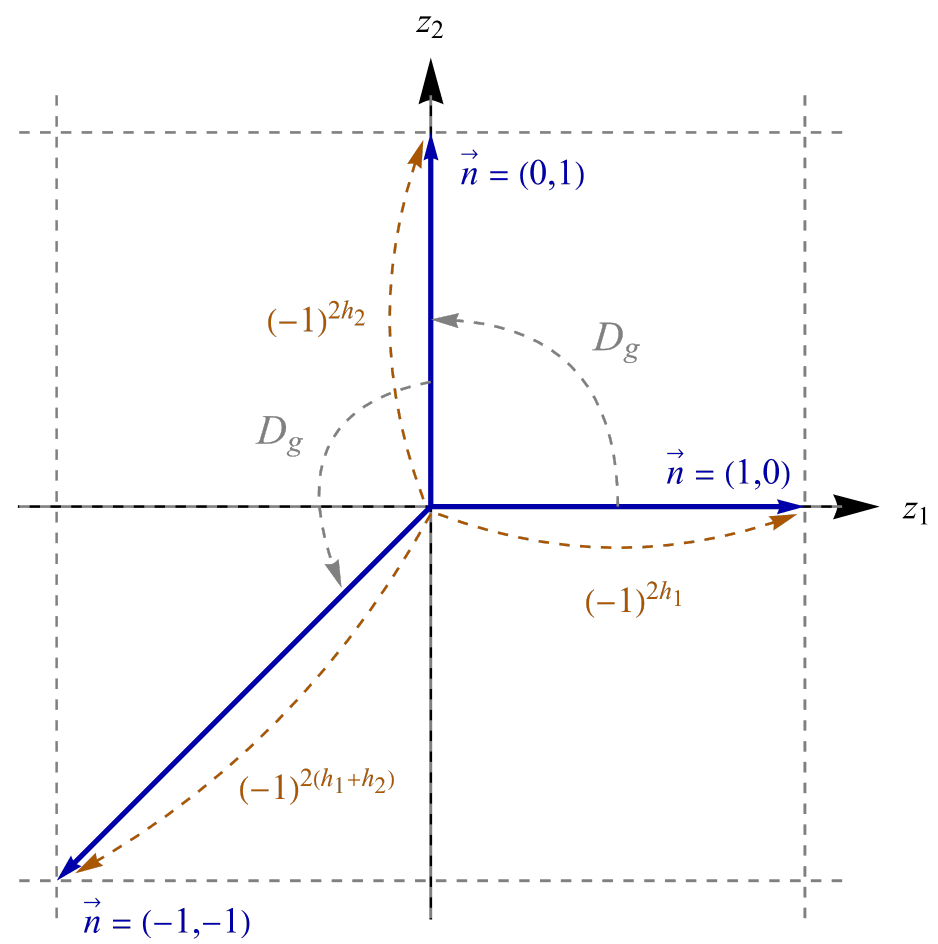}
    \caption{$T^3/\Z_3$}
    \label{fig:spin-structures-T3Z3}
\end{subfigure}
\hspace{0.1\textwidth}
\begin{subfigure}{0.4\textwidth}
    \includegraphics[width=\textwidth]{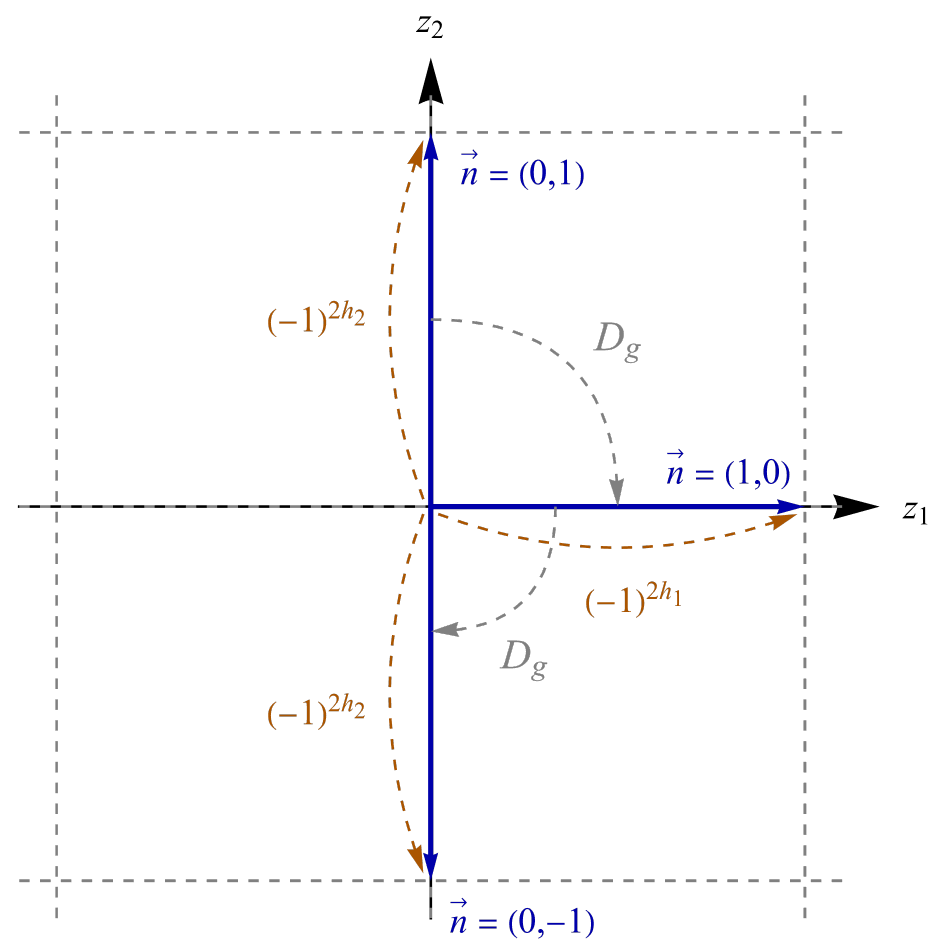}
    \caption{$T^3/\Z_4$}
    \label{fig:spin-structures-T3Z4}
\end{subfigure}
\caption{Visualising the constraints on spin structures that are preserved under the diffeomorphisms $\D[g]$ for the $\Z_3$ \eq{eq:Z3-action} and $\Z_4$ \eq{eq:Z4-action} actions on $T^3$. The action on a fermionic field under a translation by a lattice vector $\vec{n}$ is given by $(-1)^{2\vec{h}\cdot\vec{n}}$. For example, since the vector $(1,0)$ is mapped to $(0,1)$ under the $\Z^3$ action $\D[g]$, a consistent spin structure must give the same action (sign) upon a translation by $(1,0)$ and $(0,1)$; thus $h_1 = h_2\,\text{mod}\,\Z$. Moreover, since $(0,1)$ is in turn mapped to $(-1,-1)\sim (1,1)$ on the covering $T^3$, a translation by $(0,1)$ must have the same effect on a spinor as a translation by $(1,1)$, i.e. $h_2 = h_1 + h_2 \,\text{mod}\,\Z$. Together, these restrictions fix $h_1 = h_2 = 0\,\text{mod}\,\Z$. Repeating the exercise with the $\Z_4$ action, one still finds $h_1 = h_2\,\text{mod}\,\Z$, but both signs remain consistent as long as they are shared by the two directions.}
\label{fig:spin-structures-T3}
\end{figure}

One can use these two 3-dimensional RFM's to visualise the spin structure consistency condition \eq{mimi}, recalling its derivation in Section \ref{sec:RFMs} (figure \ref{fig:spin-structures-T3}), by omitting the subspace left invariant by $\D[g]$ (in both cases the $z_3$ direction). The condition \eqref{swer2} can also be understood from these pictures: the action $\D[g]^3$ leaves all lattice vectors invariant, $\D[g]^3\,\vec{z} = \vec{z}$; the only effect of this action is therefore a translation by the lattice vector $(0,0,1) = 3\,\bvec[g]$, which must be compatible with the corresponding action $(-1)^{2h_3}$. This must then be the sign of $\mathcal{D}_\mathbf{g}^3$.  

Finally, let us compare our method with a result known in the literature. We will use \eq{eq:VC-invariant-subspace} to compute the Casimir energy for M-theory on the interval $S^1/\Z_2$ and compare the result with equations (2.12) and (2.13) of \cite{Fabinger:2000jd}\footnote{Equation (2.13) in \cite{Fabinger:2000jd} is written in terms of the Jacobi theta function $\theta_4(0,i\tau)$ and only argued to be convergent and positive. However, one can compute it analytically using Poisson resummation and a Mellin transform---just as in the derivation of our general formula---to find \eq{eq:Fabinger-Horava}.},
\begin{align}
    \Vcas^{\text{M-theory on }S^1/\Z_2} = -\frac{\mathcal{J}}{L^{10}}\,,
    && \mathcal{J} = \frac{\Gamma(\frac{11}{2})}{2\pi^{11/2}}\frac{2^{11}-1}{2^{13}}\zeta(11) \approx 1.206\cdot 10^{-2} \,.
    \label{eq:Fabinger-Horava} 
\end{align}
Strictly speaking, this case is outside of the realm of validity of our method, since the $\mathbb{Z}_2$ leading to the Horava-Fabinger interval $S^1/\mathbb{Z}_2$ has fixed points (it acts by sending $z\rightarrow-z$). Therefore, this is \emph{not} an RFM, but rather, a Riemann-flat orbifold (away from the singularities). In our notation, this would correspond to $\D[g]=-1$, $\bvec[g]=0$. Nonetheless we may still apply our formula \eq{eq:VC-invariant-subspace}, summing over the two elements of this $\Z_2$, to find precisely the result \eqref{eq:Fabinger-Horava}. In fact this contribution comes uniquely from the identity term, since our formula is proportional to the size of the invariant subspace, which for the non-trivial element of this $\Z_2$ action corresponds to the two individual fixed points $z\in\{0,\frac12\}$ that have zero volume. More accurately, the twisted sector only contributes through the divergences associated with the orbifold (and so, in this case, it is a twisted sector in the standard sense). However since the twisted sectors correspond to a Horava-Witten $E_8$ brane and its antibrane in this example, we know the twisted sector contribution vanishes. It is also important to note that $L$ used here denotes the size of the interval, which is related to the size of the covering $S^1$ as $R=2L$. 

We are now in a position to compute the Casimir potential for our 7d solution of M-theory on $T^4$ discussed in Section \ref{sec:warm-up-dS7}. It is simply  
\begin{equation} 
    \Vcas^{\text{M-theory on }T^4} = -\frac{\Gamma\left(\frac{11}{2}\right)}{2\pi^{11/2}}\sqrt{G}\sum^{\sim}_{\substack{\vec{n}\in\mathbb{Z}^4}} \frac{\Tr{\mathbf{44}}{\mathbf{I}} + \Tr{\mathbf{84}}{\mathbf{I}} - \Tr{\mathbf{128}}{\mathbf{I}}\, e^{2\pi i\,\vec{h}\cdot\vec{n}} }{\vert\vec{n}\vert^{11}} \,,
    \label{eq:Casimir-T4}
\end{equation}
since there is no quotienting and thus the only element in the $\gamma\in\Gamma$ sum is the identity (covering $T^4$); we also include the sum over the representations in M-theory, i.e. the graviton $(\mathbf{44})$, the 3-form $(\mathbf{84})$ and the gravitino $(\mathbf{128})$, which for the identity are simply their respective dimensions. For the specific symmetric point in moduli space where the lattice $\Lambda$ defining the $T^4$ is the root lattice of $C_4$ with its standard inner product, written in a basis in which it takes the form
\begin{equation}
    \mathbf{G} = R^2 \begin{pmatrix}
        2 & -1 & 0 & 1 \\
        -1 & 2 & -1 & 0 \\
        0 & -1 & 2 & -1 \\
        1 & 0 & -1 & 2 
    \end{pmatrix}
    \label{eq:C4-metric-T4}
\end{equation}
and choosing boundary conditions for the gravitino that are twisted around every direction, $\vec{h}=\{\frac12,\frac12,\frac12,\frac12\}$, the Casimir energy is
\begin{equation}
    \Vcas^{\text{M-theory on }T^4} = -128\cdot\frac{945}{64\pi^5}\cdot 2R^4\sum^{\sim}_{\substack{\vec{n}\in\mathbb{Z}^4}} \frac{1 - (-1)^{n_1+n_2+n_3+n_4}}{\vert\vec{n}\vert^{11}} \approx -\frac{8.827}{R^7} \,,
    \label{eq:Casimir-energy-density-T4}
\end{equation}
which is precisely the Casimir contribution to the potential \eqref{f09} in our $dS_7$ example of Section \ref{sec:warm-up-dS7}.
Note that we chose a basis that does not correspond to the inner-product matrix in the $C_4$ root-basis\footnote{Recall that for a non-simply laced group such as $C_4$ the inner-product matrix for the root lattice does not directly match its Cartan matrix.},

\begin{equation}
    \begin{pmatrix}
        2 & -1 & 0 & 0 \\
        -1 & 2 & -1 & 0 \\
        0 & -1 & 2 & -2 \\
        0 & 0 & -2 & 4 
    \end{pmatrix} \,,
\end{equation}
but is rather related to it through the transformation
\begin{equation}
    \begin{pmatrix}
        z_1 \\ z_2 \\ z_3 \\ z_4 
    \end{pmatrix}
    \to \left(
        \begin{array}{*4{C{1em}}}
            1 & 0 & 0 & 1 \\
            0 & 1 & 0 & 1 \\
            0 & 0 & 1 & 1 \\
            0 & 0 & 0 & 1
        \end{array}
        \right)\begin{pmatrix}
        z_1 \\ z_2 \\ z_3 \\ z_4 
    \end{pmatrix} \,.
\end{equation}
We will see in Section \ref{sec:dS5-maximum} that it is the form \eqref{eq:C4-metric-T4} that naturally appears at special points in the $T^6/\Z_8$ moduli space---of course, the lattice is still the $C_4$ root lattice.

\subsection{Casimir branes and Atkin-Lehner symmetry}\label{sec:alsym}

The main goal of this paper is to achieve Casimir-de Sitter vacua. In this section, which is somewhat out of this main line of development, we wish to emphasize a very interesting phenomenon that we observed while deriving \eq{eq:VC-invariant-subspace}. As explained in the discussion right after this equation, the expression for the integrated Casimir energy contains a factor $\hat{\delta}_{\vec{h}}$ which evaluates to zero whenever a certain condition on the $\vec{h}$ vector of spin structures  (equation \eq{eq:eta-condition}) is not satisfied. Within our context, this can only happen for fermionic fields\footnote{Whenever one considers non-trivial bundles for discrete symmetries in an RFM, this will also happen for bosons. However, it can never happen for the zero mode of the graviton, which is necessarily uncharged under all internal symmetries.}. Therefore, we have found a mechanism that will ensure the vanishing of (one-loop, twisted) contributions to the Casimir energy of fermions, even in the absence of supersymmetry---we would like to understand this in some detail.

First, consider Figure \ref{fig:CBrane-T3Z4-fermion}, which shows a plot of the Casimir energy where $\hat{\delta}_{\vec{h}}$ vanishes. We see that there are two Casimir branes at different locations, so the energy density is non-trivial. The vanishing vacuum energy is achieved by a detailed cancellation of the tensions between these two branes. This is reminiscent of what happens at tree level in supersymmetric orientifold compactifications---but now, it is happening at one-loop, and in a fully non-supersymmetric context. 

The apparent symmetry of Figure \ref{fig:CBrane-T3Z4-fermion} suggests that to explain this vanishing we should look for a translation symmetry under which the integrand of \eq{csum} is odd.  Indeed, a  shift $\vec{z}\to\vec{z}+\vec{e}$ in the integrand can be compensated by a shift of the summation variable $\vec{n}$ whenever
\begin{equation} 
    (\mathbf{I}-\mathbf{D}_\gamma)\cdot \vec{e}\in\mathbb{Z}^k \,,
    \label{c0}
\end{equation}
at the cost of a phase factor 
\begin{equation} e^{2\pi i\,[(\mathbf{I} - \D[g])^T\vec{h}]\cdot\vec{e}}.\label{e3323}\end{equation}
If the integral is not to vanish identically, for any solution to \eq{c0} the corresponding phase \eq{e3323} must be trivial. In other words, we must have
\begin{equation}
    \vec{h}\cdot  (\mathbf{I}-\mathbf{D}_\gamma)\, \vec{e}\in\mathbb{Z},\quad \text{whenever}\quad (\mathbf{I}-\mathbf{D}_\gamma)\,\vec{e}\in\mathbb{Z}^k \,.
    \label{rerer}
\end{equation}
If \eq{eq:eta-condition} holds, so that $\hat{\delta}_{\vec{h}}=1$, then $\vec{h}$ can be replaced by $\vec{\eta}\in\mathbb{Z}^k$ and \eq{rerer} holds automatically. Conversely, equation \eq{rerer} is the statement that $\vec{h}$ is in the dual lattice to the lattice of vectors of the form $(\mathbf{I}-\mathbf{D}_\gamma)\, \vec{e}\in\mathbb{Z}^k$. This lattice is primitive in $\mathbb{Z}^k$, and therefore, by arguments similar to those leading to \eq{mam}, we can conclude that the dual lattice is spanned by vectors of the form
\begin{equation} 
    (\mathbf{I} - \D)^T\vec{\eta} +\vec{\xi}_{\text{Inv}},\quad \vec{\eta}\in\mathbb{Z}^k,\quad \D^T\, \vec{\xi}_{\text{Inv}}=\vec{\xi}_{\text{Inv}}\,.
\end{equation}
The statement that $\vec{h}$ is in the dual lattice is then  \eq{eq:eta-condition}. Thus, $\hat{\delta}_{\vec{h}}$ only vanishes when one can find a non-trivial vector $\vec{e}$. 

In summary, what happens is the following: We have an explicit formula for the Casimir energy, equation \eq{csum}, which gives us (a contribution to) the one-loop vacuum energy as an integral of some function (the 11-dimensional energy density $\hat{\rho}(\vec{z})$) on a domain (the RFM),
\begin{equation} 
    V_{\text{Cas}}\sim \int_{\text{RFM}} \hat{\rho}(\vec{z})\,dV\,,
\end{equation}
and we have found an ``anomalous'' symmetry of the integrand, by which we merely mean that it transforms with a non-trivial phase,
\begin{equation} 
    \hat{\rho}(\vec{z}+\vec{e})= e^{i\varphi(\vec{e})}\, \hat{\rho}(\vec{z})\,,
\end{equation}
and which is also a symmetry of the integration region (the RFM, since it is flat).
As a result, the integral $V_{\text{Cas}}$ vanishes due to cancellations between different regions of the integration domain. Phrased in this way, the phenomenon we have found is quite reminiscent of Atkin-Lehner symmetry \cite{atkin1970hecke,Moore:1987ue,Dienes:1990qh}, a proposed mechanism to achieve vanishing vacuum energy at higher loops in perturbative string theory. In Atkin-Lehner symmetry, one writes the one-loop contribution to the vacuum energy in the worldsheet as the usual integral over the fundamental domain $\mathcal{F}$ \cite{Ibanez:2012zz},
\begin{equation} 
    V^{\text{1-loop}}\sim \int_{\mathcal{F}}\, \mathcal{Z}(\tau) \, \frac{d^2\tau}{\tau_2} \,.
    \label{onelop}
\end{equation}
The idea of Atkin-Lehner symmetry is that one may find special worldsheet CFT's where the theory is symmetric with respect to additional torus diffeomorphsims not included in $SL(2,\mathbb{Z})$ (for instance, $\tau\rightarrow -1/(2\tau)$), but where this symmetry is anomalous, so that the partition function transforms with a non-trivial phase, 
\begin{equation}  
    \mathcal{Z}\left(-\frac{1}{2\tau}\right)= e^{i\mathcal{A}}\, \mathcal{Z}(\tau)\,.
    \label{als}
\end{equation}
As a result of \eq{als} and the fact that $\tau\rightarrow -1/(2\tau)$ is a symmetry of the integration domain $\mathcal{F}$, the integral vanishes.

To our knowledge, Atkin-Lehner symmetry has only ever been discussed in perturbative string theory and, in fact, explicit examples of worldsheet CFT's realizing it have only been constructed with a two-dimensional target space \cite{Moore:1987ue,Dienes:1990qh,Gannon:1992su}. Yet this mechanism is totally analogous to what we have just described for Casimir energies, under the correspondence
\begin{center}\renewcommand{\arraystretch}{1.2}
    \begin{tabular}{c|c}
\textbf{Worldsheet Atkin-Lehner symmetry}&\textbf{RFM Casimir energy symmetry}\\\hline
Torus worldsheet partition function $\mathcal{Z}(\tau)$& Higher-dim. Casimir energy density $\hat{\rho}(\vec{z})$\\
$SL(2,\mathbb{Z})$&RFM defining group $\mathcal{B}$\\
$SL(2,\mathbb{Z})$ fundamental domain& Riemann-flat manifold (integration region)\\
Atkin-Lehner symmetry $\mathcal{Z}(\tau)\rightarrow e^{i\mathcal{A}}\, \mathcal{Z}(\tau')$& Symmetry $\hat{\rho}(\vec{z}+\vec{e})= e^{i\varphi(\vec{e})}\, \hat{\rho}(\vec{z})$
\end{tabular}\end{center}
Thus, in some sense RFM's exhibit what may be the first example of a \emph{spacetime} Atkin-Lehner-like symmetry. This suggests that the mechanism of \cite{Moore:1987ue} is not intrinsically stringy, and similar phenomena are also under the reach of EFT. One important difference with respect to the worldsheet Atkin-Lehner symmetry is that, in the original formulation, the mechanism ensures the vanishing of the complete one-loop vacuum energy, while the field-theoretic version we have found in RFM's only works for some contributions (twisted sectors) to the vacuum energy, and for fermionic fields only (or more generally, for fields charged under internal symmetries with a non-trivial bundle on the RFM). Nevertheless, it seems possible to e.g. engineer QFT's with vanishing one-loop vacuum energy when compactified on appropriate RFM's.
Perhaps thinking along these lines can provide a new way to engineer solutions with an anomalously small vacuum energy.  We hope to return to these very interesting questions in the near future.

\subsection{Numerical evaluation of Casimir energies}
\label{sec:numerical-sums}
Although we have now a fully explicit, analytic formula to compute Casimir energies (i.e. the tension of Casimir branes), there is still the matter of evaluating \eq{eq:VC-invariant-subspace} explicitly in concrete examples. In general, there is no analytic expression for sums like \eq{eq:VC-invariant-subspace} \cite{borwein2013lattice}, so we must restort to numerical methods. 
A sum like \eq{eq:VC-invariant-subspace} is convergent, but the speed of convergence can be very slow depending on $s_\gamma$ and $\vec{\beta}_\gamma$. Furthermore, we will need to evaluate these sums for  high-dimensional lattices, up to dimension 7 for the examples in \secref{sec:dS4-maxima}. Moreover, to scan for $dS$ maxima we will need to evaluate the sum many times, for different values of the moduli. Under these circumstances a direct attack, by first truncating the sum, evaluating it, and then performing the integral numerically, will quickly become computationally prohibitive\footnote{We tried anyway, just to be sure, and it is indeed very bad.}.  

However, we are in luck: a sum like \eq{eq:VC-invariant-subspace} is a higher-dimensional analog of the sums employed to compute lattice energies of ionic crystals, for which efficient numerical methods have been developed. So to evaluate the sums efficiently, we will use the technique of Ewald resummation \cite{ewaldOG}, which is standard in computational chemistry (see e.g. \cite{TOUKMAJI199673}). The idea of the technique is as follows: Suppose one wants to evaluate a sum like that appearing in \eq{eq:VC-invariant-subspace}, schematically of the form
\begin{equation}\sum_{\vec{n}\neq-\vec{c}} \frac{e^{2\pi i \vec{h}\cdot\vec{n}}}{\vert \vec{n}+\vec{c}\vert^{2s}}.\label{ewaldOG}\end{equation}
If the quantity we are summing over was a smooth function of $\vec{n}$, we could do Poisson resummation and use the Fourier-converted sum to quickly compute the tails of the sum in position space. However, the function $\vert \vec{n}+\vec{c}\vert^{-2s}$ has a pole at $\vec{n}+\vec{c}=\vec{0}$, and therefore, a direct application of Poisson resummation will not yield convergent results. Instead, Ewald resummation proceeds by splitting the sum as

\begin{equation} 
    \sum_{\vec{n}\neq-\vec{c}} \frac{e^{2\pi i \vec{h}\cdot\vec{n}}}{\vert \vec{n}+\vec{c}\vert^{2s}}=   \sum_{\vec{n}\neq-\vec{c}}  \frac{F(\vec{n})\,e^{2\pi i \vec{h}\cdot\vec{n}}}{\vert \vec{n}+\vec{c}\vert^{2s}} +  \sum_{\vec{n}} \frac{(1-F(\vec{n}))\, e^{2\pi i \vec{h}\cdot\vec{n}}}{\vert \vec{n}+\vec{c}\vert^{2s}}\,,
    \label{mhg}
\end{equation}
where the function $F(\vec{n})$ is chosen such that 
\begin{equation}
    \lim_{\vec{y}\rightarrow-\vec{c}}\frac{1-F(\vec{y})}{\vert\vec{y}+\vec{c}\vert^{2s}}\quad \text{is finite, and }\,  F(\vec{y})\,\rightarrow0\,\text{ as $\vert\vec{y}\vert\rightarrow\infty$ faster than a polynomial} \,.
    \label{conds}
\end{equation}
Under these circumstances, the first sum in \eq{mhg} can be evaluated directly and converges quickly in real space. The second sum now is over all $\vec{n}$, and the conditions \eq{conds} ensure that it can be evaluated by Poisson resummation. As the function being summed over is smooth, its Fourier coefficients will decay quickly, and the momentum space sum will also converge rapidly.  Which precise form of $F(\vec{n})$ is best depends on the particular sum being evaluated;  the Coulomb case $s=1/2, k=3$ most often studied in chemistry is customarily tackled with $F(\vec{n})\propto  \text{erfc}(\alpha\vert\vec{n}\vert)$ \cite{TOUKMAJI199673}. Exactly the more general sum \eq{ewaldOG} was studied in \cite{NIJBOER1957309}, but only in three dimensions. We have extended these techniques to arbitrary dimension (details\footnote{Shortly after we  independently derived these results, \cite{buchheit2024} appeared, which uses similar ideas to achieve efficient evaluation of Ewald sums. The final expressions of \cite{buchheit2024} are equivalent to ours; the only practical difference is that the numerical implementation provided in \cite{buchheit2024} becomes significantly slow for higher-dimensional sums, due to details on how the sample points for the sum are selected in the numerical code. We produced an independent implementation which is more efficient for higher-dimensional cases, and was used for all results in this paper.} can be found in Appendix \ref{app:ewy}). The general result, applied to the integrand in \eq{csum}, reads as follows:
\begin{align}&\sum_{\vec{n}+\vec{c}\neq0}\frac{e^{2\pi i \vec{h}\cdot\vec{n}}}{\vert \vec{n}+\vec{c}\vert^{2s}}=\nonumber\\&\sum_{\vec{n}+\vec{c}\neq0}\frac{e^{2\pi i \vec{h}\cdot\vec{n}}}{\vert \vec{n}+\vec{c}\vert^{2s}} \frac{\Gamma(s,\alpha \vert \vec{n}+\vec{c}\vert^{2})}{\Gamma(s)}\,\, +\,\, \frac{\pi^{2s-\frac{k}{2}}}{\Gamma(s)\sqrt{G}} \sum_{\vec{k}-\vec{h}\neq0}\, \Gamma \left(\frac{k}{2}-s,\frac{\pi ^2 \vert\vec{h}-\vec{k}\vert_D^2}{\alpha }\right) \frac{e^{-2\pi i(\vec{h}-\vec{k})\cdot\vec{c}}}{ \vert\vec{h}-\vec{k}\vert_D^{k-2s}}\nonumber\\&+ \delta_{\vec{h},\vec{0}}\, \frac{\pi^{\frac{k}{2}}\alpha^{s-k/2}}{\sqrt{G}\, \Gamma(s)\left(s-\frac{k}{2}\right)}-\frac{\alpha^s\, e^{-2\pi i \vec{h}\cdot\vec{c}}}{\Gamma(s+1)}\chi_{\mathbb{Z}^k}(\vec{c}).\label{ewald0} \end{align}
In practice, we will evaluate these sums numerically, including a hard cutoff in momentum and position sums, and using the freedom in choosing $\alpha$ to ensure the value of fastest convergence (which will depend on the parameters of the sum and the cutoff in the sums, see Appendix \ref{app:num}). 
 In these expressions,
\begin{equation} \Gamma(s,z)\equiv \int_z^\infty t^{s-1}e^{-t}\,dt\end{equation}
is the incomplete $\Gamma$ function, and $\chi_{\mathbb{Z}^k}(\vec{c})$ is the characteristic function of the integers (it is 1 when all components of $\vec{c}$ are integers, and zero otherwise). Since $\vec{c}\in[0,1)^k$ as discussed above, this can only happen when $\vec{c}=\vec{0}$.  The norms in the second sum have a $_D$ subscript to indicate that they are dual lattice sums, where the inner products are to be taken with respect to $G^{-1}$, the natural metric in the dual space, as befits to a momentum space sum. Finally, $\alpha>0$ is a free parameter. The actual value of the sum does not depend on $\alpha$, but the relative contribution of the different terms do; $\alpha$ may then be chosen to optimize the speed of convergence of the overall sum. Since the incomplete $\Gamma$ function has the asymptotics
\begin{align}\Gamma(s,z)&\approx \Gamma(s)-\frac{z^s}{s}+\mathcal{O}(z^{s+1}),\quad \vert\vec{z}\vert\ll 1,\nonumber\\\Gamma(s,z)&\approx e^{-z}\, z^{s-1}\left[1+\mathcal{O}(z^{-1})\right]\quad \vert\vec{z}\vert\gg 1,\end{align}
in the limit $\alpha\rightarrow0$ the momentum space sum is completely suppressed, and one recovers the original expression in \eq{ewaldOG}. On the other hand, when $\alpha$ is very large, the position space sum is negligible, and the sum is dominated by the momentum space term. Notice that, in the regime of interest $2s>k$, the terms $\vert \vec{h}-\vec{k}\vert_D$ in the momentum sum diverge badly, so the regulating effect is due to the oscillating phases $e^{-2\pi i(\vec{h}-\vec{k})\cdot\vec{c}}$, which introduce large cancellations whenever $\vec{c}\neq0$. When $\vec{c}=\vec{0}$ or $\vec{h}=\vec{0}$, the terms in the second line of \eq{ewald0} are crucial for convergence, as is the $\Gamma$ function factor.

All numerical results for Casimir energies in this paper were computed via the Ewald formula \eq{ewald0}. While these are not really necessary for lower-dimensional examples, such as the two or three-dimensional examples at the end of the previous subsection where brute force suffices, an efficient implementation is a key requirement for the higher-dimensional sums that we will encounter in Sections \ref{sec:dS5-maximum} and \ref{sec:dS4-maxima}.

Finally, there is a nice physical interpretation of the Ewald result \eq{ewald0}, when particularized to Casimir sums of the form \eq{eq:VC-invariant-subspace}.  In that expression, the dual space sum (second term) in \eq{ewald0} is over the lattice $\Lambda^\parallel$, dual to $\Xi$,  which contains precisely the momenta $\vec{k}$ for which the plane wave $e^{2\pi i\,\vec{k}\cdot\vec{y}}$ is well-defined as a function on the RFM. These vectors are then shifted by $e^{2\pi i\, \vec{h}\cdot\vec{y}}$, which implements twisted boundary conditions. In other words, the second sum in \eq{ewald0} is a sum over the KK spectrum of the RFM, and $\vert\vec{h}-\vec{k}\vert$ is precisely the mass of the corresponding KK mode. Since $2s=D$, the total spacetime dimension, we have that $k-2s=-d$, where $d$ is the number of non-compact spacetime dimensions, and the second term in \eq{ewald0} can be recast as a regularized version of the sum
\begin{equation}\sum_{\text{KK spectrum}} m_{\text{KK}}^d,\end{equation}
of zero-point energies of KK modes, over the whole KK spectrum. We recognize the standard, UV divergent, computation of vacuum energies as a sum of one-loop diagrams/zero-point energies over the particle spectrum of the lower-dimensional theory.  Doing this sum directly in lower-dimensional EFT leads to divergences that have to be regularized and removed e.g. by imposing higher-dimensional Lorentz invariance in the decompactification limit. This is the standard sum of one-loop vacuum bubble diagrams commonly employed to compute Casimir energies \cite{Arkani-Hamed:2007ryu} in momentum space, and is dual to the position space point-splitting method we used here to regularize the Casimir stress-energy tensor in Subsection \ref{sec:casimir-massless-fields}. The Ewald expression \eq{ewald0} then provides a regularization for both terms, and includes a parameter $\alpha$ that smoothly interpolates between a purely momentum (at $\alpha\rightarrow\infty$) or a purely position sum (at $\alpha\rightarrow0$). One can then regard the $\alpha$-independence of the result as a consequence of the fact that physical results are independent of the regularization scheme. The advantage of the position space point-splitting method over the momentum one is that it also produces an expression for the full higher-dimensional backreacted stress-energy tensor, as stressed in \secref{sec:warm-up-dS7}, which means that the higher-order corrections can in principle be computed and evaluated explicitly, as discussed there. 

\section{A \texorpdfstring{$dS_5$}{dS5} maximum in M-theory}
\label{sec:dS5-maximum}
After setting up all the groundwork, we are finally ready to come to the main point of this paper: an explicit de Sitter maximum solution which is under theoretical control.  The vacuum we have in mind is a compactification of M-theory on $dS_5\times \mathcal{F}_6$, where $\mathcal{F}_6$ is a particular Riemann-flat manifold, threaded by $G_4$-flux. The five-dimensional scalar potential will arise from just two sources: a Casimir and a $G_4$-flux piece,
\begin{equation}
    V^{(5d)} = \Vcas + V_{G_4} \,,
    \label{totpot}
\end{equation}
where the fluxes will be carefully chosen to ensure that $V^{(5d)}$ has a saddle point in all directions---including the volume. We will first present the manifold $\mathcal{F}_6$, then study its Casimir potential using the formulae in \secref{sec:Casimir-on-RFMs}, followed by a detailed study of the flux potential, and a determination of the maximum parameters. Finally, we will study corrections to the solution, describing in which sense they are small. 

\subsection{The Riemann-flat manifold \texorpdfstring{$\mathcal{F}_6$}{F6}}
\label{subsec:dS5-maximum-RFM}
To construct our de Sitter maximum, we will compactify M-theory on the RFM $\mathcal{F}_6$, described as follows. Consider a $T^6$ parametrized by coordinates $\vec{z}=(z_1,z_2,z_3,z_4,z_5,z_6)$, subject to the identifications
\begin{equation}
    z_i\,\sim\,z_i+1 \,.
\end{equation}
We will quotient by the $\mathbb{Z}_8$ action generated by the affine transformation
\begin{equation}
    \vec{z}\,\rightarrow\iota_{\mathbf{g}}(\vec{z})=\D[g]\,\vec{z}+\bvec[g] \,, 
    \quad\text{with}\quad 
    \D[g]\equiv\left(
        \begin{array}{*6{C{1em}}}
             0 & 0 & 0 & -1 & 0 & 0 \\
             1 & 0 & 0 & 0 & 0 & 0 \\
             0 & 1 & 0 & 0 & 0 & 0 \\
             0 & 0 & 1 & 0 & 0 & 0 \\
             0 & 0 & 0 & 0 & 1 & 0 \\
             0 & 0 & 0 & 0 & 0 & 1 \\
        \end{array}
        \right)\,,\quad \bvec[g]=\left(\begin{array}{c}0\\0\\0\\0\\0\\ \frac18\end{array}\right) \,.
    \label{Z8-rfmdef}
\end{equation}
This action has no fixed points on $T^6$, since the vector $\bvec[g]$ lies in the invariant subspace of $\D[g]$. The matrix $\D[g]$ is of order 8, and the manifold $\mathcal{F}_6$ is defined as the quotient
\begin{equation}
    \mathcal{F}_6\equiv T^6/\mathbb{Z}_8 \,.
    \label{cqw}
\end{equation}
We must now endow $\mathcal{F}_6$ with a Riemann-flat metric. As discussed in \secref{sec:RFMs}, they all come from Riemann-flat metrics on the covering $T^6$ that are invariant under $\D[g]$. The most general possibility is
\begin{align}
    ds^2 &= G_{ij}dz^i\, dz^j \,, \nonumber \\ 
    \mathbf{G} &=\left(
    \begin{array}{cccccc}
         (\gamma +2) R_2^2 & -(\gamma +1) R_2^2 & 0 & (\gamma +1) R_2^2 & 0 & 0 \\
         -(\gamma +1) R_2^2 & (\gamma +2) R_2^2 & -(\gamma +1) R_2^2 & 0 & 0 & 0 \\
         0 & -(\gamma +1) R_2^2 & (\gamma +2) R_2^2 & -(\gamma +1) R_2^2 & 0 & 0 \\
         (\gamma +1) R_2^2 & 0 & -(\gamma +1) R_2^2 & (\gamma +2) R_2^2 & 0 & 0 \\
         0 & 0 & 0 & 0 & \beta  R_1^2 & \alpha  R_1^2 \\
         0 & 0 & 0 & 0 & \alpha  R_1^2 & \beta^{-1} R_1^2 \\
    \end{array}
\right) \,. 
    \label{eq:Z8-generator}
\end{align}
There are five moduli, parametrized here as $\{R_1,R_2,\alpha,\beta,\gamma\}$---this means that 16 of the 21 moduli of $T^6$ have been fixed by the quotient \eq{cqw}. Since we are looking for a dS maximum, we do not need to stabilize all these moduli; a saddle point, where the first derivative of the potential vanishes, will suffice. We will now identify a subspace of the moduli space, given by specific values $\{\alpha,\beta,\gamma\}$, where the partial derivatives along these directions vanish automatically due to symmetry arguments (see also \cite{Parameswaran:2024mrc,Chen:2014fqa});  we will not have to worry about them henceforth. To do this, notice that an affine transformation $\vec{z}\,\rightarrow f(\vec{z})=\mathbf{A}\,\vec{z}+\vec{b}$ of the parent $T^6$ will descend to a well-defined diffeomorphism of $\mathcal{F}_6$ if the points $\vec{z}$ and $\mathbf{A}\,\vec{z}+\vec{b}$ of $T^6$ are in the same $\mathbb{Z}_8$ orbit generated by \eq{Z8-rfmdef}, or in other words, if
\begin{equation} 
    f(\iota_\mathbf{g}(\vec{z}))=\iota_{\mathbf{g}}^{n_1}(f(\vec{z})) \,.
    \label{normy}
\end{equation}
for some $n_1\in\mathbb{Z}$. 
This is the condition that the affine transformation $f(\vec{z})$ is in the normalizer of the $\mathbb{Z}_8$ generated by $\iota_{\mathbf{g}}(\vec{z})$\footnote{The error that affected previous versions of the paper was here;  we included condition \eq{blorp1}, but neglected \eq{blorp2} in the analysis. Although it is correct for symmetries fixing the moduli $\alpha$ and $\gamma$, there is no symmetry fixing $\alpha, \beta$ simultaneously, as was erroneously claimed in the previous versions of this paper. As a result, we are forced to stabilize $\beta$ numerically, which is what the rest of this section does.}. Working this out explicitly, one obtains
\begin{equation} 
    \mathbf{A}\,\D[g]\, \vec{z} + \mathbf{A}\,\vec{b}_g+\vec{b}= \D[g]^{n_1}\,\mathbf{A}\, \vec{z} + \D[g]^{n_1}\,\vec{b}+ \left(\sum_{k=0}^{n_1-1}\D[g]^k\,\bvec[g]\right)+\vec{n} \,,
\end{equation}
or equivalently, since this should hold for any $\vec{z}$,
\begin{align} 
    \mathbf{A}\cdot \D[g]\cdot \mathbf{A}^{-1}&=\D[g]^{n_1} \,,\label{blorp1}\\ 
    (\mathbf{I}-\D[g]^{n_1})\,\vec{b}&=\sum_{k=0}^{n_1-1}\D[g]^k\,\bvec[g]-\mathbf{A}\,\bvec[g]+\vec{n} \,.\label{blorp2}
\end{align}
In the particular case where $\D[g]\,\vec{b}=\vec{b}$, \eq{blorp2} simplifies to
\begin{equation}
    \mathbf{A}\,\bvec[g]= n_1\,\bvec[g]+\vec{n} \,.
\end{equation}
For $\D[g]$ as in \eq{Z8-rfmdef}, one possibility is 
\begin{equation}
    \mathbf{A}=\left(
    \begin{array}{*6{C{1em}}}
         1 & 0 & 0 & 0 & 0 & 0 \\
         0 & 1 & 0 & 0 & 0 & 0 \\
         0 & 0 & 1 & 0 & 0 & 0 \\
         0 & 0 & 0 & 1 & 0 & 0 \\
         0 & 0 & 0 & 0 & -1 & 0 \\
         0 & 0 & 0 & 0 & 0 & 1 \\
    \end{array}
\right),\label{az4}\end{equation}
a matrix of order 2 that in fact commutes with $\D[g]$ and therefore satisfies the above with $n_1=1$. The matrix \eq{az4} therefore descends to a well-defined diffeomorphism on $\mathcal{F}_6$. Its action on the metric 
\begin{equation}\mathbf{G}\,\rightarrow\mathbf{A}^T\cdot\mathbf{G}\cdot\mathbf{A}=\left(
\begin{array}{cccccc}
 (\gamma +2) R_2^2 & -(\gamma +1) R_2^2 & 0 & (\gamma +1) R_2^2 & 0 & 0 \\
 -(\gamma +1) R_2^2 & (\gamma +2) R_2^2 & -(\gamma +1) R_2^2 & 0 & 0 & 0 \\
 0 & -(\gamma +1) R_2^2 & (\gamma +2) R_2^2 & -(\gamma +1) R_2^2 & 0 & 0 \\
 (\gamma +1) R_2^2 & 0 & -(\gamma +1) R_2^2 & (\gamma +2) R_2^2 & 0 & 0 \\
 0 & 0 & 0 & 0 & \beta^{-1}  R_1^2 & -\alpha  R_1^2 \\
 0 & 0 & 0 & 0 & -\alpha  R_1^2 & \beta R_1^2 \\
\end{array}
\right)\end{equation}
maps $\alpha\,\rightarrow-\alpha$, leaving $\beta$ and $\gamma$ invariant. Hence, at $\alpha=0$ it becomes a $\mathbb{Z}_2$ isometry. The component of the gradient $\partial_\alpha V$ transforms as a vector under this $\mathbb{Z}_2$, and therefore, if the $G_4$ flux choice respects this isometry (something that will be ensured in Subsection \ref{subsec:dS5-maximum-flux-potential}), we will have $\partial_\alpha V=0$ as desired. 

We can run a somewhat similar argument with
\begin{equation}
    \mathbf{A}=\left(
    \begin{array}{*6{C{1em}}}
         1 & -1 & 1 & 0 & 0 & 0 \\
         1 & -1 & 0 & 1 & 0 & 0 \\
         1 & 0 & -1 & 1 & 0 & 0 \\
         0 & 1 & -1 & 1 & 0 & 0 \\
         0 & 0 & 0 & 0 & 1 & 0 \\
         0 & 0 & 0 & 0 & 0 & 1 \\
    \end{array}
    \right) \,,
    \label{aD4}
\end{equation}
which satisfies \eq{blorp1} with $n_1=3$. In this case, \eq{blorp2} is not satisfied; however, in the Casimir potential \eq{eq:VC-invariant-subspace}, the twisted sector terms depend only on the metric restricted to the invariant subspace (which is spanned by the last two coordinates), while the identity term has $\bvec[g]=0$ and hence satisfies \eq{blorp2}. Hence, although \eq{aD4} does not descend to a well-defined isometry on $\mathcal{F}_6$, it acts in the Casimir and flux potentials as if it was. In particular, this transformation acts on $\mathbf{G}$ by sending $\gamma\rightarrow-\gamma$. Since \eq{aD4} is of order 2, at $\gamma=0$ this is a $\mathbb{Z}_2$ isometry under which $\partial_\gamma V$ picks up a sign. It follows that, as long as fluxes are chosen appropriately to respect these symmetries, at the locus of moduli space given by
\begin{equation} \mathbf{G}= R^2_1\, \left(
    \begin{array}{*6{C{1.3em}}}
         0 & 0 & 0 & 0 & 0 & 0 \\
         0 & 0 & 0 & 0 & 0 & 0 \\
         0 & 0 & 0 & 0 & 0 & 0 \\
         0 & 0 & 0 & 0 & 0 & 0 \\
         0 & 0 & 0 & 0 & $\beta^{-1}$ & 0 \\
         0 & 0 & 0 & 0 & 0 & $\beta$
    \end{array}\right)
    + R^2_2\, \left(
    \begin{array}{*6{C{1em}}}
         2 & -1& 0 & 1 & 0 & 0 \\
         -1 & 2& -1 & 0 & 0 & 0 \\
         0 & -1 & 2 & -1 & 0 & 0 \\
         1 & 0 & -1 & 2 & 0 & 0 \\
         0 & 0 & 0 & 0 & 0 &0 \\
         0 & 0 & 0 & 0 & 0 &0
    \end{array}
    \right) \,,
    \label{ewmmm}
\end{equation}
the potential is automatically a saddle for $\{\alpha,\gamma\}$. In fact, after a change of basis where the first four coordinates are transformed as
\begin{equation}
    \left(\begin{array}{c}z_1\\z_2\\z_3\\z_4\\\end{array}\right)\,\rightarrow \left(
    \begin{array}{*4{C{1em}}}
         1 & 0 & 0 & -1 \\
         0 & 1 & 0 & -1 \\
         0 & 0 & 1 & -1 \\
         0 & 0 & 0 & 1 \\
    \end{array}
    \right)\cdot\left(\begin{array}{c}z_1\\z_2\\z_3\\z_4\\\end{array}\right) \,,
\end{equation}
the upper left block of the term proportional to $R_2$ in \eq{ewmmm} becomes precisely the Cartan matrix of the $C_4$ algebra.
Thus, the special locus of moduli space we have found is such that the parent $T^6$ defining lattice is $C_4(R_2)\oplus\Lambda_R$---the $C_4$ root lattice, rescaled by a factor of $R_2$, plus a rectangular lattice $\Lambda_R$\footnote{The $D_4$ and $C_4$ root lattices are identical, so all the discussion here can be equivalently phrased in terms of the $D_4$ root lattice. In this picture the transformation \eq{ewmmm} is not part of the Weyl group of $D_4$; rather, an element in this Weyl group needs to be combined with an outer automorphism of the $D_4$ algebra of order 2 to obtain \eq{ewmmm}. The two pictures are of course equivalent.}. The transformation \eq{aD4} is precisely one of the reflections in the Weyl group of the $C_4$ lattice.  Describing the root lattice in terms of an ambient $\mathbb{R}^4$ space with basis $\{e_1,e_2,e_3,e_4\}$ equipped with standard inner product, the $C_4$ roots are vectors of the form $\pm (e_i-e_j)$ plus $\pm 2e_i$. The Weyl group acts by permuting the roots and flipping the sign of any number of them \cite{Carter1972}. In terms of these variables, \eq{aD4}
corresponds simply to the $\mathbb{Z}_2$ transformation that sends
\begin{equation}
    e_1\,\rightarrow e_1 \,,
    \quad e_2\,\leftrightarrow\,e_4 \,,
    \quad e_3\,\rightarrow\,-e_3 \,.
\end{equation}
In short, the reason why this special locus of moduli space exists is because the $T^6$ lattice becomes a root lattice for some simple Lie algebra, and has a large isometry group corresponding to the Weyl group of the lattice. This technique to find saddle points of the potential by searching for root lattices will also be helpful in other setups, such as those of \secref{sec:dS4-T7-example}.

For the question of Casimir energies it will be important to specify the spin structure on $\mathcal{F}_6$. As explained in \secref{sec:RFMs}, the choices of spin structure are labelled by a vector $\vec{h}$ whose components are all 0 or 1/2 modulo 1, and encode the periodicity of fermions along the six cycles of $T^6$. However, not every spin structure on $T^6$ descends to a well-defined spin structure on $\mathcal{F}_6$; the allowed ones are determined by solving equations \eq{mimi} and \eq{swer2}. Since we are quotienting by a group of even order, the spin lift of the generator $\mathcal{D}_\mathbf{g}$ satisfies $\mathcal{D}_\mathbf{g}^8 = \mathbf{I}$. For $\mathcal{F}_6$, there are then 2 possibilities for $\vec{h}$ compatible with the isometries discussed above, 
\begin{equation} 
    \left(0,0,0,0,0,0\right) 
    \quad \text{or} \quad 
    \left(\frac12,\frac12,\frac12,\frac12,0,0\right) \,.
\end{equation}
In other words, we may only choose the spin structure on the $C_4$ lattice (it is fixed to be periodic on the $\mathbb{Z}^2$),
and we will choose $(\frac12,\frac12,\frac12,\frac12,0,0)$, where fermions are antiperiodic along all coordinates of $C_4$, since the alternative will not yield a dS saddle.
As we will see in \secref{sec:dS4-maxima}, most cyclic RFM's do not allow for an antiperiodic spin structure on the subspace where the point action acts transitively; in fact, out of all the possibilities discussed in Section \ref{sec:dS4-maxima}, only the one used in this section and those closely related to it (by which we mean that they have the same $\mathbb{Z}_8$ block as the one discussed here) allow for antiperiodic boundary conditions. These turn out to be crucial to achieve a saddle point, which explains our particular choice \eq{Z8-rfmdef} in this section. Relatedly, any $SL(6,\mathbb{Z})$ transformation with the same block decomposition (in the sense of \secref{sec:RFMs}) as $\D[g]$ can be brought, via an $SL(6,\mathbb{Z})$ change of basis (is conjugate) to either $\D[g]$ or the closely related transformation 
\begin{equation}\left(
    \begin{array}{*6{C{1em}}}
         0 & 0 & 0 & 1 & 0 & 0 \\
         1 & 0 & 0 & 0 & 0 & 0 \\
         0 & 1 & 0 & 0 & 0 & 0 \\
         0 & 0 & 1 & 0 & 1& 0 \\
         0 & 0 & 0 & 0 & 1 & 0 \\
         0 & 0 & 0 & 0 & 0 & 1 \\
    \end{array}
\right).\end{equation}
Using this instead of $\D[g]$ in \eq{Z8-rfmdef} does not allow for an antiperiodic spin structure in the first four components; as a result we do not study it either. The only other freedom is in choosing the shift vector $\bvec[g]$. For the example we are discussing there is more than one inequivalent choice of vector $\bvec[g]$. As explained in Section \ref{sec:RFMs}, the choices of $\bvec[g]$ are in one-to-one correspondence with solutions to the equation
\begin{equation}
    \D[g]\,\bvec[g]=\bvec[g]\quad\text{mod}\,\mathbb{Z} \,,
\end{equation}
which in our case has two inequivalent solutions,
\begin{equation}
    \bvec[g]=\left(0,0,0,0,\frac{a}{8},\frac{b}{8}\right)\quad\text{or}\quad \left(\frac12,\frac12,\frac12,\frac12,\frac{a}{8},\frac{b}{8}\right) \,,
    \label{eq:Z8-shift-vector}
\end{equation}
where $a,b$ are integers constrained such that the first multiple of $\bvec[g]$ to be a lattice vector is precisely $8\,\bvec[g]$.  Thus, we see that \eq{Z8-rfmdef} provides only a particularly simple choice of $\bvec[g]$. We have also explored all others, and they are either as good as this one or worse in terms of Casimir energy and the size of the space, as explained in Subsection \ref{subsec:dS5-maximum-casimir-potential}. 

All of the above explains our choice of $\mathcal{F}_6$ via \eq{Z8-rfmdef} and spin structure on the parent $T^6$. We just need to study the scalar potential as a function of $(R_1,R_2,\beta)$. Our last task in this subsection is to perform a convenient change of variables. The volume of the covering $T^6$ in our locus of moduli space is 
\begin{equation}
    \text{Vol}(T^6)= 2\, R_2^4 \,R_1^2 \,.
\end{equation}
The volume of the RFM is reduced by a factor of 8 from this. We will introduce new variables $(x,R)$ defined by
\begin{equation} R^6\equiv R_2^4\, R_1^2,\quad x\equiv\frac{R_2}{R_1}.\label{xdef}\end{equation}
The variable $R$ is essentially the volume modulus of $\mathcal{F}_6$ (we have $\text{Vol}(\mathcal{F}_6)=\frac14\, R^6$), while $x$ measures the aspect ratio between the $C_4$ and $\mathbb{Z}^2$ factors of the RFM discussed above. The inverse relation reads
\begin{equation} 
    R_1=\frac{R}{x^{2/3}},\quad R_2= R\, x^{1/3} \,.
    \label{rinverse}
\end{equation}

\subsection{Casimir energy}
\label{subsec:dS5-maximum-casimir-potential}

We must now calculate the Casimir energy term $\Vcas$ in \eq{totpot}, using the formula \eq{eq:VC-invariant-subspace}. In five dimensions, the Casimir energy is a moduli-dependent quantity with units of length$^{-5}$ which is obtained from a one-loop calculation of the massless fields. Since the only dimensionful modulus in \eq{xdef} is $R$, dimensional analysis forces a Casimir term of the form
\begin{equation}
    \Vcas = -\frac{\mathcal{C}(x,\beta)}{4\, R^5} \,,
    \label{casdef}
\end{equation}
where we have packed all non-trivial dependence in a Casimir function $\mathcal{C}(x,\beta)$. We have also introduced an overall minus sign in the definition of \eq{casdef} to take into account the fact that Casimir energies of periodic fields are usually negative; hence, we expect $\mathcal{C}(x,\beta)$ to be a positive function. In this section we will study $\mathcal{C}(x,\beta)$ in some detail, but it is important to keep in mind that the actual physical energy contains an additional minus sign.

Recall that the Casimir energy $\Vcas$ on $\mathcal{F}_6 = T^6/\Z_8$ is given by
\begin{align}
    \Vcas = -\frac{\Gamma(s)}{2\pi^{s}}\cdot \int_{\mathcal{F}_6} d^{6}z ~\sqrt{G}~\sum_{j=0}^{7}\sum_{\vec{n}\in\Z^6} \frac{\TrB{\D[g]^j} - \TrF{\D[g]^j}\cdot e^{2\pi i \Vec{h}\cdot\Vec{n}}}{|(\mathbf{I} - \D[g]^j)\,\vec{z} - j\,\bvec[g] + \vec{n}|^{2s}} \,, 
\end{align}
with $s=D/2=11/2$ for M-theory. 
\begin{figure}[!htb]
\centering
\begin{subfigure}{0.49\textwidth}
    \includegraphics[width=\textwidth]{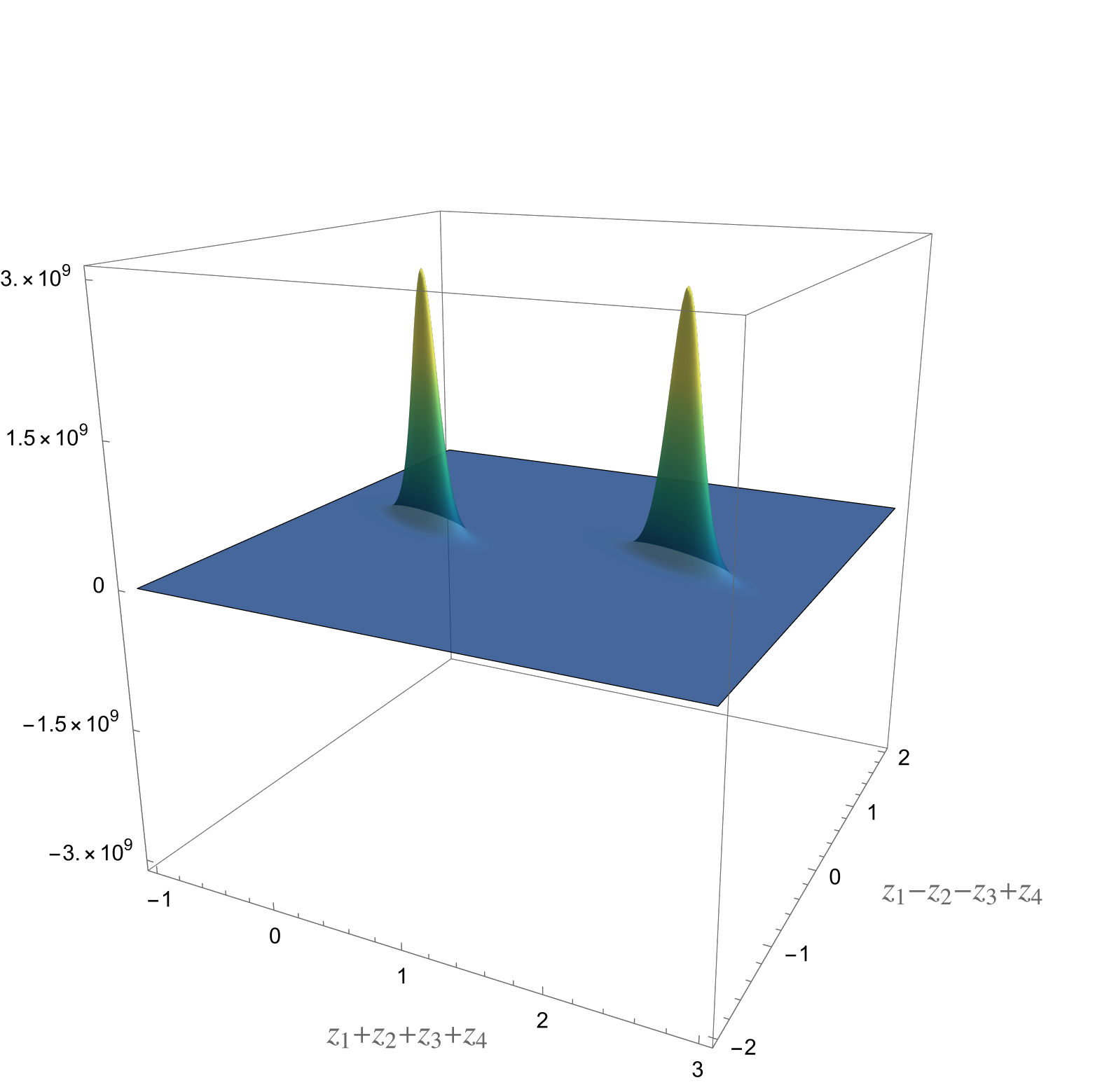}
\end{subfigure}
\hfill
\begin{subfigure}{0.49\textwidth}
    \includegraphics[width=\textwidth]{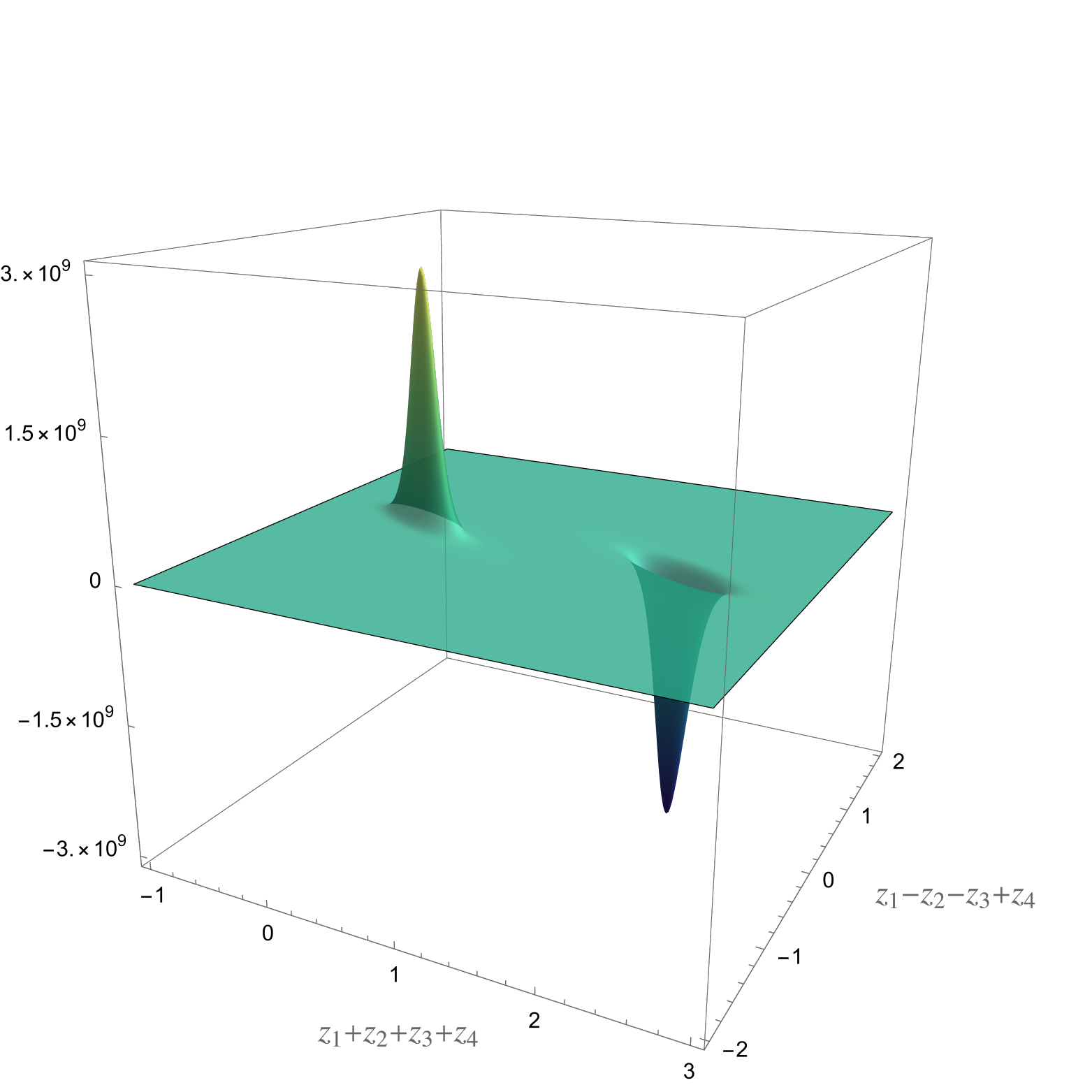}
\end{subfigure}
        
\caption{Integrands of $\mathcal{E}(\D[g])$ \eq{csum} for a bosonic and a fermionic degree of freedom, respectively, with the fermionic boundary conditions being fixed by our choice of spin structure $\vec{h} = (\frac12,\frac12,\frac12,\frac12,0,0)$ for a reference point in moduli space with $x=\beta=1$. By rotating the axes appropriately and plotting on the plane $(z_1+z_2+z_3+z_4\,,\,z_1-z_2-z_3+z_4)$, we can see both Casimir branes, one at $(0,0,0,0)$ and one at $(\frac12,\frac12,\frac12,\frac12)$ on the fibre $T^4$ that wrap the base $T^2$. While for bosons both Casimir branes have positive tension, for fermions they have opposite tension and cancel each other out upon integration over the covering $T^6$.}
\label{fig:Casimir-branes-Z8}
\end{figure}
The contribution of each group element (apart from the identity) can be seen as the energy of a number of codimension-4 Casimir branes, localised on the $T^4$ fibre and wrapping the base $T^2$, corresponding to the subspaces invariant under $\D$, i.e. $\D\,\vec{z} = \vec{z}~\text{mod}~\Z^6$ (see figure \ref{fig:Casimir-branes-Z8}). In total, these non-trivial elements contribute as 32 codimension-4 branes.
The identity element contributes as a space-filling Casimir brane, wrapping the whole of $\mathcal{F}_6$.
Given the block diagonal structure of the invariant metric \eq{ewmmm}, the norm splits into two pieces
\begin{align}
    \Vcas = -\frac{\Gamma(s)}{2\pi^{s}}\cdot \frac{2R_1^2 R_2^4}{|\Z_8|}\cdot \sum_{j=0}^{7}\int_{T^6} d^{4}z ~\sum_{\substack{\vec{n}\in\Z^4 \\ \vec{m}\in\Z^2}} \frac{\TrB{\D[g]^j} - \TrF{\D[g]^j}\cdot e^{2\pi i \Vec{h}\cdot(\Vec{n},\vec{m})}}{\Big[R_2^2\,|(\mathbf{I} - \D[g]^j)\,\vec{z} + \vec{n}|^2 + R_1^2\,|\vec{m} - j\,\bvec[g]|^2 \Big]^{s}} \,.
\end{align}
We also split the 6d vector $\vec{n}$ into a 4d vector $\vec{n}$ along the fibre and a 2d vector $\vec{m}$ along the base. Substituting in the order of $|\Z_8| = 8$ and performing the change of variables \eq{rinverse}, we have 
\begin{align}
    \Vcas = -\frac{\Gamma(s)}{2\pi^{s}}\cdot \frac{x^{\frac{4s}{3}}}{4\cdot R^{2s-6}}\cdot \sum_{j=0}^{7}\int_{T^6} d^{4}z ~\sum_{\substack{\vec{n}\in\Z^4 \\ \vec{m}\in\Z^2}} \frac{\TrB{\D[g]^j} - \TrF{\D[g]^j}\cdot e^{2\pi i \Vec{h}\cdot(\Vec{n},\vec{m})}}{\Big[x^2\,|(\mathbf{I} - \D[g]^j)\,\vec{z} + \vec{n}|^2 + |\vec{m} - j\,\bvec[g]|^2 \Big]^{s}} \,.
\end{align}
For the choice of fermionic boundary conditions given by $\vec{h} = (\frac12,\frac12,\frac12,\frac12,0,0)$, we find a potential of the form \eq{casdef}, with 
\begin{align}
    \mathcal{C}(x,\beta) = \frac{\Gamma(s)}{2\pi^{s}}\cdot x^{\frac{22}{3}} \sum_{j=0}^{7}\int_{T^6} d^{4}z ~\sum_{\substack{\vec{n}\in\Z^4 \\ \vec{m}\in\Z^2}} \frac{\TrB{\D[g]^j} - \TrF{\D[g]^j}\cdot e^{\pi i (n_1 + n_2 + n_3 + n_4)}}{\Big[x^2\,|(\mathbf{I} - \D[g]^j)\,\vec{z} + \vec{n}|^2 + \beta^{-1}m_1^2 + \beta\big(m_2 - \frac{j}{8}\big)^2\Big]^{s}} \,.
    \label{eq:Casimir-T6Z8}
\end{align}
Note that the Casimir potential is a sum over the $8$ elements of $\Z_8$. As we show in Section \ref{sec:Casimir-on-RFMs}, the terms corresponding to non-trivial elements $\D[g]^j$, $j\in\{1,...,7\}$, collapse into a 2d sum over the base $T^2$ corresponding to the invariant sublattice under $\D[g]$, 
\begin{align}
    \mathcal{C}_{\rm twisted}(x,\beta) 
    &= \frac{\Gamma(s-2)}{2\pi^{s-2}}\cdot x^{\frac{10}{3}}\sum_{j=1}^{7}\sum_{\vec{m}\in\Z^2} \frac{\TrB{\D[g]^j}}{\Big[\beta^{-1}m_1^2 + \beta\big(m_2 - \frac{j}{8}\big)^2 \Big]^{s-2}} \nonumber \\ 
    &= x^{\frac{10}{3}}\cdot \begin{cases}
        3.08\times 10^6\,\beta^{-\frac{7}{2}}  &,\, \beta\lesssim 8 \\
        4.15\times 10^5\,\beta^{-\frac{5}{2}}  &,\, \beta\to\infty 
        \end{cases} \,. 
    \label{eq:casimir-Z8-twisted-analytical}
\end{align}
Note that the fermionic contribution drops out due to the factor $\hat{\delta}_{\vec{h}}$ in \eq{eq:VC-invariant-subspace}, since there is no solution to equation \eq{eq:eta-condition} with $\vec{\eta}\in\Z^6$. 
The traces of each element in the graviton, 3-form and gravitino representations were obtained using the formulas in Appendix \ref{ap:traces}.\footnote{The generator of our $\Z_8$ action \eq{eq:Z8-generator}, embedded as an $SO(9)$ matrix, has eigenvalue arguments $\vec{\theta} = \{0,0,\frac{\pi}{4},\frac{3\pi}{4}\}$. 
Note that, since our quotient group has even order, the choice of spin lift $\mathcal{D}_\mathbf{g}$ does not affect the sign of $\mathcal{D}_\mathbf{g}^8 = \mathbf{I}$, which forces us to set $h_6=0$ through \eq{swer2} as argued above.}
\begin{table}[ht]
    \centering
    \renewcommand{\arraystretch}{1.2}
    \begin{tabular}{c|cccccccc}
         k & 0 & 1 & 2 & 3 & 4 & 5 & 6 & 7  \\ \hline 
        $\Tr{\mathbf{44}}{\D[g]^j}$ & 44 & 14 & 12 & 14 & 4 & 14 & 12 & 14  \\
        $\Tr{\mathbf{84}}{\D[g]^j}$ & 84 & 10 & 20 & 10 & $-4$ & 10 & 20 & 10  \\
        $\Tr{\mathbf{128}}{\D[g]^j}$ & 128 & $16\sqrt{2}$ & $-32$ & $-16\sqrt{2}$ & 0 & $-16\sqrt{2}$ & $-32$ & $16\sqrt{2}$ 
    \end{tabular}
    \label{tab:my_label}
\end{table}

The twisted terms contribute with a very large numerical coefficient, which is going to be crucial for our dS solution; this large coefficient benefits from the fact that the gravitino, whose contribution comes with an overall minus sign, ends up not contributing at all through the twisted terms. Moreover, in order to take advantage of this, we need a solution with $\beta$ as small as possible. Recalling the overall minus sign in the Casimir energy \eq{casdef}, we conclude that by going to the quotient RFM, the energy becomes very large and negative, allowing for large dS solutions.

On the other hand, the term corresponding to the identity element $\D[g]^0 = \mathbf{I}$ cannot be simplified---the whole lattice remains invariant under this element---and is a full 6d infinite lattice sum that we must compute numerically,
\begin{equation}
    \mathcal{C}_\mathbf{I}(x,\beta) = \frac{\Gamma(s)}{2\pi^{s}}\cdot 128\cdot x^{\frac{22}{3}}  ~\sum_{\substack{\vec{n}\in\Z^4 \\ \vec{m}\in\Z^2}} \frac{1 - e^{\pi i (n_1 + n_2 + n_3 + n_4)}}{\Big[x^2\,|\vec{n}|^2 + \beta^{-1}m_1^2 + \beta\,m_2^2 \Big]^{s}} \,.
    \label{eq:Casimir-dS5-Id}
\end{equation}
Note that the identity contribution, unlike the twisted terms, is an even function of $\log\beta$, i.e. $\mathcal{C}_\mathbf{I}(x,\beta) = \mathcal{C}_\mathbf{I}(x,\beta^{-1})$, and thus it must have a critical point at $\beta=1$. To study its asymptotics we can therefore focus on $\beta\to\infty$,
\begin{align}
    \mathcal{C}_\mathbf{I}(x,\beta) 
    &= \frac{\Gamma(s)}{2\pi^{s}}\cdot 128\cdot x^{\frac{22}{3}}\beta^{s}  ~\sum_{\substack{\vec{n}\in\Z^4 \\ \vec{m}\in\Z^2}} \frac{1 - e^{\pi i (n_1 + n_2 + n_3 + n_4)}}{\Big[\beta\,x^2\,|\vec{n}|^2 + m_1^2 + \beta^2\,m_2^2 \Big]^{s}} \nonumber \\
    &\approx 4.414\,x^{-\frac{11}{3}} + 5.116\,x^{-\frac{8}{3}}\beta^{\frac{1}{2}} \quad\text{as}\quad \beta\to\infty \,. 
    \label{eq:Casimir-dS5-Id-largebeta}
\end{align}

Since the flux potential will not depend on $\beta$---as we shall explicitly see in the next section---a critical point for this modulus must arise from the Casimir potential alone. Given the asymptotic behaviours of each contribution, we expect it to come from balancing the identity against the twisted terms and to be near $\beta\sim 45.3\,x^{3/2}$.

Nonetheless, the function $\mathcal{C}(x,\beta)$ depends on the two moduli $(x,\beta)$ in a complicated way, so that it is best to solve this as a numerical 2-modulus problem by using the analytical solution for $R$ to define a potential $V_R(x,\beta)$ that takes into account the flux contribution. We do this in Section \ref{subsec:dS5-maximum}, after introducing the flux potential.

\subsection{Flux potential}
\label{subsec:dS5-maximum-flux-potential}

We will now determine the flux piece $V_{G_4}$ of the scalar potential \eq{totpot}. In our setup, this comes entirely from M-theory $G_4$ flux. The $G_4$ kinetic term is proportional to the Hodge norm $\vert G_4\vert^2$, so we will compute that first. We must take into account that $G_4$ is quantized,
\begin{equation}
    \int_{\mathcal{C}_4}G_4\in\mathbb{Z}\,,\quad\forall\, \mathcal{C}_4\in H_4(\mathcal{F}_6,\mathbb{Z}) \,. 
    \label{uf}
\end{equation}
Using Poincar\'{e} duality, this can be equivalently rewritten as 
\begin{equation}
    \int_{\mathcal{F}_6}\alpha\wedge G_4\in\mathbb{Z}\,,\quad\forall\, \alpha\in H^2(\mathcal{F}_6,\mathbb{Z}) \,. 
    \label{uf2}
\end{equation}
If we choose a basis $\{\alpha_i\}$ of $H^2(\mathcal{F}_6,\mathbb{Z})$, we may satisfy \eq{uf} and \eq{uf2} by expanding $G_4$ in terms of the Hodge-dual forms $\{*\alpha_i\}$ (which will not, in general, have quantized periods) as
\begin{equation} 
    G_4=\sum_{i,j} n^i\, \mathcal{G}_{ij}^{-1} *\alpha_j\,,\quad n^i\in\mathbb{Z} \,,
\end{equation}
where we have defined 
\begin{equation}
    \mathcal{G}_{ij}\equiv \int_{\mathcal{F}^6} \alpha_i\wedge *\alpha_j \,.
    \label{kinterm}
\end{equation}
With these definitions, we have 
\begin{equation} 
    \int_{\mathcal{F}^6}\vert G_4\vert^2=\sum_{i,j} \mathcal{G}_{ij}^{-1}\, n^in^j\,.
    \label{fluxpoty}
\end{equation}
The only task left is to determine the Hodge norm acting on 2-forms \eq{kinterm}. To do this, we must construct a basis of $H^2(\mathcal{F}_6,\mathbb{Z})$. Because of the natural projection map $p:\, T^6\rightarrow\mathcal{F}_6$, we may work instead in the covering $T^6$. We have
\begin{equation} 
    \int_{\mathcal{F}^6} \alpha_i\wedge *\alpha_j=  \frac18\int_{T^6} p^*(\alpha_i)\wedge *p^*(\alpha_j)\,,
    \label{kinterm2}
\end{equation}
where the factor of $1/8$ appears because the fundamental class of $\mathcal{F}_6$ pulls back to 8 times the fundamental class of $T^6$, so the Hodge duals in $T^6$ are rescaled by a factor of $1/8$ with respect to Hodge duals in $\mathcal{F}_6$ (in other words, $T^6$ is an eightfold covering of $\mathcal{F}_6$, which can be identified with a subset of $T^6$ with 1/8 its volume). The pullback $p^*(\alpha_i)\in H^2(T^6,\mathbb{Z})$ is represented by a 2-form on $T^6$ invariant under the action \eq{Z8-rfmdef}. Out of the fifteen 2-cycles of $T^6$, only a three-dimensional subspace is invariant under \eq{Z8-rfmdef}, generated by the 2-forms
\begin{align}
    \omega_1&\equiv dz_1\wedge dz_2+dz_1\wedge dz_4 + dz_2\wedge dz_3+dz_3\wedge dz_4 - dz_1\wedge dz_3 - dz_2\wedge dz_4\,,\nonumber\\ \omega_2&\equiv dz_1\wedge dz_2+dz_1\wedge dz_4 + dz_2\wedge dz_3+dz_3\wedge dz_4 -2 dz_1\wedge dz_3 -2 dz_2\wedge dz_4\,,\nonumber\\ \omega_3&\equiv dz_5\wedge dz_6 \,.
    \label{invariant2forms}  
\end{align}
Furthermore, the forms $\omega_i$ are all invariant under \eq{az4} and transform under \eq{aD4} as
\begin{equation}\omega_1\,\rightarrow\,\omega_1,\quad \omega_2\,\rightarrow\,-\omega_2,\quad \omega_3\,\rightarrow\,\omega_3,\end{equation}
This means that if the $*\omega_2$ piece of $G_4$-flux vanishes, the flux potential will respect both symmetries used in Subsection \ref{subsec:dS5-maximum-RFM} to ensure a saddle along $\alpha$ and $\gamma$. 

The pullback $p^*H^2(\mathcal{F}_6,\mathbb{Z})$ is a sublattice of the invariant subspace generated by the $\omega_i$. To determine this sublattice, we can recall that classes in $H^2(X,\mathbb{Z})$ are in one-to-one correspondence (via the Chern class) with $U(1)$ bundles over $X$. These, in turn, are fully characterized by the field-strength $F$ of the gauge bundle (an element of $H^2(X,\mathbb{R})$) and by the holonomies 
\begin{equation} 
    W(\ell)=\exp\left(2\pi i\int_\ell A\right)
    \label{wl}
\end{equation}
of the gauge field $A$ evaluated on any closed line $\ell$\,\footnote{From a math point of view, this characterization comes from the long exact sequence $\ldots\rightarrow H^1(X,U(1))\rightarrow H^2(X,\mathbb{Z})\rightarrow  H^2(X,\mathbb{R})\rightarrow\ldots$ arising from the short exact sequence $1\rightarrow\mathbb{Z}\rightarrow\mathbb{R}\rightarrow\mathbb{R}/\mathbb{Z}=U(1)\rightarrow1$. The map sending holonomies to the corresponding Chern class is an example of a Bockstein homomorphism, see e.g \cite{Davighi:2020vcm} for some more details.}. The quantity $W(\ell)$ must be a well-defined function from the space of closed lines of $X$ to $U(1)$, and this condition can be used to determine a basis of $H^2(X,\mathbb{Z})$ in terms of differential forms.   We will first illustrate how this is done for $T^6$. A gauge field in $\mathbb{R}^6$ can be expanded into 1-forms $A=\vec{A}\cdot d\vec{z}$ and there are no identifications since all bundles on $\mathbb{R}^6$ are trivial. The holonomies \eq{wl} can be written as integrals of a field-strength and are all well-defined. When passing to the quotient $T^6\equiv\mathbb{R}^6/\mathbb{Z}^6$, we must impose the well-definedness of $W(\ell)$ for any 1-cycle in the torus. These are parametrized as $\ell=\vec{z}_0+t \,\vec{n}$ where $t\in[0,1)$ and $\vec{n}\in\mathbb{Z}^6$. Without loss of generality, we will restrict to gauge fields where the components are linear, so that $\vec{A}=\mathbf{F}\cdot\vec{z}$ with $\mathbf{F}$ antisymmetric\footnote{So that it is represented as $\frac12\mathbf{F}_{ij} dz^i\wedge dz^j$ in form notation.}. The holonomy is 
$W(\ell)=\exp\left(2\pi \,i\,  \vec{n}\cdot\mathbf{F}\cdot\vec{z}_0\right)$, and for it to be a well-defined function on the torus it must be invariant under shifts by lattice vectors $\vec{z}_0\rightarrow\vec{z}_0+\vec{m}$, or in other words that
\begin{equation} 
    \vec{n}\cdot \mathbf{F}\cdot \vec{m}\in\mathbb{Z} \,,
\end{equation}
which is the familiar quantization condition. The only difference in the case of $\mathcal{F}_6$ is that the $\mathbb{Z}_8$ quotient introduces additional identifications, $\vec{z}_0\rightarrow \D[g]^j\,\vec{z}_0 + j\, \bvec[g]$, and additional 1-cycles, parametrized as
\begin{equation}
    \ell_1=(1-t)\vec{z}_0+ t\, (\D[g]^j\,\vec{z}_0 + j\,\bvec[g])\quad t\in\,[0,1) \,.
\end{equation}
The corresponding condition is that the quantity
\begin{equation} 
    [(\D[g]^l\,\vec{z}_0 + l\,\bvec[g]+\vec{n})]\cdot \mathbf{F}\cdot [(\D[g]^j-\mathbf{I})\,\vec{z}_0 + j\,\bvec[g]] \,,
    \label{cond}
\end{equation}
is independent of $k,l,\vec{n}$. Taking $\mathbf{F}$ to be invariant under the $\mathbb{Z}_8$, and using that $\bvec[g]$ is in our case invariant under $\D[g]$, turns the above into the condition
\begin{equation} 
    \vec{n}\cdot \mathbf{F}\cdot \bvec[g]\in\mathbb{Z} \,,
    \label{cond2}
\end{equation}
which is what we need to determine the sublattice $p^*H^2(\mathcal{F}_6,\mathbb{Z})$: it is generated by $\omega_1,\omega_2$ and $8\, \omega_3$. Notice that the forms $\omega_1,\omega_2$ are fully localized along the $T^4$ fiber, while $\omega_3$ lies along the base; the result we just obtained then matches an intuitive picture where $\mathcal{F}_6$ is identified with a fundamental domain for the $\mathbb{Z}_8$ action; since it acts freely on cycles along the fiber, forms which were integer-quantized and located along the fiber still have integer periods. However, the 2-cycle along the base is replaced by a domain with 1/8 its volume, so proper quantization requires us to rescale $\omega_3$ by a factor of 8.

We can now determine the metric $\mathcal{G}_{ij}$ using \eq{kinterm2}, which gives $\mathcal{G}_{ij}=\text{Vol}(\mathcal{F}_6)\, \kappa_{i}\,\delta_{ij}$. The orthogonality is due to the fact that $\omega_1,\omega_2$ have different eigenvalues\footnote{The more general result for $\gamma\neq0$ is $\left(
\begin{array}{cc}
     \frac{2 \left(\gamma ^2+2\right)}{\left(\gamma ^2-2\right)^2 R_2^4} & -\frac{8 \gamma }{\left(\gamma ^2-2\right)^2 R_2^4} \\
     -\frac{8 \gamma }{\left(\gamma ^2-2\right)^2 R_2^4} & \frac{4 \left(\gamma ^2+2\right)}{\left(\gamma ^2-2\right)^2 R_2^4} \\
\end{array}
\right)$. We see that a linear, off-diagonal term proportional to $\gamma$ appears, which is due to the fact that \eq{aD4} is no longer an isometry, and the invariant subspaces no longer have to be orthogonal. Thus, when an $\omega_2$ flux piece is turned on, the potential no longer has a saddle for $\gamma=0$ in the $\gamma$ direction.} with respect to \eq{aD4}. We have
\begin{equation}
    \kappa_1 = \frac{1}{R_2^4} \,,
    \quad \kappa_2 = \frac{2}{R_2^4} \,,
    \quad \kappa_3 = \frac{64}{R_1^4} \,.
\end{equation}
Using \eq{fluxpoty}, we finally arrive to our flux potential,
\begin{equation} 
    \int_{\mathcal{F}^6}\vert G_4\vert^2 = \frac{1}{R^2}\left(4 n_1^2 x^{4/3}+\frac{n_3^2}{16 x^{8/3}}\right) \,.
    \label{fluxpoty2}
\end{equation}
where we have set the piece proportional to $*\omega_2$ to zero already. As we will see in the next subsection, the factor of 16 in the denominator of \eq{fluxpoty2}, which comes from the effect of the $\mathbb{Z}_8$ quotient involved in defining $\mathcal{F}_6$, is an example of yet another boon of RFM's: the quotient decreases the flux quantum allowing for finer choices of flux that lead to more vacua in the parent torus. 

\subsection{The \texorpdfstring{$dS_5$}{dS5} maximum}
\label{subsec:dS5-maximum}

We can now combine the results of Subsections \ref{subsec:dS5-maximum-casimir-potential} and \ref{subsec:dS5-maximum-flux-potential} to obtain the scalar potential. The M-theory flux potential for the reduction on $\mathcal{F}_6$ is (see Appendix \ref{app:fluxQ})
\begin{equation} 
    V_{G_4}=\frac{(2\pi^2)^{\frac13}}{2\ell_{11}^3}\int_{\mathcal{F}_6} \vert G_4\vert^2= \frac{(2\pi^2)^{\frac13}}{2R^2\ell_{11}^3}\left(4 n_1^2 x^{4/3}+\frac{n_3^2}{16 x^{8/3}}\right) \,.
\end{equation}
We see that $V_{G_4}$ has dimensions of length$^{-5}$, as befits a five-dimensional energy density. We will now set $\ell_{11}=1$ and add the Casimir term \eq{casdef}, to construct the total scalar potential \eq{totpot} explicitly as a function of the moduli,
\begin{equation} 
    V^{(5d)} = \frac{2^{\frac{10}{3}}}{R^{10}}\left[ \frac{(2\pi^2)^{\frac13}}{2R^2}\left(4 n_1^2 x^{4/3}+\frac{n_3^2}{16 x^{8/3}}\right) -\frac{\mathcal{C}(x,\beta)}{4\, R^5} \right] \,.
    \label{eq:dS5-Z8-scalar-potential}
\end{equation}
We also included an overall Weyl rescaling factor, required to go to Einstein frame in 5d. All that is left to do is to find extrema of this potential---denoting by $\mathcal{C}_x$ ($\mathcal{C}_\beta$) the first derivative of $\mathcal{C}(x,\beta)$ with respect to $x$ ($\beta$), the extremum conditions are
\begin{subequations}
    \begin{align}
        R^3 &= \frac{\mathcal{C}(x,\beta)}{(2\pi^2)^{1/3}}\cdot\frac{10x^{8/3}}{64n_1^2\,x^4 + n_3^2} \,, \label{eq:dS5-maximum-Req} \\
        \frac{x\,\mathcal{C}_x(x,\beta)}{\mathcal{C}(x,\beta)} &= \frac{10}{3}\cdot\frac{32 n_1^2 \,x^4 - n_3^2}{64n_1^2 ~x^4 + n_3^2} \,,
        \label{eq:dS5-maximum-xeq}
    \end{align}    
\end{subequations}
and $\mathcal{C}_\beta(x,\beta) = 0$. Note that along $\beta$ an extremum of the full potential is necessarily an extremum of the Casimir function $\mathcal{C}(x,\beta)$, but along $x$ the Casimir contribution balances against the flux potential. We want to find the solution with the biggest possible $R$, since this will give us the most control. It is clear from the $R$ equation \eq{eq:dS5-maximum-Req} that we want the largest possible $\mathcal{C}(x)$ and the smallest allowed fluxes---this is where the large contribution of the twisted sector becomes key. 

Since we do not have simple analytical expressions for $\mathcal{C}(x,\beta)$ and its derivatives, we will evaluate the potential \eqref{eq:dS5-Z8-scalar-potential} at its extremum along the $R$ direction, using \eqref{eq:dS5-maximum-Req}, and find numerically extrema of the potential
\begin{equation}
    V_R(x,\beta) \equiv V^{(5d)}\big|_{R_{\rm saddle}} = \frac{2^{\frac{43}{3}}}{5^5} \frac{(2\pi)^{5/3}}{\mathcal{C}(x,\beta)^4}\left(4 n_1^2 x^{4/3}+\frac{n_3^2}{16 x^{8/3}}\right)^5 \,,
\end{equation}
for the remaining moduli $(x,\beta)$. When $n_1=0$ and $n_3=1$, we find a maximum for which the vacuum energy is  $V_{\text{saddle}} = 4.43\cdot 10^{-8} \,\ell_5^{-5}$ (see Table \ref{tab:dS5-Z8-flux-solutions-new}). This is a factor of $10^{2}$ better than the na\"ive $T^4$ solution in \secref{sec:warm-up-dS7}\footnote{And still, it is not as good as one might have expected. The dS saddle we found arises as a balance of a large twisted sector and a small untwisted one, to the effect that the actual saddle is at a not-so-large $R$. Moreover, the dependence of $\Vcas$ on $\beta$ forces us to turn on $n_3$ flux in order to find a saddle, which leads to smaller $R$. In future work we will explore situations where the saddle comes from balancing two twisted sectors and more favourable conditions for the fluxes, where one may expect much larger results.}.
\begin{table}[h]
    \centering
    \renewcommand{\arraystretch}{1.2}
    \begin{tabular}{|c|c|c|c|c|c|c|c|}
        \hline
        $R$ & $x$ & $\beta$ & $\frac18\beta^{\frac12} R_1$ & $\beta^{-\frac12} R_1$ & $\sqrt{2}\,R_2$ & $\mathrm{vol}(\mathcal{F}_6)$ & $V_{\rm saddle}$ \\ \hline 
        4.68 & 0.247 & 6.78 & 3.87 & 4.56 & 4.15 & $2.61\times 10^3$ & $4.43\times 10^{-8}$ \\ \hline
    \end{tabular}
    \caption{Saddle point for $(n_1,n_3)=(0,1)$. We also list the length of the smallest curves in the base and fiber of $\mathcal{F}_6$, in 11-dimensional Planck units, as well as the volume of the space; the vacuum energy is given in 5-dimensional Planck units.}
    \label{tab:dS5-Z8-flux-solutions-new}
\end{table}

Note that the moduli $R_1$ and $R_2$ are the radii of the covering $T^6$, and as such, they do not directly measure the diameters of closed curves in $\mathcal{F}_6$. To get these, notice that the smallest 1-cycles of $\mathcal{F}_6$ will be either entirely contained in the $T^4$ fiber or in the $T^2$ base. Since the $T^4$ fiber is the $C_4$ Cartan torus rescaled by $R_2$, and the root lattice of $C_4$ is even, physical 1-cycles along the fiber have a minimal length of $\sqrt{2}\, R_2$. Similarly, $\beta^{-1/2}\,R_1$ and $\beta^{1/2}\,R_1$ are the sizes of the torus base---but the physical size of the latter is eight times smaller due to the $\mathbb{Z}_8$ quotient. Therefore, the sizes of physical closed curves on $\mathcal{F}_6$ are
\begin{equation} 
    \frac{\beta^{\frac12}R_1}{8}\,,\quad \beta^{-\frac12}R_1\quad \text{and}\quad \sqrt{2}\, R_2\,,
\end{equation}
which are the parameters quoted in Table \ref{tab:dS5-Z8-flux-solutions-new}.

Although the value of the vacuum energy is improved with respect to the solution in \secref{sec:warm-up-dS7}, the size of the torus is still Planckian---which again raises issues of control. We now turn to a systematic discussion of these issues. 

\subsection{Characterization and control of the solutions}
\label{sec:ctrl}

\vspace{1em}
\begin{flushright}
    \textit{“A common mistake that people make when trying to design something\\ completely foolproof is to underestimate the ingenuity of complete fools.”} \\ \vspace{1em}
    Douglas Adams, \textit{A Hitchhiker's Guide to the Galaxy} 
\end{flushright}
\vspace{1em}

The solution with $n_3=1$ that we found has a small vacuum energy of $10^{-8}$ in five-dimensional Planck units, and radii which are roughly Planckian in size. 
We will now provide some more details about this solution, and discuss at length just how under control it is.

We will first determine the spectrum of light fields and masses around this solution. To compute these masses, we need the properly normalized kinetic terms of the moduli in RFM's. Since these are all quotients of tori, the kinetic terms from dimensional reduction of the $(D=d+k)$--dimensional theory on $T^k$ are enough. 

Just like in \secref{sec:RFMs}, we will parametrize the moduli of $T^k$ by the $k\times k$ inner product matrix $\mathbf{G}$ that describes the inner product when the torus lattice is just $\mathbb{Z}^k$, an $n\times n$ matrix $\mathbf{M}$ whose columns are the vectors spanning the torus lattice. The kinetic terms of the moduli coming from the Einstein-Hilbert action on $T^k$ are
\cite{Etheredge:2022opl}
\begin{equation}
    \frac{1}{2\kappa_d^2}\int d^dx\, \sqrt{-g_d}\left[\frac{1}{(d-2)} (\partial\log\sqrt{\text{det}\,\mathbf{G}})^2+ \frac{1}{4}\text{Tr}((\mathbf{G}^{-1}\partial\mathbf{G})^2)\right] \,.
\end{equation}
One can check that this gives e.g. the standard metric on the upper half-plane for the case of $T^2$, agreeing e.g. with (4.9) of \cite{Etheredge:2022opl}
and Appendix A of \cite{Castellano:2023jjt}
for the canonical normalization of the volume modulus of $T^k$, which is precisely $\sqrt{\text{det}\,\mathbf{G}}$. Using these, it is possible to determine the mass of the moduli fields $\{R,x,\gamma,\alpha,\beta\}$ from the eigenvalues of the Hessian normalized with respect to the proper kinetic term. They are 
\begin{equation}
       \{-83.3 \,,\, -5.98 \,,\, -359 \,,\, 6.68\,,\, -22.1\}\cdot H_0^2 \,,
    \label{eq:dS5-masses}
\end{equation}
so that our solution is a stable minimum along one direction and an unstable maximum along the remaining four.  The masses acquired by the five moduli are of the order of the Hubble scale $H_0\approx 8.61\cdot 10^{-5}\, M_{\rm Pl}^{(5)}$ \eqref{hradius}. 

On top of these geometric moduli, we also get two KK vectors from the base $T^2$, and additional fields from the dimensional reduction of $C_3$, the RR axion. The free part of the $\mathcal{F}_6$ cohomology is 
\begin{equation}
    \begin{array}{c|ccccccc}
        p&0&1&2&3&4&5&6\\\hline 
        H^p(\mathcal{F}_6,\mathbb{R})&\mathbb{R}&2\mathbb{R}&3\mathbb{R}&4\mathbb{R}&3\mathbb{R}&2\mathbb{R}&\mathbb{R}
    \end{array}
\end{equation}
so from dimensional reduction of $C_3$ we get another five vectors and five axions, four from $C_3$ periods and another one from the five-dimensional dual of $C_3$ with no legs on $T^6$. All of these fields are massless to a first approximation; the axions constitute compact directions of the moduli space, which receive a non-perturbative potential due to M2 and M5 brane instantons \cite{Harvey:1999as}. We will study this potential in detail in the next section; while we have not computed their exact minimum, the smallness of the potential together with the compact range of the scalars ensures that they give a negligible correction to the solution, and can be treated in practice as a moduli space. The shift symmetry of the axions ensures that there are no perturbative corrections \cite{Dine:1986vd}. In some cases, one can use the parity symmetry of M-theory to ensure that the exact $C_3$-axion potential will have a minimum at zero; it is not possible to do this for all five components of the axion fields in this case, due to the $G_4$ flux that breaks some of these parity symmetries. 

The fact that we have a dS solution with massless vector fields means that we could check the Festina Lente bound \cite{Montero:2019ekk}. This is a proposed dS Swampland condition setting a lower bound on the mass of charged states under any $U(1)$ gauge field, stating that the masses of all charged states must satisfy
\begin{equation}m^2\gtrsim \frac{g\,q}{\sqrt{G}} H,\label{fl}\end{equation}
in terms of the Hubble scale, the gauge coupling $g$, the integer quantized charge $q$ of the particle, and Newton's constant $G$. Although the arguments supporting the bound do not work for dS maxima \cite{Montero:2019ekk,Montero:2021otb}, it is perhaps interesting that \eq{fl} is satisfied nonetheless. For the two KK photons, the content of the FL bound is equivalent to the statement $m_{KK}\gtrsim V^{1/2}$ \cite{Montero:2021otb}; as we have seen, this is satisfied in our vacuum. For the vectors obtained from $C_3$, we know that the charged states obtained from M2 and M5 branes saturate the Weak Gravity Conjecture, so $m\sim g M_p$ and \eq{fl} becomes the statement $g/\sqrt{G}\gtrsim H$. In terms of 11-dimensional quantities, we have
\begin{equation}
    \frac{g}{\sqrt{G}}\sim \frac{\text{Vol}(\omega_2)}{\ell_{11}^3}\sim \frac{R^2}{\ell_{11}^3}\gg H\,,
\end{equation}
for electric $C_3$ vectors (those that come from reduction of $C_3$ with two legs on $\mathcal{F}_6$), where $\text{Vol}(\omega_2)$ is the volume of the 2-cycle corresponding to the $C_3$ vector field under consideration. Therefore, FL is satisfied in this case. We should also consider magnetic vectors, obtained from dimensional reduction of the dual potential $C_6$ on a five-cycle of $\mathcal{F}_6$. In this case, the electrically charged object is an M5 brane wrapping this same five-cycle, leading to
\begin{equation}\frac{g}{\sqrt{G}}\sim \frac{\text{Vol}(\omega_5)}{\ell_{11}^6}\sim \frac{R^5}{\ell_{11}^6}\gg H,\end{equation}
so FL is satisfied for these gauge fields as well.

Finally, there are no light fermions in our solution, since $\mathcal{F}_6$ is the quotient of a $T^6$ where we have chosen an antiperiodic spin structure around four of the six 1-cycles. This means that the Dirac operator has no zero modes, nor does the Rarita-Schwinger field since the tangent bundle of $T^6$ is flat. As a result, our five-dimensional 5d theory is purely bosonic, even though we started with a maximally supersymmetric theory in 11 dimensions.

Having described the EFT, we now turn to the values of the parameters involved in it. With such a small vacuum energy, the dS radius is
\begin{equation}
    H_0^{-1}=\sqrt{\frac{6}{V_{\text{Saddle}}}}\approx 1.16\cdot 10^4 \,\ell_5 \,,
    \label{hradius}
\end{equation}
so that the smallest closed curve $\tfrac18\beta^{\frac12}R_1 \approx 4.6\cdot 10^{-3} H_0^{-1}$, and the solution is scale separated (like the AdS example of \cite{Luca:2022inb}).
The ratio of the 5d and 11d Planck lengths is
\begin{equation}
    \frac{\ell_{11}}{\ell_5} = \frac{R^2}{4^{1/3}}\approx 13.78 \,, 
\end{equation}
and the five-dimensional EFT is valid up to, roughly, $7.3\cdot 10^{-2}\, M_P^{(5)}$.

We now turn to the question of just how under control the solution is. The average energy density of the 11-dimensional solution, obtained by dividing the 5d vacuum energy by the volume of the internal manifold, is
\begin{equation} 
    \rho_{11d} = 8.40\cdot 10^{-6}\, \ell_{11}^{-11} \,,
\end{equation}
which arises as a balance of a flux and Casimir terms each of which is, on average,
\begin{equation}
    \rho_{\text{Cas}}\approx -3.36\cdot10^{-5}\, \ell_{11}^{-11}\,, 
    \quad \rho_{G_4}\approx 4.20\cdot10^{-5}\, \ell_{11}^{-11} \,.
    \label{avbar}
\end{equation}
We see that the 11-dimensional energy densities are very small; in fact, this small number coincides with the $\epsilon$ of the iterative approach to solving Einstein's equations discussed in Section \ref{sec:warm-up-dS7}, which means that the leading-order curvatures go like $10^{-5}$, too. Taken together, this means that any higher-derivative correction to the action of the schematic form 
\begin{equation}
    \int P(R_{\mu\nu\rho\sigma}, \vert G_4\vert), 
    \label{hder} 
\end{equation}
where $P$ is some polynomial,
will be suppressed by powers of $10^{-5}$. While we do not know the coefficients in front of these higher derivative corrections (save for a few cases \cite{Grimm:2017okk,Grimm:2017pid,Hyakutake:2005rb,Hyakutake:2007sm,Hyakutake:2006aq}), we expect them to become smaller and smaller as corrections get larger; in fact, if we assume that 11d supergravity has a cutoff $\Lambda$ (which we would naturally expect to appear at or around the 11d Planck scale), then \cite{Caron-Huot:2022ugt} would imply that coefficients of higher-derivative operators cannot be more than $\mathcal{O}(1)$ in units of the cutoff to ensure unitarity of the S-matrix (although, strictly speaking, we are outside of the regime of validity of this result, since M-theory is not weakly coupled at the cutoff scale); as a result, the solution we found seems quite safe under na\"ive higher-derivative and classical corrections, at least at the homogeneous level. In this regard, it is quite important that the RFM does not contain a non-trivial tadpole equation that would force higher-derivative terms to be a relevant part of the classical solution, avoiding an important subtlety present in standard flux compactifications \cite{Sethi:2017phn}.

We must also analyze whether our solution has inhomogeneities that are strong enough to invalidate the above analysis; for instance, if the internal profile on $\mathcal{F}_6$ of the energy density has very steep gradients, these will show up in corrections like \eq{hder}, and will affect the estimate of $\epsilon$, which was roughly speaking the maximum value of the energy density on the internal space. To leading order, the flux term yields a homogeneous 11-dimensional energy density, so we do not have to worry about it. However, this is not so for the Casimir term; as we saw in Section \ref{sec:casimir-RFMs}, the energy density coalesces around the location of the Casimir branes. In our case, we therefore expect a Casimir energy profile that is homogeneous along the directions of the base $T^2$, but inhomogeneous on the fiber $T^4$. From \eq{csum}, we can estimate the inhomogeneities in the Casimir energy by evaluating the energy density at $\{\frac12,\frac12,\frac12,\frac12,0,0\}$, a maximum where a Casimir brane is located, which is
\begin{equation}
    \hat{\rho}(\tfrac12,\tfrac12,\tfrac12,\tfrac12,0,0)\approx 3.37\cdot 10^{-5} \,\ell_{11}^{-11} \,,
\end{equation}
which is approximately the same as the average value \eq{avbar}. This reflects the fact that at this point in moduli space, the Casimir branes are not significantly enhanced and providing large contributions, which we can deduce from \eqref{eq:casimir-Z8-twisted-analytical} since $\langle\beta\rangle\approx 6.78$ at the saddle point.
For the same reason, the average gradient of the Casimir energy, which we can estimate with a value of $\hat{\rho}\approx 3.30\times 10^{-5} \ell_{11}^{-11}$ far away from a Casimir brane --- of the same order of magnitude as the average as well. Thus, the gradients are very small in Planck units and, in particular, much smaller than $1/R$, so the effective field theory is under control. Had this not been the case, all the KK modes of the internal space (including KK copies of $T^6$ moduli frozen by the $\mathbb{Z}_8$ quotient) would have been excited, ruining the solution. In other words, we have shown that internal space inhomogeneities do not ruin the na\"ive scale separation we obtained\footnote{This is something that we cannot do conclusively for some supersymmetric solutions, such as DGKT \cite{DeWolfe:2005uu}, due to the presence of singularities in the internal space. All of that is avoided here, since the Casimir branes are smooth, with a core profile that we have determined explicitly.}.

What about higher loop corrections? Since our solution relies on an unexpectedly large Casimir coefficient, one might wonder whether higher-order loops can be even larger, thereby destroying the solution\footnote{We thank A. Hebecker for bringing up this point.}. On dimensional grounds, the loop expansion of the Casimir energy (in  11d Planck units) is of the form
\begin{equation}V_{\text{Cas}} \sim\frac{1}{4R^5}\left[ \mathcal{C}+ \sum_{l=1}^\infty \mathcal{C}_i \left(\frac{\ell_{11}}{R}\right)^{9l}\right],\label{Casex}\end{equation}
where the power of $\ell_{11}$ is determined by the fact that the loop expansion in M-theory is in powers of Newton's constant $G\sim \ell_{11}^9$. Our Casimir coefficient, $\mathcal{C}$, is just the leading one in an infinite series. The higher-order Casimir coefficients $\mathcal{C}_i$ will come from divergent loop integrals requiring regularization, unlike $\mathcal{C}$, since at higher loops there are unprotected higher-derivative terms in the M-theory action that can contribute to the vacuum energy. The question is then whether the higher-order effects in \eq{Casex} can spoil our solution. However,  even though the Casimir coefficient is large, the overall contribution of the Casimir energy is quite small in 11-dimensional Planck units, since $R\sim 4.68\, \ell_{11}$. This is why we obtained a Hubble radius as large as \eq{hradius};  it was necessary in order to argue stability against classical corrections, and it  means that the loop counting parameter in  \eq{Casex} is
\begin{equation} 
    \left(\frac{\ell_{11}}{R}\right)^{9}\sim 10^{-6}\,.
    \label{lp}
\end{equation}
This is comparable to the classical parameter $\epsilon$ controlling classical corrections, and it means that, in order to overcome the classical contribution $\mathcal{C}\sim 10^{3}$ to the Casimir energy, the higher Casimir coefficients would have to satisfy
\begin{equation} 
    \mathcal{C}_i \gtrsim 10^{3+6l} \,.
    \label{rer3r}
\end{equation}
While we have not computed the coefficients $\mathcal{C}_i$, the estimate \eq{rer3r} is reassuring. It means that, even if the $\mathcal{C}_i$ are enhanced in the same way that $\mathcal{C}_0$ was, they still have to beat a factor  \eq{lp} of loop suppression to significantly affect the solution.  In other words, since $\rho_{11d}$ is small in Planck units, the configuration is quite classical, and quantum corrections are expected to be smaller still. Nevertheless, it would be nice to check the above via a smart estimate of the coefficients $\mathcal{C}_i$, for which explicit expressions as sums over the KK spectrum can be obtained.

An important point is that our solution is a perturbative expansion around  a static, classical solution of the equations of motion (the RFM), and as such, we expect corrections to be convergent when computed around the true static vacuum. This contrasts with what happens in more general, time-dependent situations \cite{Callan:1986bc,Angelantonj:2002ct,Dudas:2004nd,Raucci:2024fnp,Sethi:2017phn}. For instance, in a non-supersymmetric model with open strings, the one-loop vacuum energy diverges when computed at constant dilaton \cite{Raucci:2024fnp,Polchinskiv2}, reflecting the fact that a constant $\phi$ is not a solution of the classical equations of motion. In our case, the one-loop calculation is finite since the RFM is a solution to the classical equations of motion. Were we to compute the higher $\mathcal{C}_i$, we would have to do it in an iterative process where the $(n+1)$-loop calculation is done around an $n$-loop corrected vacuum. For instance, the two-loop Casimir energy will diverge unless computed at the precise point of moduli space where we found a one-loop maximum. While they certainly complicate a detailed treatment, these are all standard issues and we do not view them as a fundamental obstacle to finding a corrected solution.

If the solution is safe from loop and higher-derivative corrections, what about non-perturbative effects? In M-theory, well-known non-perturbative effects come from Euclidean M2 and M5 branes wrapping cycles of $\mathcal{F}_6$. Using the formulae of Appendix \ref{app:fluxQ}, we can estimate the size of these effects as
\begin{align} 
    e^{-S_{\text{M2}}}\sim e^{-T_{\text{M2}} \big(\beta^{\frac12}\frac{R_1R_2^2}{4}\big)}\approx\, e^{-180} \,,
    && e^{-S_{\text{M5}}}\sim e^{-T_{\text{M5}}\frac{R^6}{4}}\approx e^{-3.04\times 10^{3}} \,,
\end{align}
so they are completely negligible. 

Since our solution is non-supersymmetric, we may also have to worry about other non-perturbative effects, such as e.g. gravitational instantons. Any instanton which may be described by 11d supergravity would have, on dimensional grounds, an action scaling as $(R\, \ell_{11})^{11}$, and therefore would be even more suppressed than the M2, M5 branes described above.

So it seems that the solution is stable against higher-derivative, classical, and quantum corrections! How does this square with the natural instinct that one should not trust a solution whose smallest physical closed curve is just $\frac18 \beta^{1/2}R_1\sim 3.87\, \ell_{11}$? Even if all the corrections we know are under control, we must also worry about unknown ones. More concretely, the analysis here has pushed 11-dimensional supergravity all the way to a few Planck lengths. We do not know if this is correct: there could be e.g. additional states, not controlled by supergravity, appearing at a mass scale $\mu\lesssim M_{11}$. If such states exist, they are necessarily non-supersymmetric, but we have no evidence that this is in fact the case---it is just a worry at this point. In fact, if $\mu$ is close to $M_{11}$, they could be thought of as ``small'' quantum-mechanical black holes. On general grounds, we expect such objects to exist in the theory. In at least one class of examples, that of $AdS_3$ quantum gravity, where there is more control, we know that there are BTZ black hole states with $\mu=1/(8G)$, and every quantum gravity in $AdS_3$ must have at least one operator other than the graviton with $\mu\leq 3/(32\, G)$ \cite{Benjamin:2019stq}. To estimate the effect of such particles, we will pretend the compactification is on a $T^6$ factorized as $T^2\times T^4$, where the radii of the first $T^2$ are  $\frac18 \beta^{1/2}R_1$, $\frac18 \beta^{-1/2}R_1$, and that of the $T^4$ is $\sqrt{2}\, R_2$. We wish to estimate the contribution to the 5d vacuum energy of an additional particle of mass $\mu$. An expression for the Casimir energy of a massive particle of mass $\mu$ on a circle was given in \cite{Arkani-Hamed:2007ryu}. Generalizing to an arbitrary torus,  this yields an expression 
\begin{equation}
    \delta V^{(5d)}= \frac{2^{\frac{10}{3}}}{R^{10}}\frac{R^6}{4} \left[\sum _{\vec{l}\in \Lambda} \frac{\left(2 \mu ^{11}\right) K_{\frac{11}{2}}(\mu  n\vert\vec{l}\vert)}{(2 \pi )^{\frac{11}{2}} (\mu  n \vert\vec{l}\vert)^{\frac{11}{2}}}\right].
    \label{e343f}
\end{equation}
where the sum runs over the defining lattice of the $T^6$. Using the parameters above, we find that even for $\mu=1$,  we have $\delta V^{(5d)}\sim 1.3\cdot 10^{-10}$, which is still a factor of $1000$ smaller than our result.  Actually, this value is not so far from the $\mu\rightarrow0$ limit, in which the above becomes just another contribution to the vacuum energy of an additional massless field, contributing an amount within a factor of 20 of our actual result. In either case, we get that  e.g. $\sim1000$ states at $\mu=1$ might have a contribution large enough to significantly impact our solution. Including the sum over spin structures reduces the estimate by an $\mathcal{O}(1)$ factor, increasing the hierarchy to $10^4$. Thus, if M-theory contained $10^4$ Planckian black hole microstates, they could overcome our solution; although this is more than what a naive extrapolation of the Bekenstein-Hawking formula would suggest, we have no way of guaranteeing this does not happen.

The comments above also apply to e.g. string states, that may arise from wrapping an M2 brane along the small circle of radius $\frac18 \beta^{1/2}R_1$. If the radius was subplanckian, we would be in the type IIA regime, and the wound M2 states would give us perturbative string states.  Since our radius is slightly superplanckian instead, we are in the M-theory regime, and the states are heavy; the associated mass scale is 
\begin{equation} 
    \sqrt{T}\sim (2\pi^2)^{1/3} \beta^{1/2}\frac{R_1}{8}\approx 10.45\,,
\end{equation}
so the wound M2 states are heavier than the black hole states we considered above, and are therefore more suppressed. In fact, since the string is so heavy, these string states will automatically be within the Schwarzschild radius, and therefore, they are part of the black hole spectrum we estimated.

In any case, there are many subtleties in estimating the effect of these Planckian states: what is the effect of including interactions, what is the lifetime of these states (they would not contribute if they are extremely short-lived), are we even allowed to run black holes in loops in the na\"ive way that we just did. Furthermore, there are other possibilities we do not know how to even begin to quantify properly. Suppose (again along one of the lines in \cite{Sethi:2017phn}) that someone comes and tells us they have discovered an exotic gravitational instanton in M-theory, with a singular core (so that its action cannot be reliably computed by supergravity), and that its action is small enough to destabilize our solution. We cannot rigorously rule out such possibility (although if true, it would have implications far beyond our solution, as it would mean the presence of unexpectedly light instantons in M-theory compactifications). The punchline is that, although things seem fine in the crude analysis we just did,  we do not have a way to control corrections such as those in equation \eq{e343f}, since we do not know the spectrum of metastable Planckian states in M-theory. To some extent, this aligns with the intuition that $\frac18 \beta^{1/2}R_1\sim 3.87\, \ell_{11}$ is dangerous. 

Absent new windows into non-perturbative M-theory, there is no way to ensure our solution is protected against these ``black hole loops''. The solution might actually be there, or it might not. On the other hand, we did achieve a huge improvement over the solution in Section \ref{sec:warm-up-dS7}. We see the RFM compactification studied here as a proof of principle, which illustrates that this not so well-known corner of the String Landscape may harbor hidden gems. Perhaps, by choosing a better RFM in one setup or another, we can push the size of the internal manifold to a regime where we feel safer from black hole and exotic states.

\subsection{The \texorpdfstring{$dS_5\times \mathcal{F}_6$}{dS5xF6} vacuum of M-theory: An executive summary}
The main point of this section---and of the whole paper---is that we achieved a $dS_5\times \mathcal{F}_6$ solution of 11-dimensional supergravity, where
\begin{equation}\mathcal{F}_6=T^6/\mathbb{Z}_8\end{equation}
is a quotient of $T^6$ by a fixed-point free $\mathbb{Z}_8$ isometry that breaks all supersymmetries. Casimir energies and $G_4$ flux balance each other to produce a classical maximum. The vacuum energy is small in Planck units, the masses of tachyons and other light fields are of order Hubble scale, the solution is scale separated, and is protected against all known higher-derivative, loop, and classical corrections. We summarize some salient features of the solution in Table \ref{resumen} for rapid reference, with pointers to parts of the paper that discuss each of these in more detail.

\begin{landscape}
\renewcommand{\arraystretch}{1.4}
\begin{table}
\centering
\begin{tabular}{ccccc}
\hline\textbf{Property}&\textbf{Value}&\textbf{Comments}&\textbf{Details in \ldots}\\\hline\hline
5d vacuum energy & $V^{(5d)}\approx4.43\cdot 10^{-8}\ell_5^{-5}$& \begin{tabular}{@{}c@{}}Quite small due to large\\ Casimir coefficient $\mathcal{C}$\end{tabular} &Subsections \ref{subsec:dS5-maximum}, \ref{sec:ctrl} \\\hline
Hubble radius& $H_0^{-1}\sim 1.16\cdot 10^4\, \ell_5 = 8.43\cdot 10^2\, \ell_{11}$&\begin{tabular}{@{}c@{}}$\ell_{11}/\ell_{5}\sim13.78$\\ (species vs. 5d Planck scales)\end{tabular}&Subsection \ref{sec:ctrl}\\\hline
Internal space volume& $\mathcal{V}= 2.61\times 10^3\,\ell_{11}^6=1.79\cdot10^{10}\,\ell_5^6$& Large volume helps with control &Subsections \ref{subsec:dS5-maximum}, \ref{sec:ctrl}\\\hline
Internal manifold radii& \begin{tabular}{@{}c@{}}$\frac18\beta^{1/2} R_1\sim 3.87\, \ell_{11}=53.3\, \ell_5,$ \\$\beta^{-1/2} R_1\sim 4.56\, \ell_{11}=62.8\, \ell_5,$ \\$\sqrt{2}\, R_2\sim 4.15\, \ell_{11}=57.2\, \ell_5$\end{tabular}&\begin{tabular}{@{}c@{}} Potential impact of black\\ hole-like states due to small $R_i$\end{tabular} &Subsections \ref{subsec:dS5-maximum}, \ref{sec:ctrl}\\\hline
Scale separation& Yes: $ \frac18\beta^{\frac12} R_1= 4.6\times 10^{-3} \, H_0^{-1}$ & Facilitates 5d EFT validity&Subsection \ref{sec:ctrl}\\\hline
Geometric moduli& \begin{tabular}{@{}c@{}}Five, with masses \\ $m_i^2=\{-83.3 \,,\, -5.98 \,,\, -359 \,,\, 6.68\,,\, -22.1\}\cdot H_0^2$ \end{tabular}&\begin{tabular}{@{}c@{}}All masses light, of order $H_0^2$;\\ complies with \cite{Ooguri:2018wrx}\end{tabular}&Subsection \ref{sec:ctrl}\\\hline
Axions & Five, with negligible masses&Generated by Euclidean M2, M5&Subsection \ref{sec:ctrl}\\\hline
Vectors&Seven: two KK photons, five from $C_3$ &Exactly massless&Subsection \ref{sec:ctrl}\\\hline
Light fermions& None&\begin{tabular}{@{}c@{}} Due to antiperiodic spin \\ structure in covering $T^6$\end{tabular}&Subsection \ref{sec:ctrl}\\\hline
\begin{tabular}{@{}c@{}}11d average\\ energy density\end{tabular}& $\rho_{11d}=8.40\cdot10^{-6}\,\ell_{11}^{-11}$&Control parameter for corrections&Subsections \ref{sec:warm-up-dS7}, \ref{sec:ctrl}\\\hline
\begin{tabular}{@{}c@{}}Flux  \& Casimir \\ contributions to $\rho_{11d}$\end{tabular} & $\rho_{\text{Cas}}\approx -3.36\cdot10^{-5}\,\ell_{11}^{-11}, \ \ \rho_{G_4}\approx 4.20\cdot10^{-5}\, \ell_{11}^{-11}$&\begin{tabular}{@{}c@{}}Each contribution larger than $\rho_{11d}$;\\ they partially cancel out\end{tabular}&Subsection \ref{sec:ctrl}\\\hline
Internal gradients& $\langle \rho_{11d}\rangle^2\sim 10^{-7}$ &\begin{tabular}{@{}c@{}} Much smaller than KK scale\end{tabular}&Subsection \ref{sec:ctrl}\\\hline
\end{tabular}
\caption{Summary of some properties of the $dS_5\times \mathcal{F}_6$ solution. Here $\ell_{11}$ is the 11-dimensional reduced Planck length, defined by $\kappa_{11}^2=\ell_{11}^9$, and $\ell_5$ is the five-dimensional reduced Planck length, analogously defined.}
\label{resumen}
\end{table}
\renewcommand{\arraystretch}{0.9}
\end{landscape}

\section{A scan for M-theory \texorpdfstring{$dS_4$}{dS4} maxima}
\label{sec:dS4-maxima}

As mentioned in \secref{sec:RFMs}, we will focus on cyclic RFM's. The reason for this is twofold---on the one hand the group $\Gamma$ for cyclic RFM's has a single generator, which makes the explicit computation of the Casimir energies more tractable and expressible in terms of that generator alone; on the other hand, it also allows us to scan a family of RFM's, since a full set of cyclic RFM's can be obtained for a given dimension. 

\subsection{Cyclic Riemann-flat manifolds in 7d}

Consider the compactification of M-theory down to 4d, which requires a 7-dimensional RFM, 
\begin{equation}
    \mathcal{F}_7 = \RFM{7} \,. 
\end{equation}
We know that M-theory on $\mathcal{F}_7$ will not admit a $dS_4$ minimum, since we are not allowed to turn on both $G_4$ and $G_7$ fluxes without causing a runaway for an axion that descends from $C_3$ \cite{dS-nogos-MMBB}. Here instead we will look for a maximum, as discussed in previous sections. 

Recall that we want the RFM to contribute in two key ways: we want it to freeze as many geometric moduli as possible and to enhance the Casimir energy contribution. While the former follows directly from the $\D[g]$ associated with the single generator $\mathbf{g}$ of the cyclic group $\Gamma$, the latter requires a choice of spin structure that avoids cancellations between boson and fermion contributions. In choosing $\mathcal{F}_7$, we must therefore look for the number of moduli and allowed spin structures of all 7d RFM's generated by a cyclic group $\Gamma$, which for RFM's with an $S^1$ base have generators of the form\footnote{As explained in \secref{sec:RFMs}, we can generalise the shift vector $\bvec[g]$ to any vector such that $\D[g]\,\bvec[g] = \bvec[g]\,\text{mod}\,\Z$ and $n\,\bvec[g]\in\Z^7\implies n\geq |A|$ \eq{eq:shift-vectors}. Here we restrict to the simplest case, as this does not affect any of the conclusions of this section.}
\begin{equation}
    \left\{\mathbf{g}: \D[g] = \begin{pmatrix}
        \mathbf{A} & 0 \\
        0 & 1 \\
    \end{pmatrix} \,, \bvec[g] = \begin{pmatrix}
        \vec{0} \\
        |\mathbf{A}|^{-1} 
    \end{pmatrix} \right\} \,,
\end{equation}
for finite order $6\times 6$ matrices $\mathbf{A}$, i.e. $\mathbf{A}^n = \mathbf{I}$ for some finite $n$. To classify these we will follow standard techniques; see \cite{Ono:1988kh} for a more detailed explanation and the results of interest in our case. The smallest such $n$ gives the order of $\mathbf{A}$, which we denote as $|\mathbf{A}|$. Any finite matrix must have a minimal polynomial $m_{\bf A}(x)$ that divides both $x^n - 1$ and its characteristic polynomial (cf. Cayley–Hamilton theorem), so that a large class of finite order matrices\footnote{These matrices correspond to the torsion elements of $GL(6,\Z)$.} $\mathbf{A}\in GL(6,\Z)$ can be obtained by factorising $x^n - 1$ into \textit{cyclotomic polynomials}\footnote{The $n^{\rm th}$ cyclotomic polynomial is the unique irreducible polynomial with integer coefficients that is a divisor of $x^n-1$ and does not divide $x^k-1$ for any $k<n$. It is given by $$\Phi_n(x) = \prod_{1\leq k\leq n} (x - e^{2\pi i\frac{k}{n}}) \,,$$ with ${\rm gcd}(k,n)=1$, i.e. $k$ and $n$ coprime.} and building matrices $\mathbf{A}$ that are direct sums of the companion matrices of these cyclotomic polynomials   \cite{MatricesFiniteOrder} (the precise statement is that any matrix of finite order and integer entries is conjugate over $SL(6,\mathbb{Q})$ to one of the matrices we construct here). We must therefore look for all possible products of cyclotomic polynomials $\Phi_n(x)$ whose degrees---given by Euler's totient function $\varphi(n)$---add up to $6$. 

Let us illustrate this with a concrete example that will be relevant in what follows. The cyclotomic polynomial $\Phi_7(x)$ has degree $6$ and has companion matrix $\mathbf{A}$ such that $\mathbf{A}^7 - \mathbf{I} = 0$, i.e. an order 7 matrix. This matrix can be used to define a 7d RFM with $S^1$ base whose quotient group $\Gamma = \Z_7$ is generated by 
\begin{equation}
    \D[g] = \left(
        \begin{array}{*6{C{1.05em}}|*1{C{1.05em}}}
         0 & 0 & 0 & 0 & 0 & -1 & 0 \\
         1 & 0 & 0 & 0 & 0 & -1 & 0 \\
         0 & 1 & 0 & 0 & 0 & -1 & 0 \\
         0 & 0 & 1 & 0 & 0 & -1 & 0 \\
         0 & 0 & 0 & 1 & 0 & -1 & 0 \\
         0 & 0 & 0 & 0 & 1 & -1 & 0 \\ \hline 
         0 & 0 & 0 & 0 & 0 & 0 & 1
        \end{array}
    \right) \,, \quad
    \bvec[g] = \left(
        \begin{array}{c}
         0 \\ 0 \\ 0 \\ 0 \\ 0 \\ 0 \\ \frac17 
        \end{array} 
    \right) \,. 
\end{equation}
The condition $\delta\mathbf{G}\cdot\D[g] = (\D[g]^T)^{-1}\cdot\delta\mathbf{G}$ \eqref{eprot0} leaves 4 invariant moduli and $(\mathbf{I} - \D[g])^T\cdot\vec{h} \in\Z^7$ \eqref{mimi} only allows for the spin structure $\vec{h} = (0,...,0,h_7)$ with $h_7\in\{0,\frac12\}$. We will come back to this example shortly. 

Since $\varphi(5) + \varphi(6) = 6$, a less direct example is a block diagonal matrix whose blocks are the companion matrices of $\Phi_5(x)$ and $\Phi_6(x)$, which has order ${\rm lcm}(5,6)=30$ and can be used to define a 7d RFM with base $S^1$ and holonomy group $\Gamma = \Z_{30}$ generated by 
\begin{equation}
    \D[g] = \left(
        \begin{array}{*4{C{1.05em}}|*2{C{1.05em}}|*1{C{1.05em}}}
         0 & 0 & 0 & -1 & 0 & 0 & 0 \\
         1 & 0 & 0 & -1 & 0 & 0 & 0 \\
         0 & 1 & 0 & -1 & 0 & 0 & 0 \\
         0 & 0 & 1 & -1 & 0 & 0 & 0 \\ \hline  
         0 & 0 & 0 & 0 & 0 & -1 & 0 \\
         0 & 0 & 0 & 0 & 1 & 1 & 0 \\ \hline  
         0 & 0 & 0 & 0 & 0 & 0 & 1 \\
        \end{array}
    \right) \,, \quad
    \bvec[g] = \left(
        \begin{array}{c}
         0 \\ 0 \\ 0 \\ 0 \\ 0 \\ 0 \\ \frac{1}{30}
        \end{array} 
    \right) \,. 
    \label{eq:Z30-RFM}
\end{equation}
Note that this is not the only order 30 matrix that we can build, since $\varphi(3)+\varphi(10) = \varphi(6) + \varphi(10) = 6$ and ${\rm lcm}(3,10) = {\rm lcm}(6,10) = {\rm lcm}(5,6) = 30$. All of these leave 4 invariant moduli and only allow for periodic spin structures, due to the constraints \eq{mimi} and \eq{swer2}. 

From all combinations of $\Phi_n(x)$ such that $\sum \varphi(n) = 6$, one can build in this way 78 finite matrices $\mathbf{A}$ that correspond to generators of finite groups $\Gamma$ and therefore cyclic RFM's (including the identity, i.e. the trivial group). We will say that cyclic RFM's of this form are of \emph{diagonal} type. In Appendix \ref{ap:7d-RFMs} we summarise all 77 non-trivial cyclic diagonal RFM's with base $S^1$, giving their generators $(\D[g],\bvec[g])$, group order, number of moduli and allowed spin structures. Generically, bigger groups tend to fix more moduli and to have more constrained spin structures. It is also important to note the structure of the generator $\D[g]$---it always contains at least two blocks (the base and the fibre), but in many cases it contains more. In order to work with a 2-moduli potential and, in particular, a Casimir potential that only depends on one modulus, we can look for enhanced symmetry points that allow us to fix some of the remaining moduli at a point that is guaranteed to be a critical point of the potential, as we did in \secref{subsec:dS5-maximum-RFM}. This will generally require the fibre to be a single block---that is the case of the $\Z_7$ example, but not of the $\Z_{30}$ that always contains three blocks. Therefore, while the $T^7/\Z_7$ compactification can be reduced to a 2-moduli problem, working with $T^7/\Z_{30}$ (which could provide a bigger enhancement of the Casimir potential) will always require the study of at least three moduli. Moreover, RFM's with more than two blocks will also have a richer structure regarding their Casimir branes (i.e. the invariant subspaces under each element in the group); for example, the Casimir energy on the $\Z_{30}$ quotient defined by \eqref{eq:Z30-RFM} can be seen as the sum over 200 codimension-five, 20 codimension-three, 10 codimension-one and 1 space-filling Casimir branes wrapping appropriate cycles of the RFM (for comparison, the $\Z_7$ quotient studied in the next subsection can be seen as the sum of a total of 42 codimension-six and 1 space-filling Casimir branes). Since the behavior of the Casimir potential depends crucially on the properties of the invariant subspaces, a proper study of these RFM's and their potential saddles requires a detailed analysis of their Casimir brane structures. 

In order for a finite matrix $\mathbf{A}$ to be a single $6\times 6$ block, the factorisation in terms of cyclotomic polynomials must contain a single $\Phi_n(x)$ of degree $6$. These matrices will then be the order-$n$ companion matrices of $\Phi_n(x)$ for each $n$ such that $\varphi(n)=6$; there are 4 such cases, $\varphi(7) = \varphi(9) = \varphi(14) = \varphi(18) = 6$, which define RFM's that are $\Z_7\,,\,\Z_9\,,\,\Z_{14}$ and $\Z_{18}$ quotients of $T^7$ (see Appendix \ref{ap:7d-RFMs}). All these RFM's have invariant metrics $G$ with $4$ moduli and, while the odd $\Z_7$ and $\Z_9$ quotients allow for the spin structures $\vec{h} = \{0,0,0,0,0,0,h_7\}$, the even ones $\Z_{14}$ and $\Z_{18}$ are further constrained by \eq{swer2} to have a periodic spin structure along every cycle. As we will see explicitly for the $\Z_7$ case in the next subsection, the periodic spin structure along the fibre does not allow for a critical point of the Casimir potential and prevents us from finding a $dS_4$ saddle point. This result can be extended to the $\Z_9$, $\Z_{14}$ and $\Z_{18}$ quotients, as we will explain at the end of the section.

\subsection{Example: \texorpdfstring{$\mathcal{F}_7 = T^7/\Z_7$}{F7}}
\label{sec:dS4-T7-example}
Let us focus on $\mathcal{F}_7 = T^7/\Z_7$. 
The most general Riemann-flat metric on $\mathcal{F}_7$, i.e. a metric on $T^7$ that is invariant under $\D[g]$, is 
\begin{align}
    \renewcommand{\arraystretch}{1.2}
    \textbf{G} = \left(\begin{array}{*6{c}|*1{c}}
         2\, R_2^2 & \gamma -R_2^2 & \alpha  & -\alpha -\gamma  & -\alpha -\gamma  & \alpha  & 0 \\
         \gamma -R_2^2 & 2\, R_2^2 & \gamma -R_2^2 & \alpha  & -\alpha -\gamma  & -\alpha -\gamma  & 0 \\
         \alpha  & \gamma -R_2^2 & 2\, R_2^2 & \gamma -R_2^2 & \alpha  & -\alpha -\gamma  & 0 \\
         -\alpha -\gamma  & \alpha  & \gamma -R_2^2 & 2\, R_2^2 & \gamma -R_2^2 & \alpha  & 0 \\
         -\alpha -\gamma  & -\alpha -\gamma  & \alpha  & \gamma -R_2^2 & 2\, R_2^2 & \gamma -R_2^2 & 0 \\ 
         \alpha  & -\alpha -\gamma  & -\alpha -\gamma  & \alpha  & \gamma -R_2^2 & 2\, R_2^2 & 0 \\ \hline
         0 & 0 & 0 & 0 & 0 & 0 & R_1^2 
    \end{array}\right) \,,
    \label{eq:Z7-metric}
\end{align}
which has $4$ moduli $\{R_1,R_2,\alpha,\gamma\}$---out of the 28 moduli of $T^7$, the $\Z_7$ quotient has fixed 24. 

The metric \eqref{eq:Z7-metric} has a rather suggestive form---at the point $\alpha=\gamma=0$ in moduli space, the upper-left block is precisely the Cartan matrix of SU(7) $(A_6)$ rescaled by $R_2^2$. As explained in detail in \secref{subsec:dS5-maximum-RFM}, one may choose a transformation in the normalizer of $\Z_7$ that becomes an isometry of the metric at this special point $\alpha=\gamma=0$, therefore guaranteeing a critical point of the potential along these directions.

Consider the transformation $\vec{z}\to\textbf{Q}\,\vec{z}$, with 
\begin{align}
    \textbf{Q} = \left(\begin{array}{*6{C{1.25em}}|*1{C{1.25em}}}
         1 & 0 & 0 & 0 & 0 & 0 & 0 \\
         1 & -1 & 1 & 0 & 0 & 0 & 0 \\
         1 & -1 & 1 & -1 & 1 & 0 & 0 \\
         1 & -1 & 1 & -1 & 1 & -1 & 0 \\
         0 & 0 & 1 & -1 & 1 & -1 & 0 \\
         0 & 0 & 0 & 0 & 1 & -1 & 0 \\ \hline
         0 & 0 & 0 & 0 & 0 & 0 & 1 \\
    \end{array}\right) \,,
\end{align}
which is a finite matrix of order 3. This transformation is in the normalizer of $\Z_7$, i.e. it leaves the $\Z_7$ group invariant. The metric \eq{eq:Z7-metric} is invariant under this transformation, i.e. $\mathbf{Q}^T\cdot\mathbf{G}\cdot\mathbf{Q} = \mathbf{G}$, on the subspace with $\alpha=\gamma=0$ and thus this becomes a $\Z_3$ isometry\footnote{In order to pick such a transformation $\mathbf{Q}$, we can focus on the metric $\mathbf{G}$ we are aiming for: the Cartan matrix of $A_6$. One can pick a transformation that only leaves the subspace with $\alpha = \gamma = 0$ invariant from the Weyl group of $A_6$, i.e. the symmetric group $S_6$, which is guaranteed to leave the Cartan matrix of $A_6$ invariant. We must also check that such a $\mathbf{Q}$ belongs to the normalizer of $\Z_7$ in order for it to be a symmetry of the RFM.}. Since the components of the gradient $(\partial_\alpha V,\partial_\gamma V)$ transform as a vector under this $\Z_3$, if our choice of $G_4$ flux respects this isometry we will necessarily have $\partial_\alpha V = \partial_\gamma V = 0$. Therefore, at points in moduli space such that 
\begin{align}
    \mathbf{G} = R_1^2 \left(\begin{array}{*6{C{1.25em}}|*1{C{1.25em}}}
            0 & 0 & 0 & 0 & 0 & 0 & 0 \\
            0 & 0 & 0 & 0 & 0 & 0 & 0 \\
            0 & 0 & 0 & 0 & 0 & 0 & 0 \\
            0 & 0 & 0 & 0 & 0 & 0 & 0 \\
            0 & 0 & 0 & 0 & 0 & 0 & 0 \\
            0 & 0 & 0 & 0 & 0 & 0 & 0 \\ \hline 
            0 & 0 & 0 & 0 & 0 & 0 & 1 \\  
        \end{array}\right) 
        + R_2^2 \left(\begin{array}{*6{C{1.25em}}|*1{C{1.25em}}}
            2 & -1 & 0 & 0 & 0 & 0 & 0 \\
            -1& 2 & -1 & 0 & 0 & 0 & 0 \\
            0 & -1 & 2 & -1 & 0 & 0 & 0 \\
            0 & 0 & -1 & 2 & -1 & 0 & 0 \\
            0 & 0 & 0 & -1 & 2 & -1 & 0 \\
            0 & 0 & 0 & 0 & -1 & 2 & 0 \\ \hline
            0 & 0 & 0 & 0 & 0 & 0 & 0 \\  
        \end{array}\right)\,,
        \label{eq:Z7-G-2-moduli}
\end{align}
the potential is automatically a saddle along all directions apart from $\{R_1,R_2\}$. 

The Casimir potential can be written in terms of 7 sums over the covering $T^7$ lattice, one for each element of $\Z_7$. Since we are dealing with a cyclic group, these elements are simply $(\D[g]^j,j\cdot\bvec[g])$, for $j=0,...,6$, with $\D[g]^0 = \mathbf{I}$.
Recall that the Casimir energy $\Vcas$ on $\mathcal{F}_7 = T^7/\Z_7$ is given by,
\begin{align}
    \Vcas = -\frac{\Gamma(s)}{2\pi^{s}}\cdot \int_{\mathcal{F}_7} d^{7}z ~\sqrt{G}~\sum_{j=0}^{6}\sum_{\vec{n}\in\Z^7} \frac{\TrB{\D[h]^j} - \TrF{\D[g]^j}\cdot e^{2\pi i \Vec{h}\cdot\Vec{n}}}{|(\mathbf{I} - \D[g]^j)\,\vec{z} - j\,\bvec[g] + \vec{n}|^{2s}} \,, 
\end{align}
with $s=11/2$ for M-theory. The contribution of each group element (apart from the identity) can be seen as the energy of seven Casimir branes localised at 
\begin{align}
    (0,0,0,0,0,0),
    &\left(\tfrac{6}{7},\tfrac{5}{7},\tfrac{4}{7},\tfrac{3}{7},\tfrac{2}{7},\tfrac{1}{7}\right),
    \left(\tfrac{5}{7},\tfrac{3}{7},\tfrac{1}{7},\tfrac{6}{7},\tfrac{4}{7},\tfrac{2}{7}\right),
    \left(\tfrac{4}{7},\tfrac{1}{7},\tfrac{5}{7},\tfrac{2}{7},\tfrac{6}{7},\tfrac{3}{7}\right),\nonumber\\
    &\left(\tfrac{3}{7},\tfrac{6}{7},\tfrac{2}{7},\tfrac{5}{7},\tfrac{1}{7},\tfrac{4}{7}\right),
    \left(\tfrac{2}{7},\tfrac{4}{7},\tfrac{6}{7},\tfrac{1}{7},\tfrac{3}{7},\tfrac{5}{7}\right),
    \left(\tfrac{1}{7},\tfrac{2}{7},\tfrac{3}{7},\tfrac{4}{7},\tfrac{5}{7},\tfrac{6}{7}\right)
\end{align} 
on the $T^6$ fibre and wrapping the base $S^1$, corresponding to the subspaces invariant under $\D[g]$, i.e. $\D[g]\,\vec{z} = \vec{z}~\text{mod}~\Z^7$. The identity element contributes as a space-filling Casimir brane, wrapping the whole of $\mathcal{F}_7$. Given the block diagonal structure of the invariant metric \eq{eq:Z7-G-2-moduli}, the norm splits into two pieces
\begin{align}
    \Vcas = -\frac{\Gamma(s)}{2\pi^{s}}\cdot \frac{\sqrt{7}R_1 R_2^6}{|\Z_7|}\cdot \sum_{j=0}^{6}\int_{T^7} d^{6}z ~\sum_{\substack{\vec{n}\in\Z^6 \\ m\in\Z}} \frac{\TrB{\D[g]^j} - \TrF{\D[g]^j}\cdot e^{2\pi i \, m\, h_7 }}{\Big[R_2^2\,|(\mathbf{I} - \D[g]^j)\,\vec{z} + \vec{n}|^2 + R_1^2\,|m - \frac{j}{7}|^2 \Big]^{s}} \,,
\end{align}
where we also split the 7d vector $\vec{n}$ into a 6d vector $\vec{n}$ along the fibre and 1d vector $\vec{m}=m$ along the base, and used the only allowed spin structure on $\mathcal{F}_7$, $\vec{h}=(0,0,0,0,0,0,h_7)$. Substituting in $|\Z_7| = 7$ and performing the change of variables 
\begin{equation}
    R_1 = \frac{R}{x^{6/7}} \,,
    \quad 
    R_2 = R~x^{1/7} \,,
\end{equation}
the Casimir potential can be written as
\begin{align}
    \Vcas = -\frac{\Gamma(s)}{2\pi^{s}}\cdot \frac{x^{\frac{66}{7}}}{\sqrt{7}\cdot R^{4}}\cdot \sum_{j=0}^{6}\int_{T^6} d^{6}z ~\sum_{\substack{\vec{n}\in\Z^6 \\ m\in\Z}} \frac{\TrB{\D[g]^j} - \TrF{\D[g]^j}\cdot e^{2\pi i \,m\, h_7}}{\Big[x^2\,|(\mathbf{I} - \D[g]^j)\,\vec{z} + \vec{n}|^2 + |m - \frac{j}{7}|^2 \Big]^{s}} \,.
\end{align}
Note that the constraint \eq{swer2} correlates our choice of $h_7$ with a choice of spin lift $\pm\mathcal{D}_\mathbf{g}$, since $\mathcal{D}_\mathbf{g}^7 = (-1)^{s_\mathbf{g}}\mathbf{I}$ is such that 
\begin{equation}
    s_\mathbf{g} = 2\cdot 7\,\vec{h}\cdot\bvec[g]\,\,\text{mod}\,\, 2\Z = 2h_7 \,\,\text{mod}\,\, 2\Z \,. 
\end{equation}
Thus the traces over bosons and fermions are
\begin{table}[h]
    \centering
    \renewcommand{\arraystretch}{1.2}
    \begin{tabular}{c|ccccccc}
         $j$ & 0 & 1 & 2 & 3 & 4 & 5 & 6  \\ \hline 
        $\TrB{\D[g]^j}$ & 128 & 2 & 2 & 2 & 2 & 2 & 2 \\
        $\TrF{\D[g]^j}$ & 128 & $\pm2$ & 2 & $\pm2$ & 2 & $\pm2$ & 2 
    \end{tabular}
    \label{tab:traces-Z7}
\end{table}

\noindent with the $(+)$ sign for $h_7=0$ and the $(-)$ sign for $h_7=\frac12$. We see immediately that choosing $h_7=0$ gives $\Vcas = 0$, reflecting the fact that supersymmetry is preserved by this choice. Indeed, four spinor components are left invariant by $\D[g]$, which therefore preserves four supercharges and results in a 4d $\mathcal{N}=2$ theory. We must therefore choose $h_7=\frac12$ in order to break this SUSY. 

Except for the identity $\D[g]^0 = \mathbf{I}$, all elements $\D[g]^j$ in the sum have the same 1-dimensional invariant subspace, and we can use \eq{eq:VC-invariant-subspace} to write their contribution to the potential as
\begin{align}
    \Vcas^{\rm twisted} &= -\frac{\Gamma(s-3)}{2\pi^{s-3}}\cdot\frac{x^{\frac{24}{7}}}{7R^4}2\sum_{j=1}^6\sum_{m\in\Z}\frac{1-(-1)^{j+m}}{(m - \frac{j}{7})^{2s-6}} \nonumber \\
    &= -\frac{\Gamma(s-3)}{2\pi^{s-3}}\cdot\frac{x^{\frac{24}{7}}}{7R^4}\cdot  \frac{2^{2s-6}-1}{2^{2s-7}} \left(7^{2s-6}-1\right) \zeta (2s-6) \,.
\end{align}
We thus find the Casimir potential 
\begin{equation}
    \Vcas = - \frac{\mathcal{C}_{\bf I}(x)}{R^4} 
    - \underbrace{\frac{\Gamma(\frac52)}{2\pi^{\frac52}}\cdot  \frac{2^{5}-1}{2^{4}}\cdot \frac{7^{5}-1}{7} \zeta (5)}_{\approx 733.1}\cdot \frac{x^{\frac{24}{7}}}{R^{4}} \,,
\end{equation}
including the identity term $\mathcal{C}_{\bf I}(x)$, 
\begin{align}
    \mathcal{C}_{\bf I}(x) = \frac{\Gamma(s)}{2\pi^{s}}\cdot \frac{128\, x^{\frac{66}{7}}}{\sqrt{7}} ~\sum_{\substack{\vec{n}\in\Z^6 \\ m\in\Z}} \frac{1 - (-1)^{m}}{\Big[x^2\,|\vec{n}|^2 + m^2 \Big]^{s}} \,.
\end{align}
Note in particular that all terms with even $m$ vanish, independently of $\vec{n}$, due to the fact that we have twisted boundary conditions for the fermions around the $S^1$ base, $\vec{h} = (0,0,0,0,0,0,\frac12)$.
Although $\mathcal{C}_{\bf I}(x)$ must be summed numerically, we can study its asymptotic behaviour as $x\to 0$ and $x\to\infty$. For large $x$, the dominant terms in the sum have $\vec{n}=0$, so that 
\begin{align}
    \mathcal{C}_{\bf I}(x) \sim \frac{\Gamma(s)}{2\pi^{s}}\cdot \frac{x^{\frac{66}{7}}}{\sqrt{7}\cdot 2^{2s-8}} ~\sum_{\substack{m\in\Z}} \frac{1}{|m + \frac{1}{2}|^{2s}} 
    = \underbrace{\frac{\Gamma(\frac{11}{2})}{2\pi^{11/2}}\cdot \frac{(2^{11} - 1)\zeta(11)}{4\sqrt{7}}}_{\approx 9.337}\, x^{\frac{66}{7}} \quad\text{as}~ x\to\infty \,.
\end{align}
On the other hand, for small $x$, the dominant terms \textit{would} correspond to $m=0$, since these would get an $x^{-11}$ enhancement---however, all these terms cancel and cannot contribute to the sum. Consequently, in this limit we find

\begin{equation}
    \mathcal{C}_{\bf I}(x) \sim \frac{186\zeta(5)}{7\pi^2} \,x^{\frac{24}{7}} 
    \approx 2.791\,x^{\frac{24}{7}} 
    \quad\text{as}~ x\to 0 \,, 
\end{equation}
which we obtain using the Mellin transform followed by Poisson resummation, as we did to compute analytically some of the sums in \secref{sec:Casimir-on-RFMs}. Although here we cannot obtain the exact analytical result, we can take the zero mode in the Poisson sum and neglect the exponentially suppressed terms to derive this asymptotic behaviour.

\begin{figure}[!htb]
\centering
\begin{subfigure}{0.49\textwidth}
    \includegraphics[width=\textwidth]{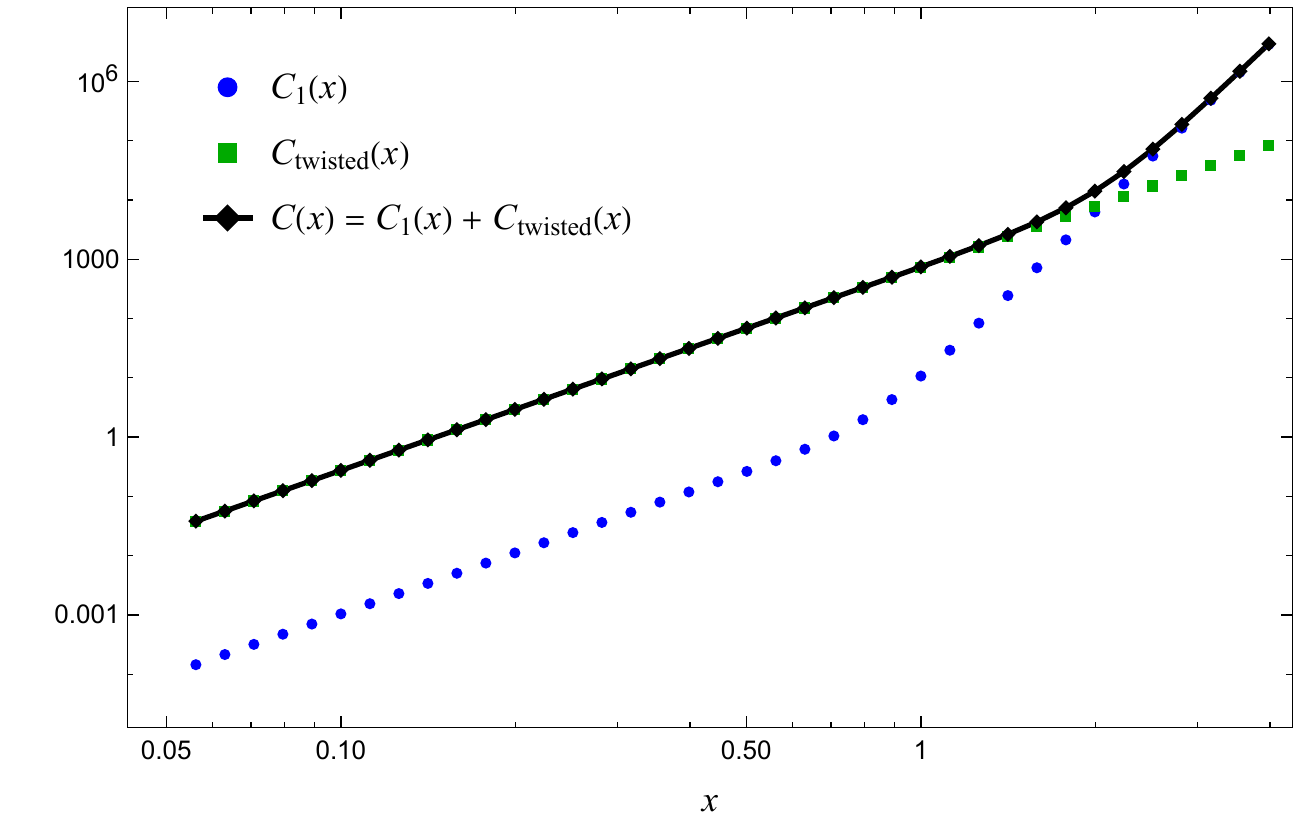}
\end{subfigure}
\hfill
\begin{subfigure}{0.49\textwidth}
    \includegraphics[width=\textwidth]{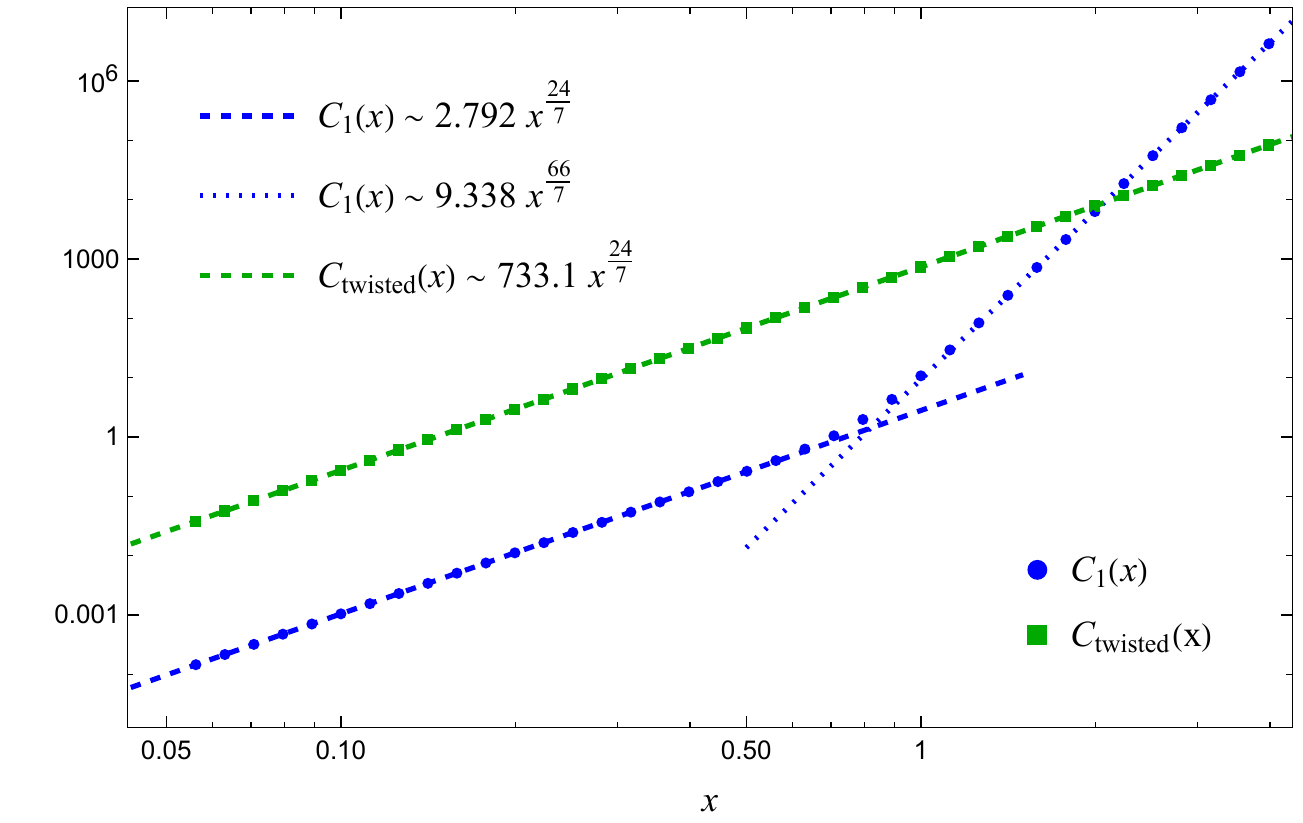}
\end{subfigure}
        
\caption{Casimir energy on $\mathcal{F}_7 = T^7/\Z^7$, evaluated numerically for the twisted and untwisted sectors. Notice that, unlike in the $\mathcal{F}_6$ example of the previous section, there is no saddle.}
\label{fig:dS4-Z7-casimir}
\end{figure}

We find that the restriction on the choice of spin structure does not allow us to break the 11-dimensional supersymmetry of M-theory and avoid the cancellations between bosons and fermions in the Casimir energy. Since these cancellations follow from the supersymmetry of the higher-dimensional theory, an antiperiodic spin structure would not be required if one started with a non-supersymmetric theory directly. However, if the starting theory is supersymmetric, one must be able to break this symmetry through the choice of compact manifold and associated spin structure---in some cases, like for $\mathcal{F}_7$, this is not possible.

Note that the corresponding equation \eqref{eq:dS5-maximum-xeq} for a general RFM $\mathcal{F}_k = T^k/\Gamma$ with an $S^1$ base takes the form\footnote{We cannot apply this expression to the compactification studied in Section \ref{sec:dS5-maximum} since $\mathcal{F}_6$ has base $T^2$.} \cite{dS-nogos-MMBB}
\begin{equation}
    \frac{x\,\mathcal{C}'}{\mathcal{C}} = \frac{d(k+4)-8}{k(k+(d-2)p)}\frac{(k-p)n_p^2\,x^2 - p\,\tilde{n}_p^2}{n_p^2\,x^2 + \tilde{n}_p^2} \,,
\end{equation}
when we restrict to a single $p$-form flux that may have legs only along the fibre $(n_p)$ or also along the base $(\tilde{n}_p)$.  For an M-theory compactification on $\mathcal{F}_7$ with $G_4$ flux, the equation becomes 
\begin{equation}
    \frac{x\,\mathcal{C}'}{\mathcal{C}} = \frac{12}{35}\cdot\frac{3n_p^2\,x^2 - 4\,\tilde{n}_p^2}{n_p^2\,x^2 + \tilde{n}_p^2} \in\left[-\frac{48}{35},\frac{36}{35}\right] \,.
    \label{eq:F7-allowed-range}
\end{equation}
Although a critical point for $\mathcal{C}(x)$ is not strictly necessary for a dS maximum, all maxima we have found  (such as the $dS_5$ solution of Section \ref{sec:dS5-maximum}) appear in the vicinity of a critical point of $\mathcal{C}(x)$. In this case, we see from \eqref{eq:F7-allowed-range} and the form of $\mathcal{C}(x)$, for which
\begin{equation}
    \frac{x\,\mathcal{C}'}{\mathcal{C}} \in \left[\frac{24}{7},\frac{66}{7}\right] = \left[\frac{120}{35},\frac{330}{35}\right] \,, 
\end{equation}
that there is no critical point for $x$. As a result, we do not expect to find a dS maximum in this setup. This argument does not depend on the precise choice of fluxes (including their proper quantisation) and therefore follows through without analysing the flux potential in detail. It will also apply to other 7d RFM's whose spin structures must be periodic around the fibre---this covers the $\Z_9$, $\Z_{14}$ and $\Z_{18}$ quotients discussed at the end of the previous subsection that are reducible to a 2-moduli problem like the $\Z_7$ quotient, and whose twisted-sector can be seen as multiple copies of the same Casimir branes (i.e. all non-trivial elements have the same invariant subspace). Thus, there does not seem to be any $dS_4$ maxima in RFM flux compactifications with Casimir energies in four dimensions. This contrasts with the maximum we found in Section   \ref{sec:dS5-maximum}, which after symmetry truncations could be reduced to a two-modulus problem.

Another important feature of the solution in Section  \ref{sec:dS5-maximum} was the antiperiodic spin structure along the fibre. The biggest order group that allows for antiperiodic spin structure along the fibre is $\Z_{24}$, which requires the study of at least 3 moduli and several sets of Casimir branes (in particular, a $\Z_{24}$ quotient will lead to 192 codimension-six, 32 codimension-four, 6 codimension-one and one space-filling Casimir branes). While cases such as these, with more than one modulus, might indeed harbour non-trivial maxima, in this paper we restrict our analysis to the setups in which, due to symmetries, the problem contains at most two moduli. It would be very interesting to study cases with three or more moduli not fixed by symmetries, a task that we leave for the future.

Since we have just argued that no diagonal cyclic RFM with periodic spin structure is likely to give us a dS saddle, what about those where an antiperiodic spin structure can be chosen? The largest holonomy group on this list of cyclic RFM's that admits a fully twisted spin structure on the fibre is $\Z_8$, for which there is an RFM that fixes all but four moduli of the $T^7$ and is defined by
\begin{align}
    \D[g] = \left(\begin{array}{*7{C{1em}}}
         0 & -1 & 0 & 0 & 0 & 0 & 0 \\
         1 & 0 & 0 & 0 & 0 & 0 & 0 \\
         0 & 0 & 0 & 0 & 0 & -1 & 0 \\
         0 & 0 & 1 & 0 & 0 & 0 & 0 \\
         0 & 0 & 0 & 1 & 0 & 0 & 0 \\
         0 & 0 & 0 & 0 & 1 & 0 & 0 \\
         0 & 0 & 0 & 0 & 0 & 0 & 1 \\
    \end{array}\right) 
\,,\quad \bvec[g] = \left(\begin{array}{c}
         0 \\
         0 \\
         0 \\
         0 \\
         0 \\
         0 \\
         \frac{1}{8} \\
    \end{array}\right) \,.
\end{align}
This is the direct sum of the $\Phi_4(x)$ and $\Phi_8(x)$ companion matrices, giving rise to a finite matrix of order lcm$(4,8)=8$. The block corresponding to $\Phi_8(x)$ of order $8$ is precisely what we used for the $T^6$ quotient studied in \secref{sec:dS5-maximum}, so in some sense it is a ``cousin'' of $\mathcal{F}_6$. We did not study this example in detail, since it is a three-modulus problem and one must identify suitable fluxes, etc. Nevertheless, we do not see an obstruction to the existence of $dS_4$ maxima, and therefore this is a promising candidate to study in the future.

\section{Conclusions}
\label{sec:conclusions} 
In this paper, we have taken a few steps in the exploration of a new class of solutions in String Theory, based on Riemann-flat manifolds (RFM's), where hopefully the existence of de Sitter minima and maxima, as well as other accelerated cosmologies, can be ascertained beyond any reasonable doubt. The landscape of RFM's offers unique opportunities for this, since it removes the need for stringy sources (orientifolds, branes, strong warping) that significantly complicate the analysis in other approaches to accelerated expansion in String Theory \cite{Kachru:2003aw,Balasubramanian:2005zx}, while allowing us to break supersymmetry and compute quantum effects very explicitly in a perturbative expansion around a flat minimum. We have described in detail a systematic framework to study Riemann-flat manifolds and determine their Casimir energies, and started a systematic exploration of their flux compactifications. 

The technical backbone of the results in this paper is the explicit formula for Casimir energies in RFM's derived in Section \ref{sec:casimir-RFMs}. We found that the Casimir energy tends to lump on certain cycles of the internal space, and can be effectively modeled as fictitious ``Casimir branes'' wrapping different loci of the RFM. This allows one to pull back all the intuition from standard string compactifications with branes \cite{Grana:2005jc}, and it is tempting to speculate that the Casimir branes may turn into actual branes in some cases via some chain of dualities. Along these lines, it is worth pointing out that in the setup of \cite{DeLuca:2021pej}, the Casimir energy also tends to lump in localized objects (there, related to small cycles of the geometry), which may suggest that the Casimir branes phenomenon is general. In practice the Casimir energy can be captured via an effective ``Casimir brane tension'', for which we provide an explicit formula that can be evaluated numerically. In some cases, the tension of the Casimir branes can be anomalously large and negative, which is helpful in attaining solutions with large volume, similarly to what happens e.g. in a Calabi-Yau with a large orientifold tadpole. 

Another very interesting phenomenon we encountered is that Casimir branes sometimes appear in pairs of exactly opposite tension, in which case their contribution to the Casimir energy exactly vanishes. As we showed in Section \ref{sec:alsym}, this phenomenon can be understood as a \emph{spacetime} version of Atkin-Lehner symmetry \cite{atkin1970hecke,Moore:1987ue,Dienes:1990qh}, since the explicit formula for Casimir energy that we obtain (arising as an integral of an 11d energy density over a compact space) is structurally analogous to the worldsheet integral of the partition function over the (super)moduli space of Riemann surfaces. Our spacetime version of Atkin-Lehner works for pure QFT, and provides us with a mechanism to engineer compactifications of both QFT's and EFT's arising from String Theory where the vacuum energy would be anomalously small, potentially giving a novel way to attack the cosmological hierarchy problem. We believe a systematic exploration of this phenomenon is important, and hope to return to it in the near future.

Using the Casimir formula, we have explored flux compactifications of M-theory to four and five dimensions, where the flux term is matched with a Casimir energy to obtain dS maximum saddles. A four-dimensional dS maximum remains a viable alternative to explain the current accelerating phase of the universe \cite{SupernovaSearchTeam:1998fmf,SupernovaCosmologyProject:1998vns}, and it might be even favoured in light of recent DESI data \cite{DESI:2024kob,DESI:2024mwx,DESI:2025fii,DESI:2025zgx}. Unfortunately, we could not find a dS maximum within the class of four-dimensional compactifications of M-theory on a 7-dimensional, cyclic RFM that we studied, which only have two moduli. However, given our results in five dimensions described below, a $dS_4$ maximum from Riemann-flat manifolds appears to be a very likely possibility to us, and we cannot refrain from emphasizing the analysis in Appendix A of \cite{Andriot:2025los} where such a scenario was found (together with a minimal matter coupling, which also arises automatically in our setup) to describe the recent DESI observations quite satisfactorily. We did not look for $dS_4$ minima (which would involve at least two fluxes pitted against the Casimir term) because they do not exist in the classical M-theory RFM landscape, as is explored in a companion paper \cite{dS-nogos-MMBB}. 

Instead, our main achievement is in five dimensions, via the very explicit solution of Section \ref{sec:dS5-maximum}. It is a $dS_5\times(T^6/\mathbb{Z}_8)$ maximum solution of M-theory, supported by a balance of $G_4$ flux and Casimir effects. The solution has a positive vacuum energy $V^{(5d)}=4.43\cdot10^{-8}$ in 5d Planck units, a Hubble radius of $H_0^{-1}\sim10^{4}\,\ell_5$, and is scale separated, with $H_0 R\sim 10^{-3}$ for $R$ a characteristic length scale of the internal manifold. The solution is only a de Sitter maximum, with four negative directions, all of which have a mass squared of order $H_0^2$. Due in part to the smallness of the vacuum energy, and in part to the fact that the internal space is Riemann-flat to leading order, our solution seems protected against classical, higher-derivative, and loop corrections. We point the reader to Section \ref{sec:dS5-maximum} and Table \ref{resumen} for more details. 

The solution we obtained is, to our knowledge, the first fully explicit example of a dS saddle point with these properties. For instance, in the solutions of \cite{Chen:2025rkb}, the superpotential can be computed reliably, but the K\"ahler potential remains uncontrolled. By looking for solutions within the regime of validity of supergravity, the full effective action can be determined, and we can provide precise numerical calculations of the parameters of the solution. We believe that being able to produce concrete numbers should be a litmus test of any would-be dS construction, as providing a very explicit solution will allow the community to explore it thoroughly. If our solution survives all corrections, then it has implications for a number of Swampland constraints. It satisfies the Refined de Sitter Conjecture \cite{Ooguri:2018wrx}, and it can be used to put bounds on the $\mathcal{O}(1)$ coefficients in the conjecture statement. Interestingly, although the mass of scalars is of order Hubble, our dS maximum is too steep to harbor eternal inflation, so our solution provides some evidence for the conjecture \cite{Rudelius:2019cfh} that eternal inflation may be in the Swampland. Another dS-specific Swampland Conjecture, the FL bound \cite{Montero:2019ekk}, is satisfied in our solution, which may therefore be regarded as giving evidence for the bound (although weak, since the bound was originally formulated for metastable dS saddles). There is also a conjecture in \cite{Andriot:2022xjh} that classical dS saddles can live in an EFT with at most four supercharges; our results suggest that quantum effects evade this conclusion. On the other hand, several works \cite{Hebecker:2018vxz,Andriot:2023wvg,Andriot:2025cyi,Friedrich:2025aec} have discussed or argued for different versions of the conjecture that cosmological or apparent horizons are in the Swampland; although it is difficult to define apparent horizons rigorously, it appears to us that the instantaneous dS horizon in our solution may qualify as apparent horizon. On the other hand, it certainly does not qualify as a cosmological horizon, since these are defined in terms of the asymptotic future. 

An important issue in any compactification obtained as a solution to a lower-dimensional EFT is that of 11-dimensional backreaction. Here, too, the RFM solutions we found are in very good position. We have explicit expressions for the 11d stress-energy tensor, coming both from Casimir energies and fluxes, and they are everywhere smooth, although inhomogeneous, in the internal space. We have checked that the inhomogeneities and general backreaction are small enough not to destroy the solution, and the explicit expressions we obtained can be used to follow up on that, constructing an iterative procedure which will converge to a full solution. We expect to come back to this interesting question in the future but, at any rate, the perturbations we obtained are so small ($10^{-5}$ in Planck units) that we expect the first-order result to be a very good approximation to the actual solution. In this aspect, our solution contrasts with standard dS minima proposals of ST \cite{Kachru:2003aw,Balasubramanian:2005zx}, or even some supersymmetric solutions like DGKT \cite{DeWolfe:2005uu}, where the presence of singular sources and strong warping makes the higher-dimensional analysis significantly harder. It shares this feature with the Casimir M-theory compactifications of \cite{DeLuca:2021pej}, with the additional difference that in an RFM the warping is very small as opposed to $\mathcal{O}(1)$. In this sense, it is similar to the $dS_3$ construction in \cite{Dong:2010pm}, where the warping variations are also very small.

Another point to emphasize is that the smallness of the cosmological constant that we find arises due to a combination of Einstein frame rescaling and the fact that the Casimir energy scales simply as $R^{-D}$, and $D=11$ in M theory. Even moderately large $R$ can yield very small numbers. In other words, to get a small cosmological constant, one does not need delicate cancellations between terms; in this case, it is simply a consequence that the volume of the internal manifold is large ($\sim 10^3$) in 11-dimensional Planck units. Using the M-theory scaling, achieving $V\sim10^{-120}$ would require $R\sim 10^{7}$ in 11d Planck units. While this estimate is certainly too na\"ive (it crucially ignores the contribution of SM fields to the cosmological constant, and uses a very na\"ive scaling of the Casimir term, ignoring its stabilization), the point that a small vacuum energy might be a natural consequence of a relatively large internal space still stands.

Even though the solution survives all known higher-derivative and loop corrections, there might be other, unknown, effects that may destabilize it. The internal manifold has a systole (the length of the smallest closed geodesic that cannot be contracted to a point) of just 3.9 eleven-dimensional Planck lengths. While we have checked that M2 and M5 instantons are negligible, loops of Planckian states, for instance, are more difficult to control. Although we have estimated that there would have to be $\sim10^4$ novel, long-lived states of Planckian mass in M-theory to significantly affect our solution, we do not know how to accurately determine whether this is the case or not. While we certainly expect M-theory to have black hole states starting at a roughly Planckian threshold, we do not really know their number, lifetime, or even how to include them in a vacuum energy calculation accurately. The only approaches which would be able in principle to access some of this information would be holography, or perhaps S-matrix unitarity arguments, but either seems currently far from achieving this. 

In short, we are inclined to believe our solution is actually there, but  it is difficult to tell with certainty\footnote{Admittedly, after so much effort, psychological factors might also play a role in our belief.}. Plus, due to the nature of the difficulties, the question is unlikely to be resolved rigorously in the near future. We believe it will be more productive to take the smallness of the vacuum energy and general properties  attained here as a  signpost encouraging us to focus on the exploration of the broader RFM Landscape, where better, more controlled models may lie, ideally finding a dS maximum that can accommodate observations. We will certainly look into this in the near future, and hope that this manuscript can entice others to join us in this exciting quest---the Earth is certainly not flat \cite{earthtround}, but the extra dimensions just might be!

\vspace{0.5cm}
\textbf{Acknowledgements:} We are indebted to 
David Andriot,
Carlo Angelantonj, 
Ivano Basile,  
Alek Bedroya,  
Jan de Boer,  
Veronica Collazuol,  
Gianguido Dall'Agata,  
Bruno de Luca,  
Alon Faraggi,
Bjorn Friedrich,  
Bernardo Fraiman,  
Arthur Hebecker,  
Luis Ib\'{a}\~{n}ez,  
Rafa{\l}  Lutowski,
Andriana Makridou,  
Fernando Marchesano,  
Jakob Moritz, 
Sonia Paban,   
Tony Padilla,  
Susha Parameswaran, 
Hector Parra de Freitas,  
Fernando Quevedo,  
Salvatore Raucci,  
Ignacio Ruiz,  
Augusto Sagnotti,  
Marco Serra,  
Gary Shiu,
Eva Silverstein,  
John Stout,  
Michelangelo Tartaglia,  
Houri-Christina Tarazi,  
Alessandro Tomasiello,  
Flavio Tonioni,  
\'{A}ngel Uranga,  
Cumrun Vafa,  
Damian Van de Heisteeg,  
Thomas Van Riet and
Irene Valenzuela for very valuable discussions and comments on the draft.
MM thanks the SCGP for hospitality and a stimulating environment during the Summer '23 workshop, where this project was initiated, and Summer '24, where it was advanced, and the KITP program ``What is String Theory?'' where an early version of these results was presented. This research was supported in part by grant NSF PHY-2309135 to the Kavli Institute for Theoretical Physics (KITP).
We also thank CERN and the Harvard Swampland
Initiative for hospitality during part of this work, and the
Aspen Center for Physics, which is supported by National Science Foundation grant PHY-2210452 and a grant from the Simons Foundation (1161654, Troyer), for providing a stimulating environment during
its completion.
The authors gratefully acknowledge the support of an Atraccion del Talento Fellowship 2022-T1/TIC-23956 from Comunidad de Madrid, which supported both authors in the early stages of this project, as well as the Spanish State Research Agency (Agencia Estatal de Investigacion) through the grants IFT Centro de Excelencia Severo Ochoa CEX2020-001007-S, PID2021-123017NB-I00, and Europa Excelencia EUR2024-153547. MM is currently supported by the RyC grant RYC2022-037545-I from AEI.

\appendix

\section{\texorpdfstring{$G_4$}{G4} flux quantization in M-theory}
\label{app:fluxQ}
The de Sitter maxima we look for have $G_4$ flux as an essential ingredient---in this appendix we work out the normalization of the kinetic term from first principles, finding agreement with \cite{Polchinskiv2,Witten:1996md,deAlwis:1996ez,deAlwis:1996hr}.
We start from the 11-dimensional M-theory action, whose bosonic part is \cite{Polchinskiv2}
\begin{equation}
    2\kappa_{11}^2\, S_{\text{M-theory}} = \int d^{11}x\,\sqrt{-g}\left[R-\frac12\vert \tilde{G}_4\vert^2\right]-\frac16\int  \tilde{G}_4\wedge \tilde{G}_4\wedge \tilde{C}_3 \,.
    \label{e332s}
\end{equation}
The 4-form $\tilde{G}_4$ does not have integer-quantized periods. The appropriate quantization condition can be determined from demanding that the Chern-Simons coupling is appropriately quantized \cite{deAlwis:1996ez,Witten:1996md}, and it is such that
\begin{equation} 
    \frac{1}{2\kappa_{11}^2}\left(\int \tilde{G}_4\right)^3=2\pi\,n,\quad n\in\mathbb{Z} \,.
\end{equation}
From this, it follows that the action \eq{e332s} can be rewritten as 
\begin{equation} 
    S_{\text{M-theory}} = \int d^{11}x\,\sqrt{-g}\left[\frac{1}{2\kappa_{11}^2} R-\frac{1}{2g_3^2}\vert G_4\vert^2\right]-\frac{1}{24\pi^2}\int  G_4\wedge G_4\wedge C_3 \,,
    \label{e332s2}
\end{equation}
in terms of a $G_4$ whose periods are quantized in multiples of $2\pi$ (so that M2-brane charges are integer-quantized), and 
\begin{equation}
    g_3^2\equiv2\cdot (2\pi^2)^{2/3}\kappa_{11}^{2/3} 
    = 2\cdot (2\pi^2)^{2/3} \,\ell_{11}^6
    \approx (3.822\, \ell_{11}^3)^2 
\end{equation}
is the M-theory 3-form gauge coupling, where in the last equality we have also introduced the 11-dimensional reduced Planck length, defined by $\kappa_{11}^2=\ell_{11}^9$. Using the formulae of \cite{Heidenreich:2015nta}, the tensions of the BPS M2 and M5 branes are
\begin{equation}
    T_{\text{M2}} = \sqrt{\frac12\frac{g_3^2}{\kappa_{11}^2}}
    = \frac{(2\pi^2)^{1/3}}{\ell_{11}^3} \,,
    \quad T_{\text{M5}} = \sqrt{\frac{1}{2}\frac{4\pi^2}{g_3^2\kappa_{11}^2}}
    =\left(\frac{\pi}{2}\right)^{\frac13}\frac{1}{\ell_{11}^{\,6}} \,.
    \label{Tm2M5}
\end{equation}
These agree with the expressions in \cite{deAlwis:1996ez}. Writing $G_4=2\pi n_4 \,\omega_4$, where $\omega_4$ is a 4-form with integer periods, the flux term of the potential becomes
\begin{equation} 
    \int \frac{1}{2g_3^2}\vert G_4\vert^2=\left(\frac{\pi}{2}\right)^{\frac23}\frac{n_4^2}{\ell_{11}^{3}} \int \vert\omega_4\vert^2 \,,
\end{equation}
which is the expression used in the main text.

\section{Lattice sums on RFM's with 1-dimensional invariant subspaces}\label{ap:lattice-sums-1d-inv-space}

Our formula \eq{eq:VC-invariant-subspace} for the tension of Casimir branes in RFM's does not generally admit an analytic expression, and we must resort to numerical methods to evaluate it. This is what happens in many simple cases, such as a torus with no quotients \cite{lambert2013closed}. In this appendix, however, we describe a particular case where the sums can be evaluated analytically and explicitly, in terms of $\zeta$ functions. This happens when the Casimir brane is one-dimensional in the compact space (i.e. when we are computing $\mathcal{E}(\gamma)$ for a $\gamma$ which has a one-dimensional invariant subspace),
\begin{equation}
    \mathcal{E}(\gamma) = -\frac{\Gamma(s_\gamma)}{2\pi^{s_\gamma}}\cdot\frac{G_\parallel^{-(s_\gamma - \frac12)}}{|\Gamma|} \sum_{n\in\Z} \frac{e^{2\pi i\,\beta\,n}}{|n-b_\gamma^\parallel|^{2s_\gamma}} \,. 
\end{equation}
In this case we can perform the sum analytically to obtain 
\begin{equation}
    \sum_{n\in\Z} \frac{e^{2\pi i\,\beta\,n}}{|n-b_\gamma^\parallel|^{2s_\gamma}} = \begin{cases}
        \zeta(2s_\gamma,b_\gamma^\parallel) + \zeta(2s_\gamma,1-b_\gamma^\parallel) & \text{if}\,\beta=0\,\text{mod}\,\Z \\ \\
        2^{2s_\gamma}\Big[\zeta(2s_\gamma,\frac{b_\gamma^\parallel}{2}) 
        - \zeta(2s_\gamma,\frac{1-b_\gamma^\parallel}{2}) \\   
        \hspace{3em} 
        + \zeta(2s_\gamma,1-\frac{b_\gamma^\parallel}{2}) 
        - \zeta(2s_\gamma,1-\frac{1-b_\gamma^\parallel}{2})
        \Big] & \text{if}\,\beta=\frac12\,\text{mod}\,\Z
    \end{cases} \,.
\end{equation}
Together with the fact that each $\gamma\in\Gamma$ has an inverse $\gamma^{-1}\in\Gamma$, whose invariant subspace is the same as that of $\gamma$,
\begin{equation}
    (\mathbf{I} - \D^{-1})\vec{n} = 0\,\text{mod}\,\Z 
    \Longleftrightarrow (\mathbf{I} - \D)\vec{n} = 0\,\text{mod}\,\Z \,,
\end{equation}
and with $b_{\gamma^{-1}}^\parallel = -b_\gamma^\parallel \sim 1 - b_\gamma^\parallel$, this implies that $\mathcal{E}(\gamma^{-1}) = (-1)^{2\beta} \mathcal{E}(\gamma)$. If $\gamma$ is an element of order 2, i.e. $\gamma^{-1} = \gamma$, the contribution with $\beta = \frac12$ vanishes; otherwise, the contributions from $\gamma$ and $\gamma^{-1}$ will add up or subtract depending on the traces weighing the sum,
\begin{equation}
    \Vcas = \sum_\mathbf{r}\sum_{\gamma\in\Gamma} \Tr{\bf r}{\D}\,\mathcal{E}(\gamma) = \frac12\sum_{\mathbf{r}}\sum_{\gamma\in\Gamma} \left(\Tr{\mathbf{r}}{\D} + (-1)^{2\beta_\mathbf{r}}\Tr{\mathbf{r}}{\D^{-1}}\right) \mathcal{E}(\gamma) \,. 
\end{equation}
For bosonic fields, which always have $\beta = 0$ and satisfy $\Tr{Boson}{\D^{-1}} = \Tr{Boson}{\D}$ (see Appendix \ref{ap:traces}), the contributions add up. The largest contribution will come from the term with smaller $b_\gamma^\parallel$, with $\zeta(2s_\gamma,\frac{1}{p})\sim p^{2s_\gamma}$ giving a good estimate of the enhancement to the Casimir potential on cyclic RFM's of order $p$. However, for fermionic fields, the result will depend on the choice of boundary conditions $\beta_\mathbf{r}$ (compatible with the allowed spin structures of the RFM); the trace of $\D^{-1}$ will be the same as that of $\D$ up to a sign (cf. Appendix \ref{ap:traces}). 

In Appendix \ref{ap:traces} we show that on a cyclic RFM of order $p = |\Gamma|$, spinor traces are related as 
\begin{equation}
    \Tr{Spinor}{\D^{-1}} = (-1)^{\sum n_i}\, \Tr{Spinor}{\D} \,,\quad\text{with}\quad p\cdot\theta_i = 2\pi n_i \,,
\end{equation}
where $\theta_i$ are the angles (eigenvalue arguments) of the rotations associated with $\D[g]$ (including the choice of spin lift). When the order of the cyclic RFM is odd, the pairity of $\sum n_i$ is the same as that of $s_\mathbf{g}$ in $\mathcal{D}_\mathbf{g}^p = (-1)^{s_\mathbf{g}}\mathbf{I}$, and correspondingly correlated with $\beta_\mathbf{r}$ through the constraint \eq{swer2} on $\vec{h}$.
For example, the $T^7/\Z^7$ we study in Section \ref{sec:dS4-T7-example} is constrained to have $2\beta_\mathbf{r} = 2h_7 = \sum n_i \,\,\text{mod}\,\,2$, so that the fermion contributions add up, rather than cancel.  

On the other hand, if the cyclic RFM has even order, $s_\mathbf{g}$ is independent of the choice of spin lift and thus of the parity of $\sum n_i$. One can then choose the spin lift with $\sum n_i$ odd, so that the traces of $\D$ and $\D[\gamma^{-1}]$ differ by a sign, while $2\beta_\mathbf{r}$ is constrained to be even---in this case, the fermion contributions will cancel pairwise. One can do this, for example, on the $\Z_{18}$ quotient (see tables in Appendix \ref{ap:7d-RFMs}).

Remarkably, if all the elements in $\Gamma$ for a cyclic RFM, except for the identity, have a one-dimensional invariant subspace---which is the case in particular for the RFM's identified in Section \ref{sec:dS4-maxima} that may be left with only two moduli in special points in moduli space---fermion fields may have no twisted sector terms at all, and only contribute through the identity element as they would on the covering torus. Whether this cancellation happens can depend on the spin structure of the RFM and its order. This interesting feature/cancellation appears in the absence of supersymmetry, and is similar in form and details to the vanishing of fiber contributions when the $\hat{\delta}_{\vec{h}}$ vanishes in the main text. Having a part of the spectrum not contributing to the vacuum energy is a quite surprising phenomenon, and variants of it could be used e.g. to provide new solutions for the cosmological constant problem, where the contribution to the vacuum energy of some fields is cancelled by variants of this mechanism. It would be very interesting to understand if there is an underlying symmetry reason for this phenomenon, and whether it can be generalized.

\section{Ewald summation}
\label{app:ewy}

In this appendix, we generalize the Ewald summation technique for sums of the form
\begin{equation}\sum_{\vec{n}+\vec{c}\neq0}\frac{e^{2\pi i \vec{h}\cdot\vec{n}}}{\vert \vec{n}+\vec{c}\vert^{2s}},\label{csumq}\end{equation}
developed in \cite{NIJBOER1957309} for the three-dimensional case, to higher-dimensional setups. We will also consider a general norm,
\begin{equation} \vert \vec{n}\vert^2= G_{ij} n^i n^j.\end{equation}
The sum we wish to regularize is
\begin{equation}\frac{e^{2\pi i \vec{h}\cdot\vec{n}}}{\vert \vec{n}+\vec{c}\vert^{2s}},\end{equation}
where there is a constant shift vector $\vec{c}$, and whose Fourier transform is

\begin{equation}\mathcal{F}\left[\frac{e^{2\pi i \vec{h}\cdot\vec{n}}}{\vert \vec{n}+\vec{c}\vert^{2s}}\right](\vec{k})\equiv \int d^k\vec{n}\, \frac{e^{2\pi i (\vec{h}-\vec{k})\cdot\vec{n}}}{\vert \vec{n}+\vec{c}\vert^{2s}}=\frac{2^{k-2s}\pi^{k/2}\Gamma\left(\frac{k}{2}-s\right)}{\Gamma(s)} \frac{e^{-2\pi i \vec{h}\cdot\vec{c}}}{\vert \vec{h}-\vec{k}\vert^{k-2s}}.\end{equation}
Following \cite{NIJBOER1957309}, we take
\begin{equation} 
    F(\vec{n})\equiv \frac{\Gamma(s,\alpha \vert \vec{n}+\vec{c}\vert^{2})}{\Gamma(s)},\quad  \Gamma(s,x)\equiv \int_x^\infty t^{s-1}e^{-t}\,dt \,.
\end{equation}
For $s=1/2$, the typical case of Coulomb interactions, the above becomes the usual $\text{erfc}$ regulator used in physical chemistry. The function
\begin{equation} f(\vec{n})\equiv 1-F(\vec{n})= \frac{\gamma(s,\alpha \vert \vec{n}+\vec{c}\vert^{2})}{\Gamma(s)}\end{equation}
has the required property that $f(\vec{0})=0$ and in fact, for small $\vec{n}+\vec{c}$ it goes as $\vert\vec{n}+\vec{c}\vert^{2s}$, so that it cancels the denominator divergence in \eq{csumq}. More precisely, we have that
\begin{equation}  
    \lim_{\vec{n}+\vec{c}\rightarrow\vec{0}}f(\vec{n})\,\frac{e^{2\pi i \vec{h}\cdot\vec{n}}}{\vert \vec{n}+\vec{c}\vert^{2s}}=\frac{\alpha^s\, e^{-2\pi i \vec{h}\cdot\vec{c}}}{\Gamma(s+1)}\,\chi_{\mathbb{Z}^k}(\vec{c}) \,,
    \label{v44}
\end{equation}
and since terms with $\vec{n}+\vec{c}=\vec{0}$ are excluded from the original sum, we will need to include a correction term equal to the negative of \eq{v44} if $\vec{c}$ is a vector of integer coordinates.

The resulting sum can then be evaluated in momentum space via Poisson resummation. The Fourier transform of the product can be evaluated directly, as
\begin{align} 
    \mathcal{F}&\left[\frac{e^{2\pi i \vec{h}\cdot\vec{n}}}{\vert \vec{n}+\vec{c}\vert^{2s}}\cdot  \frac{\gamma(s,\alpha \vert \vec{n}+\vec{c}\vert^{2})}{\Gamma(s)}\right](\vec{k})\equiv\int d^k\vec{n}\, e^{-2\pi i\vec{k}\cdot\vec{n}}\frac{e^{2\pi i \vec{h}\cdot\vec{n}}}{\vert \vec{n}+\vec{c}\vert^{2s}}\cdot  \frac{\gamma(s,\alpha \vert \vec{n}+\vec{c}\vert^{2})}{\Gamma(s)} \\
    &=\frac{1}{\Gamma(s)}\int d^k\vec{n}\,\, \frac{e^{2\pi i(\vec{h}-\vec{k})\cdot\vec{n}}}{\vert \vec{n}+\vec{c}\vert^{2s}} \gamma(s,\alpha \vert \vec{n}+\vec{c}\vert^{2}) =\frac{e^{-2\pi i(\vec{h}-\vec{k)}\cdot\vec{c}}}{\Gamma(s)}\int d^k\vec{n'}\, \frac{e^{2\pi i(\vec{h}-\vec{k})\cdot\vec{n'}}}{\vert \vec{n'}\vert^{2s}} \gamma(s,\alpha \vert \vec{n'}\vert^{2}) \,. \nonumber
 \end{align} 
Next, we find a similarity transformation diagonalizing the metric $\mathbf{G}$,
\begin{equation} 
    \mathbf{G}=\mathbf{M}^T\mathbf{M}\,,\quad \vec{m}=\mathbf{M}\, \vec{n'}\,,
\end{equation}
so that the norms become the standard ones, and we define the variable 
\begin{equation} 
    \vec{q}\equiv \mathbf{G}^{-1}(\vec{h}-\vec{k})\,,
\end{equation}
which turns the inner product in the Fourier transform phase into the metric one. We then have (after a change of variables to spherical coordinates in the second line, with $z$ axis antiparallel to the vector $\vec{q}$, and with $\mathcal{S}_{k-2}$ the volume of the $(k-2)$-dimensional sphere)
\begin{align}
    \mathcal{F}&\left[\frac{e^{2\pi i \vec{h}\cdot\vec{n}}}{\vert \vec{n}+\vec{c}\vert^{2s}}\cdot  \frac{\gamma(s,\alpha \vert \vec{n}+\vec{c}\vert^{2})}{\Gamma(s)}\right](\vec{k})=\frac{e^{-2\pi i(\vec{h}-\vec{k)}\cdot\vec{c}}}{\Gamma(s)\sqrt{G}}\int d^k\vec{m}\, \frac{e^{2\pi i((\mathbf{M}^{-1})^T(\vec{h}-\vec{k}))\cdot\vec{m}}}{\vert \vec{m}\vert^{2s}} \gamma(s,\alpha \vert \vec{m}\vert^{2})\nonumber\\&= \frac{e^{-2\pi i(\vec{h}-\vec{k)}\cdot\vec{c}}}{\Gamma(s)\sqrt{G}}\mathcal{S}_{k-2}\int_0^\infty dr\, \frac{\gamma(s,\alpha r^2)}{r^{2s-k+1}}\int_0^\pi d\theta\, (\sin\theta)^{k-2} e^{-2\pi i \vert\vec{q}\vert r\cos\theta}\nonumber\\&=\frac{(2\pi)^{\frac{k}{2}}e^{-2\pi i(\vec{h}-\vec{k)}\cdot\vec{c}}}{\Gamma(s)\sqrt{G}}\frac{1}{(2\pi \vert\vec{q}\vert)^{k-1-2s}}\int_0^\infty dr\, \frac{\gamma(s,\alpha r^2)}{(2\pi \vert\vec{q}\vert r)^{2s-\frac{k}{2}}}J_{\frac{k-2}{2}}(2\pi \vert\vec{q}\vert r)\nonumber\\&=\frac{(2\pi)^{\frac{k}{2}}e^{-2\pi i(\vec{h}-\vec{k)}\cdot\vec{c}}}{\Gamma(s)\sqrt{G}}\frac{1}{(2\pi \vert\vec{q}\vert)^{k-2s}}\int_0^\infty d\tau\, \frac{\gamma\left(s,\frac{\alpha}{(2\pi \vert\vec{q}\vert)^2} \tau^2\right)}{\tau^{2s-\frac{k}{2}}}J_{\frac{k-2}{2}}(\tau)\nonumber\\&= \frac{2^{k-2s}\pi^{\frac{k}{2}}}{\Gamma(s)} \Gamma \left(\frac{k}{2}-s,\frac{\pi ^2 \vert\vec{q}\vert^2}{\alpha }\right) \frac{e^{-2\pi i(\vec{h}-\vec{k})\cdot\vec{c}}}{\sqrt{G}\, (2\pi \vert\vec{q}\vert)^{k-2s}} \,. 
    \label{we4} 
\end{align} 
In the above derivation, we have used one of the definitions of the Bessel $J$ function,
\begin{equation} 
    J_\nu(t)= \frac{t^\nu}{(2\pi)^{\nu+1}}\mathcal{S}_{2\nu}\int_0^\pi e^{-it\cos\theta} \sin^{2\nu}\theta\, d\theta \,.
\end{equation}
The chain of manipulations in \eq{we4} is only valid when $\vert \vec{q}\vert\neq0$; since $\vec{h}$ is a vector whose components are all in $[0,1)$, vanishing $\vert\vec{q}\vert$ can only happen when $\vec{h}=0$ and $\vec{k}=0$. Then, we obtain a term
\begin{equation} 
    \frac{\mathcal{S}_{k-2}}{\Gamma(s)\sqrt{G}}\int_0^\infty dr\, \frac{\gamma(s,\alpha r^2)}{r^{2s-k+1}}\int_0^\pi d\theta\, (\sin\theta)^{k-2}=\frac{\pi^{\frac{k}{2}}\alpha^{s-k/2}}{\sqrt{G}\,\Gamma(s)\left(s-\frac{k}{2}\right)} \,. 
\end{equation}
Therefore, the Ewald summation formula we will use is
\begin{align}
    &\sum_{\vec{n}+\vec{c}\neq0}\frac{e^{2\pi i \vec{h}\cdot\vec{n}}}{\vert \vec{n}+\vec{c}\vert^{2s}}=\nonumber\\&\sum_{\vec{n}+\vec{c}\neq0}\frac{e^{2\pi i \vec{h}\cdot\vec{n}}}{\vert \vec{n}+\vec{c}\vert^{2s}} \frac{\Gamma(s,\alpha \vert \vec{n}+\vec{c}\vert^{2})}{\Gamma(s)}\,\, +\,\, \frac{\pi^{2s-\frac{k}{2}}}{\Gamma(s)\sqrt{G}} \sum_{\vec{k}-\vec{h}\neq0}\, \Gamma \left(\frac{k}{2}-s,\frac{\pi ^2 \vert\vec{h}-\vec{k}\vert_D^2}{\alpha }\right) \frac{e^{-2\pi i(\vec{h}-\vec{k})\cdot\vec{c}}}{ \vert\vec{h}-\vec{k}\vert_D^{k-2s}}\nonumber\\&+ \delta_{\vec{h},\vec{0}}\, \frac{\pi^{\frac{k}{2}}\alpha^{s-k/2}}{\sqrt{G}\, \Gamma(s)\left(s-\frac{k}{2}\right)}-\frac{\alpha^s\, e^{-2\pi i \vec{h}\cdot\vec{c}}}{\Gamma(s+1)}\chi_{\mathbb{Z}^k}(\vec{c}) \,.
    \label{ewald} 
\end{align}

In these expressions, $\vert \vec{v}\vert_{D}^2\equiv G_{ij}^{-1} v^i v^j$ is the canonical norm in momentum space. The first term in the second line is there to take into account the special case $\vec{q}=0$ that only happens when $\vec{h}=0$; in this case, the term $\vec{k}=0$ is not included in the second sum, and the $\delta_{\vec{h},\vec{0}}$ function term of the second line appears instead. The second term in the second line is the correction \eq{v44} that only appears if the vector $\vec{c}$ has all integer coordinates; the function $\chi_{\mathbb{Z}^k}(\vec{c})$ is 1 in this case, and zero otherwise.

The expression \eq{ewald} is exact for any $\alpha$; in practice, we will truncate both the sum over $\vec{n}$ and the one over $\vec{k}$, and choose an optimal value of $\alpha$ to maximize the speed of convergence. As a first check, when $s=3/2$ the above formula does become the expression of \cite{NIJBOER1957309}\footnote{Modulo an additional term, related to the last term of \eq{ewald}, that appears in \cite{NIJBOER1957309} because that reference always excludes the contribution from $\vec{n}=0$ to the sum, which we do include here. Also, when $\vec{h}=0$, the $\vec{k}=0$ term of the second sum should be replaced by its limiting value, which is convergent.}. We have also checked numerically the validity of \eq{ewald} for several concrete sums that can be found in the literature, including
 \begin{align}
    &\sum_{\vec{n}\neq\vec{0}} \frac{(-1)^m}{m^2+mn+n^2}=-\frac{4\pi\,\ln 2}{3\sqrt{3}},\quad \sum_{\vec{n}\neq\vec{0}} \frac{(-1)^{m+n}}{m^2+5n^2}=-\frac{\pi}{\sqrt{5}}\ln\left(1+\sqrt{5}\right),& \text{\cite{zucker1975}}\nonumber\\ &\sum_{\vec{n}\in\mathbb{R}^3}\frac{(-1)^{n_1+n_2+n_3}}{\left[\left(n_1+\frac16\right)^2+\left(n_2+\frac16\right)^2+\left(n_3+\frac16\right)^2\right]^s}=12^s\beta(2s-1),&\text{\cite{borwein1998pi}}
 \nonumber\\ &\sum_{\vec{n}\in\mathbb{R}^8-\{\vec{0}\}}\frac{(-1)^{\sum_i n_i}}{\vert\vec{n}\vert^{2s}}=-16\zeta(s)\eta(s-3),\quad \sum_{\vec{n}\in\mathbb{R}^8}\frac{1}{\vert\vec{n}+\vec{1}/2\vert^{2s}}=\frac{256}{2^s}\zeta(s-3)\lambda(s),&\text{\cite{zucker1974exact}}
 \end{align}
 which provide examples in 2, 3, and 8 dimensions. In our numerical implementation, the sums are performed by first determining the set of vectors of norm smaller than the cutoff, and performing the sum only after this set of vectors has been fully identified and stored in memory. This provides a large speedup over implementations such as that in \cite{buchheit2024} (which uses a similar approach to ours, and appeared as we were evaluating these sums), where the approach is to sum over a large square box of points. This means one does not need to spend computer resources identifying the set of points for the sum, but it comes at a hefty price, since many points of very large norm are included, whose contribution to the sum is negligible, but where nevertheless one spends some time evaluating the integrand. This is particularly important in larger dimension where there are many points in the unit cube that are not contained in the unit ball for a generic norm.
 
 Furthermore, for larger values of $s$ and completely alternating sums (where the phase is oscillating quickly in the unit cell), evaluating the sum at small $\alpha$ and truncating the direct sum to just a single term served to produce outstanding results very quickly. This is because for large $s$, the contribution from points away from the origin decays very quickly, all the more so for alternating sums.  
 
 \section{Numerical estimate}\label{app:num}
 For the numerical evaluation of the sum \eq{ewald}, it is also convenient to have an estimate of when the second term becomes small. In practice, the sum will yield accurate results as long as the terms which have been neglected are beyond the exponential tail of the incomplete $\Gamma$ function. Defining $\lambda\equiv \vert\vec{h}-\vec{k}\vert^2$, we want that
 \begin{equation} 
    \frac{\Gamma\left(\frac{k}{2}-s,\frac{\pi^2\lambda}{\alpha}\right)}{\lambda^{k/2-s}}\leq \frac{\Gamma(k/2)}{2\pi^{k/2}}\frac{\epsilon}{\lambda^{\frac{k-1}{2}}} \,.
    \label{fre}
\end{equation}
 The quantity dividing $\epsilon$ on the right-hand side times the prefactor is the area of a sphere of radius $\sqrt{\lambda}$, so the denominator ensures that upon summing over all points of norm $\sqrt{\lambda}$, their total contribution is less than or equal to $\epsilon$. Thus, $\epsilon$ directly controls the precision of the sum; setting $\epsilon=0.1$ will produce results at least accurate to the first digit after the decimal point, and so on.

The above can be rearranged to 
 \begin{equation} 
    \frac{\Gamma\left(\frac{k}{2}-s,\frac{\pi^2\lambda}{\alpha}\right)}{\left(\frac{\pi^2}{\alpha}\lambda\right)^{1/2-s}}\leq \frac{\Gamma(k/2)}{2\pi^{k/2}}\, \left(\frac{\pi^2}{\alpha}\right)^{s-1/2} \epsilon \,,
    \label{gfre}
\end{equation}
and using the fact that the incomplete $\Gamma$ function is upper bounded by its leading asymptote,
 \begin{equation}
    \frac{\Gamma(a,z)}{z^{a}}\leq \frac{e^{-z}}{z} \,,
\end{equation}
 the bound \eq{fre} will be satisfied if 
 \begin{equation}
    \frac{e^{-\frac{\pi^2\lambda}{\alpha}}}{\frac{\pi^2\lambda}{\alpha}}\leq  \frac{\Gamma(k/2)}{2\pi^{k/2}}\, \left(\frac{\pi^2}{\alpha}\right)^{s-1/2}\, \epsilon\,,
    \label{fre3}
\end{equation}
 which results in
 \begin{equation} \lambda\geq  \frac{\alpha}{\pi^2}\, W\left(\frac{2\pi^{k/2}}{\Gamma(k/2)\,\epsilon}\left(\frac{\alpha}{\pi^2}\right)^{s-1/2}\right).\label{damn}\end{equation}
 where $W$ is the Lambert function. We use \eq{damn} in order to determine the cutoff value in the momentum space sums as a function of $\alpha$ in the numerical implementation.

\section{Alternative derivations  of the Casimir formula for RFM's}\label{directspace}
In this appendix, we recover the result \eq{eq:VC-invariant-subspace} of the main text in two different ways.
\subsection{Direct evaluation in position space}

We will first show how \eq{eq:VC-invariant-subspace} may be recovered by a ``direct'' method, in which the integrals are evaluated without using additional analytical tools. Our starting point is again \eq{csum},  which we reproduce here for convenience:
\begin{equation} 
    \int_{[0,1]^k}dV \sum_{n\in\mathbb{Z}^k}\frac{e^{2\pi i \vec{h} \cdot\vec{n}}}{\vert (\mathbf{I}-\mathbf{D}_\gamma)\,\vec{z}+\vec{b}_\gamma+\vec{n}\vert^{2s}} \,.
    \label{csump}
\end{equation}
To evaluate this sum, notice that $\mathbf{D}_\gamma$ is an $SL(n,\mathbb{Z})$ transformation, that maps the lattice $\mathbb{Z}^k$ to itself. As a result, the image of the lattice $\mathbb{Z}^k$ by $(\mathbf{I}-\mathbf{D}_\gamma)$ is a sublattice of $\mathbb{Z}^k$,
\begin{equation}\Lambda^\perp\equiv(\mathbf{I}-\mathbf{D}_\gamma)\,\mathbb{Z}^k\subset\mathbb{Z}^k.\end{equation}
We will denote its dimension by $k''$. $\Lambda^\perp$ is a sublattice of another lattice, $\Sigma^\perp$, defined as the intersection of $\Lambda^\perp\otimes\mathbb{R}$ with $\mathbb{Z}^k$. We can similarly define the lattice of vectors invariant under $\mathbf{D}_\gamma$\footnote{Notice that this is different from the definition of $\Lambda^\parallel$ in the rest of the paper, which uses the transpose of $\mathbf{D}_\gamma$. The definition in terms of $\mathbf{D}_\gamma$ is used in this subsection of this appendix only.}, 
\begin{equation}
    \Lambda^\parallel\equiv\{\vec{n}\in\mathbb{Z}^k\,\vert\,  (\mathbf{I}-\mathbf{D}_\gamma)\vec{n}=\vec{0}\}\,.
\end{equation}
This is a lattice of dimension $k'=k-k''$. The lattices $\Lambda^\parallel$ and $\Lambda^\perp$ are orthogonal with respect to the inner product $G$. Their sum is a lattice 
\begin{equation} \Lambda\equiv\Lambda^\parallel\oplus \Lambda^\perp\supset \mathbb{Z}^k\label{lat0}\end{equation}
of dimension $k$, and hence, it is a sublattice of $\mathbb{Z}^k$ of finite index. The group 
\begin{equation}\Gamma_\Lambda\equiv \mathbb{Z}^k/\Lambda\end{equation}
is a finite abelian group. Each $\gamma\in \Gamma_\Lambda$ can be represented by a vector $\vec{\gamma}$ of integer coordinates in a particular unit cell of $\Lambda$ (say, the Brillouin zone), with the group law being vector addition modulo $\Lambda$. With these provisions, any vector $\vec{n}\in\mathbb{Z}^k$ can be written as
\begin{equation}\vec{n} =\vec{k} + \vec{l}+\vec{\gamma},\end{equation}
where  $\vec{k}\in\Lambda^\perp$ and $\vec{l}\in\Lambda^\parallel$. 
As a result, the sum over $\vec{n}$ in \eq{csump} can be rewritten as
\begin{equation}
    \sum_{\substack{\vec{k}\in\Lambda^\perp\\\vec{l}\in\Lambda^\parallel,\gamma\in\Gamma_\Lambda}}\int_{[0,1]^k}  dV\frac{e^{2\pi i \vec{h} \cdot[\vec{k} + \vec{l}+\vec{\gamma}]}}{\vert (\mathbf{I}-\mathbf{D}_\gamma)\,\vec{x}+\vec{k}+\vec{b}_\gamma+\vec{l}+\vec{\gamma}\vert^{2s}} \,.
    \label{csum2}
\end{equation}
Now recall the consistency condition \eq{mimi}, discussed in Section \ref{sec:RFMs}, that
\begin{equation}  
    \vec{h} \cdot(\mathbf{I}-\mathbf{D}_\gamma)\, \vec{l}=0\,\text{mod}\,\,\mathbb{Z} \,.
\end{equation}
Since $\vec{k}=(\mathbf{I}-\mathbf{D}_\gamma)\, \vec{l}$ for some $\vec{l}$ by virtue of being an element in $\Lambda^\perp$, the $e^{2\pi i\, \vec{h}\cdot\vec{k}}$ term in the numerator in \eq{csum2} is trivial. In fact, writing $\vec{x}=\vec{x}^\perp+\vec{x}^\parallel$ where $\vec{x}^{\parallel,\perp}$ are the orthogonal projections of $\vec{x}$ onto $\Lambda^\parallel\otimes\mathbb{R}$ and  $\Lambda^\perp\otimes\mathbb{R}$, respectively, the whole integrand is independent of $\vec{x}^{\parallel}$ since this is annihilated by $(\mathbf{I}-\mathbf{D}_\gamma)$. The integral over $\vec{x}_\parallel$ can be carried out directly and gives an overall factor of $V_\parallel$, where $V_\parallel$ is the volume of the unit cell of the $\Lambda^\parallel$ lattice. The range of the variable $(\mathbf{I}-\mathbf{D}_\gamma)\,\vec{x}_\perp$ is that of the unit cell of $\Lambda^\perp$, so the combined variable
\begin{equation}
    \vec{y}\equiv (\mathbf{I}-\mathbf{D}_\gamma)\,\vec{x}_\perp+\vec{k}
\end{equation}
runs over all of $\mathbb{R}^{k''}$. This means that, after changing variables to $\vec{y}$ (and including the resulting Jacobian factor $\vert \text{det} (\mathbf{I}-\mathbf{D}_\gamma)\vert^{-1}$), the sum over $\vec{k}$ and the integral over $\vec{x}_\perp$ can be combined into a single integral over $\mathbb{R}^{k''}$,
\begin{equation}
    \frac{V_\parallel}{\vert \text{det} (\mathbf{I}-\mathbf{D}_\gamma)\vert} \sum_{\substack{\vec{l}\in\Lambda^\parallel\\\gamma\in\Gamma_\Lambda}}  \int_{\mathbb{R}^{k''}} dV \frac{e^{2\pi i \vec{h} \cdot(\vec{\gamma}+\vec{l})}}{\vert \vec{y}+\vec{b}_\gamma+\vec{l}+\vec{\gamma}\vert^{2s}} \,.
    \label{csum3}
\end{equation}
Notice that $\vert \text{det} (\mathbf{I}-\mathbf{D}_\gamma)\vert$ counts precisely the number of elements of the finite abelian group $\Gamma_\Sigma=\Lambda^\perp/\Sigma^\perp$.  The integral over $\vec{y}$ can be carried out now by elementary methods. First, by a shift of the $\vec{y}$ variable, we can replace
\begin{equation} \vec{y}+\vec{b}_\gamma+\vec{l}+\vec{\gamma}\,\rightarrow\,  \vec{y}+\vec{b}^\parallel_\gamma+\vec{l}+\vec{\gamma}^\parallel,\end{equation}
where $\vec{v}^\parallel$ denotes the projection of $\vec{v}$ onto $\Lambda^\parallel\otimes\mathbb{R}$. Then, the norm splits as
\begin{equation} \vert \vec{y}+\vec{b}^\parallel_\gamma+\vec{l}+\vec{\gamma}^\parallel\vert^2=\vert\vec{y}\vert^2+\alpha^2,\quad \alpha^2\equiv \vert\vec{b}^\parallel_\gamma+\vec{l}+\vec{\gamma}^\parallel\vert^2.\end{equation}
The resulting integral can be evaluated directly by performing a change of coordinates to make the metric diagonal, and using spherical coordinates, to obtain
\begin{equation} 
    \int_{\mathbb{R}^{k''}} \frac{dV}{(\vert\vec{y}\vert^2+\alpha^2)^s}= \frac{2\pi^{\frac{k''}{2}}}{\Gamma(k''/2)}\int_0^\infty \frac{y^{k''-1}\, dy}{(y^2+\alpha^2)^s}= \frac{2\pi^{\frac{k''}{2}}}{\Gamma(k''/2)}=\pi^{\frac{k-k'}{2}}\frac{\Gamma(s')}{\Gamma(s)}\, \frac{1}{\alpha^{2s'}} \,,
    \label{csum4}
\end{equation}
where we have used the variables $k'=k-k''$ and $s'=s+(k'-k)/2$, as in the main text. We can already recognize the numerical prefactor of \eq{eq:VC-invariant-subspace}. Putting everything together, we get that \eq{csum} equals
\begin{equation} 
    \pi^{\frac{k-k'}{2}}\frac{\Gamma(s')}{\Gamma(s)}\, \frac{ V_\parallel}{\vert \text{det} (\mathbf{I}-\mathbf{D}_\gamma)\vert}  \sum_{\substack{\vec{l}\in\Lambda^\parallel\\\gamma\in\Gamma_\Lambda}} \frac{e^{2\pi i \vec{h} \cdot(\vec{\gamma}+\vec{l})}}{ \vert\vec{b}^\parallel_\gamma+\vec{l}+\vec{\gamma}^\parallel\vert^{2s'}} \,.
    \label{csum5}
\end{equation}
Now, any vector in $\Sigma^\perp$ but not in $\Lambda^\perp$ will yield a $\vec{\gamma}$ with $\vec{\gamma}^\parallel=0$. For any such element, the argument of the sum in \eq{csum} only depends on $\vec{\gamma}$ via the phase of the numerator. Denoting the set of all such $\gamma$ by $\Gamma_\Sigma$, the sum over $\Gamma_\Lambda$ in \eq{csum5} can be performed first over these, yielding
\begin{equation} 
    \sum_{\gamma\in\Gamma_\Sigma} e^{2\pi i\,\vec{h}\cdot\vec{\gamma}}= \vert \text{det} (\mathbf{I}-\mathbf{D}_\gamma)\vert \,\hat{\delta}_{\vec{h}} \,,
\end{equation}
where we have used that the cardinality of $\Gamma_\Sigma$ equals $ \vert \text{det} (\mathbf{I}-\mathbf{D}_\gamma)\vert$, as explained above, and defined
\begin{equation} \hat{\delta}_{\vec{h}}\equiv 1\quad\text{iff $\vec{h}\cdot\vec{l}\in\mathbb{Z},\, \forall \vec{l}\in\Sigma^\perp$, and $0$ otherwise.}\end{equation}
The condition for $\hat{\delta}_{\vec{h}}=1$ is precisely that $\vec{h}^\perp$, the orthogonal projection of $\vec{h}$ onto $\Sigma^\perp\otimes\mathbb{R}$, coincides with the same projection for some lattice vector $\vec{n}\in\mathbb{Z}^{k''}$. This is equivalent to the existence of some $\vec{\eta}\in\mathbb{Z}^k$ such that
\begin{equation} (\mathbf{I}-\mathbf{D}_\gamma)^T\cdot\vec{h}=(\mathbf{I}-\mathbf{D}_\gamma)^T\cdot\vec{\eta},\label{r133}\end{equation}
which is precisely the consistency condition \eq{betadef}. As in the main text, \eq{r133} means that $\vec{h}=\vec{\beta}+\vec{\eta}$, where $\vec{\beta}$ is in $\Lambda^\parallel\otimes\mathbb{Q}$.   Finally, we can combine $\vec{w}\equiv\vec{l}+\vec{\gamma}^\parallel$. All vectors $\vec{w}$ have the property that $\vec{w}\cdot\vec{l}\in\mathbb{Z}$ for all $\vec{l}\in\mathbb{Z}^k$, and they all lie in $\Lambda^\parallel\otimes\mathbb{Q}$. Therefore they span the projection of the dual lattice to $\Lambda^\parallel$ onto itself. Denoting this lattice as $\Xi$, relabelling $V_\parallel\rightarrow\sqrt{G_\parallel}$ in terms of the induced metric, and noticing that
\begin{equation}
    \vec{h}\cdot\vec{w}=\vec{\beta}\cdot\vec{w} +\vec{\eta}\cdot\vec{w}= \vec{\beta}\cdot\vec{w} \,\,\,\text{mod}\,\,\,\mathbb{Z} 
    \label{csum6}
\end{equation}
for all $\vec{w}\in\Xi$, the Casimir energy becomes
\begin{equation} \sqrt{G_\parallel}\, \pi^{\frac{k-k'}{2}}\frac{\Gamma(s')}{\Gamma(s)}\,   \sum_{\vec{w}\in\, \Xi} \frac{e^{2\pi i \vec{\beta} \cdot\vec{w}}}{ \vert\vec{w}+\vec{b}^\parallel_\gamma\vert^{2s'}},\label{csumfin}\end{equation}
which agrees with expression \eq{eq:VC-invariant-subspace} in the main text.

\subsection{RFM Casimir formula from Ewald expressions}
A second cross-check of \eq{eq:VC-invariant-subspace} comes from using the Ewald resummation expressions of Subsection \ref{sec:numerical-sums} and Appendix \ref{app:ewy}. We now start by ignoring momentarily the integral in \eq{csump}, and focusing on the sum inside the integral first. This sum can be tackled by Ewald methods, and indeed the Ewald resummation formula \eq{ewald0} applies directly provided that we identify
\begin{equation}
    \vec{c}=(\mathbf{I}-\mathbf{D}_\gamma)\,\vec{z} +\vec{b}_\gamma \,.
\end{equation}
One key advantage of the expression \eq{ewald0} is that after Ewald resummation, the integral over $\vec{z}$ can be performed directly in the momentum sum (just like in the Mellin transform derivation of Subsection \ref{casformula}), since the dependence on $\vec{z}$ is purely oscillatory. The integral one needs to evaluate is
\begin{align}
    \sqrt{G} &\int_{[0,1]^k} d^kz\, e^{-2\pi i (\vec{h}-\vec{k})\cdot [(\mathbf{I}-\mathbf{D}_\gamma)\,\vec{z}+\vec{b}_\gamma]}=  \sqrt{G}\, e^{-2\pi i (\vec{h}-\vec{k})\cdot\vec{b}_\gamma} \int_{[0,1]^k}  d^kz\, e^{-2\pi i (\vec{h}-\vec{k})\cdot [(\mathbf{I}-\mathbf{D}_\gamma)\,\vec{z}]}&\nonumber\\&= \sqrt{G}\, e^{-2\pi i (\vec{h}-\vec{k})\cdot\vec{b}_\gamma} \prod_{l=1}^k\frac{e^{2\pi i \vec{q}_l}-1}{2\pi i \vec{q}_l},\quad \vec{q}\equiv-(\mathbf{I}-\mathbf{D}_\gamma)^T (\vec{h}-\vec{k}) \,,
\end{align}
where $\vec{q}_l$ is the $l$-th component of the vector $\vec{q}$.  Further simplification is not possible without additional assumptions on the vector $\vec{h}$. However, the consistency condition on spin structures \eq{mimi}, discussed in Section \ref{sec:RFMs}, that
\begin{equation}  
    \vec{h} \cdot(\mathbf{I}-\mathbf{D}_\gamma)\,\vec{l}\in\mathbb{Z},\quad \forall\, \vec{l}\in\mathbb{Z}^k \,,
\end{equation}
implies that $(\mathbf{I}-\mathbf{D}_\gamma)^T\cdot \vec{h}$ is a vector of integer coordinates, and therefore, that $\vec{q}$ is as well. Since
\begin{equation}\frac{e^{2\pi i \vec{q}_l}-1}{2\pi i \vec{q}_l}\,\rightarrow\delta_{q_l,0}\quad\text{for}\quad q_l\in\mathbb{Z},\end{equation}
the integrals simplify to a delta function $\delta_{\vec{q},\vec{0}}$ imposing that the sum is restricted to the sublattice satisfying
\begin{equation*} (\mathbf{I}-\mathbf{D}_\gamma)^T(\vec{h}- \vec{k})=0,\quad\Rightarrow\quad \vec{k}=\vec{\eta}+ \vec{\kappa},\end{equation*}
where $\vec{\kappa}$ is a generic vector of the lattice $\Lambda^\parallel$ defined as the intersection of $\mathbb{Z}^k$ with the invariant subspace of  $\mathbf{D}_\gamma^T$, and $\vec{\eta}\in\mathbb{Z}^k$ is any particular vector of integer coordinates satisfying
\begin{equation} 
    (\mathbf{I}-\mathbf{D}_\gamma)^T\vec{\eta}=(\mathbf{I}-\mathbf{D}_\gamma)^T\vec{h}\,.
    \label{betadef}
\end{equation}
If there is no $\vec{\eta}\in\mathbb{Z}^k$ satisfying this equation, then the sum vanishes identically. We will remember this by including in the final result a factor $\hat{\delta}_{\vec{h}}$ in the sum which evaluates to 1 when \eq{betadef} is satisfied, and to 0 otherwise. Notice that the vector $\vec{\beta}\equiv \vec{h}-\vec{\eta}$ lies in $\Lambda^\parallel\otimes\mathbb{Q}$. We can similarly define $\vec{b}_\gamma^\parallel$ as the projection of $\vec{b}_\gamma$ onto the subspace invariant under $\mathbf{D}_\gamma$. We have that
\begin{equation}
    (\vec{\beta}-\vec{k})\cdot\vec{b}_\gamma= (\vec{\beta}-\vec{k})\cdot\vec{b}^\parallel_\gamma \,.
\end{equation}

Taking these into account, the integral over expression \eq{ewald0} takes the form
\begin{align} &\mathcal{S}^\alpha_{\text{position}}+ \frac{\pi^{2s-\frac{k}{2}}}{\Gamma(s)} \sum_{\substack{\vec{\kappa}\in\Lambda^\parallel\\\kappa-\vec{\beta}\neq\vec{0}}}\, \Gamma \left(\frac{k}{2}-s,\frac{\pi ^2 \vert\vec{\beta}-\vec{\kappa}\vert_D^2}{\alpha }\right) \frac{e^{-2\pi i (\vec{\beta}-\vec{\kappa})\cdot\vec{b}^\parallel_\gamma}}{ \vert\vec{\beta}-\vec{\kappa}\vert_D^{k-2s}}\nonumber\\&+ \delta_{\vec{h},\vec{0}}\, \frac{\pi^{\frac{k}{2}}\alpha^{s-k/2}}{ \Gamma(s)\left(s-\frac{k}{2}\right)}-\frac{\sqrt{G}\, \alpha^s}{\Gamma(s+1)}\delta_{(\mathbf{D}_\gamma,\vec{b}_\gamma)},\label{ewald1} \end{align}
where by $\mathcal{S}^\alpha_{\text{position}}$ denotes the integral over $\vec{x}$ of the first term in \eq{ewald0}, which we did not evaluate explicitly. Notice that the last term in \eq{ewald1}, which comes from the integral in the last term of \eq{ewald0}, is only present for the identity contribution $\mathbf{D}_\gamma=\mathbf{I},\,\vec{b}_\gamma=\vec{0}$, where the integral over $\vec{x}$ is trivial anyway. For any other element of $\mathcal{B}$, the $\chi_{\mathbb{Z}^k}(\vec{c})$ term localizes the last term in \eq{ewald0} to a zero-measure set of the unit cube, so we will drop it from now on. 

Denoting $k'\equiv\text{dim}(\Lambda^\parallel)$, and writing $\vec{\kappa}=\mathbf{V}\cdot \vec{l}$ where $\mathbf{V}$ is a matrix whose columns form a basis of $\Lambda^\parallel$, and $\vec{l}\in\mathbb{Z}^r$, we can rewrite \eq{ewald1} as a sum over $\vec{l}$. Using the definitions\footnote{ In general, given a rectangular matrix $\mathbf{V}$, there is more than one pseudoinverse $\mathbf{V}^{-1}$. Since $\vec{b}^\parallel_\gamma$ lies in the invariant space of $D_\gamma$, we will choose the particular pseudoinverse such that the columns of $(V^{-1})^T$ generate the invariant subspace of $D_\gamma$.}  
\begin{equation} 
    \vert \vec{v}\vert_I\equiv \vec{v}^T\cdot \mathbf{G}_I\cdot \vec{v}\,,\quad \mathbf{G}_I\equiv \mathbf{V}^T\,\mathbf{G}_D\mathbf{V}\,,\quad \vec{d}\equiv \mathbf{V}^{-1}\,\vec{\beta}\,,\quad \vec{v}_\gamma\equiv \mathbf{V}^{T}\,\vec{b}^\parallel_\gamma\,,\quad s'=\frac{k'-k}{2}+s \,,
\end{equation}
equation \eq{ewald1} can be rewritten as
\begin{align} 
    &\mathcal{S}^\alpha_{\text{position}}+ \frac{\pi^{2s'-k' +\frac{k}{2}}}{\Gamma(s)} \sum_{\substack{\vec{l}\in\mathbb{Z}^{k'}\\\vec{l}-\vec{d}\neq\vec{0}}}\, \Gamma \left(\frac{k'}{2}-s',\frac{\pi ^2 \vert\vec{d}-\vec{l}\vert_I^2}{\alpha }\right) \frac{e^{-2\pi i (\vec{d}-\vec{l})\cdot\vec{v}_\gamma}}{ \vert\vec{d}-\vec{l}\vert_I^{k'-2s'}}+ \delta_{\vec{d},\vec{0}}\, \frac{\pi^{\frac{k}{2}}\alpha^{s'-k'/2}}{ \Gamma(s)\left(s'-\frac{k'}{2}\right)}\,.
    \label{ewald2} 
\end{align}
In particular, if we multiply the whole sum by a common factor $\frac{\Gamma(s)}{\pi^{k/2}}\frac{\pi^{k'/2}}{\Gamma(s')}\frac{1}{\sqrt{G_I}}$, where $G_I\equiv\text{det}(\mathbf{G}_I)$, we obtain
\begin{align} 
    &\frac{\Gamma(s)}{\pi^{k/2}}\frac{\pi^{k'/2}}{\Gamma(s')}\frac{1}{\sqrt{G_I}}\, \mathcal{S}^\alpha_{\text{position}}+ \frac{\pi^{2s'-\frac{k'}{2}}}{\Gamma(s')\, \sqrt{G_I}} \sum_{\substack{\vec{l}\in\mathbb{Z}^{k'}\\\vec{l}-\vec{d}\neq\vec{0}}}\, \Gamma \left(\frac{k'}{2}-s',\frac{\pi ^2 \vert\vec{d}-\vec{l}\vert_I^2}{\alpha }\right) \frac{e^{-2\pi i (\vec{d}-\vec{l})\cdot\vec{v}_\gamma}}{ \vert\vec{d}-\vec{l}\vert_I^{k'-2s'}}\nonumber\\&+ \delta_{\vec{d},\vec{0}}\, \frac{\pi^{\frac{k'}{2}}\alpha^{s'-k'/2}}{\sqrt{G_I}\, \Gamma(s')\left(s'-\frac{k'}{2}\right)}\,.
    \label{ewald2ymedio} 
\end{align}
The last two terms now look exactly like the momentum piece of a Ewald sum \eq{ewald0}, in $k'$ dimensions instead of $k$, and with different metric and shift vectors. According to \eq{ewald0}, the momentum piece gives the exact result as $\alpha\rightarrow\infty$, where the position piece becomes negligible. Since $\mathcal{S}^\alpha$, being an integral of a similar position term, also becomes neglibile as $\alpha\rightarrow\infty$, \eq{ewald2} must equal the Ewald sum, which means that for any $\alpha$ we can replace $\mathcal{S}^\alpha$ with the position term. Putting everything together, adding the prefactor in \eq{precas} coming from the volume of the sphere and the propagator derivatives, and the extra sign due to our definition of the Casimir coefficient, the Casimir energy $\mathcal{E}$ of \eq{csum} evaluates to 
\begin{align} 
    \mathcal{E}(\gamma)=&-\hat{\delta}_{\vec{h}}\,\frac{\Gamma(s')}{2\pi^{s'}}\cdot
    \frac{\sqrt{G_\parallel}}{\vert\Gamma\vert} \sum_{\vec{m}}\frac{e^{2\pi i \vec{d}\cdot\vec{m}}}{\vert \vec{m}+\vec{v}_\gamma\vert_\parallel^{2s'}}\,,
    \label{ewald3} 
\end{align}
where $\vert\cdot\vert_\parallel^2$ is the norm with respect to $\mathbf{G}_\parallel=\mathbf{G}_I^{-1}$, and the definitions of the various quantities appearing in \eq{ewald3} are as follows: let  as above $\mathbf{V}$ be matrix whose columns form a basis of the invariant subspace of $\mathbb{Z}^k$ with respect to $\mathbf{D}^T_\gamma$, of dimension $k'$. Then
\begin{equation} \mathbf{G}_\parallel= \mathbf{V}^{-1} \,\mathbf{G}\, (\mathbf{V}^{-1})^T,\quad \vec{v}_\gamma= \mathbf{V}^{T} \cdot\vec{b}_\gamma^\parallel,\quad \vec{d}\equiv \mathbf{V}^{-1}\vec{\beta},\quad s'=\frac{k'-k}{2}+s.\end{equation}
where $\vec{b}_\gamma^\parallel$ is the projection of $\vec{b}_\gamma$ onto the invariant subspace of $\mathbf{D}_\gamma$, and $\vec{\beta}$ is defined in \eq{betadef}. In short, the integral over $\vec{z}$ localizes the sum to a lower-dimensional sublattice, but the resulting sum is still like that of \eq{ewaldOG}, and can be evaluated numerically in an efficient way via the Ewald formula \eq{ewald0}. Finally, \eq{ewald3} can also be written in terms of the original metric as 
\begin{equation}
    \mathcal{E}(\gamma) = -\hat{\delta}_{\vec{h}}\,\frac{\Gamma(s')}{2\pi^{s'}}\cdot
    \frac{\sqrt{G_\parallel}}{\vert\Gamma\vert} \sum_{\vec{\xi}\in\,\Xi} \frac{e^{2\pi i \,\Vec{\beta}\cdot\vec{\xi}}}{|\vec{\xi} + \bvec^\parallel|^{2s'}} \label{ewald4} 
\end{equation}
where $\Xi$ is the lattice spanned by the columns of $(\mathbf{V}^{-1})^T$.  As explained above, these can be taken to lie in the invariant subspace of $\gamma$. By the same argument as that around equation \eq{eq:eta-condition}, the lattice spanned by these is precisely the projection of the ambient lattice onto the $\mathbf{D}_\gamma$ invariant subspace. Equation \eq{ewald4} is our final result, and it agrees with \eq{eq:VC-invariant-subspace} of the main text. In short, it turns out that performing the $\vec{z}$ integral in the sum \eq{csum} turns it into another sum of the same form, which can be evaluated efficiently via Ewald methods. 

\section{Lorentz group traces for Casimir energies}
\label{ap:traces}
In this appendix we collect some information regarding the calculation of traces of several matrices that appear in the calculation of Casimir energies in the main text. As described there, the Casimir energy involves traces over representations of a certain group action acting on the Hilbert space of the massless particles in $d$ dimensions. The little group acting on $D$-dimensional massless particles is $SO(D-2)$, which decomposes as
\begin{equation}
    SO(D-2)\to SO(k)\times SO(d-2) \,,
\end{equation}
as we compactify the theory on a $k$-dimensional manifold down to $d=D-k$ dimensions. The geometric action we use to obtain a Riemann-flat manifold as a quotient of $T^k$ is in the $SO(k)$ factor, but it is useful to work in terms of $SO(D-2)$ matrices. Hence, we will embed the $SO(k)$ matrix in the vector representation as a block-diagonal $SO(D-2)$ matrix, with an additional $(d-2)\times(d-2)$ block representing the identity. 

In this paper, we focus on M-theory compactifications, and thus start with a $9\times 9$ matrix $\mathbf{M}$ representing an element of $SO(9)$ in the vector representation. To compute Casimir energies we must compute the trace of the matrices that represent this element in the representations of $SO(9)$ that correspond to the fields of $11$-dimensional supergravity---the graviton $g_{\mu\nu}$, transforming in the symmetric-traceless $\mathbf{44}$ representation; the 3-form $C_{\mu\nu\rho}$, transforming in the antisymmetric, three-index $\mathbf{84}$ representation; and the gravitino, transforming in the spin-$3/2$ $\mathbf{128}$ representation. Notice that the little group accurately counts the physical polarizations (degrees of freedom) for each field, and that the total number of bosonic d.o.f. equals the number of fermionic d.o.f., as befits a supersymmetric theory.

In principle, the matrix $\mathbf{M}$ must be orthogonal. However, because all the formulas below will involve traces, and the trace of a matrix is invariant under similarity transformations, $\mathbf{M}$ can be replaced by $\mathbf{X} \mathbf{M}\mathbf{X}^{-1}$ for any non-singular matrix $\mathbf{X}$. In particular, we may replace the orthogonal matrices by non-orthogonal representatives of themselves (for instance, by the $SL(n,\mathbb{Z})$ matrix $\mathbf{D}$ that implements the action in the lattice basis), which is what we do in the main text.

How do we relate the trace of $\mathbf{M}$ to the trace of $\mathbf{M}^\text{Sym.}$, the representative in the symmetric tensor representation relevant for the graviton? If $v^a$ and $w^a$ both transform in the vector representation of $SO(n)$, one can construct the tensor representation whose objects are two-index tensors $T^{ab}$, and for which a basis can be constructed via bivectors $v^a w^b$. The $SO(n)$ action on the tensor representation is
\begin{equation} 
    v^aw^b\,\rightarrow M^a_cM^b_d v^c w^d \,,
\end{equation}
and the pairing to a dual basis is given by the tensor $\delta^a_c\delta^b_d$. Hence, the trace in the tensor representation is simply
\begin{equation} 
    \text{Tr}(\mathbf{M}^\text{Tens.}) =\delta^c_a\delta^d_b M^a_cM^b_d = (\text{Tr}(\mathbf{M}))^2 \,.
\end{equation}
The tensor representation is reducible, and it decomposes as a sum of symmetric traceless, antisymmetric, and singlet representations. The corresponding projection operators are
\begin{align}
    \mathbb{P}^{\text{Sym.}}\, T^{ab} &=\frac12(T^{ab}+T^{ba})-\frac{T}{n}\delta^{ab} \,, \nonumber \\ 
    \mathbb{P}^{\text{ASym.}}\, T^{ab} &=\frac12(T^{ab}-T^{ba}) \,, \nonumber \\ 
    \mathbb{P}^{\text{Sing}.}\, T^{ab} &=\frac{T}{n}\delta^{ab} \,.
\end{align}
One can check that each of these squares to themselves, as a projection operator must, and that the sum of the three equals the identity. The trace in the symmetric tensor representation is just the trace in the tensor representation of the projected matrix, 
\begin{equation}
    \Tr{}{\mathbf{M}^{\text{Sym.}}}=\frac12\left(\Tr{}{\mathbf{M}}^2 + \Tr{}{\mathbf{M}^2}\right)-1 \,,
\end{equation}
where the $-1$ comes from the singlet contribution which is proportional to $(1/n)\text{Tr}(\mathbf{M}^T\mathbf{M})=1$ since $\mathbf{M}$ is (similar to) an orthogonal matrix.
As a check, when $\mathbf{M}=\mathbf{I}$ we get 
\begin{equation}
    \Tr{}{\mathbf{I}^{\text{Sym.}}} = \frac{n^2}{2}\left(\frac12-\frac{1}{n}\right)\cdot n= \frac{n(n+1)}{2}-1 \,,
\end{equation}
which is the correct dimension for the symmetric traceless representation. 

For the three-index antisymmetric representation, the corresponding projector acts as
\begin{equation}
    \mathbb{P}^{\text{3-Antisym.}}\, T^{abc} = \frac16\left(T^{abc}+T^{bca}+T^{cab}-T^{bac}-T^{acb}-T^{cba}\right) \,,
\end{equation}
and, consequently, the trace is
\begin{equation}
    \Tr{}{\mathbf{M}^\text{3-Antisym.}} = \frac16\left(\Tr{}{\mathbf{M}}^3 - 3\,\Tr{}{\mathbf{M}^2}\cdot \Tr{}{\mathbf{M}} + 2\,\Tr{}{\mathbf{M}^3}\right) \,.
\end{equation}
Evaluating this for the identity matrix gives
\begin{equation}
    \Tr{}{\mathbf{I}^\text{3-Antisym.}} = \frac16(n^3-3n^2+2n)=\frac{n(n-1)(n-2)}{6}=\binom{n}{3} \,,
\end{equation}
which is the correct dimension for the three-index antisymmetric representation. This is a special case of a more general formula \cite{vanRitbergen:1998pn} encoding the character of a fully antisymmetric representation\footnote{Note a typo in the rhs of (49) of \cite{vanRitbergen:1998pn}, the term $\text{Ch}_R(lF)$ should be divided by $l$ \cite{weyl1946classical,Schellekens:1986xh}.} via a generating function,
\begin{equation} 
    \sum_{k=0}^\infty x^k\Tr{}{\mathbf{M}^{k\text{-Antisym.}}}= \prod_{l=1}^\infty\exp\left(-(-x)^l\frac{\text{Tr}(\mathbf{M}^l)}{l}\right) \,.
    \label{eq:character-traces}
\end{equation}

Finally, to evaluate the traces in the gravitino (Rarita-Schwinger) representation, we may realize it as the tensor product of a vector with a spinor $\rho_a^\alpha$, where $a$ is as before a vector index and $\alpha$ is a spinor index, and then substract a spinor with opposite chirality, 
\begin{equation} 
    \mathbb{P}^{\text{R.S}}\, \rho_a^\alpha = \rho_a^\alpha - \frac{1}{n^2}\delta^{ab}(\gamma_b)^\alpha_{\beta} \rho_a^\beta\,,
\end{equation}
where the relative factor of $1/n^2$ comes from the property $\delta^{ab}(\gamma_b)^\alpha_{\beta}(\gamma_a)^\beta_{\alpha}=n^2$ of $\gamma$ matrices. Taking traces on both sides of this equation, one obtains
\begin{equation} 
    \Tr{}{\mathbf{M}^{\text{R.S.}}} = \left(\Tr{}{\mathbf{M}} - 1 \right)\Tr{}{\mathbf{M}^\text{Spinor}} \,,
    \label{rs}
\end{equation}
and the only thing left to do is to express $\Tr{}{\mathbf{M}^\text{Spinor}}$ in terms of $\Tr{}{\mathbf{M}}$. To do this, we can employ the fact that, for $n=2k+1$ odd, the spinor representation $\mathbf{2^{\lfloor\frac{k}{2}\rfloor}}$ satisfies 
\begin{equation} \mathbf{2^{k}}\otimes \mathbf{2^{k}}= \sum_{l=1}^k\mathbf{\binom{n}{l}},\end{equation}
i.e. it becomes a sum of fully antisymmetrized representations. The dimensions also agree, as can be checked using Pascal's identity. Combining this with \eq{eq:character-traces}, we obtain an expression for the spinor traces, which for the particular case of interest $n=9$ becomes
\begin{align}
    \Tr{}{\mathbf{\mathbf{M}}^\text{Spinor}}^2
    &= 1 + \Tr{}{\mathbf{M}} + \frac12 (\Tr{}{\mathbf{M}}^2 - \Tr{}{\mathbf{M}^2}) \nonumber  \\ 
    & + \frac{1}{6} (\Tr{}{\mathbf{M}}^3 - 3\,\Tr{}{\mathbf{M}} \Tr{}{\mathbf{M}^2} + 2\,\Tr{}{\mathbf{M}^3}) \\
    & + \frac{1}{24} (\Tr{}{\mathbf{M}}^4 - 6\,\Tr{}{\mathbf{M}}^2 \Tr{}{\mathbf{M}^2} + 3\,\Tr{}{\mathbf{M}^2}^2 + 8\,\Tr{}{\mathbf{M}} \Tr{}{\mathbf{M}^3} - 
    6\,\Tr{}{\mathbf{M}^4}), \nonumber
    \label{sptr}
\end{align}
and together with \eq{rs} fully determines the traces in the gravitino representation, up to a sign ambiguity in taking the square root in \eq{sptr}. This ambiguity is physical, and corresponds to the two lifts (related by a sign) that a given $SO$ element has in Spin. With these expressions, we get $\Tr{}{\mathbf{I}^{\text{R.S.}}}=128$, which is the correct result for the Rarita-Schwinger representation.

All these formulae can be checked by evaluating them explicitly for $SO(9)$ matrices in Cartan form, in terms of Cartan angles. We can write any $SO(9)$ element in the form
\begin{align}
    \mathbf{M} = \begin{pmatrix}
        \cos\theta_1  & \sin\theta_1 & 0 & 0 & 0 & 0 \\
        -\sin\theta_1 & \cos\theta_1 & 0 & 0 & 0 & 0 \\
        0             & 0                & \ddots & \vdots & \vdots & \vdots  \\
        0             & 0            & \dotsm & \cos\theta_4  & \sin\theta_4 & 0 \\
        0             & 0            & \dotsm & -\sin\theta_4 & \cos\theta_4 & 0 \\
        0 & 0 & \dotsm & 0 & 0 & 1 
    \end{pmatrix}
    = \exp\left(i \theta_\alpha E^3_\alpha \right) \,,
\end{align}
where $E^3_\alpha$ is a diagonal generator that is a linear combination of the Cartan generators of $SO(9)$, i.e. $E^3_\alpha = |\alpha|^{-2}(\alpha\cdot H)$, with $\alpha$ a root of $SO(9)$ and $H_i,\, i=1,...,4$ the Cartan generators.
Since we can decompose each representation into eigenstates $|\mu\rangle$ of $H_i$ labelled by a weight vector $\vec{\mu}$, once we know each decomposition in terms of its weights we can take the trace in some representation $R$ as
\begin{align}
    \Tr{R}{\mathbf{M}} = \sum_{\vec{\mu}\in R} \exp\left(i\, \vec{\theta}\cdot\vec{\mu}\right) \,,
    \label{eq:trace-formula}
\end{align}
where $\vec{\mu}\in R$ includes all weight vectors $\vec{\mu}$ in the representation $R$ (e.g. there are $9$ such vectors in the fundamental representation and $44$ in the rank-2 symmetric traceless representation). We then find that the traces computed in this way match the ones computed using the character formulas above. 
For example, the weights of the vector representation correspond to all possible permutations of $(\pm 1,0,0,0)$ plus $(0,0,0,0)$, which is indeed a 9-dimensional representation. Using \eq{eq:trace-formula} we find
\begin{equation}
    \Tr{\mathbf{9}}{\mathbf{M}} = 2\sum_{i=1}^4 \cos\theta_i + 1 \,, 
\end{equation}
which matches the trace of $\mathbf{M}$. Taking instead the spinor representation of $SO(9)$, corresponding to weight vectors $\{\pm\frac12,\pm\frac12,\pm\frac12,\pm\frac12\}$ (which does add up to $2^4 = 16$ d.o.f.), we find
\begin{equation}
    \Tr{\text{Spinor}}{\mathbf{M}} = \prod_{i=1}^{4} 2\cos\frac{\theta_i}{2} \,.
    \label{eq:spinor-trace-weights}
\end{equation}
For spinor representations we should keep in mind that the transformation $\mathbf{M}$ has two spin lifts $\pm\mathcal{M}$ related by a sign. If we choose this set of $\theta_i$ to define $+\mathcal{M}$, then $-\mathcal{M}$ corresponds to any choice of $\theta_i' = \theta_i + 2\pi \eta_i$ such that $\sum\eta_i \in(2\Z+1)$.
Indeed, we find that
\begin{align}
    \Tr{\text{Spinor}}{-\mathcal{M}} 
    &= \prod_{i=1}^{4} 2\cos\left(\frac{\theta_i'}{2} + \pi\eta_i\right) \nonumber \\ 
    &= (-1)^{\sum\eta_i}\prod_{i=1}^{4} 2\cos\frac{\theta_i'}{2}
    = (-1)^{\sum\eta_i}\Tr{\text{Spinor}}{\mathcal{M}} \,,
    \label{eq:Spinor-trace-signs}
\end{align}
with the trace differing by a sign whenever $\sum\eta_i$ is odd.

When we consider cyclic RFM's, as we do in this paper, all elements in the group $\Gamma$ are powers of the generator, i.e. $\Gamma = \{\mathbf{I},\D[g],...,\D[g]^{p-1}\}$, where $p=|\Gamma|$ is the order of the finite group $\Gamma$. Correspondingly the angles $\theta_i^{(j)}$ for a given element $\D[g]^j$ are just multiples of the angles $\theta_i^{\mathbf{g}}$ of the generator, $\theta_i^{(j)} = j\cdot\theta_i^{\mathbf{g}}$, which gives the traces in the vector representation
\begin{equation}
    \Tr{\mathbf{9}}{\D[g]^j} = 2\sum_{i=1}^4 \cos(j\cdot\theta_i^{\mathbf{g}}) + 1 \,.
\end{equation}
Since $\D[g]^p = \mathbf{I}$, we have that $p\,\theta_i = 2\pi n_i$, with $n_i\in\Z$, and the trace is indeed $9$, the dimension of the vector representation. For spinor representations, 
\begin{align}
    \Tr{\text{Spinor}}{\D[g]^p} &= \prod_{i=1}^{4} 2\cos(\pi\,n_i) 
    = (-1)^{\sum n_i}\cdot 16 \,,
\end{align}
whose sign depends on the choice of angles $\theta_i$, i.e. on the choice of spin lift $\pm\mathcal{D}_\mathbf{g}$. Changing the spin lift by shifting $\theta_i\to \theta_i + 2\pi\eta_i$, with $\sum\eta_i$ odd, as in \eq{eq:Spinor-trace-signs}, corresponds to a shift $n_i\to n_i + p\,\eta_i$ and thus $\sum n_i\to\sum n_i + p\sum\eta_i$; this changes the spinor trace by a factor $(-1)^{p\sum\eta_i}$ which is $1$ for $p$ even and $(-1)$ for $p$ odd. Note the relation to our constraint \eq{swer2}---for even $p$, $s_\mathbf{g}$ does not depend on the specific choice of spin lift and thus the phase vector $\vec{h}$ is constrained in the directions not orthogonal to $\bvec[g]$.

Finally taking the representations of M-theory, the traces of a given element $\D[g]^j\in\Gamma$ in the case of cyclic groups are
\begin{equation}
    \Tr{\mathbf{44}}{\D[g]^j}  = \sum_{a=1}^4 \bigg[2\cos(2 j\,\theta_a^\mathbf{g}) + 2\cos (j\,\theta_a^\mathbf{g}) + 1 + 2\sum_{\substack{b=1\\j\neq i}}^4 \cos (j\,\theta_a^\mathbf{g})\cos (j\theta_b^\mathbf{g}) \bigg]
\end{equation}
for the graviton, 
\begin{align}
    \Tr{\mathbf{84}}{\D[g]}  &= \sum_{a=1}^4 \left[1 + 6\cos (j\,\theta_a^\mathbf{g}) + 2\sum_{\substack{b=1\\j\neq i}}^4 \cos (j\,\theta_a^\mathbf{g})\cos (j\,\theta_b^\mathbf{g})\right.\nonumber\\&\quad\left. +\frac43\sum_{\substack{b,c=1\\b\neq c\neq a}}^4 \cos (j\,\theta_a^\mathbf{g})\cos (j\,\theta_b^\mathbf{g})\cos (j\,\theta_c^\mathbf{g})\right] \,, 
\end{align}
for the three-form, and 
\begin{align}
    \Tr{\mathbf{128}}{\D[g]^j} &= \left(2\sum_{a=1}^4 \cos(j\,\theta_a^{\mathbf{g}})\right)\prod_{a=1}^{4} 2\cos\left(j\,\frac{\theta_a^\mathbf{g}}{2}\right) \,,
\end{align}
for the gravitino, in terms of the eigenvalue arguments $\theta_i^\mathbf{g}$ of the generator $\D[g]$ and keeping in mind the choice of spin lift $\mathcal{D}_\mathbf{g}$.

Note that the traces over bosonic representions are invariant under the change $j\to p-j$ and thus the traces of $\D[g]^j$ and $\D[g]^{p-j} = (\D[g]^j)^{-1}$ are the same. However, for the gravitino representation (as would also be in the spinor representation) we have 
\begin{equation}
    \Tr{\mathbf{128}}{\D[g]^{p-j}} = (-1)^{\sum n_i}\left(2\sum_{a=1}^4 \cos(j\,\theta_a^{\mathbf{g}})\right)\prod_{a=1}^{4} 2\cos\left(j\,\frac{\theta_a^\mathbf{g}}{2}\right) \,,\quad\text{with}\quad p\cdot\theta_i = 2\pi n_i \,. 
\end{equation}
The traces in the gravitino representation will then be the same \textit{up to a sign}, which is determined by the eigenvalues of the generator $\D[g]$. Note that this sign is only affected by the specific choice of spin lift when $p$ is odd, but not when $p$ is even. 

\section{Cyclic 7d RFM's with \texorpdfstring{$S^1$}{S1} base}
\label{ap:7d-RFMs}
This appendix contains a list of all the Riemann-flat manifolds of cyclic holonomy that we analyzed for the work in Section \ref{sec:dS4-maxima}. While this list is not exhaustive, there is at least one representative for each $\mathbb{Q}$-equivalence class of $SL(7,\mathbb{Z})$ matrices. As explained in the main text, we could not find a dS saddle point in any of these, although we only studied carefully the examples with two moduli only. The data listed includes: the explicit expression for the generator $\vec{z}\,\rightarrow \D[g]\,\vec{z}+\bvec[g]$, the order of the group $\Gamma$, the number of moduli that the RFM has, and the possible choices for the vector $\vec{h}$ of spin structures. Each entry $h_i$ not fixed to zero can take values $0$ (corresponding to periodic boundary conditions) or $1/2$ (corresponding to antiperiodic boundary conditions).

{
\begin{table}
    \centering
    \scriptsize
    \begin{tabular}{|C{12em}C{4em}C{2em}C{4em}C{11em}|}
        \hline
        $\D[g]$ & $\bvec[g]$ & $|\Gamma|$ & Moduli & \parbox[c][2.5em][c]{1em}{$\vec{h}$} \\  
        \hline & & & & \\ $\left(
\begin{array}{*7{C{1em}}}
 0 & -1 & 0 & 0 & 0 & 0 & 0 \\
 1 & -1 & 0 & 0 & 0 & 0 & 0 \\
 0 & 0 & 0 & 0 & 0 & -1 & 0 \\
 0 & 0 & 1 & 0 & 0 & 1 & 0 \\
 0 & 0 & 0 & 1 & 0 & -1 & 0 \\
 0 & 0 & 0 & 0 & 1 & 1 & 0 \\
 0 & 0 & 0 & 0 & 0 & 0 & 1 \\
\end{array}
\right)$ & $\left(
\begin{array}{c}
 0 \\
 0 \\
 0 \\
 0 \\
 0 \\
 0 \\
 \frac{1}{30} \\
\end{array}
\right)$ & $30$ & 4 & $(0,0,0,0,0,0,0)$ \\ & & & & \\ \hline & & & & \\ $\left(
\begin{array}{*7{C{1em}}}
 0 & -1 & 0 & 0 & 0 & 0 & 0 \\
 1 & 1 & 0 & 0 & 0 & 0 & 0 \\
 0 & 0 & 0 & 0 & 0 & -1 & 0 \\
 0 & 0 & 1 & 0 & 0 & 1 & 0 \\
 0 & 0 & 0 & 1 & 0 & -1 & 0 \\
 0 & 0 & 0 & 0 & 1 & 1 & 0 \\
 0 & 0 & 0 & 0 & 0 & 0 & 1 \\
\end{array}
\right)$ & $\left(
\begin{array}{c}
 0 \\
 0 \\
 0 \\
 0 \\
 0 \\
 0 \\
 \frac{1}{30} \\
\end{array}
\right)$ & $30$ & 4 & $(0,0,0,0,0,0,0)$ \\ & & & & \\ \hline & & & & \\ $\left(
\begin{array}{*7{C{1em}}}
 0 & 0 & 0 & -1 & 0 & 0 & 0 \\
 1 & 0 & 0 & -1 & 0 & 0 & 0 \\
 0 & 1 & 0 & -1 & 0 & 0 & 0 \\
 0 & 0 & 1 & -1 & 0 & 0 & 0 \\
 0 & 0 & 0 & 0 & 0 & -1 & 0 \\
 0 & 0 & 0 & 0 & 1 & 1 & 0 \\
 0 & 0 & 0 & 0 & 0 & 0 & 1 \\
\end{array}
\right)$ & $\left(
\begin{array}{c}
 0 \\
 0 \\
 0 \\
 0 \\
 0 \\
 0 \\
 \frac{1}{30} \\
\end{array}
\right)$ & $30$ & 4 & $(0,0,0,0,0,0,0)$ \\ & & & & \\ \hline & & & & \\ $\left(
\begin{array}{*7{C{1em}}}
 0 & -1 & 0 & 0 & 0 & 0 & 0 \\
 1 & -1 & 0 & 0 & 0 & 0 & 0 \\
 0 & 0 & 0 & 0 & 0 & -1 & 0 \\
 0 & 0 & 1 & 0 & 0 & 0 & 0 \\
 0 & 0 & 0 & 1 & 0 & 0 & 0 \\
 0 & 0 & 0 & 0 & 1 & 0 & 0 \\
 0 & 0 & 0 & 0 & 0 & 0 & 1 \\
\end{array}
\right)$ & $\left(
\begin{array}{c}
 0 \\
 0 \\
 0 \\
 0 \\
 0 \\
 0 \\
 \frac{1}{24} \\
\end{array}
\right)$ & $24$ & 4 & $\left(0,0,h_3,h_3,h_3,h_3,0\right)$ \\ & & & & \\ \hline & & & & \\ $\left(
\begin{array}{*7{C{1em}}}
 0 & -1 & 0 & 0 & 0 & 0 & 0 \\
 1 & 1 & 0 & 0 & 0 & 0 & 0 \\
 0 & 0 & 0 & 0 & 0 & -1 & 0 \\
 0 & 0 & 1 & 0 & 0 & 0 & 0 \\
 0 & 0 & 0 & 1 & 0 & 0 & 0 \\
 0 & 0 & 0 & 0 & 1 & 0 & 0 \\
 0 & 0 & 0 & 0 & 0 & 0 & 1 \\
\end{array}
\right)$ & $\left(
\begin{array}{c}
 0 \\
 0 \\
 0 \\
 0 \\
 0 \\
 0 \\
 \frac{1}{24} \\
\end{array}
\right)$ & $24$ & 4 & $\left(0,0,h_3,h_3,h_3,h_3,0\right)$ \\ & & & & \\ \hline & & & & \\ $\left(
\begin{array}{*7{C{1em}}}
 0 & -1 & 0 & 0 & 0 & 0 & 0 \\
 1 & 0 & 0 & 0 & 0 & 0 & 0 \\
 0 & 0 & 0 & 0 & 0 & -1 & 0 \\
 0 & 0 & 1 & 0 & 0 & -1 & 0 \\
 0 & 0 & 0 & 1 & 0 & -1 & 0 \\
 0 & 0 & 0 & 0 & 1 & -1 & 0 \\
 0 & 0 & 0 & 0 & 0 & 0 & 1 \\
\end{array}
\right)$ & $\left(
\begin{array}{c}
 0 \\
 0 \\
 0 \\
 0 \\
 0 \\
 0 \\
 \frac{1}{20} \\
\end{array}
\right)$ & $20$ & 4 & $\left(h_1,h_1,0,0,0,0,0\right)$ \\ & & & & \\ \hline & & & & \\ $\left(
\begin{array}{*7{C{1em}}}
 0 & -1 & 0 & 0 & 0 & 0 & 0 \\
 1 & 0 & 0 & 0 & 0 & 0 & 0 \\
 0 & 0 & 0 & 0 & 0 & -1 & 0 \\
 0 & 0 & 1 & 0 & 0 & 1 & 0 \\
 0 & 0 & 0 & 1 & 0 & -1 & 0 \\
 0 & 0 & 0 & 0 & 1 & 1 & 0 \\
 0 & 0 & 0 & 0 & 0 & 0 & 1 \\
\end{array}
\right)$ & $\left(
\begin{array}{c}
 0 \\
 0 \\
 0 \\
 0 \\
 0 \\
 0 \\
 \frac{1}{20} \\
\end{array}
\right)$ & $20$ & 4 & $\left(h_1,h_1,0,0,0,0,0\right)$ \\ & & & & \\ \hline \end{tabular}

	\end{table}\begin{table}
    \centering
    \scriptsize
    \begin{tabular}{|C{12em}C{4em}C{2em}C{4em}C{11em}|}
        \hline
        $\D[g]$ & $\bvec[g]$ & $|\Gamma|$ & Moduli & \parbox[c][2.5em][c]{1em}{$\vec{h}$} \\  
        \hline & & & & \\ $\left(
\begin{array}{*7{C{1em}}}
 0 & 0 & 0 & 0 & 0 & -1 & 0 \\
 1 & 0 & 0 & 0 & 0 & 0 & 0 \\
 0 & 1 & 0 & 0 & 0 & 0 & 0 \\
 0 & 0 & 1 & 0 & 0 & 1 & 0 \\
 0 & 0 & 0 & 1 & 0 & 0 & 0 \\
 0 & 0 & 0 & 0 & 1 & 0 & 0 \\
 0 & 0 & 0 & 0 & 0 & 0 & 1 \\
\end{array}
\right)$ & $\left(
\begin{array}{c}
 0 \\
 0 \\
 0 \\
 0 \\
 0 \\
 0 \\
 \frac{1}{18} \\
\end{array}
\right)$ & $18$ & 4 & $(0,0,0,0,0,0,0)$ \\ & & & & \\ \hline & & & & \\ $\left(
\begin{array}{*7{C{1em}}}
 0 & -1 & 0 & 0 & 0 & 0 & 0 \\
 1 & -1 & 0 & 0 & 0 & 0 & 0 \\
 0 & 0 & 0 & 0 & 0 & -1 & 0 \\
 0 & 0 & 1 & 0 & 0 & -1 & 0 \\
 0 & 0 & 0 & 1 & 0 & -1 & 0 \\
 0 & 0 & 0 & 0 & 1 & -1 & 0 \\
 0 & 0 & 0 & 0 & 0 & 0 & 1 \\
\end{array}
\right)$ & $\left(
\begin{array}{c}
 0 \\
 0 \\
 0 \\
 0 \\
 0 \\
 0 \\
 \frac{1}{15} \\
\end{array}
\right)$ & $15$ & 4 & $\left(0,0,0,0,0,0,h_7\right)$ \\ & & & & \\ \hline & & & & \\ $\left(
\begin{array}{*7{C{1em}}}
 0 & 0 & 0 & 0 & 0 & -1 & 0 \\
 1 & 0 & 0 & 0 & 0 & 1 & 0 \\
 0 & 1 & 0 & 0 & 0 & -1 & 0 \\
 0 & 0 & 1 & 0 & 0 & 1 & 0 \\
 0 & 0 & 0 & 1 & 0 & -1 & 0 \\
 0 & 0 & 0 & 0 & 1 & 1 & 0 \\
 0 & 0 & 0 & 0 & 0 & 0 & 1 \\
\end{array}
\right)$ & $\left(
\begin{array}{c}
 0 \\
 0 \\
 0 \\
 0 \\
 0 \\
 0 \\
 \frac{1}{14} \\
\end{array}
\right)$ & $14$ & 4 & $(0,0,0,0,0,0,0)$ \\ & & & & \\ \hline & & & & \\ $\left(
\begin{array}{*7{C{1em}}}
 0 & -1 & 0 & 0 & 0 & 0 & 0 \\
 1 & 0 & 0 & 0 & 0 & 0 & 0 \\
 0 & 0 & 0 & 0 & 0 & -1 & 0 \\
 0 & 0 & 1 & 0 & 0 & 0 & 0 \\
 0 & 0 & 0 & 1 & 0 & 1 & 0 \\
 0 & 0 & 0 & 0 & 1 & 0 & 0 \\
 0 & 0 & 0 & 0 & 0 & 0 & 1 \\
\end{array}
\right)$ & $\left(
\begin{array}{c}
 0 \\
 0 \\
 0 \\
 0 \\
 0 \\
 0 \\
 \frac{1}{12} \\
\end{array}
\right)$ & $12$ & 4 & $\left(h_1,h_1,0,0,0,0,0\right)$ \\ & & & & \\ \hline & & & & \\ $\left(
\begin{array}{*7{C{1em}}}
 0 & -1 & 0 & 0 & 0 & 0 & 0 \\
 1 & -1 & 0 & 0 & 0 & 0 & 0 \\
 0 & 0 & 0 & 0 & 0 & -1 & 0 \\
 0 & 0 & 1 & 0 & 0 & 0 & 0 \\
 0 & 0 & 0 & 1 & 0 & 1 & 0 \\
 0 & 0 & 0 & 0 & 1 & 0 & 0 \\
 0 & 0 & 0 & 0 & 0 & 0 & 1 \\
\end{array}
\right)$ & $\left(
\begin{array}{c}
 0 \\
 0 \\
 0 \\
 0 \\
 0 \\
 0 \\
 \frac{1}{12} \\
\end{array}
\right)$ & $12$ & 4 & $(0,0,0,0,0,0,0)$ \\ & & & & \\ \hline & & & & \\ $\left(
\begin{array}{*7{C{1em}}}
 0 & -1 & 0 & 0 & 0 & 0 & 0 \\
 1 & 1 & 0 & 0 & 0 & 0 & 0 \\
 0 & 0 & 0 & 0 & 0 & -1 & 0 \\
 0 & 0 & 1 & 0 & 0 & 0 & 0 \\
 0 & 0 & 0 & 1 & 0 & 1 & 0 \\
 0 & 0 & 0 & 0 & 1 & 0 & 0 \\
 0 & 0 & 0 & 0 & 0 & 0 & 1 \\
\end{array}
\right)$ & $\left(
\begin{array}{c}
 0 \\
 0 \\
 0 \\
 0 \\
 0 \\
 0 \\
 \frac{1}{12} \\
\end{array}
\right)$ & $12$ & 4 & $(0,0,0,0,0,0,0)$ \\ & & & & \\ \hline & & & & \\ $\left(
\begin{array}{*7{C{1em}}}
 -1 & 0 & 0 & 0 & 0 & 0 & 0 \\
 0 & -1 & 0 & 0 & 0 & 0 & 0 \\
 0 & 0 & 0 & 0 & 0 & -1 & 0 \\
 0 & 0 & 1 & 0 & 0 & 0 & 0 \\
 0 & 0 & 0 & 1 & 0 & 1 & 0 \\
 0 & 0 & 0 & 0 & 1 & 0 & 0 \\
 0 & 0 & 0 & 0 & 0 & 0 & 1 \\
\end{array}
\right)$ & $\left(
\begin{array}{c}
 0 \\
 0 \\
 0 \\
 0 \\
 0 \\
 0 \\
 \frac{1}{12} \\
\end{array}
\right)$ & $12$ & 6 & $\left(h_1,h_2,0,0,0,0,0\right)$ \\ & & & & \\ \hline \end{tabular}

	\end{table}\begin{table}
    \centering
    \scriptsize
    \begin{tabular}{|C{12em}C{4em}C{2em}C{4em}C{11em}|}
        \hline
        $\D[g]$ & $\bvec[g]$ & $|\Gamma|$ & Moduli & \parbox[c][2.5em][c]{1em}{$\vec{h}$} \\  
        \hline & & & & \\ $\left(
\begin{array}{*7{C{1em}}}
 1 & 0 & 0 & 0 & 0 & 0 & 0 \\
 0 & -1 & 0 & 0 & 0 & 0 & 0 \\
 0 & 0 & 0 & 0 & 0 & -1 & 0 \\
 0 & 0 & 1 & 0 & 0 & 0 & 0 \\
 0 & 0 & 0 & 1 & 0 & 1 & 0 \\
 0 & 0 & 0 & 0 & 1 & 0 & 0 \\
 0 & 0 & 0 & 0 & 0 & 0 & 1 \\
\end{array}
\right)$ & $\left(
\begin{array}{c}
 0 \\
 0 \\
 0 \\
 0 \\
 0 \\
 0 \\
 \frac{1}{12} \\
\end{array}
\right)$ & $12$ & 6 & $\left(h_1,h_2,0,0,0,0,0\right)$ \\ & & & & \\ \hline & & & & \\ $\left(
\begin{array}{*7{C{1em}}}
 1 & 0 & 0 & 0 & 0 & 0 & 0 \\
 0 & 1 & 0 & 0 & 0 & 0 & 0 \\
 0 & 0 & 0 & 0 & 0 & -1 & 0 \\
 0 & 0 & 1 & 0 & 0 & 0 & 0 \\
 0 & 0 & 0 & 1 & 0 & 1 & 0 \\
 0 & 0 & 0 & 0 & 1 & 0 & 0 \\
 0 & 0 & 0 & 0 & 0 & 0 & 1 \\
\end{array}
\right)$ & $\left(
\begin{array}{c}
 0 \\
 0 \\
 0 \\
 0 \\
 0 \\
 0 \\
 \frac{1}{12} \\
\end{array}
\right)$ & $12$ & 8 & $\left(h_1,h_2,0,0,0,0,0\right)$ \\ & & & & \\ \hline & & & & \\ $\left(
\begin{array}{*7{C{1em}}}
 0 & -1 & 0 & 0 & 0 & 0 & 0 \\
 1 & 0 & 0 & 0 & 0 & 0 & 0 \\
 0 & 0 & 0 & -1 & 0 & 0 & 0 \\
 0 & 0 & 1 & 0 & 0 & 0 & 0 \\
 0 & 0 & 0 & 0 & 0 & -1 & 0 \\
 0 & 0 & 0 & 0 & 1 & 1 & 0 \\
 0 & 0 & 0 & 0 & 0 & 0 & 1 \\
\end{array}
\right)$ & $\left(
\begin{array}{c}
 0 \\
 0 \\
 0 \\
 0 \\
 0 \\
 0 \\
 \frac{1}{12} \\
\end{array}
\right)$ & $12$ & 6 & $\left(h_1,h_1,h_3,h_3,0,0,0\right)$ \\ & & & & \\ \hline & & & & \\ $\left(
\begin{array}{*7{C{1em}}}
 0 & -1 & 0 & 0 & 0 & 0 & 0 \\
 1 & 0 & 0 & 0 & 0 & 0 & 0 \\
 0 & 0 & 0 & -1 & 0 & 0 & 0 \\
 0 & 0 & 1 & 1 & 0 & 0 & 0 \\
 0 & 0 & 0 & 0 & 0 & -1 & 0 \\
 0 & 0 & 0 & 0 & 1 & 1 & 0 \\
 0 & 0 & 0 & 0 & 0 & 0 & 1 \\
\end{array}
\right)$ & $\left(
\begin{array}{c}
 0 \\
 0 \\
 0 \\
 0 \\
 0 \\
 0 \\
 \frac{1}{12} \\
\end{array}
\right)$ & $12$ & 6 & $\left(h_1,h_1,0,0,0,0,0\right)$ \\ & & & & \\ \hline & & & & \\ $\left(
\begin{array}{*7{C{1em}}}
 0 & -1 & 0 & 0 & 0 & 0 & 0 \\
 1 & -1 & 0 & 0 & 0 & 0 & 0 \\
 0 & 0 & 0 & -1 & 0 & 0 & 0 \\
 0 & 0 & 1 & 0 & 0 & 0 & 0 \\
 0 & 0 & 0 & 0 & 0 & -1 & 0 \\
 0 & 0 & 0 & 0 & 1 & 0 & 0 \\
 0 & 0 & 0 & 0 & 0 & 0 & 1 \\
\end{array}
\right)$ & $\left(
\begin{array}{c}
 0 \\
 0 \\
 0 \\
 0 \\
 0 \\
 0 \\
 \frac{1}{12} \\
\end{array}
\right)$ & $12$ & 6 & $\left(0,0,h_3,h_3,h_5,h_5,0\right)$ \\ & & & & \\ \hline & & & & \\ $\left(
\begin{array}{*7{C{1em}}}
 0 & -1 & 0 & 0 & 0 & 0 & 0 \\
 1 & -1 & 0 & 0 & 0 & 0 & 0 \\
 0 & 0 & 0 & -1 & 0 & 0 & 0 \\
 0 & 0 & 1 & 0 & 0 & 0 & 0 \\
 0 & 0 & 0 & 0 & 0 & -1 & 0 \\
 0 & 0 & 0 & 0 & 1 & 1 & 0 \\
 0 & 0 & 0 & 0 & 0 & 0 & 1 \\
\end{array}
\right)$ & $\left(
\begin{array}{c}
 0 \\
 0 \\
 0 \\
 0 \\
 0 \\
 0 \\
 \frac{1}{12} \\
\end{array}
\right)$ & $12$ & 4 & $\left(0,0,h_3,h_3,0,0,0\right)$ \\ & & & & \\ \hline & & & & \\ $\left(
\begin{array}{*7{C{1em}}}
 0 & -1 & 0 & 0 & 0 & 0 & 0 \\
 1 & -1 & 0 & 0 & 0 & 0 & 0 \\
 0 & 0 & 0 & -1 & 0 & 0 & 0 \\
 0 & 0 & 1 & -1 & 0 & 0 & 0 \\
 0 & 0 & 0 & 0 & 0 & -1 & 0 \\
 0 & 0 & 0 & 0 & 1 & 0 & 0 \\
 0 & 0 & 0 & 0 & 0 & 0 & 1 \\
\end{array}
\right)$ & $\left(
\begin{array}{c}
 0 \\
 0 \\
 0 \\
 0 \\
 0 \\
 0 \\
 \frac{1}{12} \\
\end{array}
\right)$ & $12$ & 6 & $\left(0,0,0,0,h_5,h_5,0\right)$ \\ & & & & \\ \hline \end{tabular}

	\end{table}\begin{table}
    \centering
    \scriptsize
    \begin{tabular}{|C{12em}C{4em}C{2em}C{4em}C{11em}|}
        \hline
        $\D[g]$ & $\bvec[g]$ & $|\Gamma|$ & Moduli & \parbox[c][2.5em][c]{1em}{$\vec{h}$} \\  
        \hline & & & & \\ $\left(
\begin{array}{*7{C{1em}}}
 -1 & 0 & 0 & 0 & 0 & 0 & 0 \\
 0 & -1 & 0 & 0 & 0 & 0 & 0 \\
 0 & 0 & 0 & -1 & 0 & 0 & 0 \\
 0 & 0 & 1 & 0 & 0 & 0 & 0 \\
 0 & 0 & 0 & 0 & 0 & -1 & 0 \\
 0 & 0 & 0 & 0 & 1 & 1 & 0 \\
 0 & 0 & 0 & 0 & 0 & 0 & 1 \\
\end{array}
\right)$ & $\left(
\begin{array}{c}
 0 \\
 0 \\
 0 \\
 0 \\
 0 \\
 0 \\
 \frac{1}{12} \\
\end{array}
\right)$ & $12$ & 6 & $\left(h_1,h_2,h_3,h_3,0,0,0\right)$ \\ & & & & \\ \hline & & & & \\ $\left(
\begin{array}{*7{C{1em}}}
 -1 & 0 & 0 & 0 & 0 & 0 & 0 \\
 0 & -1 & 0 & 0 & 0 & 0 & 0 \\
 0 & 0 & 0 & -1 & 0 & 0 & 0 \\
 0 & 0 & 1 & -1 & 0 & 0 & 0 \\
 0 & 0 & 0 & 0 & 0 & -1 & 0 \\
 0 & 0 & 0 & 0 & 1 & 0 & 0 \\
 0 & 0 & 0 & 0 & 0 & 0 & 1 \\
\end{array}
\right)$ & $\left(
\begin{array}{c}
 0 \\
 0 \\
 0 \\
 0 \\
 0 \\
 0 \\
 \frac{1}{12} \\
\end{array}
\right)$ & $12$ & 6 & $\left(h_1,h_2,0,0,h_5,h_5,0\right)$ \\ & & & & \\ \hline & & & & \\ $\left(
\begin{array}{*7{C{1em}}}
 1 & 0 & 0 & 0 & 0 & 0 & 0 \\
 0 & -1 & 0 & 0 & 0 & 0 & 0 \\
 0 & 0 & 0 & -1 & 0 & 0 & 0 \\
 0 & 0 & 1 & 0 & 0 & 0 & 0 \\
 0 & 0 & 0 & 0 & 0 & -1 & 0 \\
 0 & 0 & 0 & 0 & 1 & 1 & 0 \\
 0 & 0 & 0 & 0 & 0 & 0 & 1 \\
\end{array}
\right)$ & $\left(
\begin{array}{c}
 0 \\
 0 \\
 0 \\
 0 \\
 0 \\
 0 \\
 \frac{1}{12} \\
\end{array}
\right)$ & $12$ & 6 & $\left(h_1,h_2,h_3,h_3,0,0,0\right)$ \\ & & & & \\ \hline & & & & \\ $\left(
\begin{array}{*7{C{1em}}}
 1 & 0 & 0 & 0 & 0 & 0 & 0 \\
 0 & -1 & 0 & 0 & 0 & 0 & 0 \\
 0 & 0 & 0 & -1 & 0 & 0 & 0 \\
 0 & 0 & 1 & -1 & 0 & 0 & 0 \\
 0 & 0 & 0 & 0 & 0 & -1 & 0 \\
 0 & 0 & 0 & 0 & 1 & 0 & 0 \\
 0 & 0 & 0 & 0 & 0 & 0 & 1 \\
\end{array}
\right)$ & $\left(
\begin{array}{c}
 0 \\
 0 \\
 0 \\
 0 \\
 0 \\
 0 \\
 \frac{1}{12} \\
\end{array}
\right)$ & $12$ & 6 & $\left(h_1,h_2,0,0,h_5,h_5,0\right)$ \\ & & & & \\ \hline & & & & \\ $\left(
\begin{array}{*7{C{1em}}}
 1 & 0 & 0 & 0 & 0 & 0 & 0 \\
 0 & 1 & 0 & 0 & 0 & 0 & 0 \\
 0 & 0 & 0 & -1 & 0 & 0 & 0 \\
 0 & 0 & 1 & 0 & 0 & 0 & 0 \\
 0 & 0 & 0 & 0 & 0 & -1 & 0 \\
 0 & 0 & 0 & 0 & 1 & 1 & 0 \\
 0 & 0 & 0 & 0 & 0 & 0 & 1 \\
\end{array}
\right)$ & $\left(
\begin{array}{c}
 0 \\
 0 \\
 0 \\
 0 \\
 0 \\
 0 \\
 \frac{1}{12} \\
\end{array}
\right)$ & $12$ & 8 & $\left(h_1,h_2,h_3,h_3,0,0,0\right)$ \\ & & & & \\ \hline & & & & \\ $\left(
\begin{array}{*7{C{1em}}}
 1 & 0 & 0 & 0 & 0 & 0 & 0 \\
 0 & 1 & 0 & 0 & 0 & 0 & 0 \\
 0 & 0 & 0 & -1 & 0 & 0 & 0 \\
 0 & 0 & 1 & -1 & 0 & 0 & 0 \\
 0 & 0 & 0 & 0 & 0 & -1 & 0 \\
 0 & 0 & 0 & 0 & 1 & 0 & 0 \\
 0 & 0 & 0 & 0 & 0 & 0 & 1 \\
\end{array}
\right)$ & $\left(
\begin{array}{c}
 0 \\
 0 \\
 0 \\
 0 \\
 0 \\
 0 \\
 \frac{1}{12} \\
\end{array}
\right)$ & $12$ & 8 & $\left(h_1,h_2,0,0,h_5,h_5,0\right)$ \\ & & & & \\ \hline & & & & \\ $\left(
\begin{array}{*7{C{1em}}}
 -1 & 0 & 0 & 0 & 0 & 0 & 0 \\
 0 & -1 & 0 & 0 & 0 & 0 & 0 \\
 0 & 0 & 0 & 0 & 0 & -1 & 0 \\
 0 & 0 & 1 & 0 & 0 & -1 & 0 \\
 0 & 0 & 0 & 1 & 0 & -1 & 0 \\
 0 & 0 & 0 & 0 & 1 & -1 & 0 \\
 0 & 0 & 0 & 0 & 0 & 0 & 1 \\
\end{array}
\right)$ & $\left(
\begin{array}{c}
 0 \\
 0 \\
 0 \\
 0 \\
 0 \\
 0 \\
 \frac{1}{10} \\
\end{array}
\right)$ & $10$ & 6 & $\left(h_1,h_2,0,0,0,0,0\right)$ \\ & & & & \\ \hline \end{tabular}

	\end{table}\begin{table}
    \centering
    \scriptsize
    \begin{tabular}{|C{12em}C{4em}C{2em}C{4em}C{11em}|}
        \hline
        $\D[g]$ & $\bvec[g]$ & $|\Gamma|$ & Moduli & \parbox[c][2.5em][c]{1em}{$\vec{h}$} \\  
        \hline & & & & \\ $\left(
\begin{array}{*7{C{1em}}}
 -1 & 0 & 0 & 0 & 0 & 0 & 0 \\
 0 & -1 & 0 & 0 & 0 & 0 & 0 \\
 0 & 0 & 0 & 0 & 0 & -1 & 0 \\
 0 & 0 & 1 & 0 & 0 & 1 & 0 \\
 0 & 0 & 0 & 1 & 0 & -1 & 0 \\
 0 & 0 & 0 & 0 & 1 & 1 & 0 \\
 0 & 0 & 0 & 0 & 0 & 0 & 1 \\
\end{array}
\right)$ & $\left(
\begin{array}{c}
 0 \\
 0 \\
 0 \\
 0 \\
 0 \\
 0 \\
 \frac{1}{10} \\
\end{array}
\right)$ & $10$ & 6 & $\left(h_1,h_2,0,0,0,0,0\right)$ \\ & & & & \\ \hline & & & & \\ $\left(
\begin{array}{*7{C{1em}}}
 1 & 0 & 0 & 0 & 0 & 0 & 0 \\
 0 & -1 & 0 & 0 & 0 & 0 & 0 \\
 0 & 0 & 0 & 0 & 0 & -1 & 0 \\
 0 & 0 & 1 & 0 & 0 & -1 & 0 \\
 0 & 0 & 0 & 1 & 0 & -1 & 0 \\
 0 & 0 & 0 & 0 & 1 & -1 & 0 \\
 0 & 0 & 0 & 0 & 0 & 0 & 1 \\
\end{array}
\right)$ & $\left(
\begin{array}{c}
 0 \\
 0 \\
 0 \\
 0 \\
 0 \\
 0 \\
 \frac{1}{10} \\
\end{array}
\right)$ & $10$ & 6 & $\left(h_1,h_2,0,0,0,0,0\right)$ \\ & & & & \\ \hline & & & & \\ $\left(
\begin{array}{*7{C{1em}}}
 1 & 0 & 0 & 0 & 0 & 0 & 0 \\
 0 & -1 & 0 & 0 & 0 & 0 & 0 \\
 0 & 0 & 0 & 0 & 0 & -1 & 0 \\
 0 & 0 & 1 & 0 & 0 & 1 & 0 \\
 0 & 0 & 0 & 1 & 0 & -1 & 0 \\
 0 & 0 & 0 & 0 & 1 & 1 & 0 \\
 0 & 0 & 0 & 0 & 0 & 0 & 1 \\
\end{array}
\right)$ & $\left(
\begin{array}{c}
 0 \\
 0 \\
 0 \\
 0 \\
 0 \\
 0 \\
 \frac{1}{10} \\
\end{array}
\right)$ & $10$ & 6 & $\left(h_1,h_2,0,0,0,0,0\right)$ \\ & & & & \\ \hline & & & & \\ $\left(
\begin{array}{*7{C{1em}}}
 1 & 0 & 0 & 0 & 0 & 0 & 0 \\
 0 & 1 & 0 & 0 & 0 & 0 & 0 \\
 0 & 0 & 0 & 0 & 0 & -1 & 0 \\
 0 & 0 & 1 & 0 & 0 & 1 & 0 \\
 0 & 0 & 0 & 1 & 0 & -1 & 0 \\
 0 & 0 & 0 & 0 & 1 & 1 & 0 \\
 0 & 0 & 0 & 0 & 0 & 0 & 1 \\
\end{array}
\right)$ & $\left(
\begin{array}{c}
 0 \\
 0 \\
 0 \\
 0 \\
 0 \\
 0 \\
 \frac{1}{10} \\
\end{array}
\right)$ & $10$ & 8 & $\left(h_1,h_2,0,0,0,0,0\right)$ \\ & & & & \\ \hline & & & & \\ $\left(
\begin{array}{*7{C{1em}}}
 0 & 0 & 0 & 0 & 0 & -1 & 0 \\
 1 & 0 & 0 & 0 & 0 & 0 & 0 \\
 0 & 1 & 0 & 0 & 0 & 0 & 0 \\
 0 & 0 & 1 & 0 & 0 & -1 & 0 \\
 0 & 0 & 0 & 1 & 0 & 0 & 0 \\
 0 & 0 & 0 & 0 & 1 & 0 & 0 \\
 0 & 0 & 0 & 0 & 0 & 0 & 1 \\
\end{array}
\right)$ & $\left(
\begin{array}{c}
 0 \\
 0 \\
 0 \\
 0 \\
 0 \\
 0 \\
 \frac{1}{9} \\
\end{array}
\right)$ & $9$ & 4 & $\left(0,0,0,0,0,0,h_7\right)$ \\ & & & & \\ \hline & & & & \\ $\left(
\begin{array}{*7{C{1em}}}
 0 & -1 & 0 & 0 & 0 & 0 & 0 \\
 1 & 0 & 0 & 0 & 0 & 0 & 0 \\
 0 & 0 & 0 & 0 & 0 & -1 & 0 \\
 0 & 0 & 1 & 0 & 0 & 0 & 0 \\
 0 & 0 & 0 & 1 & 0 & 0 & 0 \\
 0 & 0 & 0 & 0 & 1 & 0 & 0 \\
 0 & 0 & 0 & 0 & 0 & 0 & 1 \\
\end{array}
\right)$ & $\left(
\begin{array}{c}
 0 \\
 0 \\
 0 \\
 0 \\
 0 \\
 0 \\
 \frac{1}{8} \\
\end{array}
\right)$ & $8$ & 4 & $\left(h_1,h_1,h_3,h_3,h_3,h_3,0\right)$ \\ & & & & \\ \hline & & & & \\ $\left(
\begin{array}{*7{C{1em}}}
 -1 & 0 & 0 & 0 & 0 & 0 & 0 \\
 0 & -1 & 0 & 0 & 0 & 0 & 0 \\
 0 & 0 & 0 & 0 & 0 & -1 & 0 \\
 0 & 0 & 1 & 0 & 0 & 0 & 0 \\
 0 & 0 & 0 & 1 & 0 & 0 & 0 \\
 0 & 0 & 0 & 0 & 1 & 0 & 0 \\
 0 & 0 & 0 & 0 & 0 & 0 & 1 \\
\end{array}
\right)$ & $\left(
\begin{array}{c}
 0 \\
 0 \\
 0 \\
 0 \\
 0 \\
 0 \\
 \frac{1}{8} \\
\end{array}
\right)$ & $8$ & 6 & $\left(h_1,h_2,h_3,h_3,h_3,h_3,0\right)$ \\ & & & & \\ \hline \end{tabular}

	\end{table}\begin{table}
    \centering
    \scriptsize
    \begin{tabular}{|C{12em}C{4em}C{2em}C{4em}C{11em}|}
        \hline
        $\D[g]$ & $\bvec[g]$ & $|\Gamma|$ & Moduli & \parbox[c][2.5em][c]{1em}{$\vec{h}$} \\  
        \hline & & & & \\ $\left(
\begin{array}{*7{C{1em}}}
 1 & 0 & 0 & 0 & 0 & 0 & 0 \\
 0 & -1 & 0 & 0 & 0 & 0 & 0 \\
 0 & 0 & 0 & 0 & 0 & -1 & 0 \\
 0 & 0 & 1 & 0 & 0 & 0 & 0 \\
 0 & 0 & 0 & 1 & 0 & 0 & 0 \\
 0 & 0 & 0 & 0 & 1 & 0 & 0 \\
 0 & 0 & 0 & 0 & 0 & 0 & 1 \\
\end{array}
\right)$ & $\left(
\begin{array}{c}
 0 \\
 0 \\
 0 \\
 0 \\
 0 \\
 0 \\
 \frac{1}{8} \\
\end{array}
\right)$ & $8$ & 6 & $\left(h_1,h_2,h_3,h_3,h_3,h_3,0\right)$ \\ & & & & \\ \hline & & & & \\ $\left(
\begin{array}{*7{C{1em}}}
 1 & 0 & 0 & 0 & 0 & 0 & 0 \\
 0 & 1 & 0 & 0 & 0 & 0 & 0 \\
 0 & 0 & 0 & 0 & 0 & -1 & 0 \\
 0 & 0 & 1 & 0 & 0 & 0 & 0 \\
 0 & 0 & 0 & 1 & 0 & 0 & 0 \\
 0 & 0 & 0 & 0 & 1 & 0 & 0 \\
 0 & 0 & 0 & 0 & 0 & 0 & 1 \\
\end{array}
\right)$ & $\left(
\begin{array}{c}
 0 \\
 0 \\
 0 \\
 0 \\
 0 \\
 0 \\
 \frac{1}{8} \\
\end{array}
\right)$ & $8$ & 8 & $\left(h_1,h_2,h_3,h_3,h_3,h_3,0\right)$ \\ & & & & \\ \hline & & & & \\ $\left(
\begin{array}{*7{C{1em}}}
 0 & 0 & 0 & 0 & 0 & -1 & 0 \\
 1 & 0 & 0 & 0 & 0 & -1 & 0 \\
 0 & 1 & 0 & 0 & 0 & -1 & 0 \\
 0 & 0 & 1 & 0 & 0 & -1 & 0 \\
 0 & 0 & 0 & 1 & 0 & -1 & 0 \\
 0 & 0 & 0 & 0 & 1 & -1 & 0 \\
 0 & 0 & 0 & 0 & 0 & 0 & 1 \\
\end{array}
\right)$ & $\left(
\begin{array}{c}
 0 \\
 0 \\
 0 \\
 0 \\
 0 \\
 0 \\
 \frac{1}{7} \\
\end{array}
\right)$ & $7$ & 4 & $\left(0,0,0,0,0,0,h_7\right)$ \\ & & & & \\ \hline & & & & \\ $\left(
\begin{array}{*7{C{1em}}}
 0 & -1 & 0 & 0 & 0 & 0 & 0 \\
 1 & -1 & 0 & 0 & 0 & 0 & 0 \\
 0 & 0 & 0 & -1 & 0 & 0 & 0 \\
 0 & 0 & 1 & -1 & 0 & 0 & 0 \\
 0 & 0 & 0 & 0 & 0 & -1 & 0 \\
 0 & 0 & 0 & 0 & 1 & 1 & 0 \\
 0 & 0 & 0 & 0 & 0 & 0 & 1 \\
\end{array}
\right)$ & $\left(
\begin{array}{c}
 0 \\
 0 \\
 0 \\
 0 \\
 0 \\
 0 \\
 \frac{1}{6} \\
\end{array}
\right)$ & $6$ & 6 & $(0,0,0,0,0,0,0)$ \\ & & & & \\ \hline & & & & \\ $\left(
\begin{array}{*7{C{1em}}}
 0 & -1 & 0 & 0 & 0 & 0 & 0 \\
 1 & -1 & 0 & 0 & 0 & 0 & 0 \\
 0 & 0 & 0 & -1 & 0 & 0 & 0 \\
 0 & 0 & 1 & 1 & 0 & 0 & 0 \\
 0 & 0 & 0 & 0 & 0 & -1 & 0 \\
 0 & 0 & 0 & 0 & 1 & 1 & 0 \\
 0 & 0 & 0 & 0 & 0 & 0 & 1 \\
\end{array}
\right)$ & $\left(
\begin{array}{c}
 0 \\
 0 \\
 0 \\
 0 \\
 0 \\
 0 \\
 \frac{1}{6} \\
\end{array}
\right)$ & $6$ & 6 & $(0,0,0,0,0,0,0)$ \\ & & & & \\ \hline & & & & \\ $\left(
\begin{array}{*7{C{1em}}}
 0 & -1 & 0 & 0 & 0 & 0 & 0 \\
 1 & 1 & 0 & 0 & 0 & 0 & 0 \\
 0 & 0 & 0 & -1 & 0 & 0 & 0 \\
 0 & 0 & 1 & 1 & 0 & 0 & 0 \\
 0 & 0 & 0 & 0 & 0 & -1 & 0 \\
 0 & 0 & 0 & 0 & 1 & 1 & 0 \\
 0 & 0 & 0 & 0 & 0 & 0 & 1 \\
\end{array}
\right)$ & $\left(
\begin{array}{c}
 0 \\
 0 \\
 0 \\
 0 \\
 0 \\
 0 \\
 \frac{1}{6} \\
\end{array}
\right)$ & $6$ & 10 & $(0,0,0,0,0,0,0)$ \\ & & & & \\ \hline & & & & \\ $\left(
\begin{array}{*7{C{1em}}}
 -1 & 0 & 0 & 0 & 0 & 0 & 0 \\
 0 & -1 & 0 & 0 & 0 & 0 & 0 \\
 0 & 0 & 0 & -1 & 0 & 0 & 0 \\
 0 & 0 & 1 & -1 & 0 & 0 & 0 \\
 0 & 0 & 0 & 0 & 0 & -1 & 0 \\
 0 & 0 & 0 & 0 & 1 & -1 & 0 \\
 0 & 0 & 0 & 0 & 0 & 0 & 1 \\
\end{array}
\right)$ & $\left(
\begin{array}{c}
 0 \\
 0 \\
 0 \\
 0 \\
 0 \\
 0 \\
 \frac{1}{6} \\
\end{array}
\right)$ & $6$ & 8 & $\left(h_1,h_2,0,0,0,0,0\right)$ \\ & & & & \\ \hline \end{tabular}

	\end{table}\begin{table}
    \centering
    \scriptsize
    \begin{tabular}{|C{12em}C{4em}C{2em}C{4em}C{11em}|}
        \hline
        $\D[g]$ & $\bvec[g]$ & $|\Gamma|$ & Moduli & \parbox[c][2.5em][c]{1em}{$\vec{h}$} \\  
        \hline & & & & \\ $\left(
\begin{array}{*7{C{1em}}}
 -1 & 0 & 0 & 0 & 0 & 0 & 0 \\
 0 & -1 & 0 & 0 & 0 & 0 & 0 \\
 0 & 0 & 0 & -1 & 0 & 0 & 0 \\
 0 & 0 & 1 & -1 & 0 & 0 & 0 \\
 0 & 0 & 0 & 0 & 0 & -1 & 0 \\
 0 & 0 & 0 & 0 & 1 & 1 & 0 \\
 0 & 0 & 0 & 0 & 0 & 0 & 1 \\
\end{array}
\right)$ & $\left(
\begin{array}{c}
 0 \\
 0 \\
 0 \\
 0 \\
 0 \\
 0 \\
 \frac{1}{6} \\
\end{array}
\right)$ & $6$ & 6 & $\left(h_1,h_2,0,0,0,0,0\right)$ \\ & & & & \\ \hline & & & & \\ $\left(
\begin{array}{*7{C{1em}}}
 -1 & 0 & 0 & 0 & 0 & 0 & 0 \\
 0 & -1 & 0 & 0 & 0 & 0 & 0 \\
 0 & 0 & 0 & -1 & 0 & 0 & 0 \\
 0 & 0 & 1 & 1 & 0 & 0 & 0 \\
 0 & 0 & 0 & 0 & 0 & -1 & 0 \\
 0 & 0 & 0 & 0 & 1 & 1 & 0 \\
 0 & 0 & 0 & 0 & 0 & 0 & 1 \\
\end{array}
\right)$ & $\left(
\begin{array}{c}
 0 \\
 0 \\
 0 \\
 0 \\
 0 \\
 0 \\
 \frac{1}{6} \\
\end{array}
\right)$ & $6$ & 8 & $\left(h_1,h_2,0,0,0,0,0\right)$ \\ & & & & \\ \hline & & & & \\ $\left(
\begin{array}{*7{C{1em}}}
 1 & 0 & 0 & 0 & 0 & 0 & 0 \\
 0 & -1 & 0 & 0 & 0 & 0 & 0 \\
 0 & 0 & 0 & -1 & 0 & 0 & 0 \\
 0 & 0 & 1 & -1 & 0 & 0 & 0 \\
 0 & 0 & 0 & 0 & 0 & -1 & 0 \\
 0 & 0 & 0 & 0 & 1 & -1 & 0 \\
 0 & 0 & 0 & 0 & 0 & 0 & 1 \\
\end{array}
\right)$ & $\left(
\begin{array}{c}
 0 \\
 0 \\
 0 \\
 0 \\
 0 \\
 0 \\
 \frac{1}{6} \\
\end{array}
\right)$ & $6$ & 8 & $\left(h_1,h_2,0,0,0,0,0\right)$ \\ & & & & \\ \hline & & & & \\ $\left(
\begin{array}{*7{C{1em}}}
 1 & 0 & 0 & 0 & 0 & 0 & 0 \\
 0 & -1 & 0 & 0 & 0 & 0 & 0 \\
 0 & 0 & 0 & -1 & 0 & 0 & 0 \\
 0 & 0 & 1 & -1 & 0 & 0 & 0 \\
 0 & 0 & 0 & 0 & 0 & -1 & 0 \\
 0 & 0 & 0 & 0 & 1 & 1 & 0 \\
 0 & 0 & 0 & 0 & 0 & 0 & 1 \\
\end{array}
\right)$ & $\left(
\begin{array}{c}
 0 \\
 0 \\
 0 \\
 0 \\
 0 \\
 0 \\
 \frac{1}{6} \\
\end{array}
\right)$ & $6$ & 6 & $\left(h_1,h_2,0,0,0,0,0\right)$ \\ & & & & \\ \hline & & & & \\ $\left(
\begin{array}{*7{C{1em}}}
 1 & 0 & 0 & 0 & 0 & 0 & 0 \\
 0 & -1 & 0 & 0 & 0 & 0 & 0 \\
 0 & 0 & 0 & -1 & 0 & 0 & 0 \\
 0 & 0 & 1 & 1 & 0 & 0 & 0 \\
 0 & 0 & 0 & 0 & 0 & -1 & 0 \\
 0 & 0 & 0 & 0 & 1 & 1 & 0 \\
 0 & 0 & 0 & 0 & 0 & 0 & 1 \\
\end{array}
\right)$ & $\left(
\begin{array}{c}
 0 \\
 0 \\
 0 \\
 0 \\
 0 \\
 0 \\
 \frac{1}{6} \\
\end{array}
\right)$ & $6$ & 8 & $\left(h_1,h_2,0,0,0,0,0\right)$ \\ & & & & \\ \hline & & & & \\ $\left(
\begin{array}{*7{C{1em}}}
 1 & 0 & 0 & 0 & 0 & 0 & 0 \\
 0 & 1 & 0 & 0 & 0 & 0 & 0 \\
 0 & 0 & 0 & -1 & 0 & 0 & 0 \\
 0 & 0 & 1 & -1 & 0 & 0 & 0 \\
 0 & 0 & 0 & 0 & 0 & -1 & 0 \\
 0 & 0 & 0 & 0 & 1 & 1 & 0 \\
 0 & 0 & 0 & 0 & 0 & 0 & 1 \\
\end{array}
\right)$ & $\left(
\begin{array}{c}
 0 \\
 0 \\
 0 \\
 0 \\
 0 \\
 0 \\
 \frac{1}{6} \\
\end{array}
\right)$ & $6$ & 8 & $\left(h_1,h_2,0,0,0,0,0\right)$ \\ & & & & \\ \hline & & & & \\ $\left(
\begin{array}{*7{C{1em}}}
 1 & 0 & 0 & 0 & 0 & 0 & 0 \\
 0 & 1 & 0 & 0 & 0 & 0 & 0 \\
 0 & 0 & 0 & -1 & 0 & 0 & 0 \\
 0 & 0 & 1 & 1 & 0 & 0 & 0 \\
 0 & 0 & 0 & 0 & 0 & -1 & 0 \\
 0 & 0 & 0 & 0 & 1 & 1 & 0 \\
 0 & 0 & 0 & 0 & 0 & 0 & 1 \\
\end{array}
\right)$ & $\left(
\begin{array}{c}
 0 \\
 0 \\
 0 \\
 0 \\
 0 \\
 0 \\
 \frac{1}{6} \\
\end{array}
\right)$ & $6$ & 10 & $\left(h_1,h_2,0,0,0,0,0\right)$ \\ & & & & \\ \hline \end{tabular}

	\end{table}\begin{table}
    \centering
    \scriptsize
    \begin{tabular}{|C{12em}C{4em}C{2em}C{4em}C{11em}|}
        \hline
        $\D[g]$ & $\bvec[g]$ & $|\Gamma|$ & Moduli & \parbox[c][2.5em][c]{1em}{$\vec{h}$} \\  
        \hline & & & & \\ $\left(
\begin{array}{*7{C{1em}}}
 -1 & 0 & 0 & 0 & 0 & 0 & 0 \\
 0 & -1 & 0 & 0 & 0 & 0 & 0 \\
 0 & 0 & -1 & 0 & 0 & 0 & 0 \\
 0 & 0 & 0 & -1 & 0 & 0 & 0 \\
 0 & 0 & 0 & 0 & 0 & -1 & 0 \\
 0 & 0 & 0 & 0 & 1 & -1 & 0 \\
 0 & 0 & 0 & 0 & 0 & 0 & 1 \\
\end{array}
\right)$ & $\left(
\begin{array}{c}
 0 \\
 0 \\
 0 \\
 0 \\
 0 \\
 0 \\
 \frac{1}{6} \\
\end{array}
\right)$ & $6$ & 12 & $\left(h_1,h_2,h_3,h_4,0,0,0\right)$ \\ & & & & \\ \hline & & & & \\ $\left(
\begin{array}{*7{C{1em}}}
 -1 & 0 & 0 & 0 & 0 & 0 & 0 \\
 0 & -1 & 0 & 0 & 0 & 0 & 0 \\
 0 & 0 & -1 & 0 & 0 & 0 & 0 \\
 0 & 0 & 0 & -1 & 0 & 0 & 0 \\
 0 & 0 & 0 & 0 & 0 & -1 & 0 \\
 0 & 0 & 0 & 0 & 1 & 1 & 0 \\
 0 & 0 & 0 & 0 & 0 & 0 & 1 \\
\end{array}
\right)$ & $\left(
\begin{array}{c}
 0 \\
 0 \\
 0 \\
 0 \\
 0 \\
 0 \\
 \frac{1}{6} \\
\end{array}
\right)$ & $6$ & 12 & $\left(h_1,h_2,h_3,h_4,0,0,0\right)$ \\ & & & & \\ \hline & & & & \\ $\left(
\begin{array}{*7{C{1em}}}
 1 & 0 & 0 & 0 & 0 & 0 & 0 \\
 0 & -1 & 0 & 0 & 0 & 0 & 0 \\
 0 & 0 & -1 & 0 & 0 & 0 & 0 \\
 0 & 0 & 0 & -1 & 0 & 0 & 0 \\
 0 & 0 & 0 & 0 & 0 & -1 & 0 \\
 0 & 0 & 0 & 0 & 1 & -1 & 0 \\
 0 & 0 & 0 & 0 & 0 & 0 & 1 \\
\end{array}
\right)$ & $\left(
\begin{array}{c}
 0 \\
 0 \\
 0 \\
 0 \\
 0 \\
 0 \\
 \frac{1}{6} \\
\end{array}
\right)$ & $6$ & 10 & $\left(h_1,h_2,h_3,h_4,0,0,0\right)$ \\ & & & & \\ \hline & & & & \\ $\left(
\begin{array}{*7{C{1em}}}
 1 & 0 & 0 & 0 & 0 & 0 & 0 \\
 0 & -1 & 0 & 0 & 0 & 0 & 0 \\
 0 & 0 & -1 & 0 & 0 & 0 & 0 \\
 0 & 0 & 0 & -1 & 0 & 0 & 0 \\
 0 & 0 & 0 & 0 & 0 & -1 & 0 \\
 0 & 0 & 0 & 0 & 1 & 1 & 0 \\
 0 & 0 & 0 & 0 & 0 & 0 & 1 \\
\end{array}
\right)$ & $\left(
\begin{array}{c}
 0 \\
 0 \\
 0 \\
 0 \\
 0 \\
 0 \\
 \frac{1}{6} \\
\end{array}
\right)$ & $6$ & 10 & $\left(h_1,h_2,h_3,h_4,0,0,0\right)$ \\ & & & & \\ \hline & & & & \\ $\left(
\begin{array}{*7{C{1em}}}
 1 & 0 & 0 & 0 & 0 & 0 & 0 \\
 0 & 1 & 0 & 0 & 0 & 0 & 0 \\
 0 & 0 & -1 & 0 & 0 & 0 & 0 \\
 0 & 0 & 0 & -1 & 0 & 0 & 0 \\
 0 & 0 & 0 & 0 & 0 & -1 & 0 \\
 0 & 0 & 0 & 0 & 1 & -1 & 0 \\
 0 & 0 & 0 & 0 & 0 & 0 & 1 \\
\end{array}
\right)$ & $\left(
\begin{array}{c}
 0 \\
 0 \\
 0 \\
 0 \\
 0 \\
 0 \\
 \frac{1}{6} \\
\end{array}
\right)$ & $6$ & 10 & $\left(h_1,h_2,h_3,h_4,0,0,0\right)$ \\ & & & & \\ \hline & & & & \\ $\left(
\begin{array}{*7{C{1em}}}
 1 & 0 & 0 & 0 & 0 & 0 & 0 \\
 0 & 1 & 0 & 0 & 0 & 0 & 0 \\
 0 & 0 & -1 & 0 & 0 & 0 & 0 \\
 0 & 0 & 0 & -1 & 0 & 0 & 0 \\
 0 & 0 & 0 & 0 & 0 & -1 & 0 \\
 0 & 0 & 0 & 0 & 1 & 1 & 0 \\
 0 & 0 & 0 & 0 & 0 & 0 & 1 \\
\end{array}
\right)$ & $\left(
\begin{array}{c}
 0 \\
 0 \\
 0 \\
 0 \\
 0 \\
 0 \\
 \frac{1}{6} \\
\end{array}
\right)$ & $6$ & 10 & $\left(h_1,h_2,h_3,h_4,0,0,0\right)$ \\ & & & & \\ \hline & & & & \\ $\left(
\begin{array}{*7{C{1em}}}
 1 & 0 & 0 & 0 & 0 & 0 & 0 \\
 0 & 1 & 0 & 0 & 0 & 0 & 0 \\
 0 & 0 & 1 & 0 & 0 & 0 & 0 \\
 0 & 0 & 0 & -1 & 0 & 0 & 0 \\
 0 & 0 & 0 & 0 & 0 & -1 & 0 \\
 0 & 0 & 0 & 0 & 1 & -1 & 0 \\
 0 & 0 & 0 & 0 & 0 & 0 & 1 \\
\end{array}
\right)$ & $\left(
\begin{array}{c}
 0 \\
 0 \\
 0 \\
 0 \\
 0 \\
 0 \\
 \frac{1}{6} \\
\end{array}
\right)$ & $6$ & 12 & $\left(h_1,h_2,h_3,h_4,0,0,0\right)$ \\ & & & & \\ \hline \end{tabular}

	\end{table}\begin{table}
    \centering
    \scriptsize
    \begin{tabular}{|C{12em}C{4em}C{2em}C{4em}C{11em}|}
        \hline
        $\D[g]$ & $\bvec[g]$ & $|\Gamma|$ & Moduli & \parbox[c][2.5em][c]{1em}{$\vec{h}$} \\  
        \hline & & & & \\ $\left(
\begin{array}{*7{C{1em}}}
 1 & 0 & 0 & 0 & 0 & 0 & 0 \\
 0 & 1 & 0 & 0 & 0 & 0 & 0 \\
 0 & 0 & 1 & 0 & 0 & 0 & 0 \\
 0 & 0 & 0 & -1 & 0 & 0 & 0 \\
 0 & 0 & 0 & 0 & 0 & -1 & 0 \\
 0 & 0 & 0 & 0 & 1 & 1 & 0 \\
 0 & 0 & 0 & 0 & 0 & 0 & 1 \\
\end{array}
\right)$ & $\left(
\begin{array}{c}
 0 \\
 0 \\
 0 \\
 0 \\
 0 \\
 0 \\
 \frac{1}{6} \\
\end{array}
\right)$ & $6$ & 12 & $\left(h_1,h_2,h_3,h_4,0,0,0\right)$ \\ & & & & \\ \hline & & & & \\ $\left(
\begin{array}{*7{C{1em}}}
 1 & 0 & 0 & 0 & 0 & 0 & 0 \\
 0 & 1 & 0 & 0 & 0 & 0 & 0 \\
 0 & 0 & 1 & 0 & 0 & 0 & 0 \\
 0 & 0 & 0 & 1 & 0 & 0 & 0 \\
 0 & 0 & 0 & 0 & 0 & -1 & 0 \\
 0 & 0 & 0 & 0 & 1 & 1 & 0 \\
 0 & 0 & 0 & 0 & 0 & 0 & 1 \\
\end{array}
\right)$ & $\left(
\begin{array}{c}
 0 \\
 0 \\
 0 \\
 0 \\
 0 \\
 0 \\
 \frac{1}{6} \\
\end{array}
\right)$ & $6$ & 16 & $\left(h_1,h_2,h_3,h_4,0,0,0\right)$ \\ & & & & \\ \hline & & & & \\ $\left(
\begin{array}{*7{C{1em}}}
 1 & 0 & 0 & 0 & 0 & 0 & 0 \\
 0 & 1 & 0 & 0 & 0 & 0 & 0 \\
 0 & 0 & 0 & 0 & 0 & -1 & 0 \\
 0 & 0 & 1 & 0 & 0 & -1 & 0 \\
 0 & 0 & 0 & 1 & 0 & -1 & 0 \\
 0 & 0 & 0 & 0 & 1 & -1 & 0 \\
 0 & 0 & 0 & 0 & 0 & 0 & 1 \\
\end{array}
\right)$ & $\left(
\begin{array}{c}
 0 \\
 0 \\
 0 \\
 0 \\
 0 \\
 0 \\
 \frac{1}{5} \\
\end{array}
\right)$ & $5$ & 8 & $\left(h_1,h_2,0,0,0,0,h_7\right)$ \\ & & & & \\ \hline & & & & \\ $\left(
\begin{array}{*7{C{1em}}}
 0 & -1 & 0 & 0 & 0 & 0 & 0 \\
 1 & 0 & 0 & 0 & 0 & 0 & 0 \\
 0 & 0 & 0 & -1 & 0 & 0 & 0 \\
 0 & 0 & 1 & 0 & 0 & 0 & 0 \\
 0 & 0 & 0 & 0 & 0 & -1 & 0 \\
 0 & 0 & 0 & 0 & 1 & 0 & 0 \\
 0 & 0 & 0 & 0 & 0 & 0 & 1 \\
\end{array}
\right)$ & $\left(
\begin{array}{c}
 0 \\
 0 \\
 0 \\
 0 \\
 0 \\
 0 \\
 \frac{1}{4} \\
\end{array}
\right)$ & $4$ & 10 & $\left(h_1,h_1,h_3,h_3,h_5,h_5,0\right)$ \\ & & & & \\ \hline & & & & \\ $\left(
\begin{array}{*7{C{1em}}}
 -1 & 0 & 0 & 0 & 0 & 0 & 0 \\
 0 & -1 & 0 & 0 & 0 & 0 & 0 \\
 0 & 0 & 0 & -1 & 0 & 0 & 0 \\
 0 & 0 & 1 & 0 & 0 & 0 & 0 \\
 0 & 0 & 0 & 0 & 0 & -1 & 0 \\
 0 & 0 & 0 & 0 & 1 & 0 & 0 \\
 0 & 0 & 0 & 0 & 0 & 0 & 1 \\
\end{array}
\right)$ & $\left(
\begin{array}{c}
 0 \\
 0 \\
 0 \\
 0 \\
 0 \\
 0 \\
 \frac{1}{4} \\
\end{array}
\right)$ & $4$ & 8 & $\left(h_1,h_2,h_3,h_3,h_5,h_5,0\right)$ \\ & & & & \\ \hline & & & & \\ $\left(
\begin{array}{*7{C{1em}}}
 1 & 0 & 0 & 0 & 0 & 0 & 0 \\
 0 & -1 & 0 & 0 & 0 & 0 & 0 \\
 0 & 0 & 0 & -1 & 0 & 0 & 0 \\
 0 & 0 & 1 & 0 & 0 & 0 & 0 \\
 0 & 0 & 0 & 0 & 0 & -1 & 0 \\
 0 & 0 & 0 & 0 & 1 & 0 & 0 \\
 0 & 0 & 0 & 0 & 0 & 0 & 1 \\
\end{array}
\right)$ & $\left(
\begin{array}{c}
 0 \\
 0 \\
 0 \\
 0 \\
 0 \\
 0 \\
 \frac{1}{4} \\
\end{array}
\right)$ & $4$ & 8 & $\left(h_1,h_2,h_3,h_3,h_5,h_5,0\right)$ \\ & & & & \\ \hline & & & & \\ $\left(
\begin{array}{*7{C{1em}}}
 1 & 0 & 0 & 0 & 0 & 0 & 0 \\
 0 & 1 & 0 & 0 & 0 & 0 & 0 \\
 0 & 0 & 0 & -1 & 0 & 0 & 0 \\
 0 & 0 & 1 & 0 & 0 & 0 & 0 \\
 0 & 0 & 0 & 0 & 0 & -1 & 0 \\
 0 & 0 & 0 & 0 & 1 & 0 & 0 \\
 0 & 0 & 0 & 0 & 0 & 0 & 1 \\
\end{array}
\right)$ & $\left(
\begin{array}{c}
 0 \\
 0 \\
 0 \\
 0 \\
 0 \\
 0 \\
 \frac{1}{4} \\
\end{array}
\right)$ & $4$ & 10 & $\left(h_1,h_2,h_3,h_3,h_5,h_5,0\right)$ \\ & & & & \\ \hline \end{tabular}

	\end{table}\begin{table}
    \centering
    \scriptsize
    \begin{tabular}{|C{12em}C{4em}C{2em}C{4em}C{11em}|}
        \hline
        $\D[g]$ & $\bvec[g]$ & $|\Gamma|$ & Moduli & \parbox[c][2.5em][c]{1em}{$\vec{h}$} \\  
        \hline & & & & \\ $\left(
\begin{array}{*7{C{1em}}}
 -1 & 0 & 0 & 0 & 0 & 0 & 0 \\
 0 & -1 & 0 & 0 & 0 & 0 & 0 \\
 0 & 0 & -1 & 0 & 0 & 0 & 0 \\
 0 & 0 & 0 & -1 & 0 & 0 & 0 \\
 0 & 0 & 0 & 0 & 0 & -1 & 0 \\
 0 & 0 & 0 & 0 & 1 & 0 & 0 \\
 0 & 0 & 0 & 0 & 0 & 0 & 1 \\
\end{array}
\right)$ & $\left(
\begin{array}{c}
 0 \\
 0 \\
 0 \\
 0 \\
 0 \\
 0 \\
 \frac{1}{4} \\
\end{array}
\right)$ & $4$ & 12 & $\left(h_1,h_2,h_3,h_4,h_5,h_5,0\right)$ \\ & & & & \\ \hline & & & & \\ $\left(
\begin{array}{*7{C{1em}}}
 1 & 0 & 0 & 0 & 0 & 0 & 0 \\
 0 & -1 & 0 & 0 & 0 & 0 & 0 \\
 0 & 0 & -1 & 0 & 0 & 0 & 0 \\
 0 & 0 & 0 & -1 & 0 & 0 & 0 \\
 0 & 0 & 0 & 0 & 0 & -1 & 0 \\
 0 & 0 & 0 & 0 & 1 & 0 & 0 \\
 0 & 0 & 0 & 0 & 0 & 0 & 1 \\
\end{array}
\right)$ & $\left(
\begin{array}{c}
 0 \\
 0 \\
 0 \\
 0 \\
 0 \\
 0 \\
 \frac{1}{4} \\
\end{array}
\right)$ & $4$ & 10 & $\left(h_1,h_2,h_3,h_4,h_5,h_5,0\right)$ \\ & & & & \\ \hline & & & & \\ $\left(
\begin{array}{*7{C{1em}}}
 1 & 0 & 0 & 0 & 0 & 0 & 0 \\
 0 & 1 & 0 & 0 & 0 & 0 & 0 \\
 0 & 0 & -1 & 0 & 0 & 0 & 0 \\
 0 & 0 & 0 & -1 & 0 & 0 & 0 \\
 0 & 0 & 0 & 0 & 0 & -1 & 0 \\
 0 & 0 & 0 & 0 & 1 & 0 & 0 \\
 0 & 0 & 0 & 0 & 0 & 0 & 1 \\
\end{array}
\right)$ & $\left(
\begin{array}{c}
 0 \\
 0 \\
 0 \\
 0 \\
 0 \\
 0 \\
 \frac{1}{4} \\
\end{array}
\right)$ & $4$ & 10 & $\left(h_1,h_2,h_3,h_4,h_5,h_5,0\right)$ \\ & & & & \\ \hline & & & & \\ $\left(
\begin{array}{*7{C{1em}}}
 1 & 0 & 0 & 0 & 0 & 0 & 0 \\
 0 & 1 & 0 & 0 & 0 & 0 & 0 \\
 0 & 0 & 1 & 0 & 0 & 0 & 0 \\
 0 & 0 & 0 & -1 & 0 & 0 & 0 \\
 0 & 0 & 0 & 0 & 0 & -1 & 0 \\
 0 & 0 & 0 & 0 & 1 & 0 & 0 \\
 0 & 0 & 0 & 0 & 0 & 0 & 1 \\
\end{array}
\right)$ & $\left(
\begin{array}{c}
 0 \\
 0 \\
 0 \\
 0 \\
 0 \\
 0 \\
 \frac{1}{4} \\
\end{array}
\right)$ & $4$ & 12 & $\left(h_1,h_2,h_3,h_4,h_5,h_5,0\right)$ \\ & & & & \\ \hline & & & & \\ $\left(
\begin{array}{*7{C{1em}}}
 1 & 0 & 0 & 0 & 0 & 0 & 0 \\
 0 & 1 & 0 & 0 & 0 & 0 & 0 \\
 0 & 0 & 1 & 0 & 0 & 0 & 0 \\
 0 & 0 & 0 & 1 & 0 & 0 & 0 \\
 0 & 0 & 0 & 0 & 0 & -1 & 0 \\
 0 & 0 & 0 & 0 & 1 & 0 & 0 \\
 0 & 0 & 0 & 0 & 0 & 0 & 1 \\
\end{array}
\right)$ & $\left(
\begin{array}{c}
 0 \\
 0 \\
 0 \\
 0 \\
 0 \\
 0 \\
 \frac{1}{4} \\
\end{array}
\right)$ & $4$ & 16 & $\left(h_1,h_2,h_3,h_4,h_5,h_5,0\right)$ \\ & & & & \\ \hline & & & & \\ $\left(
\begin{array}{*7{C{1em}}}
 0 & -1 & 0 & 0 & 0 & 0 & 0 \\
 1 & -1 & 0 & 0 & 0 & 0 & 0 \\
 0 & 0 & 0 & -1 & 0 & 0 & 0 \\
 0 & 0 & 1 & -1 & 0 & 0 & 0 \\
 0 & 0 & 0 & 0 & 0 & -1 & 0 \\
 0 & 0 & 0 & 0 & 1 & -1 & 0 \\
 0 & 0 & 0 & 0 & 0 & 0 & 1 \\
\end{array}
\right)$ & $\left(
\begin{array}{c}
 0 \\
 0 \\
 0 \\
 0 \\
 0 \\
 0 \\
 \frac{1}{3} \\
\end{array}
\right)$ & $3$ & 10 & $\left(0,0,0,0,0,0,h_7\right)$ \\ & & & & \\ \hline & & & & \\ $\left(
\begin{array}{*7{C{1em}}}
 1 & 0 & 0 & 0 & 0 & 0 & 0 \\
 0 & 1 & 0 & 0 & 0 & 0 & 0 \\
 0 & 0 & 0 & -1 & 0 & 0 & 0 \\
 0 & 0 & 1 & -1 & 0 & 0 & 0 \\
 0 & 0 & 0 & 0 & 0 & -1 & 0 \\
 0 & 0 & 0 & 0 & 1 & -1 & 0 \\
 0 & 0 & 0 & 0 & 0 & 0 & 1 \\
\end{array}
\right)$ & $\left(
\begin{array}{c}
 0 \\
 0 \\
 0 \\
 0 \\
 0 \\
 0 \\
 \frac{1}{3} \\
\end{array}
\right)$ & $3$ & 10 & $\left(h_1,h_2,0,0,0,0,h_7\right)$ \\ & & & & \\ \hline \end{tabular}

	\end{table}\begin{table}
    \centering
    \scriptsize
    \begin{tabular}{|C{12em}C{4em}C{2em}C{4em}C{11em}|}
        \hline
        $\D[g]$ & $\bvec[g]$ & $|\Gamma|$ & Moduli & \parbox[c][2.5em][c]{1em}{$\vec{h}$} \\  
        \hline & & & & \\ $\left(
\begin{array}{*7{C{1em}}}
 1 & 0 & 0 & 0 & 0 & 0 & 0 \\
 0 & 1 & 0 & 0 & 0 & 0 & 0 \\
 0 & 0 & 1 & 0 & 0 & 0 & 0 \\
 0 & 0 & 0 & 1 & 0 & 0 & 0 \\
 0 & 0 & 0 & 0 & 0 & -1 & 0 \\
 0 & 0 & 0 & 0 & 1 & -1 & 0 \\
 0 & 0 & 0 & 0 & 0 & 0 & 1 \\
\end{array}
\right)$ & $\left(
\begin{array}{c}
 0 \\
 0 \\
 0 \\
 0 \\
 0 \\
 0 \\
 \frac{1}{3} \\
\end{array}
\right)$ & $3$ & 16 & $\left(h_1,h_2,h_3,h_4,0,0,h_7\right)$ \\ & & & & \\ \hline & & & & \\ $\left(
\begin{array}{*7{C{1em}}}
 -1 & 0 & 0 & 0 & 0 & 0 & 0 \\
 0 & -1 & 0 & 0 & 0 & 0 & 0 \\
 0 & 0 & -1 & 0 & 0 & 0 & 0 \\
 0 & 0 & 0 & -1 & 0 & 0 & 0 \\
 0 & 0 & 0 & 0 & -1 & 0 & 0 \\
 0 & 0 & 0 & 0 & 0 & -1 & 0 \\
 0 & 0 & 0 & 0 & 0 & 0 & 1 \\
\end{array}
\right)$ & $\left(
\begin{array}{c}
 0 \\
 0 \\
 0 \\
 0 \\
 0 \\
 0 \\
 \frac{1}{2} \\
\end{array}
\right)$ & $2$ & 22 & $\left(h_1,h_2,h_3,h_4,h_5,h_6,0\right)$ \\ & & & & \\ \hline & & & & \\ $\left(
\begin{array}{*7{C{1em}}}
 1 & 0 & 0 & 0 & 0 & 0 & 0 \\
 0 & -1 & 0 & 0 & 0 & 0 & 0 \\
 0 & 0 & -1 & 0 & 0 & 0 & 0 \\
 0 & 0 & 0 & -1 & 0 & 0 & 0 \\
 0 & 0 & 0 & 0 & -1 & 0 & 0 \\
 0 & 0 & 0 & 0 & 0 & -1 & 0 \\
 0 & 0 & 0 & 0 & 0 & 0 & 1 \\
\end{array}
\right)$ & $\left(
\begin{array}{c}
 0 \\
 0 \\
 0 \\
 0 \\
 0 \\
 0 \\
 \frac{1}{2} \\
\end{array}
\right)$ & $2$ & 18 & $\left(h_1,h_2,h_3,h_4,h_5,h_6,0\right)$ \\ & & & & \\ \hline & & & & \\ $\left(
\begin{array}{*7{C{1em}}}
 1 & 0 & 0 & 0 & 0 & 0 & 0 \\
 0 & 1 & 0 & 0 & 0 & 0 & 0 \\
 0 & 0 & -1 & 0 & 0 & 0 & 0 \\
 0 & 0 & 0 & -1 & 0 & 0 & 0 \\
 0 & 0 & 0 & 0 & -1 & 0 & 0 \\
 0 & 0 & 0 & 0 & 0 & -1 & 0 \\
 0 & 0 & 0 & 0 & 0 & 0 & 1 \\
\end{array}
\right)$ & $\left(
\begin{array}{c}
 0 \\
 0 \\
 0 \\
 0 \\
 0 \\
 0 \\
 \frac{1}{2} \\
\end{array}
\right)$ & $2$ & 16 & $\left(h_1,h_2,h_3,h_4,h_5,h_6,0\right)$ \\ & & & & \\ \hline & & & & \\ $\left(
\begin{array}{*7{C{1em}}}
 1 & 0 & 0 & 0 & 0 & 0 & 0 \\
 0 & 1 & 0 & 0 & 0 & 0 & 0 \\
 0 & 0 & 1 & 0 & 0 & 0 & 0 \\
 0 & 0 & 0 & -1 & 0 & 0 & 0 \\
 0 & 0 & 0 & 0 & -1 & 0 & 0 \\
 0 & 0 & 0 & 0 & 0 & -1 & 0 \\
 0 & 0 & 0 & 0 & 0 & 0 & 1 \\
\end{array}
\right)$ & $\left(
\begin{array}{c}
 0 \\
 0 \\
 0 \\
 0 \\
 0 \\
 0 \\
 \frac{1}{2} \\
\end{array}
\right)$ & $2$ & 16 & $\left(h_1,h_2,h_3,h_4,h_5,h_6,0\right)$ \\ & & & & \\ \hline & & & & \\ $\left(
\begin{array}{*7{C{1em}}}
 1 & 0 & 0 & 0 & 0 & 0 & 0 \\
 0 & 1 & 0 & 0 & 0 & 0 & 0 \\
 0 & 0 & 1 & 0 & 0 & 0 & 0 \\
 0 & 0 & 0 & 1 & 0 & 0 & 0 \\
 0 & 0 & 0 & 0 & -1 & 0 & 0 \\
 0 & 0 & 0 & 0 & 0 & -1 & 0 \\
 0 & 0 & 0 & 0 & 0 & 0 & 1 \\
\end{array}
\right)$ & $\left(
\begin{array}{c}
 0 \\
 0 \\
 0 \\
 0 \\
 0 \\
 0 \\
 \frac{1}{2} \\
\end{array}
\right)$ & $2$ & 18 & $\left(h_1,h_2,h_3,h_4,h_5,h_6,0\right)$ \\ & & & & \\ \hline & & & & \\ $\left(
\begin{array}{*7{C{1em}}}
 1 & 0 & 0 & 0 & 0 & 0 & 0 \\
 0 & 1 & 0 & 0 & 0 & 0 & 0 \\
 0 & 0 & 1 & 0 & 0 & 0 & 0 \\
 0 & 0 & 0 & 1 & 0 & 0 & 0 \\
 0 & 0 & 0 & 0 & 1 & 0 & 0 \\
 0 & 0 & 0 & 0 & 0 & -1 & 0 \\
 0 & 0 & 0 & 0 & 0 & 0 & 1 \\
\end{array}
\right)$ & $\left(
\begin{array}{c}
 0 \\
 0 \\
 0 \\
 0 \\
 0 \\
 0 \\
 \frac{1}{2} \\
\end{array}
\right)$ & $2$ & 22 & $\left(h_1,h_2,h_3,h_4,h_5,h_6,0\right)$ \\ & & & & \\ \hline \end{tabular}

	\end{table}
} 

\newpage
\bibliographystyle{utphys}
\bibliography{references}

\end{document}